\documentclass{aa}  
\usepackage{caption}
\usepackage{multirow}
\usepackage{graphicx}
\usepackage{tikz}
\usetikzlibrary{tikzmark, calc}
\usepackage{hyperref}
\usepackage{enumitem}
\usepackage{amsmath}
\usepackage{subcaption}
\hypersetup{
    colorlinks=true,
    linkcolor=blue,
    filecolor=magenta,      
    urlcolor=blue,
    citecolor=blue,
    pdftitle={Overleaf Example},
    pdfpagemode=FullScreen,
    }
\usepackage{calc}  
\newlength{\imageheight}
\usepackage{booktabs}  

\usepackage{txfonts}
\newcommand{\HI}{H\,{\sc i}}

\begin{document}

   \title{MeerKAT view of Hickson Compact Groups:}

   \subtitle{I. Data description and release}
   \titlerunning{Data description and release}
   \authorrunning{Ianjamasimanana et al.}

\author{R. Ianjamasimanana\inst{1},
      L. Verdes-Montenegro\inst{1},
      A. Sorgho\inst{1},
      K. M. Hess\inst{2,3,1},
      M. G. Jones\inst{4},
      J. M. Cannon\inst{9},
      J. M. Solanes\inst{6,7},
      M. E. Cluver\inst{8},
      J. Mold\'on\inst{1},
      B. Namumba\inst{1},
      J. Rom\'an\inst{14},
      I. Labadie-García\inst{1},
      C.C. de la Casa\inst{1},
      S. Borthakur\inst{13},
      J. Wang\inst{5},
      R. Garc\'ia-Benito\inst{1},
      A. del Olmo\inst{1},
      J. Perea\inst{1},
      T. Wiegert\inst{1},
      M. Yun\inst{15},
      J. Garrido\inst{1},
      S. Sanchez-Exp\'osito\inst{1},
      A. Bosma\inst{10},
      E. Athanassoula\inst{10},
      G. I. G. J\'ozsa\inst{11,12},
      T.H. Jarrett\inst{16, 17},
      C.K. Xu\inst{18, 19}
      O.M. Smirnov\inst{12, 20} 
      }

\institute{\inst{1} Instituto de Astrof\'isica de Andaluc\'ia (CSIC), Glorieta de la Astronom\'ia s/n, 18008 Granada, Spain\\
          \email{ianja@iaa.es}\\
    \inst{2} Department of Space, Earth and Environment, Chalmers University of Technology, Onsala Space Observatory, 43992 Onsala, Sweden\\
    \inst{3} ASTRON, the Netherlands Institute for Radio Astronomy, Oude Hoogeveensedijk 4,7991 PD Dwingeloo, The Netherlands\\
    \inst{4} Steward Observatory, University of Arizona, 933 North Cherry Avenue, Rm. N204, Tucson, AZ 85721-0065, USA\\
    \inst{5} Kavli Institute for Astronomy and Astrophysics, Peking University, Beijing 100871, People’s Republic of China\\
    \inst{6} Departament de F\'isica Qu\`antica i Astrof\'isica, Universitat de Barcelona, C. Mart\'i i Franqu\`es 1, 08028, Barcelona, Spain\\
    \inst{7} Institut de Ci\`encies del Cosmos (ICCUB), Universitat de Barcelona., C. Mart\'i i Franqu\`es 1, 08028, Barcelona, Spain\\
    \inst{8} Centre for Astrophysics and Supercomputing, Swinburne University of Technology, Hawthorn, VIC 3122, Australia\\
    \inst{9} Department of Physics \& Astronomy, Macalester College, 1600 Grand Avenue, Saint Paul, MN 55105, USA\\
    \inst{10} Aix Marseille Univ, CNRS, CNES, LAM, Marseille, France\\
    \inst{11} Max-Planck Institut fur Radioastronomie, Auf dem H\"ugel 69, 53121, Bonn, Germany\\
    \inst{12} Centre for Radio Astronomy Techniques and Technologies (RATT), Department of Physics and Electronics, Rhodes University, PO Box 94, Grahamstown 6140, South Africa\\
    \inst{13} School of Earth and Space Exploration, Arizona State University, 781 Terrace Mall, Tempe, AZ, 85287, USA\\
    \inst{14} Departamento de F\'isica de la Tierra y Astrof\'isica, Universidad Complutense de Madrid, E-28040 Madrid, Spain\\
    \inst{15} Department of Astronomy, University of Massachusetts, Amherst, MA 01003, USA\\
    \inst{16} Department of Astronomy, University of Cape Town, Rondebosch, Cape Town 7700, South Africa\\
    \inst{17} Western Sydney University, Locked Bag 1797, Penrith South DC, NSW 1797, Australia\\
    \inst{18} Chinese Academy of Sciences South America Center for Astronomy, National Astronomical Observatories, CAS, Beijing, People’s Republic of China\\
    \inst{19} National Astronomical Observatories, Chinese Academy of Sciences (NAOC), Beijing, People’s Republic of China\\
    \inst{20} South African Radio Astronomy Observatory, Cape Town, 7925, South Africa 
         }

   \date{Received November 14, 2024; accepted February 03, 2025}

\abstract
{Hickson Compact Groups (HCGs) are dense gravitationally-bound collections of 4-10 galaxies ideal for studying gas and star formation quenching processes. }
{We aim to understand the transition of HCGs from possessing complex \HI\ tidal structures (so-called phase 2 groups) to a phase where galaxies have lost most or all their \HI\ (phase 3). We also seek to detect diffuse \HI\ gas that was previously missed by the Very Large Array (VLA).}
{We observed three phase 2 and three phase 3 HCGs with MeerKAT and reduced the data using the Containerized Automated Radio Astronomy Calibration (CARACal) pipeline. We produced data cubes, moment maps, integrated spectra, and  compared our findings with previous VLA and Green Bank Telescope (GBT) observations.}
{Compared with previous VLA observations, MeerKAT reveals much more extended tidal features in phase 2 and some new high surface brightness features in phase 3 groups. However, no diffuse \HI\ component was found in phase 3 groups. We also detected many surrounding galaxies for both phase 2 and phase 3 groups, most of which are normal disk galaxies.}
{The difference between phase 2 and phase 3 groups is still substantial, supporting previous finding that the transition between the two phases must be 
abrupt.}

   \keywords{galaxies:evolution --
                galaxies:groups --
                galaxies:interactions --
                galaxies: ISM
               }

   \maketitle
%

\section{Introduction}
Hickson Compact Groups (HCGs) are systems of typically four to ten groups of galaxies in close proximity to each other. 
They were first catalogued by Paul Hickson in \citeyear{1982ApJ...255..382H}. On large scales, they are located in low-density environment \citep{1995AJ....109.1476P}. 
Their galaxy members are characterised by low-velocity dispersion \citep[$\mathrm{\sim 200~km~s^{-1}},$][]{1992ApJ...399..353H} but are separated by small distances like in the centre of clusters, making HCGs an ideal case to study gas and star formation quenching processes. According to the evolutionary sequence proposed by \citet{2001A&A...377..812V}, 
in phase 1, galaxies in HCGs have the majority of their neutral atomic hydrogen (\HI) gas in the disk of the galaxies. 
In phase 2, gravitational interactions between member galaxies are thought to be the main driver in  disrupting gas morphology and kinematics, displacing approximately 30\% to 60\% of the gas from the disks, and forming structures such as tidal tails and bridges. Phase 3 HCGs are divided into two subcategories. In phase 3a, most or all the gas is found in the form of intra-group clouds or tidal tails. In phase 3b, the \HI\ in the groups forms a large 
cloud with a single peaked global profile and a single velocity gradient. However, \citet{2023A&A...670A..21J} questioned the existence of this latter phase, 
as subsequent studies found that only HCG 49, among the 72 HCGs analysed by \citet{2001A&A...377..812V}, fit this classification.  
\citet{2023A&A...670A..21J} made a minor modification to the original evolutionary sequence of HCGs proposed by \citet{2001A&A...377..812V}. 
The first modification concerns the threshold at which groups are classified as being phase 2 or phase 3. \citet{2023A&A...670A..21J} 
redefined phase 2 groups as those where 25\% to 75\% of the \HI\ was associated with features not related to the disk of the galaxies. 
If more than 75\% of the \HI\ was associated with extended features, they classified the groups as phase 3. The second modification involved 
the elimination of the phase 3b classification and the introduction of phase 3c. They used this new designation scheme to specify HCGs that 
would have been classified as phase 1, but where only one galaxy among the group members was detected in \HI. In fact, phase 3c groups are thought to be systems at a 
very late evolutionary stage but have recently acquired a new member galaxy. In this paper, we will use the modified classification 
scheme by \citet{2023A&A...670A..21J}. \\

The \HI\ content of HCGs have been the subject of many studies in the literature. \citet{1987ApJS...63..265W} observed 51 HCGs using 
the NRAO 91 m and NAIC 305 m telescopes. They found that HCGs contain, on average, half as much \HI\ as loose groups of similar optical morphology and luminosity types. 
Their \HI\ mass is lower than the expected \HI\ mass from their optical morphology and luminosity types. \citet{1997A&A...325..473H} 
used the Effelsberg 100-m telescope to observe 54 HCGs at 21 cm wavelength. They detected \HI\ for 41 HCGs. 
The detection rate increased with the number of spiral galaxies in the groups. \citet{1997A&A...325..473H} expanded their study by 
incorporating the sample from \citet{1987ApJS...63..265W}, resulting in a total of 75 HCGs, including 61 detections and 14 upper limits. 
\citet{1997A&A...325..473H} observed many HI-deficient groups by comparing 
their integrated \HI\ mass with their blue luminosity. \citet{2001A&A...377..812V} used single-dish data for 72 HCGs to examine their overall HI contents. In addition, they used high-resolution VLA mapping for 16 of these groups to investigate their \HI\ spatial distributions and kinematics in more details. They found that, on average, HCGs contain only 40\% of the expected \HI\ mass based on the optical luminosity and morphological types of the member galaxies. Although the more deficient groups showed higher X-ray detection rate than the less deficient ones, suggesting the importance of phase transformation for the missing \HI, the results were not statistically significant; subsequent analysis showed no significant correlation between \HI\ deficiency and X-ray emission \citep{2008MNRAS.388.1245R}. \citet{2001A&A...377..812V} also found that the individual 
member galaxies have higher deficiency than the group as a whole, indicating the importance of gas removal processes from galaxies in HCGs. High angular resolution VLA observations showed that 70\% of the spiral galaxies had disturbed morphology and kinematics. However, they did not find a one-to-one 
correlation between the amount of \HI\ mass in tidal features and \HI\ deficiency. This may indicate a rapid evolutionary sequence once the gas is dislodged from the member galaxies. 
Our data will be crucial in investigating this scenario. \\
   
Our paper is motivated by the GBT observations of 
\citet{2010ApJ...710..385B} and the VLA-based analysis of \citet{2023A&A...670A..21J}. \citet{2010ApJ...710..385B} observed the \HI\ spectrum of 26 galaxies, in which only one group was undetected. They detected diffuse \HI\ emission, spread 
over 1000 $\mathrm{km~s^{-1}}$, which was missed by previous observations. They quantified the newly observed \HI\ emission in terms of excess mass, 
which was defined as the difference between the mass measured by the GBT and the mass measured by the VLA ($M_{\mathrm{excess}}=M_{\mathrm{GBT}}-M_{\mathrm{VLA}}$). The excess gas mass 
fraction ($M_{\mathrm{excess}}=M_{\mathrm{GBT}}-M_{\mathrm{VLA}}/M_{\mathrm{GBT}}$) was found to vary between 5\% to 80\%, and higher for more \HI\ deficient groups. 
The result suggested that at least part of the missing \HI\ was in the form of a faint, diffuse \HI\ component, spread over a large velocity range. 
Despite the detection of the diffuse component, there is still a significant discrepancy between the observed \HI\ mass and the expected values.
Thus, the question of where the remaining \HI\ in HCGs is located 
remains unanswered. \\
   
If the rest of the gas in HCGs is heated, we would expect to detect hot intra-group X-ray gas in \HI\ deficient groups. However, previous observations showed 
that either most of the detected X-ray gas was confined within the individual member galaxies \citep{2017MNRAS.464..957H} or the properties of the detected
hot intra-group gas among the studied \HI\ deficient sample were very diverse and not correlated with the \HI\ properties \citep{2008MNRAS.388.1245R}. 
In addition, being \HI\ deficient did not necessarily imply the presence of hot intergalactic medium (IGM) gas. For example, HCG 30 is one of the most \HI\ 
deficient group, yet no hot IGM has been observed \citep{2008MNRAS.388.1245R}. Previous observations also ruled out thermal evaporation as the likely dominant 
mechanisms to remove \HI\ gas in HCGs \citep{2008MNRAS.388.1245R}. As suggested by \citet{2017MNRAS.464..957H}, the missing \HI\ is either too diffuse to be detected 
or has been ionised and is in the form of a diffuse X-ray gas that is also undetected. The increased sensitivity of new radio telescopes is promising in detecting diffuse
\HI\ component that might have been missed by previous instruments like the VLA or the Westerbork Synthesis Radio Telescope (WSRT). For example, using the 
Five hundred meter Aperture Spherical Telescope (FAST), \citet{2023ApJ...944..102W} detected faint, extended, diffuse \HI\ emission around the tidally interacting 
NGC 4631 group that was missed by the WSRT. The column density of the diffuse \HI\ was still above the critical column density for photo-ionisation and thermal evaporation.
By combining data from the Karoo Array Telescope (KAT-7), WSRT, and The Arecibo Legacy Fast ALFA (ALFALFA) survey, \citet{2017MNRAS.464..957H} also detected 
faint \HI\ emission in HCG 44, far from the group centre. Thus, we expect that part of the missing \HI\ in HCGs is in the form of extended, diffuse \HI\ that 
was missed by previous observations. \\
   
In an effort to recover the missing \HI\ in HCGs and understand more about the survival of \HI\ in groups, we observed 6 HCGs at intermediate and advanced 
evolutionary stages with MeerKAT. The main scientific goals of the proposal were to:
\begin{enumerate}[label=\alph*)]
    \item understand the transition of HCGs from a complex of \HI\ tidal structures to the most extreme phase, where galaxies seem to have entirely lost their \HI,
    \item determine the distance from the core at which the \HI\ can survive in groups, and assess if magnetic fields aid in this survival within the harsh Intra-group Medium (IGrM),
    \item examine the effects of potential encounters with intruder galaxies located outside the field of view of the VLA,
    \item investigate the previously unexplored role of intra-group gas in the accelerated transition of galaxies from an active to quiescent state.
\end{enumerate}   
This paper describes the processing and general description of the data; specific goals as listed above will be addressed in future papers. 
This paper is organised as follows. We introduce the individual groups in section~\ref{sample}. We describe the observations and data reduction in section~\ref{observation}. 
We present the results in section~\ref{results}. We summarise the findings and give conclusion in section~\ref{summary}.
\\

\textbf{Reproducibility}: This paper is fully reproducible. We used the Snakemake workflow management system \citet{snakemake} to organise our scripts in terms of "rules" that contain the names of inputs and shell 
commands to produce different data products, figures, and tables. A single command was then used to execute the different rules sequentially and produce the final PDF paper. Snakemake is an advanced workflow management system to facilitate reproducibility and encourage transparency in scientific research by automating the execution of complex data analysis steps. This aligns with the FAIR principles (Findable, Accessible, Interoperable, and Reusable) and Open Science policies.  

\section{Sample}\label{sample}
To have a coherent view of the evolutionary sequence of \HI\ in HCGs, we have observed 6 HCGs, in which 3 are in phase 2 and 3 are in phase 3. 
We have omitted groups in phase 1 due to their similarities with galaxies in isolation. We have selected phase 2 groups 
since the mechanisms responsible for gas removal, dispersion or depletion are mostly active during this evolutionary stage. In addition, we have included phase 3 groups 
as they represent the most advanced stage of evolution, allowing us to investigate the ultimate fate of the \HI\ gas. We list the properties of the groups in Table~\ref{table:sample1} and summarise previous findings below. 

\begin{table*}
    \centering
    \caption{\label{table:sample1}Group parameters}
     \resizebox{0.85\textwidth}{!}{%
    \begin{tabular}{lccccclll}
    \toprule \toprule
    HCG & RA & Dec & Distance & Phase & $cz$ & members & Morphology &  \HI\ content \\
     & & & [Mpc] & & [$\mathrm{km~s^{-1}}$] &  & &\\
    \midrule
    \multirow{4}{*}{16} & \multirow{4}{*}{02:09:31.3} & \multirow{4}{*}{$-$10:09:30} & \multirow{4}{*}{49} & \multirow{4}{*}{2} & \multirow{4}{*}{3957} & HCG16a & SBab & Deficient\\
                        &  & &  &  &  & HCG16b & Sab & Deficient\\
                        &  & &  &  &  & HCG16c & Im & Normal \\
                        &  & &  &  &  & HCG16d & Im & Normal \\
                        &  & &  &  &  & NGC 848 & Im & Normal \\\\
    \multirow{5}{*}{31} & \multirow{5}{*}{05:01:38.3} & \multirow{5}{*}{$-$04:15:25} & \multirow{5}{*}{53} & \multirow{5}{*}{2} & \multirow{5}{*}{4039} & HCG31a& Sdm&Normal\\
                        &  & &  &  &  & HCG31b & Sm & Normal\\
                        &  & &  &  &  & HCG31c & Im & Deficient \\
                        &  & &  &  &  & HCG31g & cI & Normal \\
                        &  & &  &  &  & HCG31q & cI & Normal \\\\
    \multirow{4}{*}{91} & \multirow{4}{*}{22:09:10.4} & \multirow{4}{*}{$-$27:47:45} & \multirow{4}{*}{92} & \multirow{4}{*}{2} & \multirow{4}{*}{7135} & HCG91a & SBc & Deficient \\
    &  & &  &  &  & HCG91b & Sc & Normal\\
    &  & &  &  &  & HCG91c & Sc & Normal\\
    &  & &  &  &  & HCG91d & SB0 & No \HI\ \\
    \midrule
    \multirow{4}{*}{30} & \multirow{4}{*}{04:36:28.6} & \multirow{4}{*}{$-$02:49:56} & \multirow{4}{*}{61} & \multirow{4}{*}{3} & \multirow{4}{*}{4617} & HCG30a & SB & No \HI\ \\
    &  & &  &  &  & HCG30b & Sa & No \HI\ \\
    &  & &  &  &  & HCG30c & SBbc & No \HI\ \\
    &  & &  &  &  & HCG30d & S0 & No \HI\ \\\\
    \multirow{4}{*}{90} & \multirow{4}{*}{22:02:05.6} & \multirow{4}{*}{$-$31:58:00} & \multirow{4}{*}{33} & \multirow{4}{*}{3} & \multirow{4}{*}{2638} & HCG90a & Sa & Deficient\\
    &  & &  &  &  & HCG90b & E0 & No \HI\ \\
    &  & &  &  &  & HCG90c & E0 & No \HI\ \\
    &  & &  &  &  & HCG90d & Im & No \HI\ \\\\
    \multirow{5}{*}{97} & \multirow{5}{*}{23:47:22.9} & \multirow{5}{*}{$-$02:19:34} & \multirow{5}{*}{85} & \multirow{5}{*}{3} & \multirow{5}{*}{6535} & HCG97a & E5 & No \HI\ \\
     &  & &  &  &  & HCG97b & Sc & Deficient \\
      &  & &  &  &  & HCG97c & Sa & No \HI\ \\
       &  & &  &  &  & HCG97d & E1 & No \HI\ \\
        &  & &  &  &  & HCG97e & S0a & No \HI\ \\
    \bottomrule
    \end{tabular}
     }
     \tablefoot{Columns: (1) HCG ID number, (2) right ascension, (3) declination, (4) distance calculated by \citet{2023A&A...670A..21J}, (5) the evolutionary phase of the group 
     as originally proposed by \citet{2001A&A...377..812V}, (6) redshift \citep{1992ApJ...399..353H}, (7) core members of the group, (8)  morphological types from the table compiled by \citet{2023A&A...670A..21J}, most of which were taken from \citet{1989ApJS...70..687H}, (9) \HI\ content based on the analysis of VLA archival data by \citet{2023A&A...670A..21J}; the deficiency for each group 
	was estimated based on the logarithmic difference between the predicted \HI\ mass from a B-band luminosity scaling relation ($L_{B}-M_{HI}$) of isolated galaxies and the observed \HI\ mass.} 
\end{table*}

\subsection{HCG 16}
The \HI\ spectrum of HCG 16 was obtained by \citet{2010ApJ...710..385B} using the GBT. Besides, HCG 16 was previously 
mapped in \HI\ with the VLA using the C and D array configurations \citep{2001A&A...377..812V, 2019A&A...632A..78J}. \citet{2019A&A...632A..78J} 
found a total mass of $\mathrm{log~}M_{\mathrm{HI}}/M_{\odot}=10.53 \pm 0.05$. This mass is higher than the mass measured by the GBT since a 
significant fraction of the \HI\ in HCG 16 was beyond the reach of the GBT field of view, which was centred at the core of the group. However, 
by weighting the VLA map with the GBT beam response, the measured \HI\ mass obtained by the GBT is about 5\% higher than that of the VLA 
\citep{2010ApJ...710..385B}. \citet{2001A&A...377..812V} found that most of the \HI\ in HCG 16 was in the form of extended features such as tidal tails and bridges \citep[see also][]{2019A&A...632A..78J}. The group members are connected by tidal features. 
The most notable one is the long tidal tail with a projected extent of $160$ kpc, stretching south-east toward NGC 848 from the core of the group. Additionally, an \HI\ 
tail was found extending from HCG~16c to a faint optical counterpart located north of HCG~16a \citep{2019A&A...632A..78J, 2021A&A...649L..14R}.    
HCG 16 was also observed in H$\alpha$ and R-band by the Survey for Ionisation 
in Neutral Gas Galaxies \citep[SINGG,][]{2006ApJS..165..307M}. The observations revealed that all four members of HCG 16 had prominent high surface brightness 
(HSB) nuclear H$\alpha$ emission, with at least three members having  minor-axis outflow. The deep \textit{Chandra} X-ray and VLA-GMRT 1.4 GHz radio continuum 
study of \citet{2014ApJ...793...73O} uncovered the presence of a ridge of diffuse X-ray emission linking the four member galaxies. Their map also showed 
similarities between the hot and cold gas structures, with the diffuse X-ray ridge generally overlapping with the \HI\ filaments, especially the brighter 
ridge connecting NGC 838 and NGC 839. However, the X-ray ridge connecting NGC 838 and NGC 835 did not match well with the \HI\ filaments. \citet{2014ApJ...793...73O} 
detected \HI\, X-ray, and radio continuum emission from the same region, indicating the presence of a multi-phase IGM in HCG 16. 
\subsection{HCG 30}
Previous VLA observations of HCG 30 failed to detect \HI\ in the core of HCG 30 \citep{2023A&A...670A..21J}. 
However, the GBT spectrum of \citet{2010ApJ...710..385B} showed faint, broad \HI\ emission, spread over the whole velocity range of the group, 
suggesting the presence of a diffuse \HI\ component in the group. 
By comparing the VLA flux with that of the GBT, \citet{2010ApJ...710..385B}   
found an excess gas mass fraction of 23\%, corresponding to an \HI\ excess mass of $\mathrm{1.45 \times 10^{8}}$ $M_{\sun}$. Despite being the most \HI\ deficient group \citep{2001A&A...377..812V}, the \textit{Chandra} 
X-ray observations of \citet{2008MNRAS.388.1245R} found no clear evidence of the presence of diffuse IGM emission. Various tests, including an examination of emission across their 
S2 and S3 CCDs and a search for radial gradients in emission did not reveal significant variation or clear spatial variations in diffuse emission. This indicated a consistency 
with the local background. In addition, previous observations failed to detect radio continuum emission in the group, and little far infrared emission (FIR) was observed in the member
galaxies \citep{2007NewAR..51...87V}. The group members also have little H$\alpha$ emission \citep{1998ApJS..117....1V}. Small amounts of H$\alpha$ emission were seen in the central 
region of the barred early-type spiral galaxy HCG~30a and HCG~30c. However, the emission in HCG~30a was contaminated by a nearby bright star, leading to a large error in the 
measured H$\alpha$  emission. The H$\alpha$ emission in the barred late-type spiral galaxy HCG~30b was found in its bulge. No H$\alpha$ emission was found in HCG 30d, a 
lenticular galaxy. Lastly, HCG~30 may have consumed all its molecular gas through star formation as evidenced by the lack of CO detection \citep{1998ApJ...497...89V}.  
\subsection{HCG~31}
HCG~31 is one of the most studied compact groups in the literature due to the peculiar characteristics of its members, such as the presence of tidal tails, a merger, and prominent starbursts. 
All members of HCG 31 were previously detected in \HI. Their \HI\ content range from 10\% to 80\% of the expected values \citep{2005A&A...430..443V}. However, the group as a whole is not deficient in \HI. The \HI\ observations by \citet{2005A&A...430..443V} revealed many tidal features, 
containing 60\% of the total HI mass in HCG 31. In addition, all member galaxies showed signs of interactions. The \HI\ analysis by \citet{2023A&A...670A..21J} using VLA data indicated that most of the \HI\ in HCG 31 was found in the IGrM, although no reliable separation of galaxies and tidal features was possible. Several members of HCG 31, namely E, F, H, and R, were 
considered to be TDGs, tidal dwarf candidates, or tidal debris \citep{2003A&A...397...99R, 2004ApJS..153..243L, 2006AJ....132..570M, 2023MNRAS.522.2655G}. 
Despite the abundance of strong dynamical processes in HCG 31, only weak diffuse X-ray emission was detected in the group 
\citep{2013ApJ...763..121D}. The CO emission in HCG 31 is also faint, with enhanced CO emission found in the overlap region between HCG~31a and HCG~31c \citep{1997ApJ...475L..21Y}. 
This deficiency in CO was mainly attributed to tidal disruption in the group, rather than a consequence of low metal abundance.  
\subsection{HCG~90}
HCG~90 was previously imaged in \HI\ by the ATCA and VLA, with only NGC 7172 detected \citep{1997ASPC..116..358O, 2023A&A...670A..21J}, while weak \HI\ emission in NGC 7176 was indicated by the GBT spectrum \citep{2010ApJ...710..385B}. However, all four members were detected in CO \citep{1989ApJ...342..735H, 1992A&A...257..455H, 1997ASPC..117..530V}.
CCD images taken with the 3.5 m New Technology Telescope at the 
European Southern Observatory (ESO) by \citet{2003ApJ...585..739W} showed that the three core members of HCG 90 are embedded in large diffuse optical light with surprisingly a very 
small amount of associated hot intra-cluster gas. The presence of the diffuse optical light was attributed to tidally stripped stars due to galaxy interactions. The amount of diffuse 
light in HCG 90 is unprecedented in comparison to either other observations or theoretical expectations involving multiple interacting systems. It has a narrow range of colour, 
consistent with an old stellar population. \citet{2015MNRAS.447.3639M} suggested that the diffuse light existed before the onset of the merger event 
in HCG~90. The \textit{Chandra} map by \citet{2013ApJ...763..121D} showed bridges of diffuse X-ray emission connecting the three core members, as well as small common envelopes 
encompassing them. However, no diffuse X-ray emission associated with the IGM was found.
\subsection{HCG 91}
HCG~91 is heavily discussed in the literature due to the anomalous 
morphologies and kinematics of its members, including the presence of extended tidal tails, bridges, clumps, a double gaseous component, disturbed velocity field, 
and asymmetric rotation curves \citep{2001MNRAS.324..859B, 2003A&A...402..865A, 2003AJ....126.2635M, 2015MNRAS.450.2593V, 2016ApJ...818..115V}. The most pronounced 
tidal feature is the extended tidal tail of the Seyfert 1 galaxy, HCG~91a, pointing toward HCG 91c, and is visible both at optical \citep{2015MNRAS.451.2793E} and 
radio wavelength \citep{2003A&A...402..865A, 2016ApJ...818..115V, 2023A&A...670A..21J}. 
Apart from HCG 91d, all other core members of HCG~91 have been detected previously in \HI. The VLA maps by \citet{2023A&A...670A..21J} showed 
prominent \HI\ tail extending to the east and curving north toward HCG~91c. In addition, HCG~91b and HCG~91c are connected by a weak extended \HI\ emission. 
Overall, the VLA data showed that HCG~91 was moderately deficient in \HI, containing 63\% of the expected amount. At the time of writing, we are not 
aware of any diffuse X-ray emission measurement reported in the literature for this group.   
\subsection{HCG 97}
Out of the five core members of HCG 97, only the approaching side of HCG 97b (IC5 359) was previously detected in \HI\ in the VLA map of \citet{2023A&A...670A..21J}, with the receding 
side having too low a signal-to-noise (S/N) to be included as real emission. \citet{2023A&A...670A..21J} suggested that HCG 97b has a disturbed morphology. 
This is in agreement with the LOFAR 144 MHz and VLA (1.4 GHz and 4.86 GHz) maps of \citet{2023MNRAS.tmp.3112H}. They found a radio tail and an extended tail at the western side of HCG 97b. 
The radio contours are compressed at the southeastern side of the galaxy, suggesting ram-pressure stripping effects. 
\citet{2023MNRAS.tmp.3112H} used the 12m Atacama Large Millimeter Array (ALMA) and the 7m Atacama Compact Array (ACA) to trace molecular gas in HCG 97b. 
They detected CO emission in the disk of HCG 97 b, but not in the tails. The CO emission shows asymmetric morphology, with the southeast part showing an 'upturn', 
consistent with the optical morphology of the disk. 
Diffuse X-ray emission was previously detected in HCG 97 but without any associated diffuse optical light, optical tidal features or any other (optical) signs of interactions \citep{1995AJ....110.1498P}. 
However, the outer stellar population of HCG 97a were found to be bluer than what was expected from an elliptical galaxy. In addition, its colour was 
consistent with the outer envelope of HCG 97d, suggesting a previous exchange of stars between the two galaxies.          

\section{Observations and data reduction}\label{observation}
The data was taken between July 31, 2021, and January 02, 2022, using the full MeerKAT array and a minimum of 61 antennas (proposal id: MKT-20101). 
The L-band receiver (856-1711.974 MHz) was used in its 32k correlator mode. The observations were centred at 1389.1322 MHz 
and a channel width of 26.123 KHz (or 5.5 $\mathrm{km~s^{-1}}$ at 1420.4 MHz) was used to divide the 856 MHz total bandwidth, 
resulting 32768 channels. The data was recorded with the four linear polarisation products of the MeerKAT antenna feeds (XX, XY, YX, and YY). 
For each observing run, the primary calibrator was observed for $\sim$8 minutes at the beginning. This was followed by a cycle of alternating 
observations between the gain calibrator ($\sim$2 minutes) and the target ($\sim$38 minutes), with the primary calibrator revisited every two hours. 
Polarisation calibrators were observed twice for $\sim$6 minutes. The total time spent per group was $\sim$6.25 hours (about 5.13 hours spent on source) or a total of 37.5 hours, 
including calibration overheads, at 8 seconds integration period for the six HCGs. Table~\ref{table:obs_prop} shows the calibrators used and the observing time spent on each target. \\
\begin{table*}
\centering
\captionsetup{justification=centering}
\caption{\label{table:obs_prop}Observational properties}
\begin{tabular}{lcccccc}
\toprule \toprule
HCG & \multicolumn{2}{c}{Observing time} & Flux cal. & Gain cal. & Polarisation cal. & Antennas\\
 & Total & On source & &  & & \\
\midrule
16 & 6.27 & 5.14 & J0408-6545 & J0240-2309 & J0521+1638 & 61\\
31 & 6.21 & 5.15 & J0408-6545 & J0503+0203 & J0521+1638 & 62\\
\multirow{1}{*}{91} & 6.26 & 5.14 & J1939-6342 & J2206-1835 & J0137+3309 & 61\\
                    &      &      &           &             & J0521+1638 & \\
\midrule
30 & 6.21 & 5.15 & J0408-6545 & J0423-0120 & J0521+1638 & 62\\
\multirow{1}{*}{90} & 6.27 & 5.15 & J1939-6342 & J2214-3835 & J0137+3309 & 63\\
                    &      &      &           &             & J0521+1638 & \\
\multirow{1}{*}{97} & 6.26 & 5.15 & J1939-6342 & J0022+0014 & J0521+1638 & 63\\
                    &   &    & J0408-6545 &  & J0137+3309 & \\
\bottomrule
\end{tabular}
\tablefoot{Columns: (1) HCG ID number, (2) total and on-source observing time in hours, (3) flux calibrators, (4) gain calibrators, (5) polarisation 
calibrators, (6) number of available MeerKAT antennas at the time of observations.}
\end{table*}
The raw data were processed using the CARACal pipeline \citep{2020ASPC..527..635J}. CARACal integrates various radio astronomical software that can be executed through a 
Stimela script \citep{2018PhDT.......215M}, enabling a user control over each step of the data processing. To produce a science-ready \HI\ data cube, the raw data 
were passed through three main steps, called workers, in CARACal: 
\begin{enumerate}
    \item flagging and cross-calibration,
    \item continuum imaging and self-calibration,
    \item continuum subtraction and line imaging.
\end{enumerate}
CARACal produces various diagnostic plots that can be visually inspected to assess the quality of the data products. Our data was processed on 
a virtual machine of the Spanish Prototype of the Square Kilometre Array (SKA) Regional Center \citep[espSRC,][]{2022JATIS...8a1004G}. 
The virtual machine has 24 CPUs and 186 GB RAM. The espSRC is hosted by the Institute of Astrophysics of Andalusia (IAA-CSIC). 
It serves as a test bed for future scientific activities with the SKA while promoting Open Science and FAIR  principles. We adopt a similar data reduction strategy as described by \citet{2023A&A...673A.146S}. However, our observations consist of a single pointing rather than a mosaic. We describe our data reduction process below. 
\subsection{Flagging and cross-calibration}
Briefly, flagging and cross-calibration were first applied to the calibrators before performing on-the-fly calibration and flagging on the target. 
The detailed steps are as follows. First, auto-correlations and shadowed antennas were flagged using the Common Astronomy Software Applications (CASA) task 
\texttt{flagdata} incorporated in CARACal. 
RFI was then flagged using \texttt{AOFlagger} \citep{2012A&A...539A..95O}, which is the default option in CARACal. Next, cross-calibration was performed. 
For the primary calibrator, time-dependent delay calibration ($K$) was first carried out, followed by gain calibration ($G$) while applying the derived $K$ term on the fly. 
After that, bandpass calibration ($B$) was run while applying the previous $K$ and $G$ terms on the fly. This sequence was repeated twice (order: KGBKGB in CARACal) before 
applying the primary calibrator delay and bandpass to the secondary calibrator (apply: KB in CARACal) and solving for gain amplitude and phase. Next, flagging was performed 
to remove spurious calibrated visibilities before solving again for the gain amplitude and phase, and additionally bootstrapping the flux scale from the primary 
calibrator (order: GAF and calmode: [ap, null, ap] in CARACal). For the target, on-the-fly calibration, applying the primary calibrator delays and bandpass, 
as well as the secondary calibrator gains, was carried out. Auto-correlations, shadowed antennas, and RFI were also flagged.                    
\subsection{Continuum imaging and self-calibration}
This step is part of the \texttt{selfcal} worker in CARACal to get continuum models and perform self-calibration. High signal-to-noise (S/N) continuum images 
and polarisation data will be described in a future paper. Here, we only describe steps relevant for \HI\ line imaging. We use WSClean \citep{2014MNRAS.444..606O} 
for imaging and Cubical \citep{2018MNRAS.478.2399K} for self-calibration. Radio continuum emission was iteratively imaged and self-calibrated. At each step, WSClean 
was allowed to clean blindly down to a 
masking threshold set by us. At each new iteration, the previous clean mask was used to allow for a deeper cleaning. Multi-scale cleaning was used, with scales 
automatically selected by the algorithm. WSclean was run on a 2 deg $\times$ 2 deg image with 1\arcsec\ pixel size using Briggs weighting of -1 with 0\arcsec\ tapering.  
Three calibration iteration was used, requiring four WSClean imaging with different clean mask threshold (\texttt{auto-mask} option for WSClean): 30, 20, 10, 
and 5 times the rms noise level. The same cleaning threshold, 0.5 (\texttt{auto-threshold} parameter for WSClean),  was used for each imaging option.     
\subsection{Continuum subtraction and line imaging}
The continuum model generated in the previous step was subtracted from the target visibilities. Additionally, doppler-tracking corrections and a first-order 
polynomial fitting to the real and imaginary parts of the line-free channels were carried out with the CASA task \texttt{mstransform} in CARACal. 
Next, any erroneous visibilities produced by \texttt{mstransform} were flagged with AOflagger \citep{2023A&A...673A.146S}. After that, \HI\ data cubes 
were produced with WSClean using different imaging parameters. Multi-scale cleaning was used using a major cycle 
clean gain of 0.2 and scales were automatically selected by WSClean. For each group, we made available cubes at 60\arcsec, as well as higher resolution cubes down to ~20 kpc linear resolution, roughly corresponding to the scale length of the optical disk. 
Cubes at intermediate resolutions are also available. The 60\arcsec\ cubes allow the study of extended emission; whereas the higher resolution cubes aid at separating galaxies from tidal tails or intra-group 
gas as previously done by \citet{2023A&A...670A..21J}. To get the final data cubes for each group, we run CARACal to obtain a cube at 98\arcsec. Then, 
we run the Source Finding Application \citep[SoFiA,][]{2015MNRAS.448.1922S, 2021MNRAS.506.3962W} outside of CARACal to get a better clean mask. 
After that, we run CARACal again using that mask to get our 
final lowest-resolution data cube. We again use SoFiA outside of CARACal to get another clean mask and use it to clean the next higher-resolution cube. We proceed like this 
until the highest resolution cube. To correct for primary beam attenuation, we used the model of \citet{2020ApJ...888...61M}. To convert the cubes from frequency to velocity, 
we used the MIRIAD task \texttt{VELSW}. Throughout this paper, we use the optical velocity definition. We also use the 60\arcsec\ cube unless stated otherwise. 
\section{Results}\label{results}
\begin{figure*}
\setlength{\tabcolsep}{0pt}
\begin{tabular}{lll}
    \includegraphics[scale=0.21]{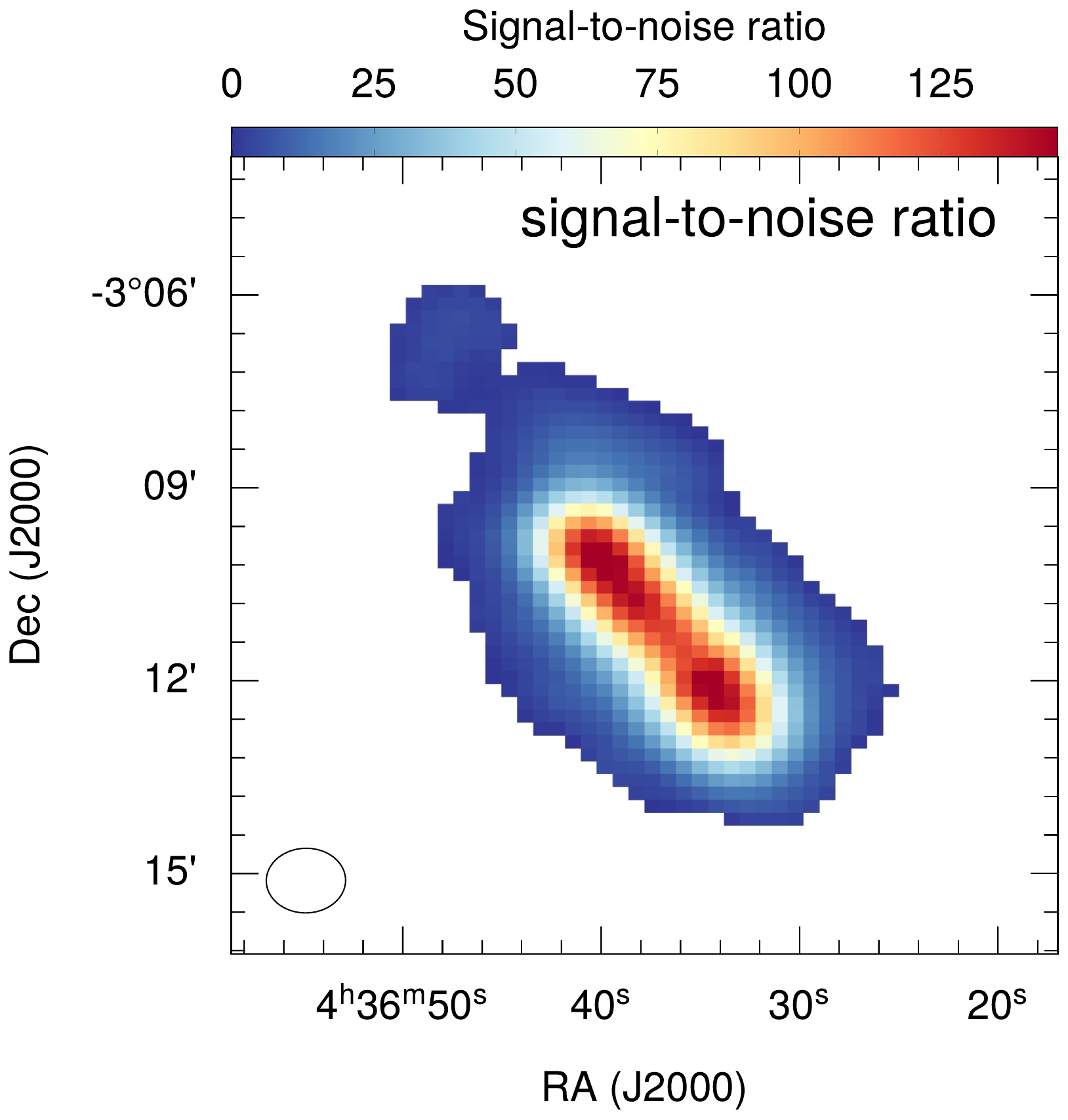}&
    \includegraphics[scale=0.21]{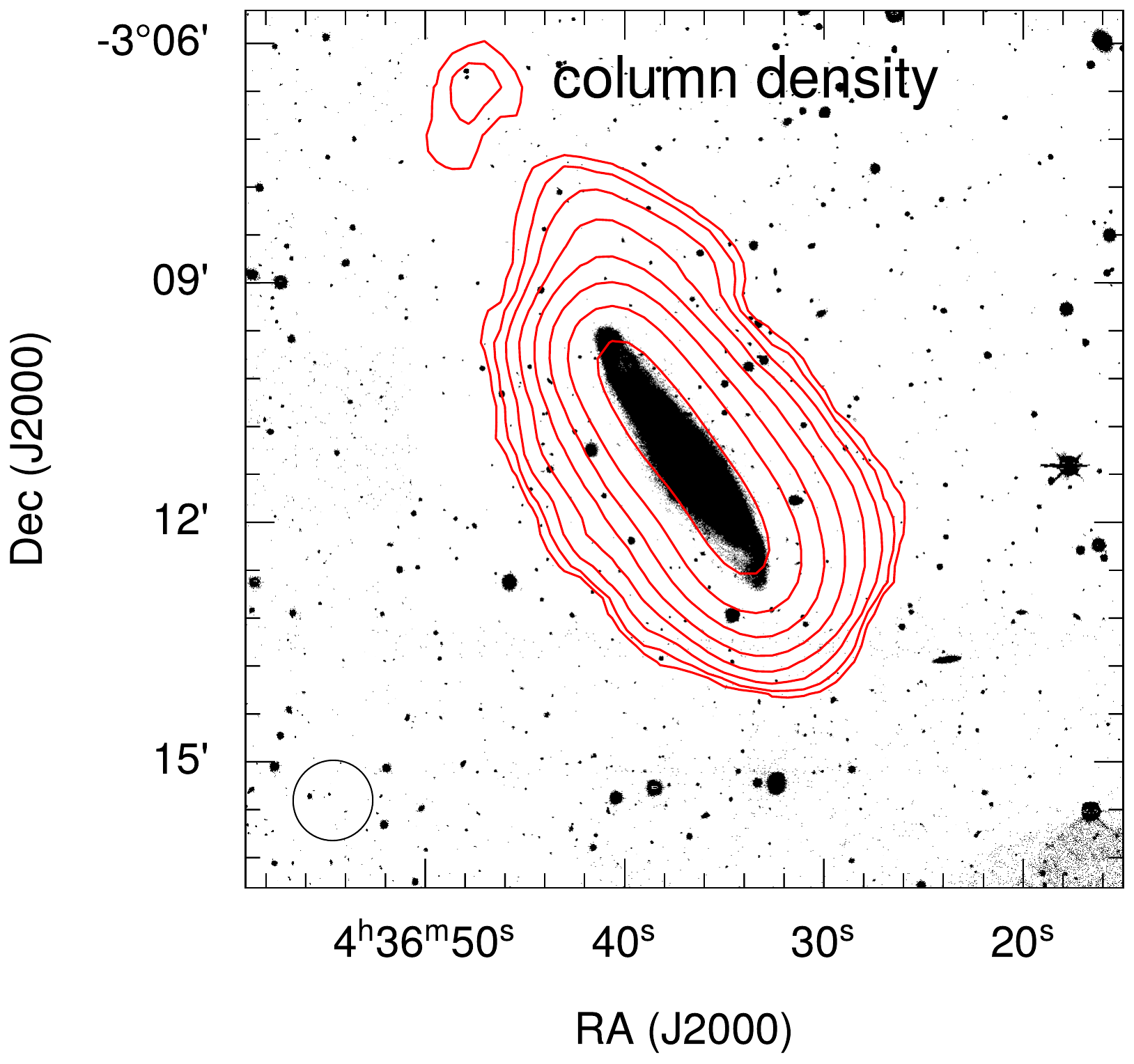}&
    \includegraphics[scale=0.21]{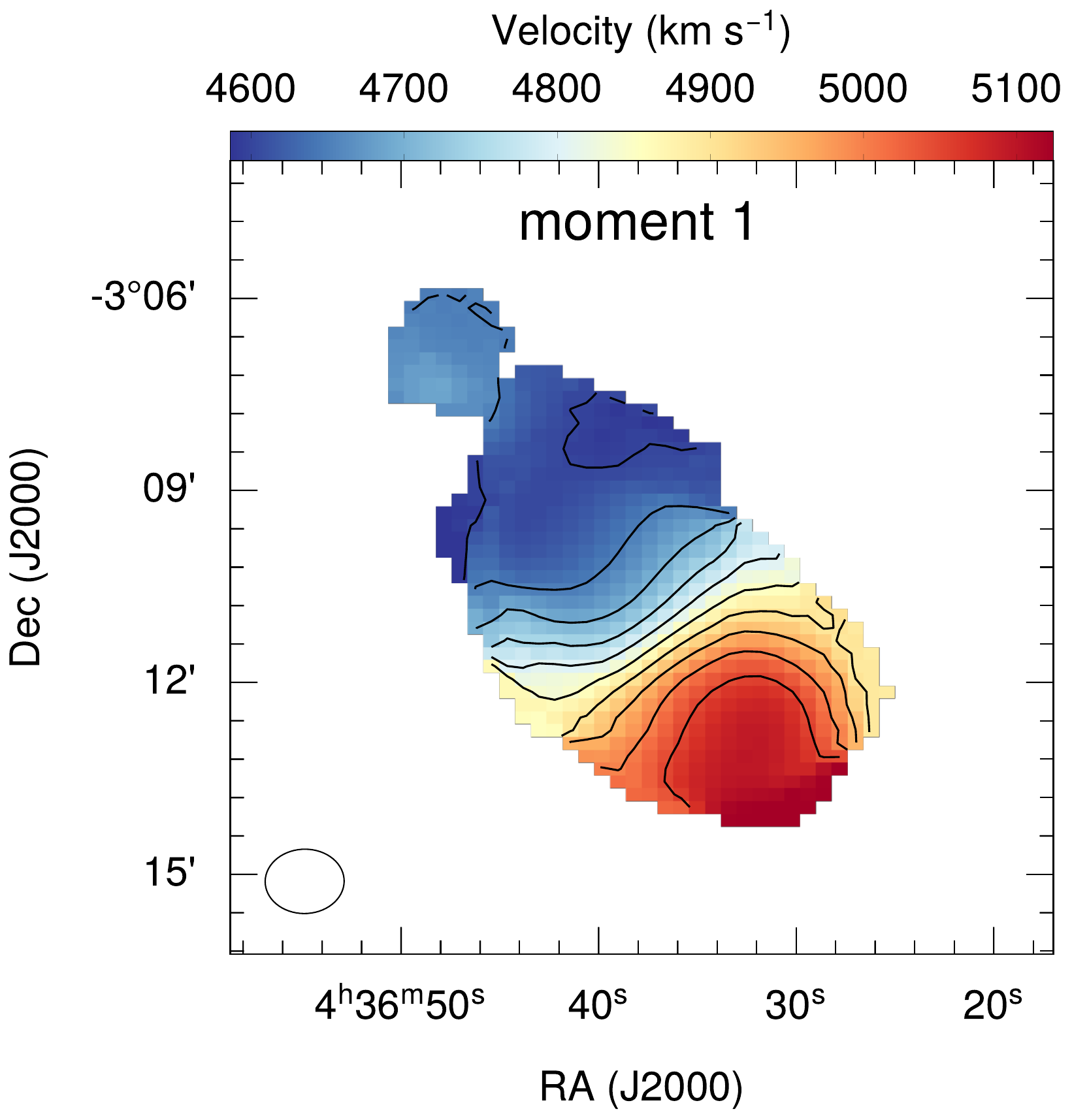}\\  
    \includegraphics[scale=0.21]{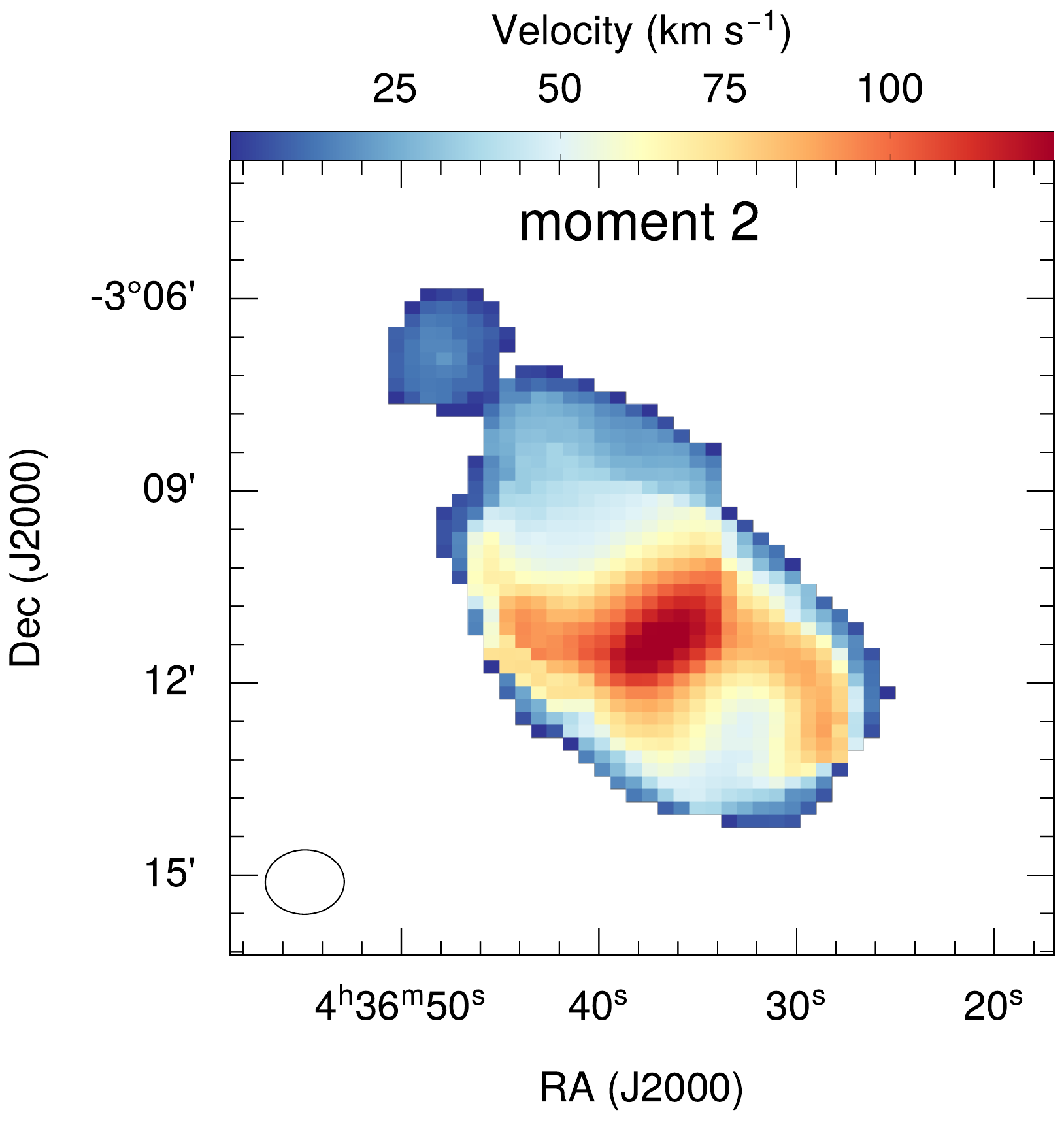}&  
    \includegraphics[scale=0.21]{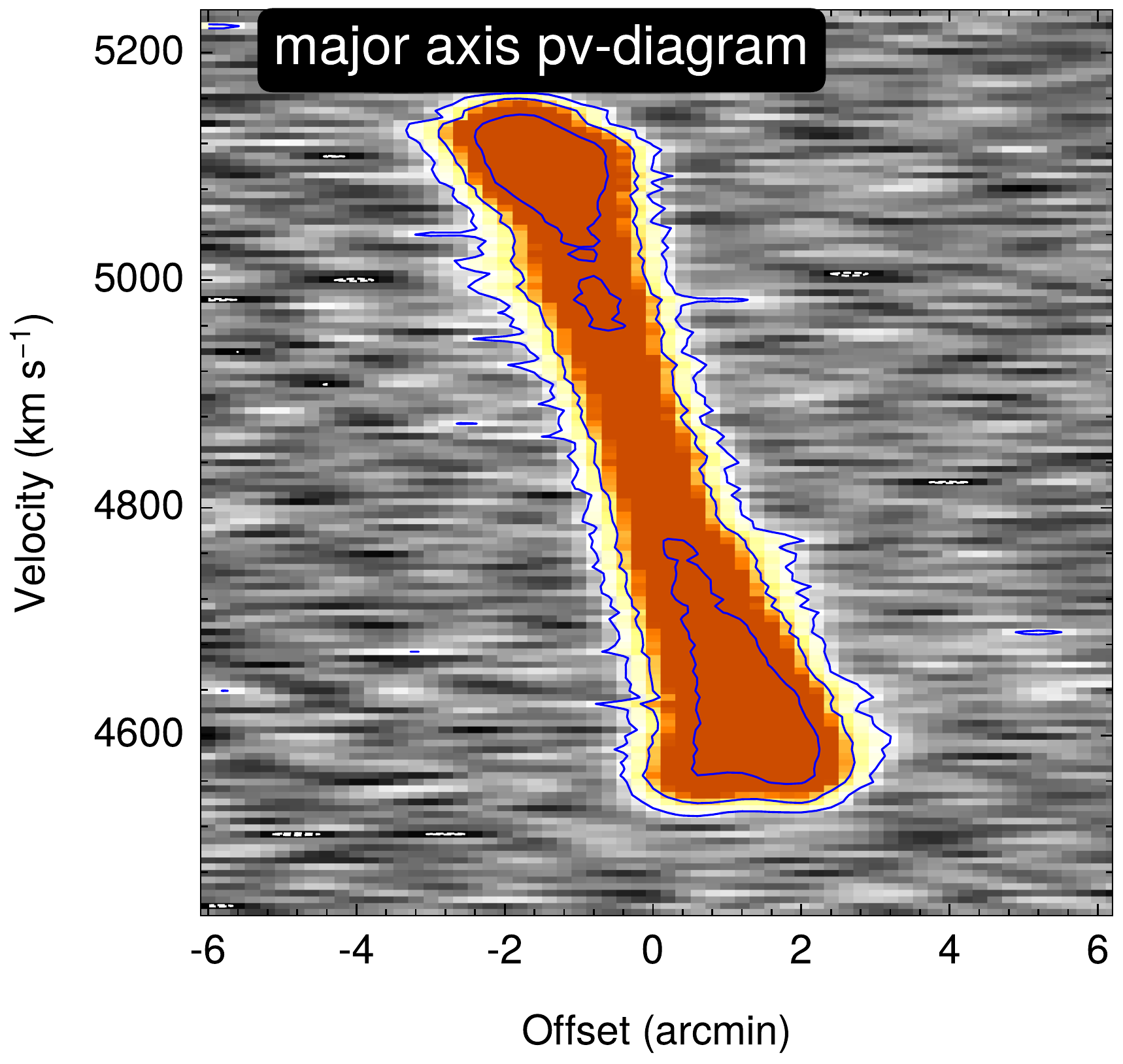}&
    \includegraphics[scale=0.21]{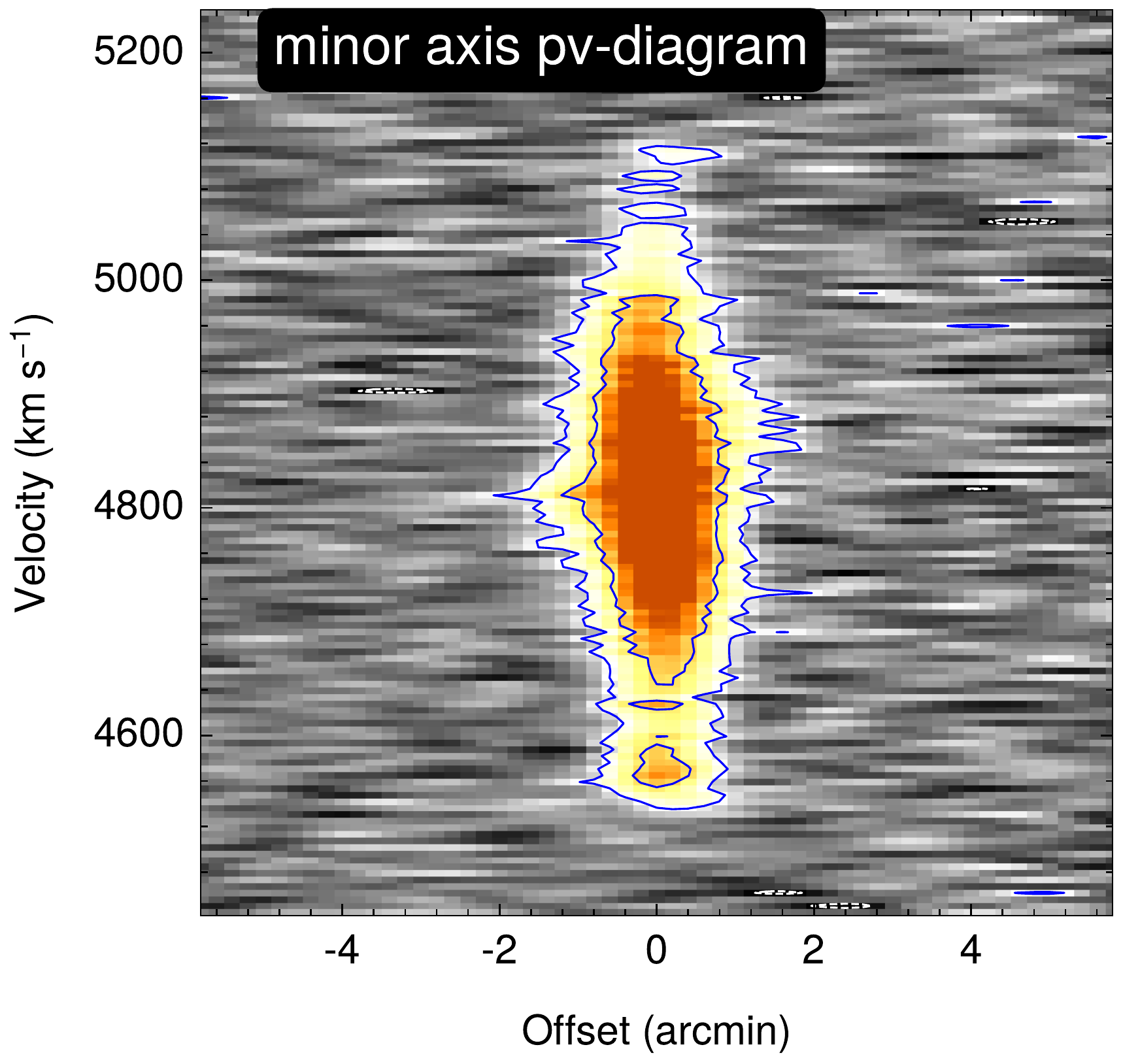}\\  
    \includegraphics[scale=0.21]{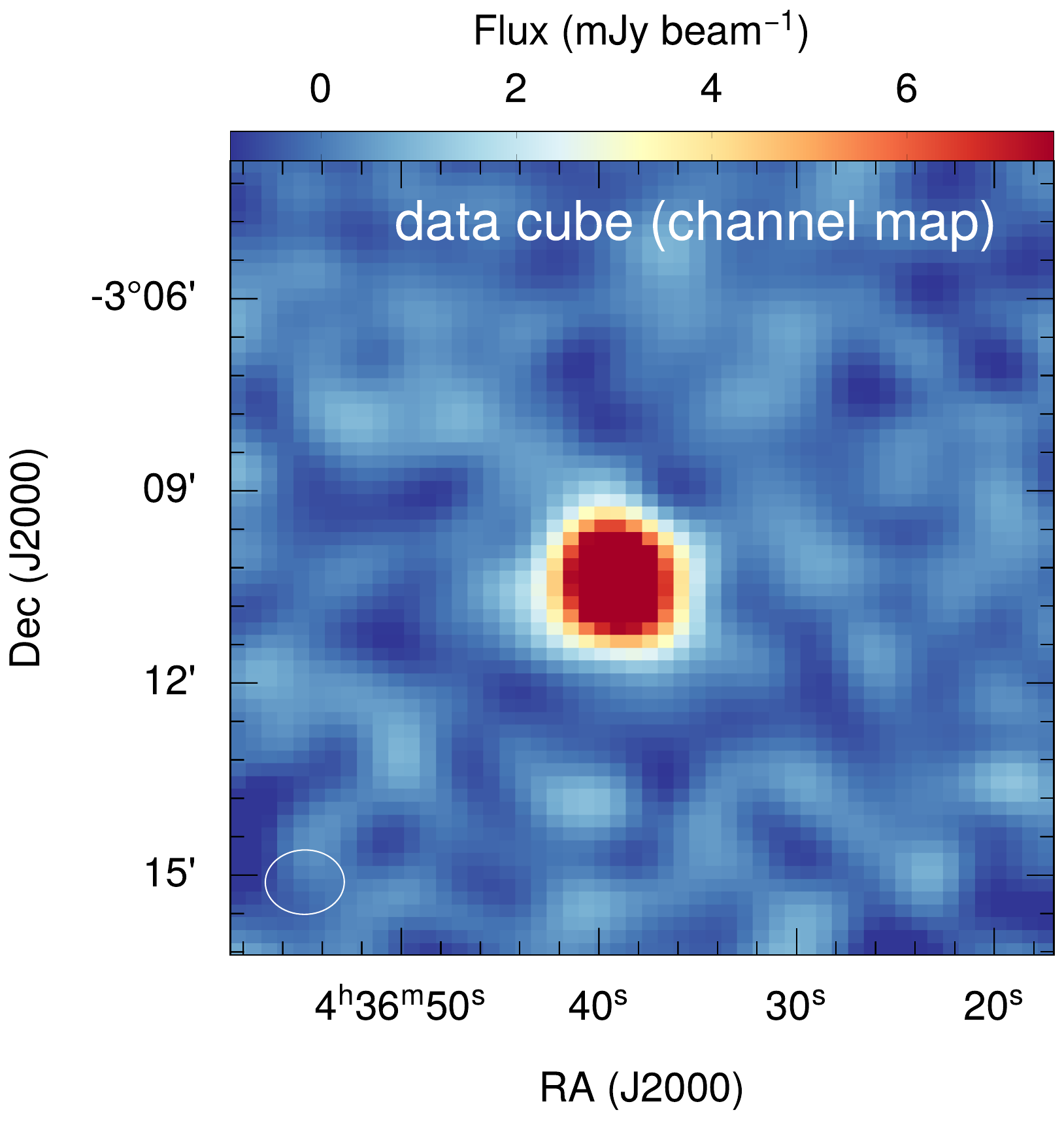}&  
    \includegraphics[scale=0.21]{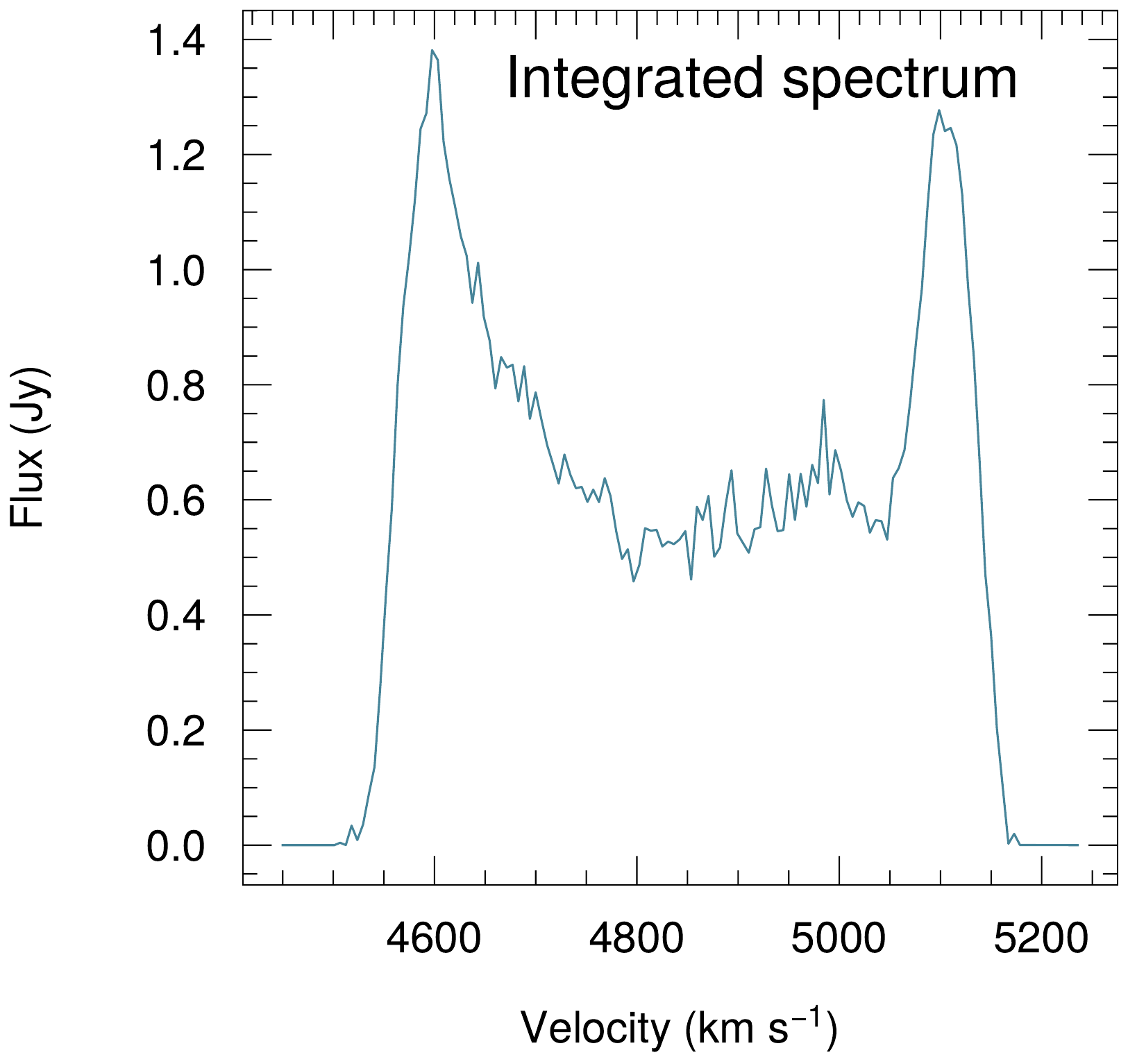}  
  \end{tabular}
  \caption{Example SoFiA data products for NGC~1622, a spiral galaxy previously detected in \HI\ located at about 381 kpc south east from the group centre. 
	From top left to bottom right panel: signal-to-noise ratio of each individual pixels, column density map and R-band DeCaLS 
  DR10 \citep{2019AJ....157..168D} optical image \textemdash~the contour levels are 
  (5.70$\times 10^{18}$, 1.140$\times 10^{19}$, 2.28$\times 10^{19}$, 4.56$\times 10^{19}$, 9.12$\times 10^{19}$, 1.82$\times 10^{20}$, 
  3.65$\times 10^{20}$, 7.30$\times 10^{20}$) $\mathrm{cm^{-2}}$, first moment map (velocity field), second moment map, major axis position velocity diagram, minor axis position velocity diagram, 
  an example channel map, and integrated spectrum from a data cube where the noise has been masked. 
  The ellipse at the bottom left corner of each plot shows the beam, 43.8\arcsec\ $\times$ 47.5\arcsec.}  
  \label{figure:data-products}
 \end{figure*}

The basic data products consist of data cubes, moment maps, and integrated spectra for the group as a whole, as well as source catalogues, cubelettes, 
moment maps, position velocity cuts, integrated spectra for each source detected by SoFiA, and 3D visualisation of each group. We used ViSL3D \citep{2024Zenodo13254756}, a new 3D visualisation software 
built upon X3D pathway  \citep{2016ApJ...818..115V} to construct the 3D view of each cube, which can be used to facilitate the separation of different \HI\ structures based on spatial and kinematical information. They are available at 
\href{https://amiga.iaa.csic.es/x3d-menu/}{https://amiga.iaa.csic.es/x3d-menu/}. For each of the groups, we provide two versions of the 
data cube: one showing the large-scale structures and one highlighting the central part of the group. 
The user can toggle between different iso-surface levels to make it easier to identify low-column density features. The technical details underlying their construction will be described in Labadie et al. (in prep.) A thorough 3D analysis of the different \HI\ features in HCGs is beyond the scope of this paper. Figure~\ref{figure:data-products} shows examples of data products for NGC~1622 at 43.8\arcsec\ $\times$ 47.5\arcsec\ resolution, a spiral galaxy previously detected in \HI\ located at about 381 kpc south east from the group centre. Below we describe the data 
for the individual groups. Overall, we observed even more extended tidal features in phase 2 groups than previously reported. For phase 3 groups, we did not detect any diffuse HI component, though we did identify some new high surface brightness features. Numerous galaxies were detected around the group centres, most notably in HCG~97 and HCG~91, 
suggesting that these groups may be embedded within larger structures. Additionally, some of the detected galaxies lack optical counterparts.  
\subsection{GBT vs MeerKAT flux}
\begin{figure*}
\begin{tabular}{ll}
    \includegraphics[scale=0.28]{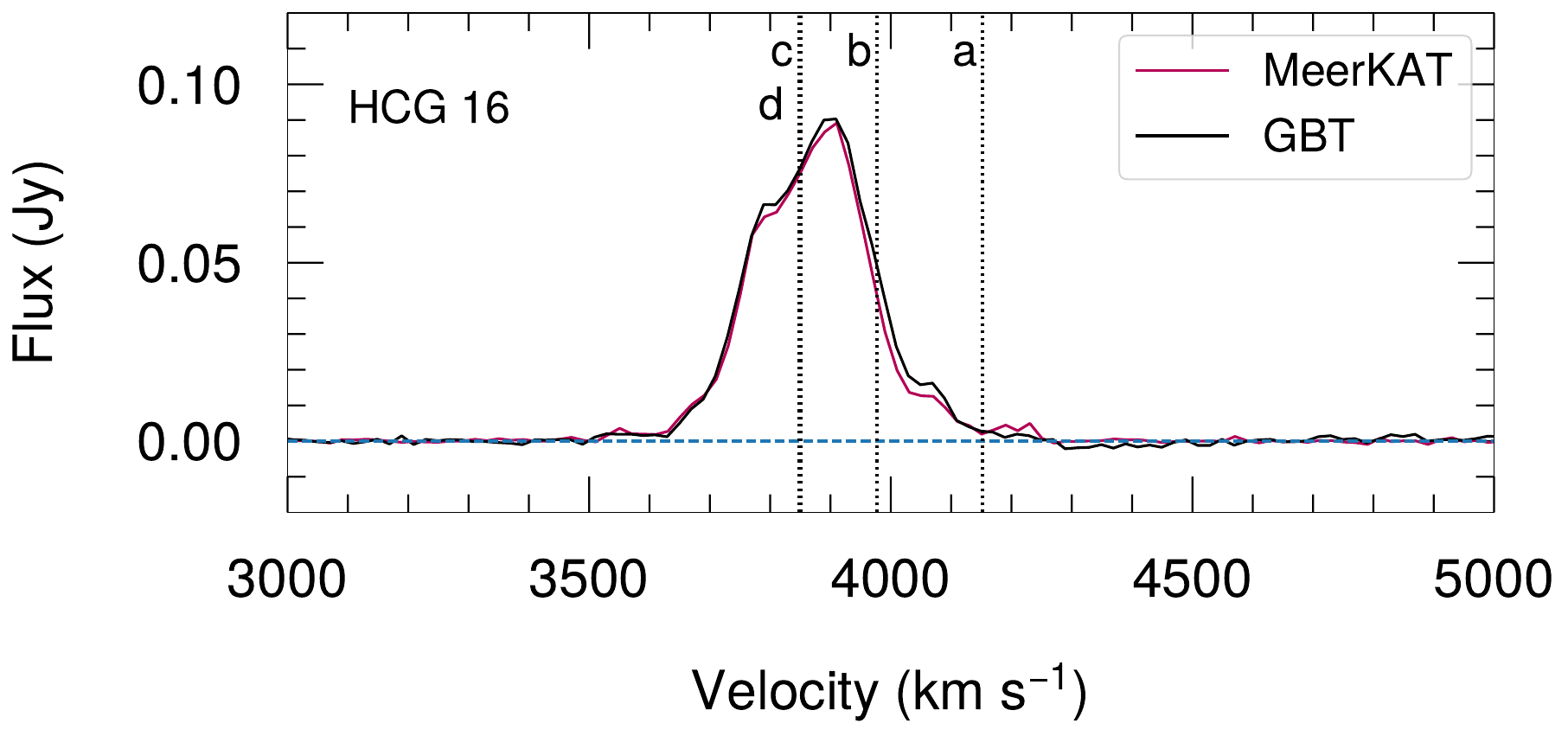}&
    \includegraphics[scale=0.28]{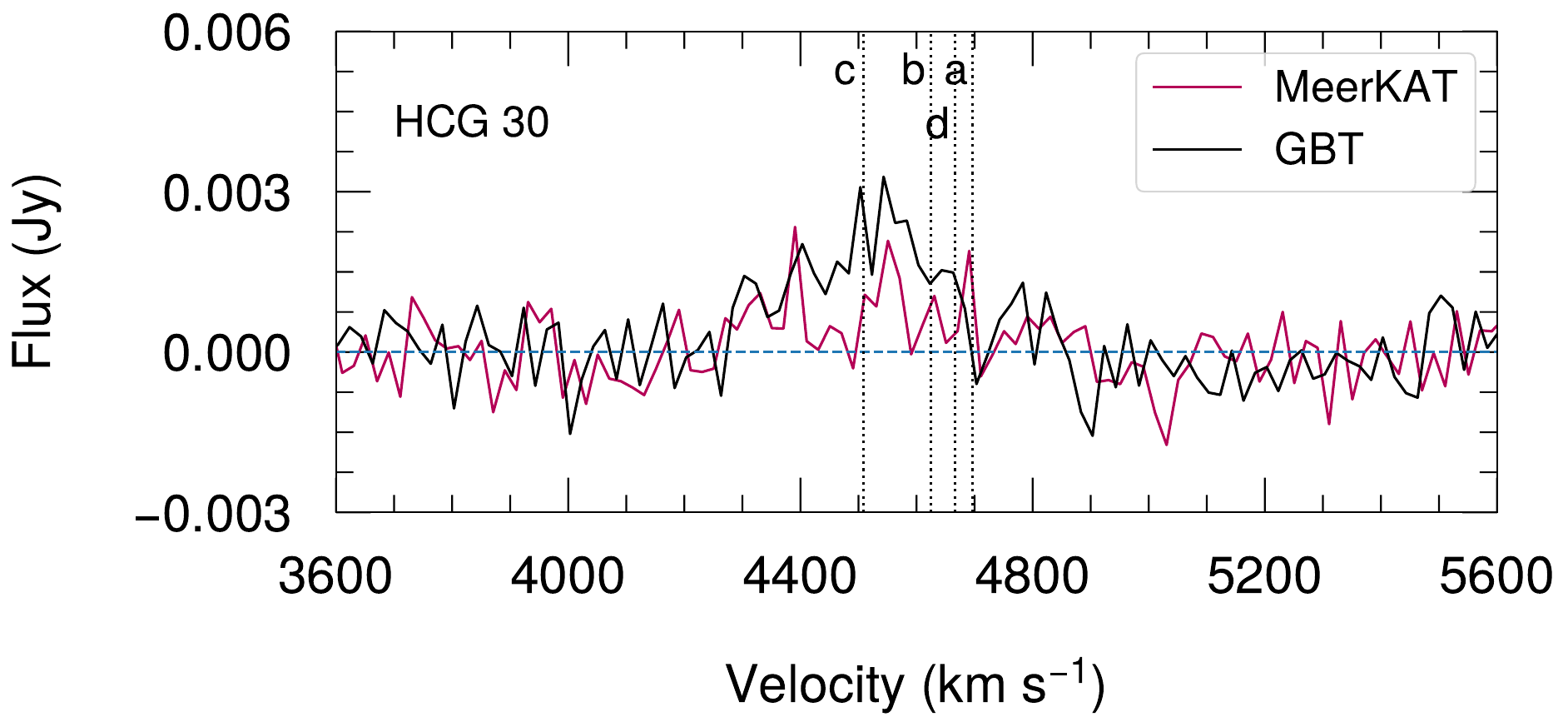}\\
    \includegraphics[scale=0.28]{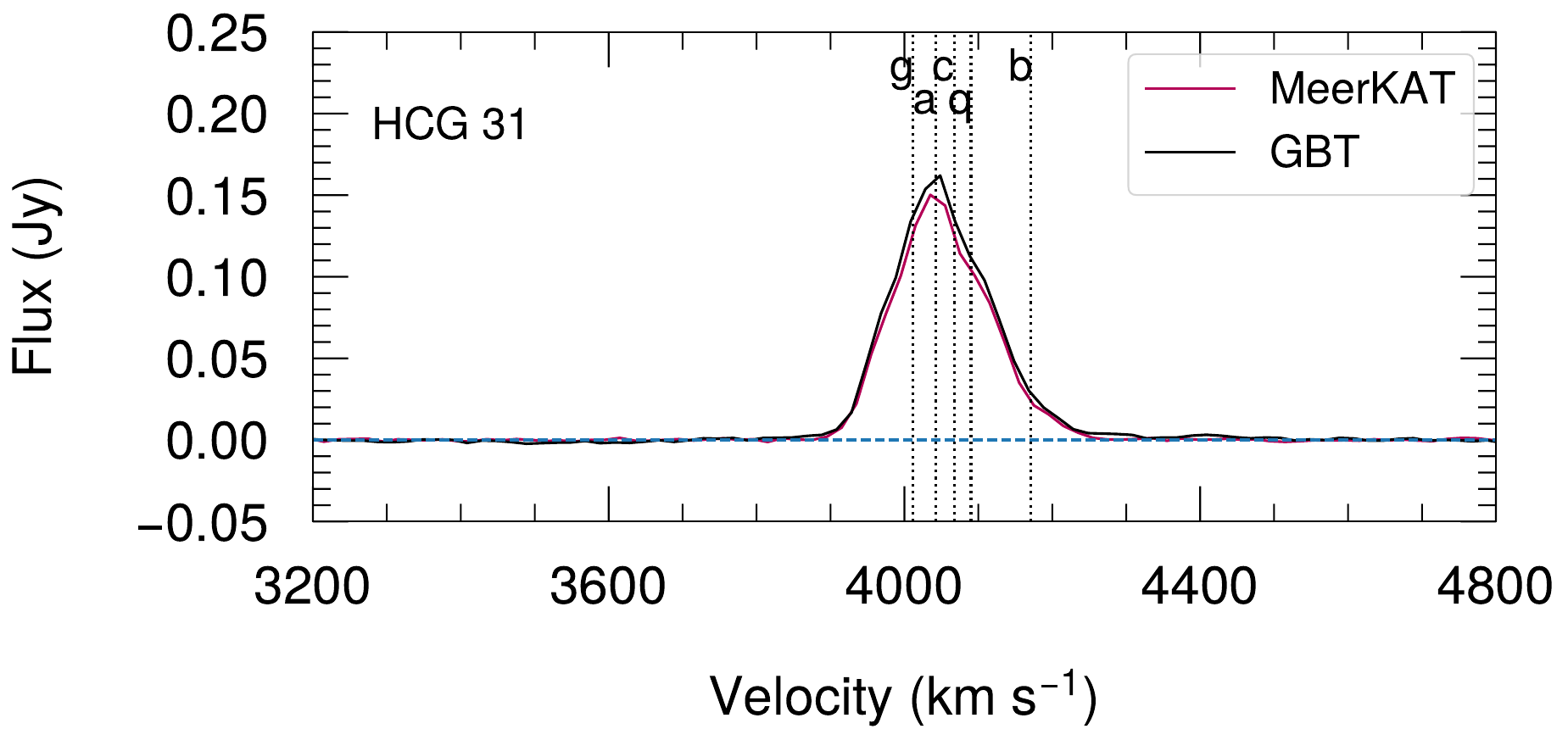}&
    \includegraphics[scale=0.28]{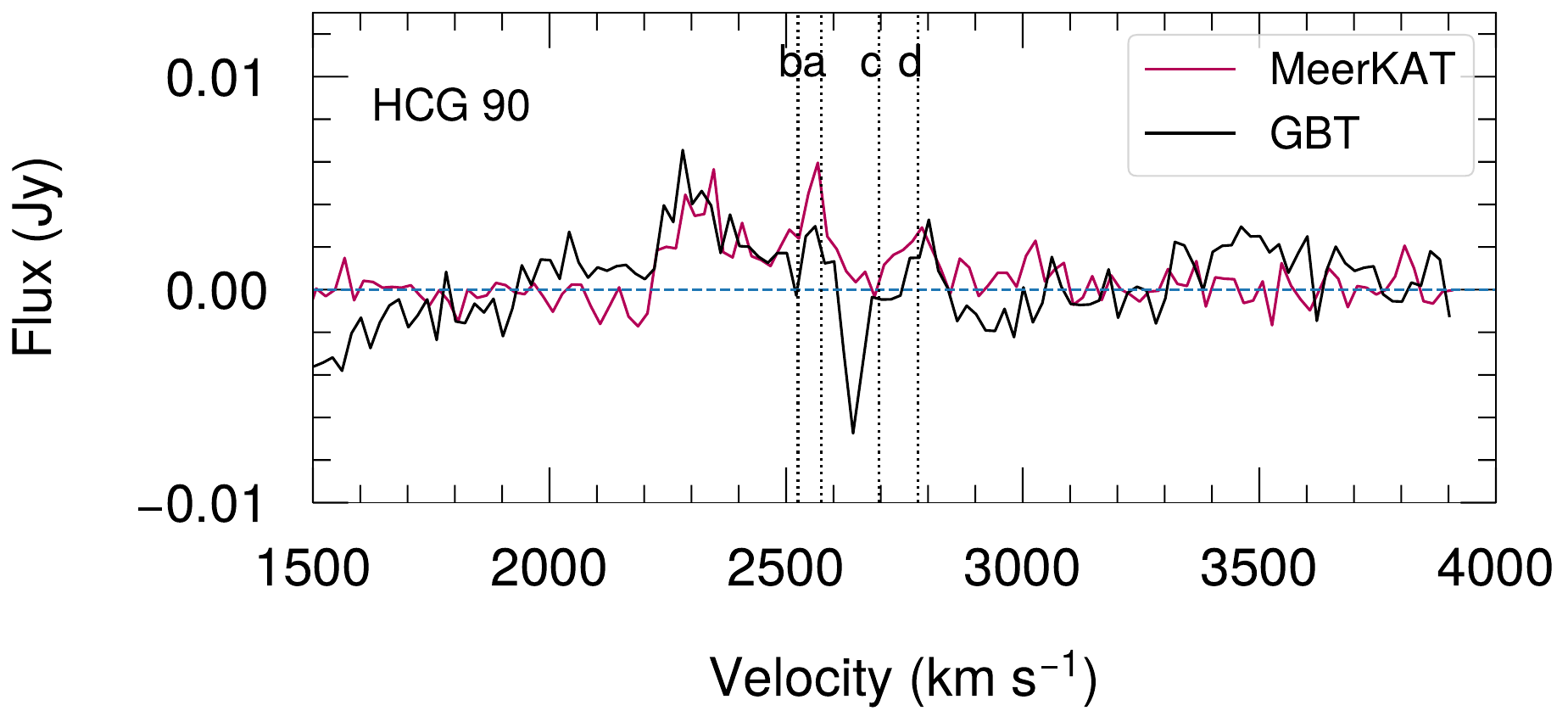}\\
    \includegraphics[scale=0.28]{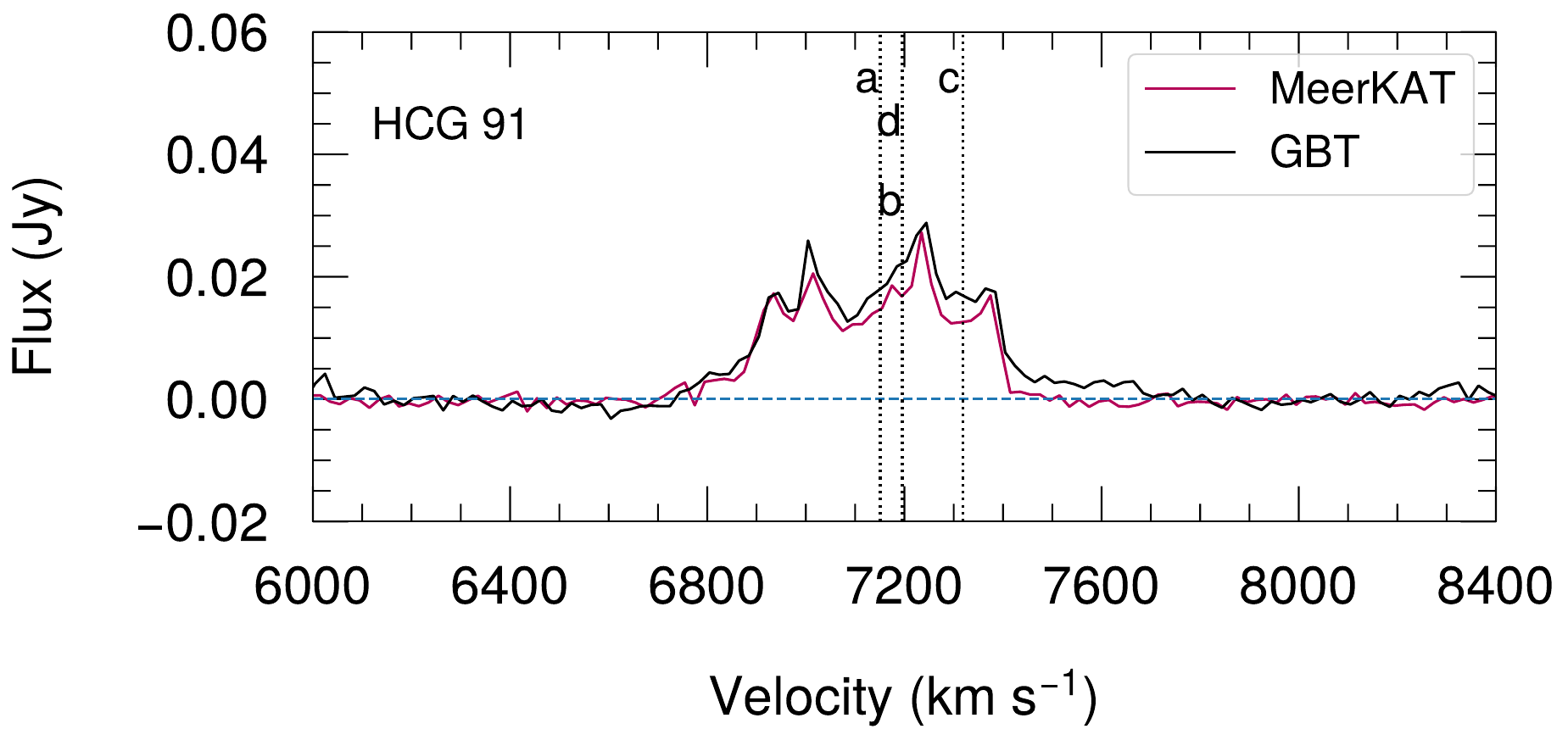}&
    \includegraphics[scale=0.28]{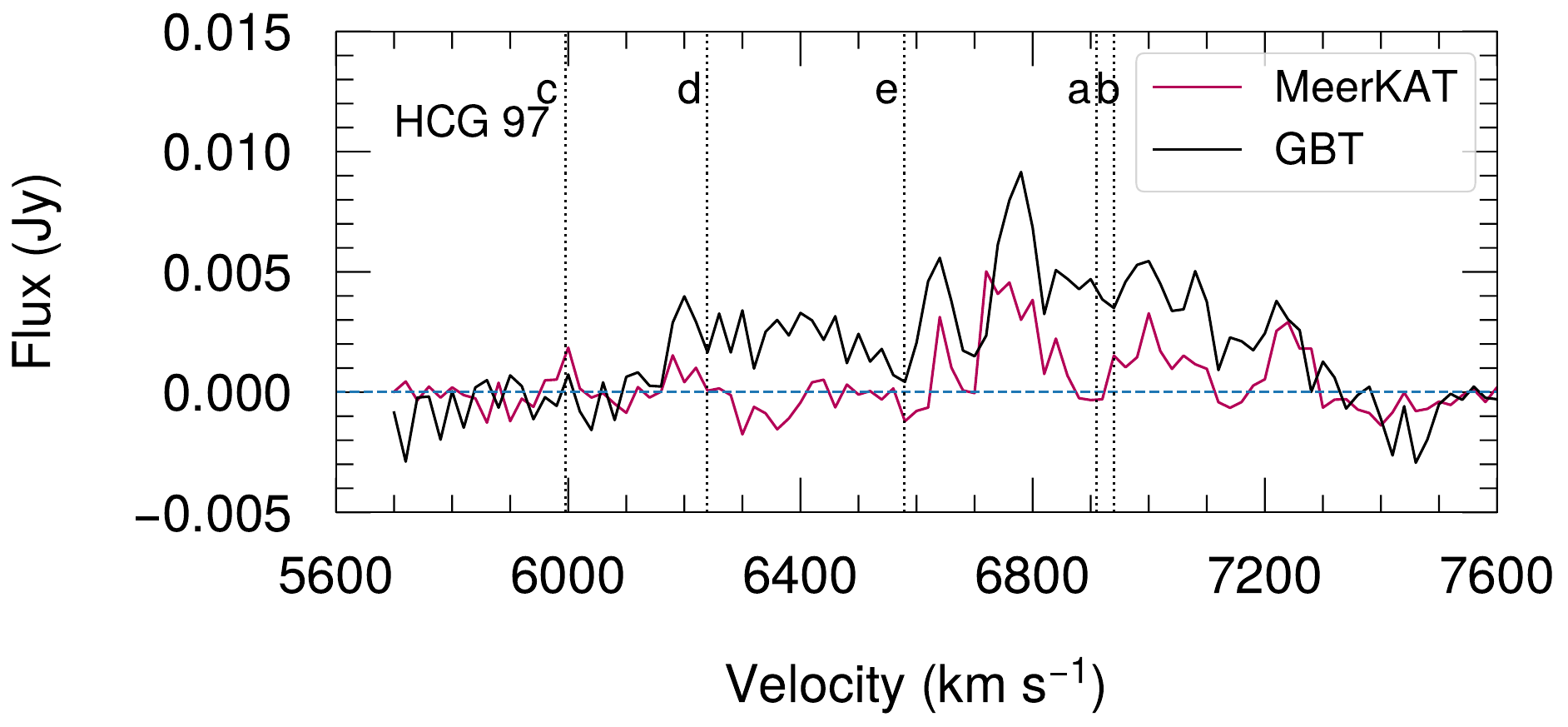}
  \end{tabular}
	\caption{GBT vs MeerKAT integrated spectrum smoothed at 20 $\mathrm{Jy~km~s^{-1}}$. The pink solid lines indicate the MeerKAT integrated spectra. The solid black lines indicate the GBT spectra of \citet{2010ApJ...710..385B}. 
    The vertical dotted lines indicate the velocities of the galaxies in the core of each group. The horizontal blue lines indicate zero intensity values to guide the eyes.}  
  \label{fig:spectrum}
 \end{figure*}

\begin{figure}
\begin{tabular}{l}
    \includegraphics[scale=0.28]{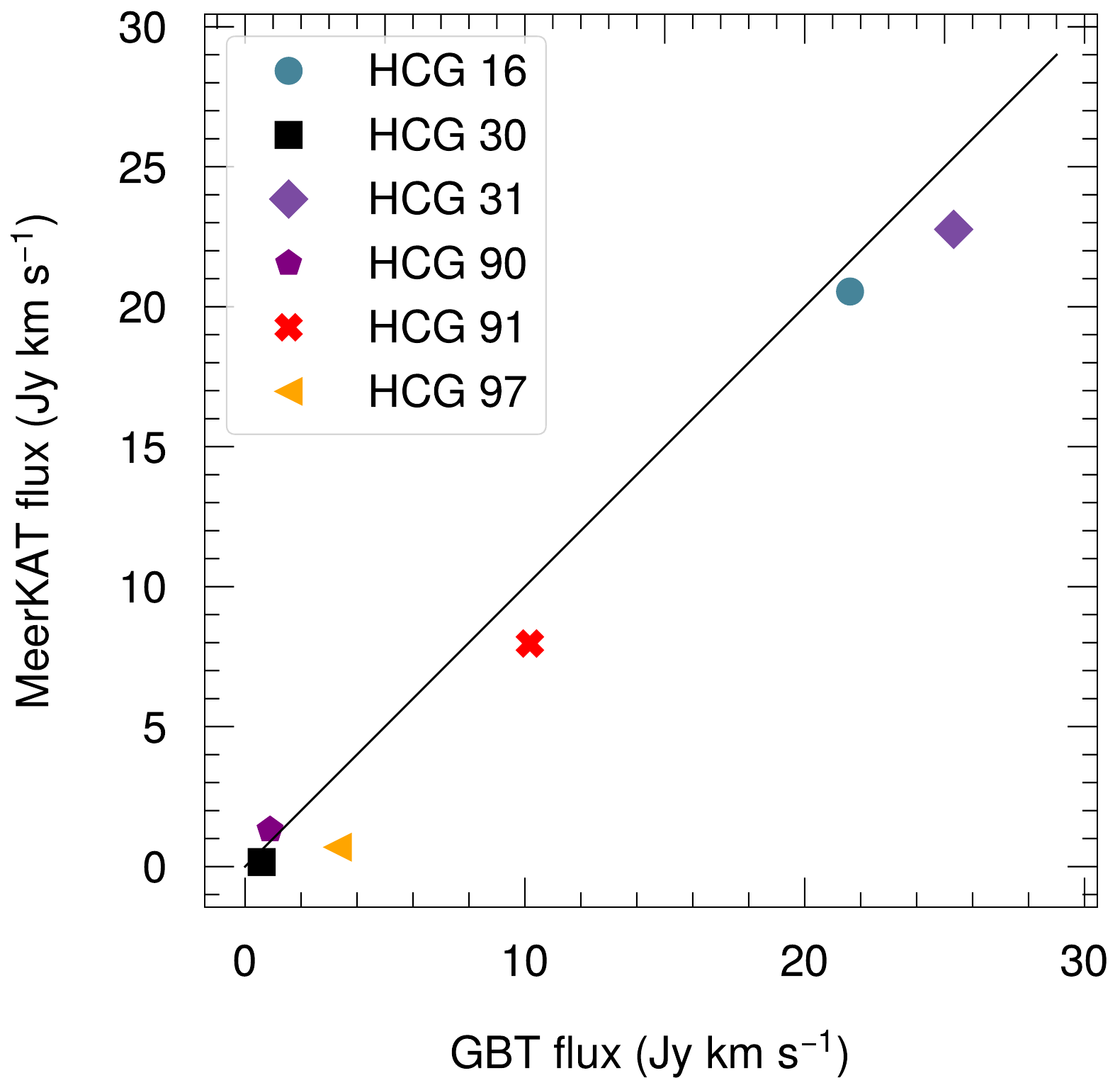}
  \end{tabular}
  \caption{GBT vs MeerKAT integrated flux measured within the GBT beam area. The solid line is a line of equality.}  
  \label{fig:flux}
 \end{figure}

To compare our flux measurements with the single dish GBT ones by \citet{2010ApJ...710..385B}, we followed the procedures outlined 
in that paper. They used the AIPS task \texttt{PATGN} to model the GBT beam response as a two-dimensional Gaussian with a FWHM of $9\overset{\arcmin}{.}1$ defined by:
\begin{equation}\label{eq:patgn}
    f(R) = C3 + (C4 - C3)~\exp[-((0.707/C5)^{2}~R^{2})],
\end{equation}
where $R$ is the radius, $C3$ is the minimum response of the beam, which is equal to zero, $C4$ is the maximum response of the beam, 
which is equal to 1, and $C5$ is the width of the distribution in arc second, which in our case, is equal to 9.1~$\times$ 60 / 2.355. 
The output of \texttt{PATGN} is a two-dimensional map containing the GBT beam correction factor. For the sake of reproducibility, 
we wrote a python function to generate the GBT beam response using equation~\ref{eq:patgn}. We multiply 
it with every channel of the MeerKAT cube. Then, we derived the integrated spectrum from the full area of the resulting cube. 
Our procedure ensures that emission outside the primary beam of the GBT is tapered to allow a reasonable comparison 
with the MeerKAT spectrum. We used the original, unsmoothed spectrum of \citet{2010ApJ...710..385B} and divided the observed brightness 
temperature by the antenna gain correction factor, 1.65 K/Jy, measured by the authors to convert it to Jy. We compare the GBT and MeerKAT integrated spectra 
in Figure~\ref{fig:spectrum}. For a better visualisation, we have smoothed the spectra from both telescopes to a common velocity resolution of 
$20~\mathrm{km~s^{-1}}$ using a boxcar kernel. For phase 2 groups, the agreement is good, the MeerKAT and GBT spectra agree well with each other in terms of shape and 
recovered flux. For phase 3 groups, the agreement is less obvious. For HCG~30, MeerKAT recovers less emission around the velocities of HCG30b, HCG30c, and HCG30d. 
For HCG~90, the GBT and MeerKAT spectra seem to agree with each other except around 3500 $\mathrm{km~s^{-1}}$ where the GBT spectrum shows positive emission. 
For HCG~97, the GBT spectrum shows emission spreading across a large velocity range. However, that of MeerKAT shows narrower emission. 
In addition, at virtually each velocity of HCG~97, GBT recovers 
more flux than MeerKAT. Figure~\ref{fig:flux} and Table~\ref{tab:gbt-percentage} compares the flux measured by MeerKAT and GBT within the GBT beam area. For phase 2 groups, there is a good agreement between the flux recovered by the GBT and MeerKAT within this area, with the largest flux difference being 22\%. 
For the groups in phase 3, GBT recovers over 70\% more flux than MeerKAT for HCG~30 and HCG~97, while for HCG~90, MeerKAT recovers 32\% more flux than GBT. This is partly due to the presence of negative fluxes in the GBT spectrum of HCG~90. Additionally, the MeerKAT spectrum around the velocity of HCG~90a peaks well above that of the GBT. Note that the total flux recovered by MeerKAT is significantly higher than that of the GBT due to the much larger field of view of MeerKAT. 
We summarise the observational parameters and derived \HI\ mass for each group in Table~\ref{table:sample2} and 
compare them with the corresponding VLA data from \citet{2023A&A...670A..21J}.
To calculate the noise level listed in the table, 
we use the noise cubes generated by SoFiA. First, we take the median value of each velocity slice across the noise cube. Then we take the median of the values from the slices.
Note that the noise has been estimated using the non-primary beam corrected data. However, the tabulated mass is calculated from the global profile of the primary beam corrected cubes using the following formula:
\begin{equation}
	M_{HI}[M_{\odot}] = 235600 \times D^2 \times \sum S_{i}\Delta v
\end{equation}
where $\sum S_{i}\Delta v $ is the sum of the flux over all channels and within the field of view of MeerKAT in units of $\mathrm{Jy~km~s^{-1}}$, and $D$ is the distance to the group.
As described by \citet{2023A&A...673A.146S}, excluding channels containing known \HI\ line emission while performing continuum subtraction using the \texttt{UVLIN} option of the CASA task 
\texttt{mstransform}, implemented in CARACaL, can introduce artefacts in the form of vertical stripes in a RA-velocity slice from the data cube. To check the possible presence of RFI or continuum residual emission in channels excluded from the visibility fit, we show a plot of the RA-velocity slice at the central declination of each HCGs cube. We describe the data products for each group below.            

\begin{table}
    \centering
    \caption{\label{tab:gbt-percentage}GBT vs MeerKAT recovered \HI\ flux}
    \begin{tabular}{c c c c}
    \toprule \toprule
    HCG & MeerKAT  & GBT & Difference \\
      & [$\mathrm{Jy~km~s^{-1}}$] & [$\mathrm{Jy~km~s^{-1}}$] &  [\%]\\
    \midrule
    16 &20.54 & 21.62 &-5    \\
    31 &22.76&25.32&-10  \\
    91 &7.97&10.18&-22  \\
    \midrule
    30 &0.16 & 0.56&-73   \\
    90 &1.31&0.89&+32  \\
    97 &0.69&3.31&-79 \\
    \bottomrule
    \end{tabular}
    \tablefoot{Difference in percentage between the integrated flux recovered by GBT and MeerKAT within the GBT beam area. The negative sign in the table 
indicates that MeerKAT recovers less flux than the GBT. The positive sign indicates that MeerKAT recovers more flux than the GBT.}
\end{table}

\begin{table*}
  \centering
  \caption{\label{table:sample2}Observational parameters and derived \HI\ mass}
  
  \resizebox{0.85\textwidth}{!}{
  \begin{tabular}{@{\extracolsep{\fill}}*{11}{c}}
  \toprule \toprule
  HCG & Distance & Pixel size & Weighting & $uv$ taper & \multicolumn{2}{c}{Beam size} & & \multicolumn{2}{c}{Lin. res.} \\
  \cline{6-7} \cline{9-10} \noalign{\vskip 2pt} 
  &  &  &  &  & MeerKAT & VLA & & MeerKAT & VLA \\
  \noalign{\vskip 3pt} 
  & [Mpc] & [\arcsec] &  & [\arcsec] & \multicolumn{2}{c}{[\arcsec]} & & \multicolumn{2}{c}{[kpc]}\\
  \midrule
  16 & 49 & 12 & 0.5 & 50 & 57.9$\times$57.5 & 38.9$\times$31.7 & & 13.5$\times$13.4 & 9.2$\times$7.5 \\
  31 & 53 & 12 & 0.5 & 50 & 59.9$\times$59.2 & 14.6$\times$12.1 & & 15.4$\times$15.2 & 3.8$\times$3.1 \\
  91 & 92 & 12 & 0.5 & 50 & 58.0$\times$56.1 & 51.3$\times$47.0 & & 25.6$\times$25.0 & 22.9$\times$21.0 \\
  \midrule
  30 & 61 & 12 & 0.5 & 50 & 60.5$\times$59.8 & 57.9$\times$44.8 & & 17.9$\times$17.7 & 17.1$\times$13.2 \\
  90 & 33 & 12 & 0.5 & 50 & 57.7$\times$56.7 & 49.1$\times$39.9 & & 9.2$\times$9.1 & 7.9$\times$6.4 \\
  97 & 85 & 12 & 0.5 & 50 & 60.2$\times$59.3 & 63.2$\times$49.5 & & 23.8$\times$23.3 & 26.0$\times$20.4 \\
  \end{tabular}
  }
  \resizebox{0.85\textwidth}{!}{
  \begin{tabular}{@{\extracolsep{\fill}}*{14}{c}}
  \toprule \toprule
  HCG & \multicolumn{2}{c}{Noise} & & \multicolumn{2}{c}{$N_{\mathrm{HI}}(3\sigma)$} & & \multicolumn{2}{c}{Flux} & & V. range & \multicolumn{2}{c}{\HI\ mass} \\
  \cline{2-3} \cline{5-6} \cline{8-9} \cline{12-13}\noalign{\vskip 2pt} 
  & MeerKAT & VLA & & MeerKAT & VLA & & MeerKAT & VLA & & & MeerKAT & VLA \\
  \noalign{\vskip 3pt} 
  & \multicolumn{2}{c}{[$\mathrm{mJy~beam^{-1}}$]} & & \multicolumn{2}{c}{[$\mathrm{10^{18}~cm^{-2}}$]} & & \multicolumn{2}{c}{[$\mathrm{Jy~km~s^{-1}}$]} & & 
  [$\mathrm{km~s^{-1}}$] & \multicolumn{2}{c}{[ $\mathrm{10^{9}~M_{\odot}}$]}\\ 
  \midrule
  16 & 0.33 & 0.41 & & 3.5 & 22.5 & & 42.2 & 39.5 & & 2811 - 5006 & 23.9 & 22.4 \\
  31 & 0.33 & 0.65 & & 3.3 & 176.1 & & 37.6 & 22.1 & & 2095 - 5986 & 24.9 & 14.7 \\
  91 & 0.33 & 0.66 & & 3.7 & 18.5 & & 55.3 & 15.8 & & 4015 - 10994 & 110.5 & 31.6 \\
  \midrule
  30 & 0.33 & 0.51 & & 3.3 & 13.3 & & 40.0 & 15.1 & & 3251 - 5733 & 35.1 & 13.3 \\
  90 & 0.32 & 0.55 & & 3.4 & 26.9 & & 40.8 & 3.4 & & 507 - 5010 & 10.5 & 0.9 \\
  97 & 0.32 & 0.44 & & 3.2 & 9.5 & & 17.7 & 7.0 & & 5232 - 8247 & 30.2 & 11.9 \\
  \bottomrule
  \end{tabular}
  }

  \tablefoot{Top: (1) HCG ID number, (2) distance derived by \citet{2023A&A...670A..21J}, (3) Pixel size of the data cubes, (4) 
  Briggs weighting parameters, (5) Gaussian taper, (6) the synthesised beam size, (7) Linear resolution. Bottom: (1) HCG ID number, 
  (2) the median noise level of the data cubes, 
  (3) the $3\sigma$ column density sensitivity limit assuming an \HI\ line width of $\mathrm{20~km~s^{-1}}$, (4) 
  total \HI\ flux, (5) Velocity range used to calculate the total \HI\ flux (6) total \HI\ mass. The VLA data is from \citet{2023A&A...670A..21J}.}
\end{table*}

\subsection{HCG 16}
\begin{figure*}
    \setlength{\tabcolsep}{0pt}
\begin{tabular}{c c c}
    \includegraphics[scale=0.215]{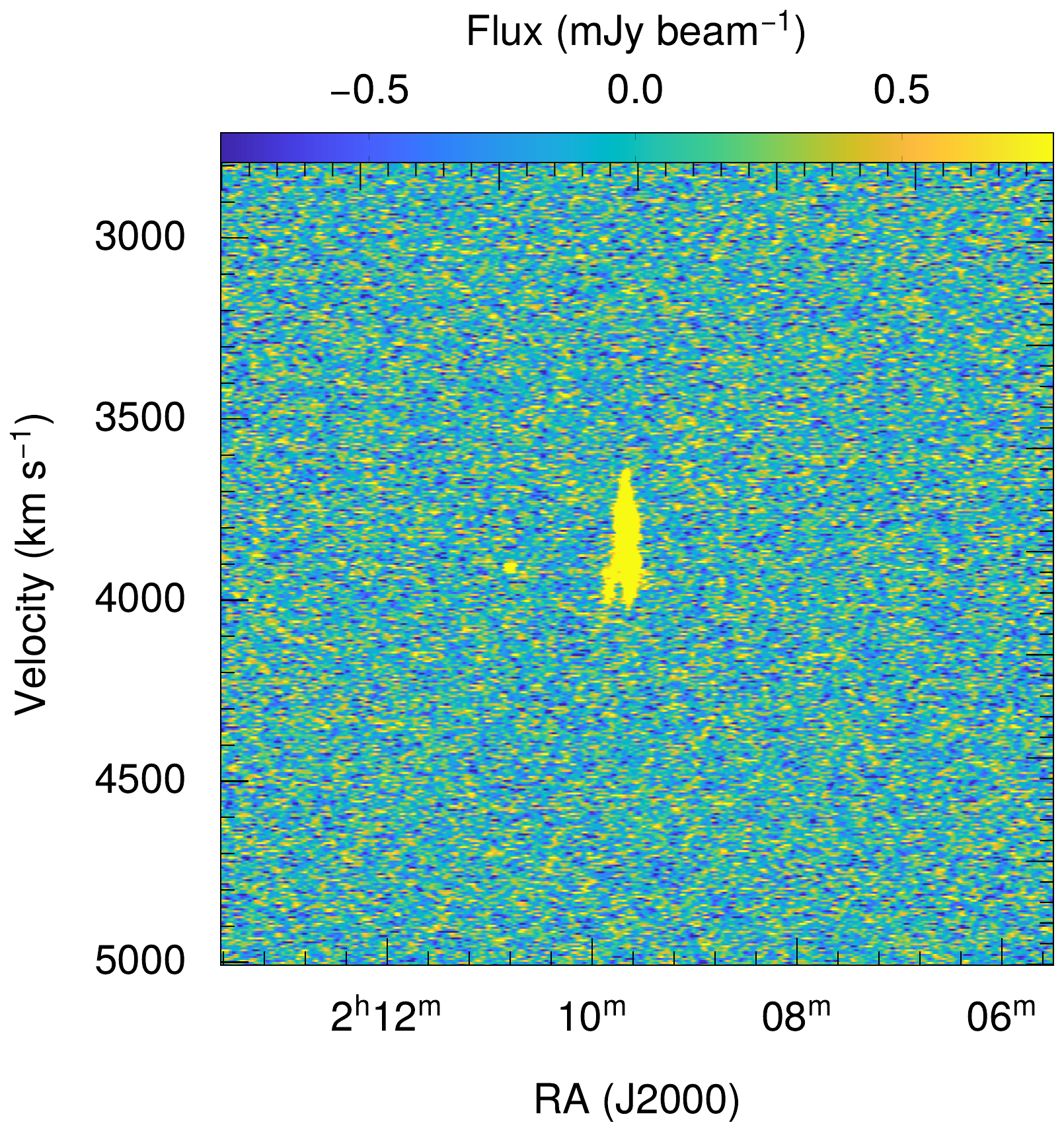} & 
    \includegraphics[scale=0.215]{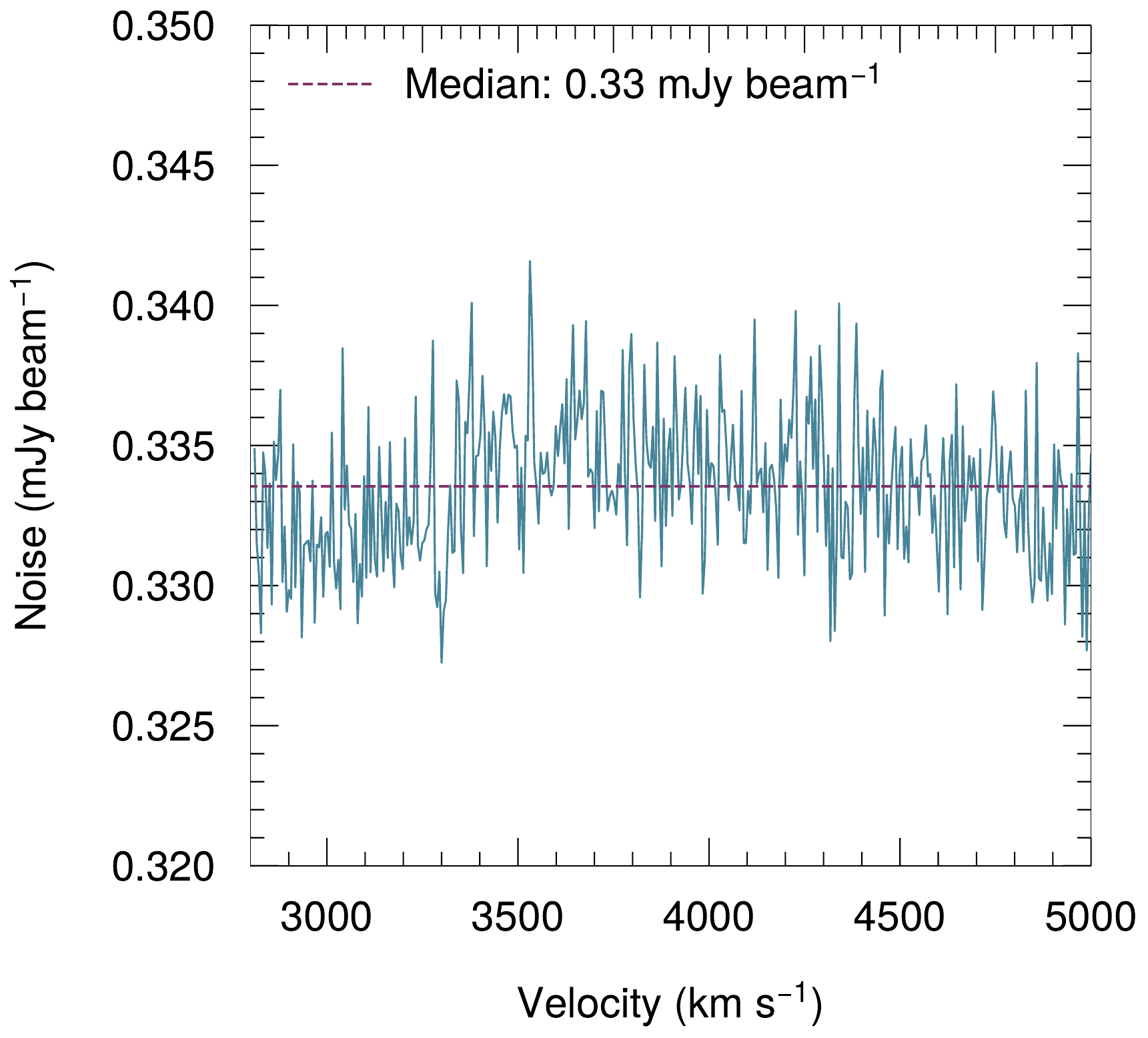} &
    \includegraphics[scale=0.215]{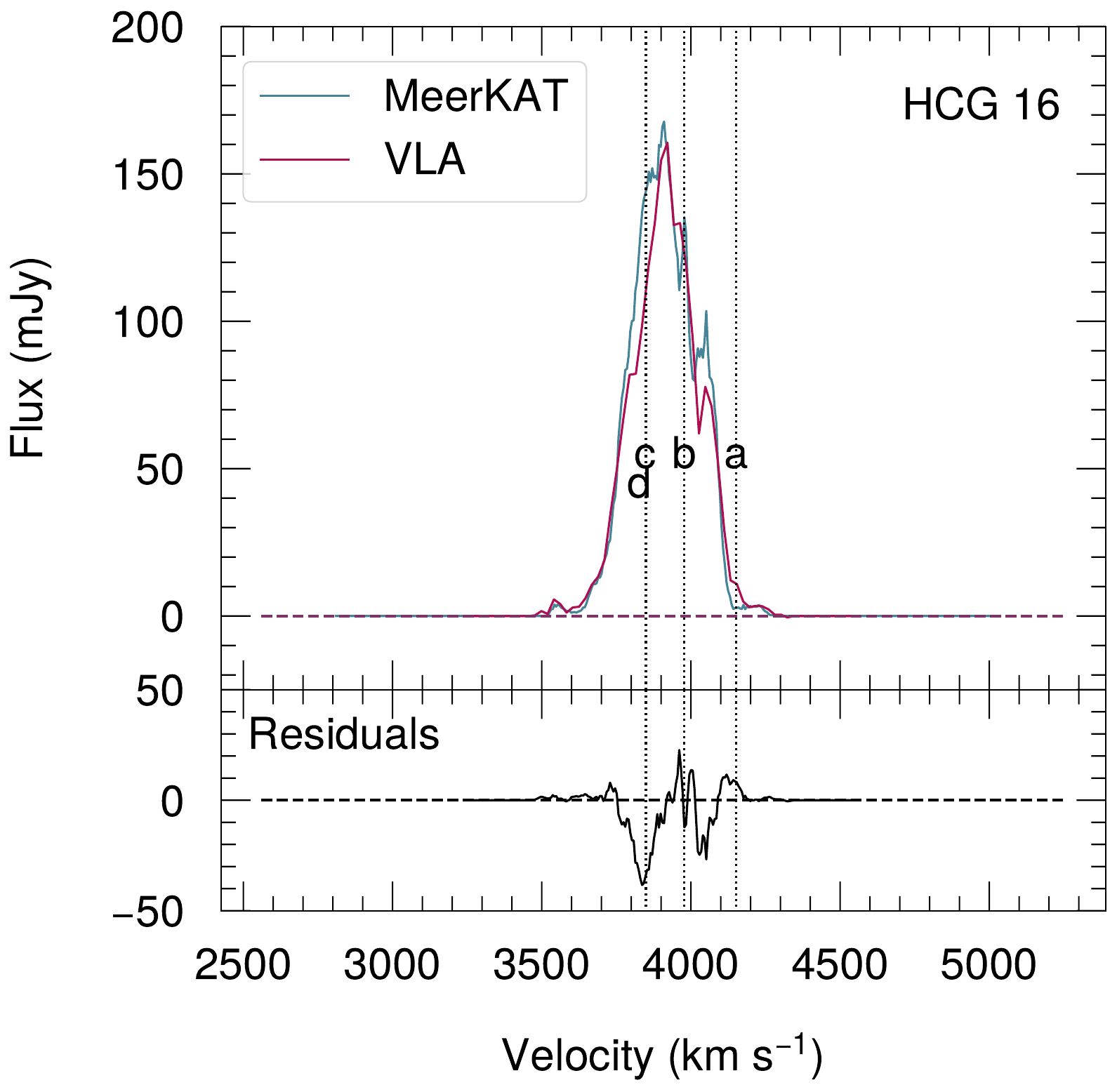} 
  \end{tabular}
  \caption{Left panel: velocity vs right ascension of HCG~16. Middle panel: median noise values of each RA-DEC slice of the non-primary beam corrected 60\arcsec data cube of 
  HCG~16 as a function of velocity. The horizontal dashed line indicates the median of all the noise values from each slice. Right panel: the blue solid lines indicates the 
  MeerKAT integrated spectrum of HCG~16; the red solid line indicates VLA integrated spectrum of the group derived by \citep{2023A&A...670A..21J}. 
  The vertical dotted lines indicate the velocities of the galaxies in the core of the group. The spectra have been extracted from area containing only genuine \HI\ emission.}  
  \label{fig:hcg16_noise}
 \end{figure*}

We show the RA-velocity slice of HCG~16 in the first panel of Figure~\ref{fig:hcg16_noise}. We found no obvious RFI or continuum residuals. 
We plot the median noise level of the data cube as a function of velocity in the second panel of the figure. The noise does not 
change much across the spectral channel.  We plot the integrated spectrum of the group in the right panel of Figure~\ref{fig:hcg16_noise}. The spectrum was 
derived from the primary beam corrected cube where the noise was blanked using a 3D mask generated by SoFiA. 
The most significant discrepancies between the VLA and MeerKAT 
spectra are observed at the velocities corresponding to galaxies HCG 16c and HCG 16d. Note that HCG 16c is the the group's most active 
star-forming galaxy and is strongly interacting with HCG~16d \citep{2023A&A...670A..21J}. The difference between the VLA and MeerKAT 
spectra can be attributed to MeerKAT's capability to detect more tidal features than the VLA. Although MeerKAT detected more \HI\ 
features as will be explained in the next section, the total mass recovered is remarkably similar to that observed by the VLA. 

\begin{figure*}
 \setlength{\tabcolsep}{0pt}
 \begin{tabular}{l l l}
     \includegraphics[scale=0.25]{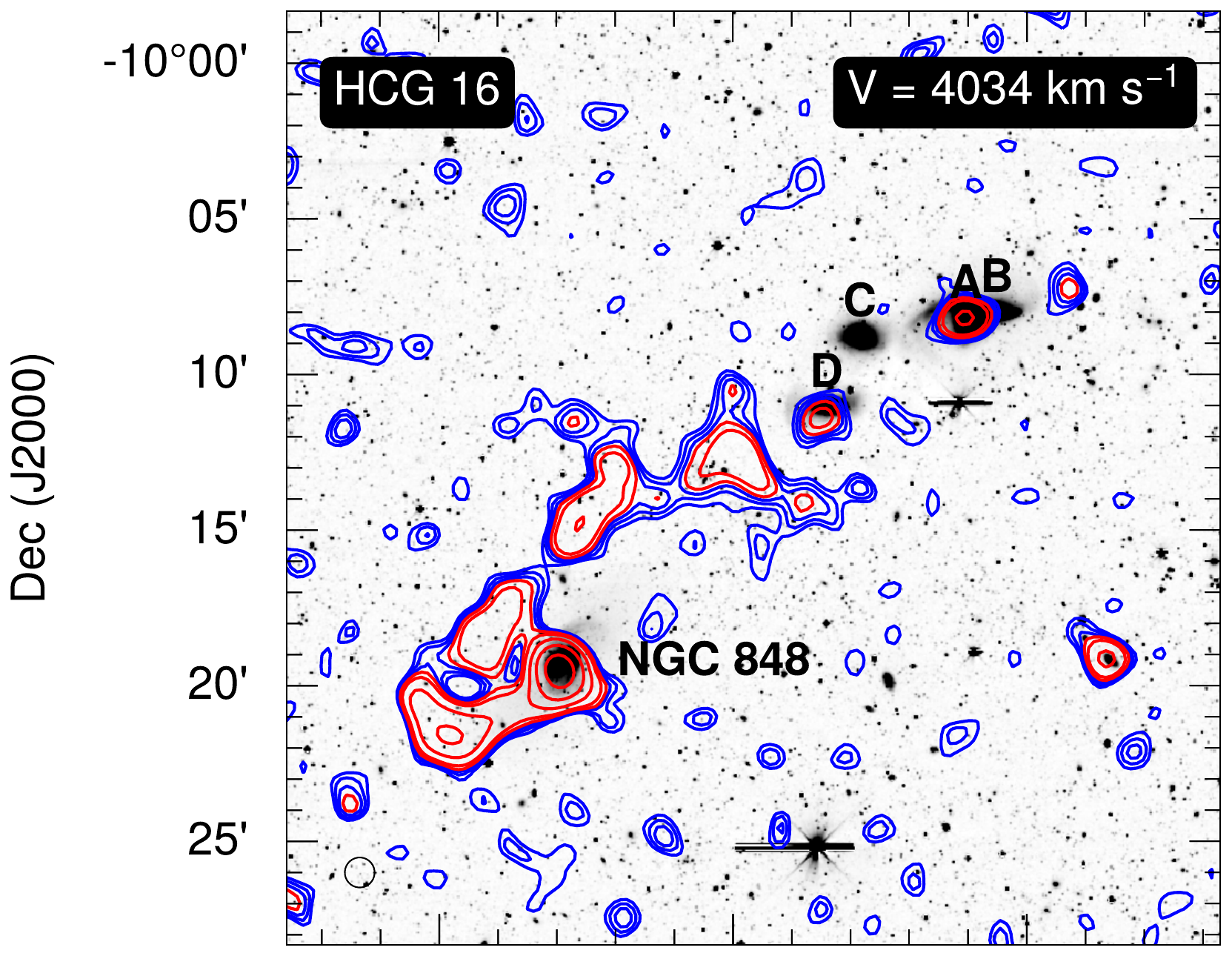} &
     \includegraphics[scale=0.25]{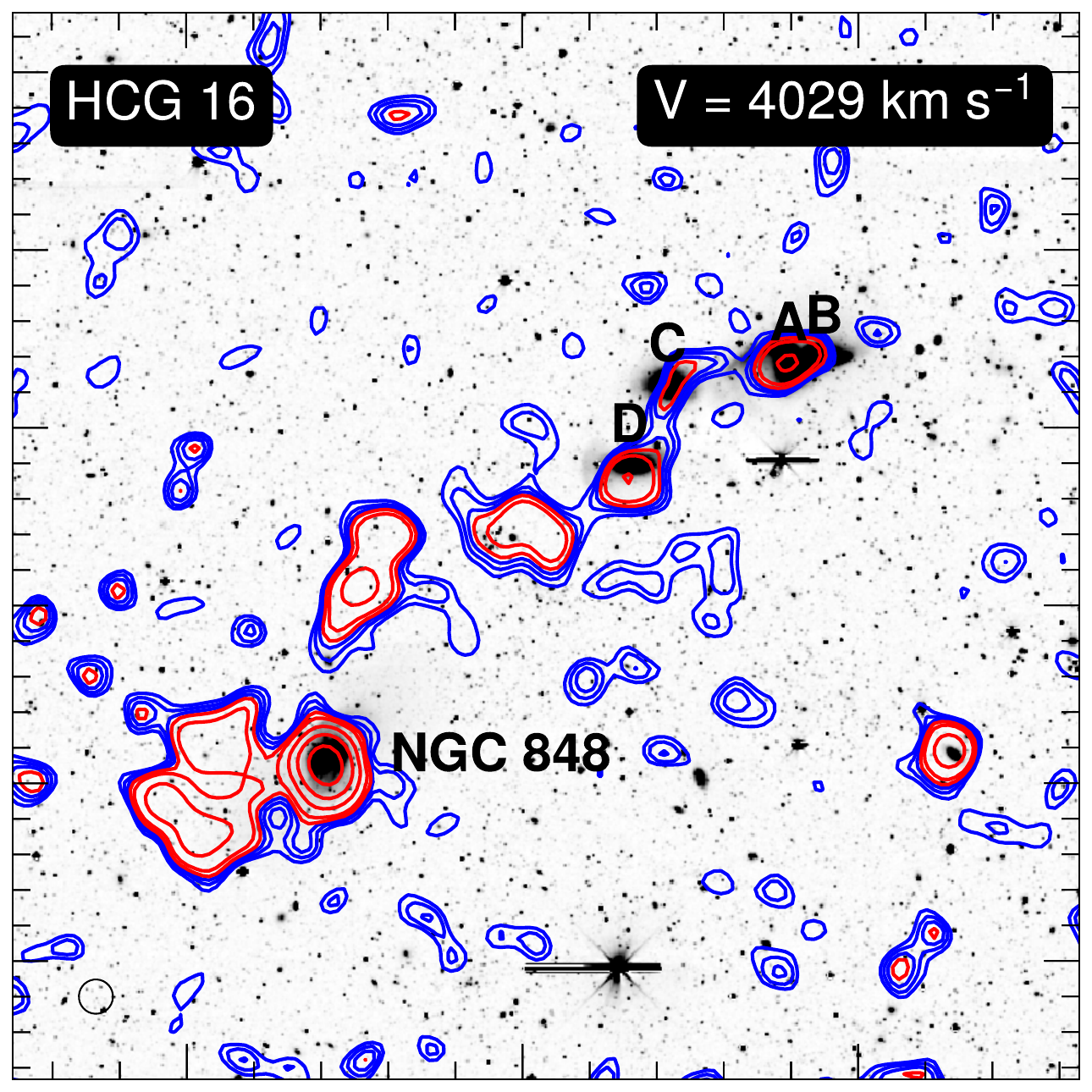} &
     \includegraphics[scale=0.25]{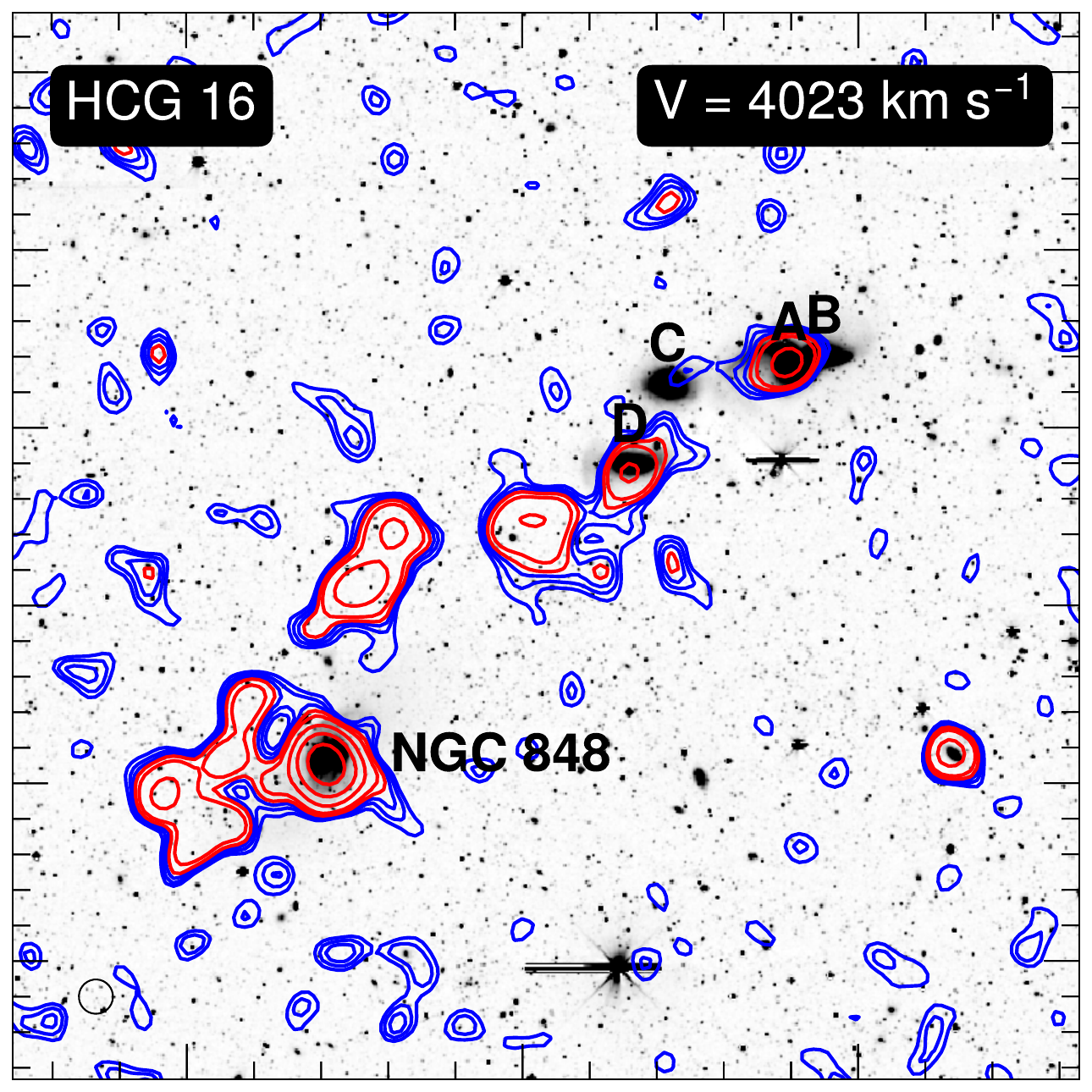} \\[-0.2cm]
     \includegraphics[scale=0.25]{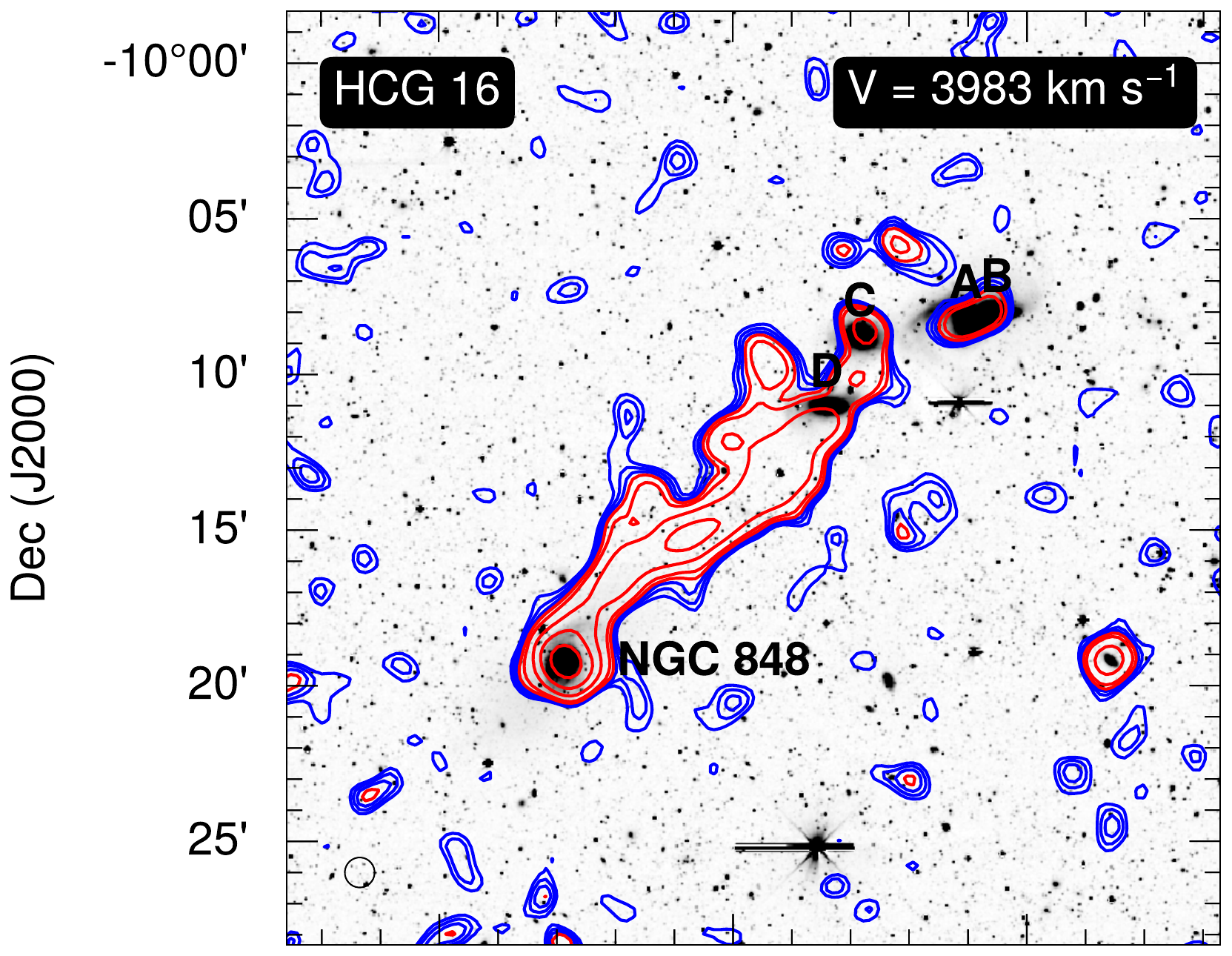} &
     \includegraphics[scale=0.25]{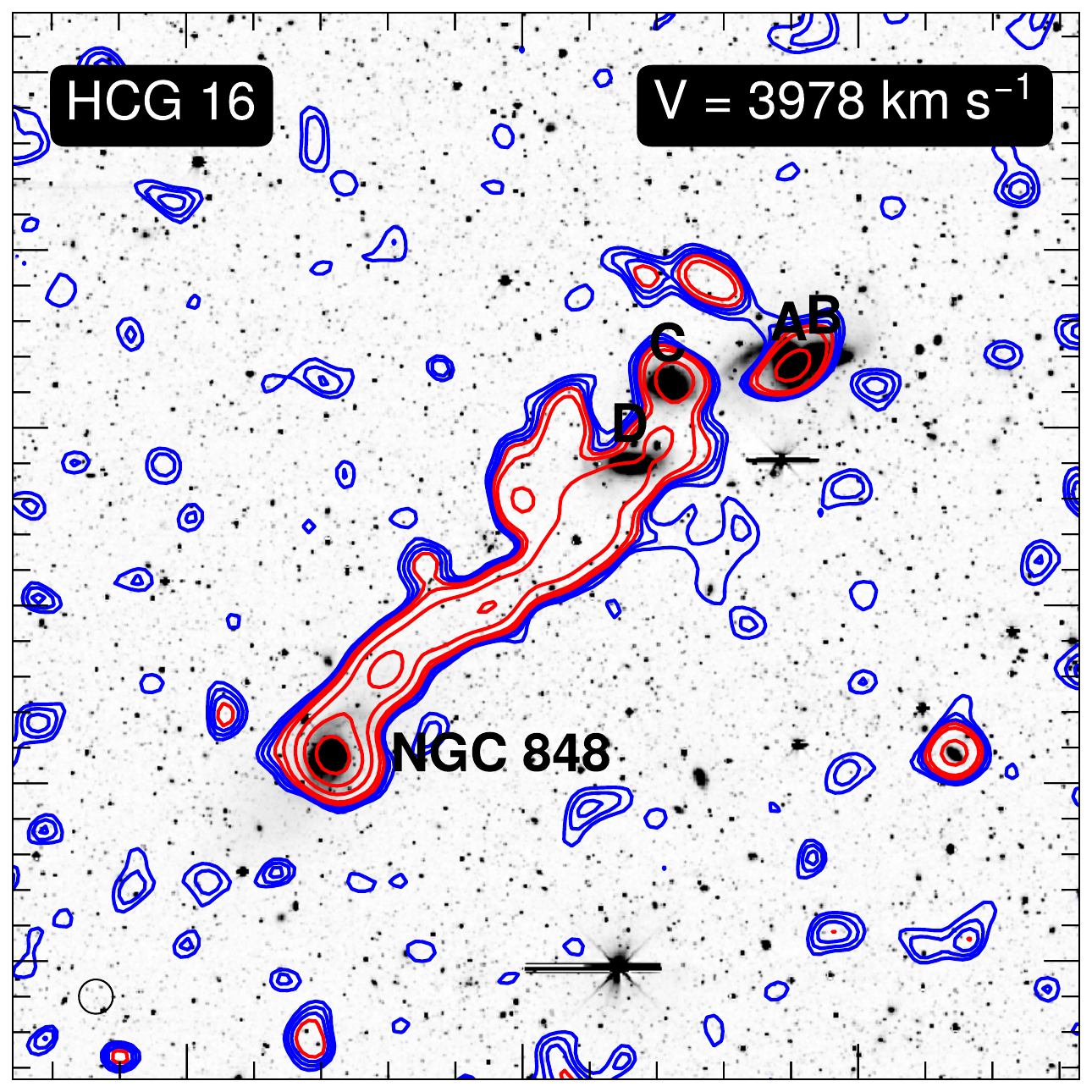} &
     \includegraphics[scale=0.25]{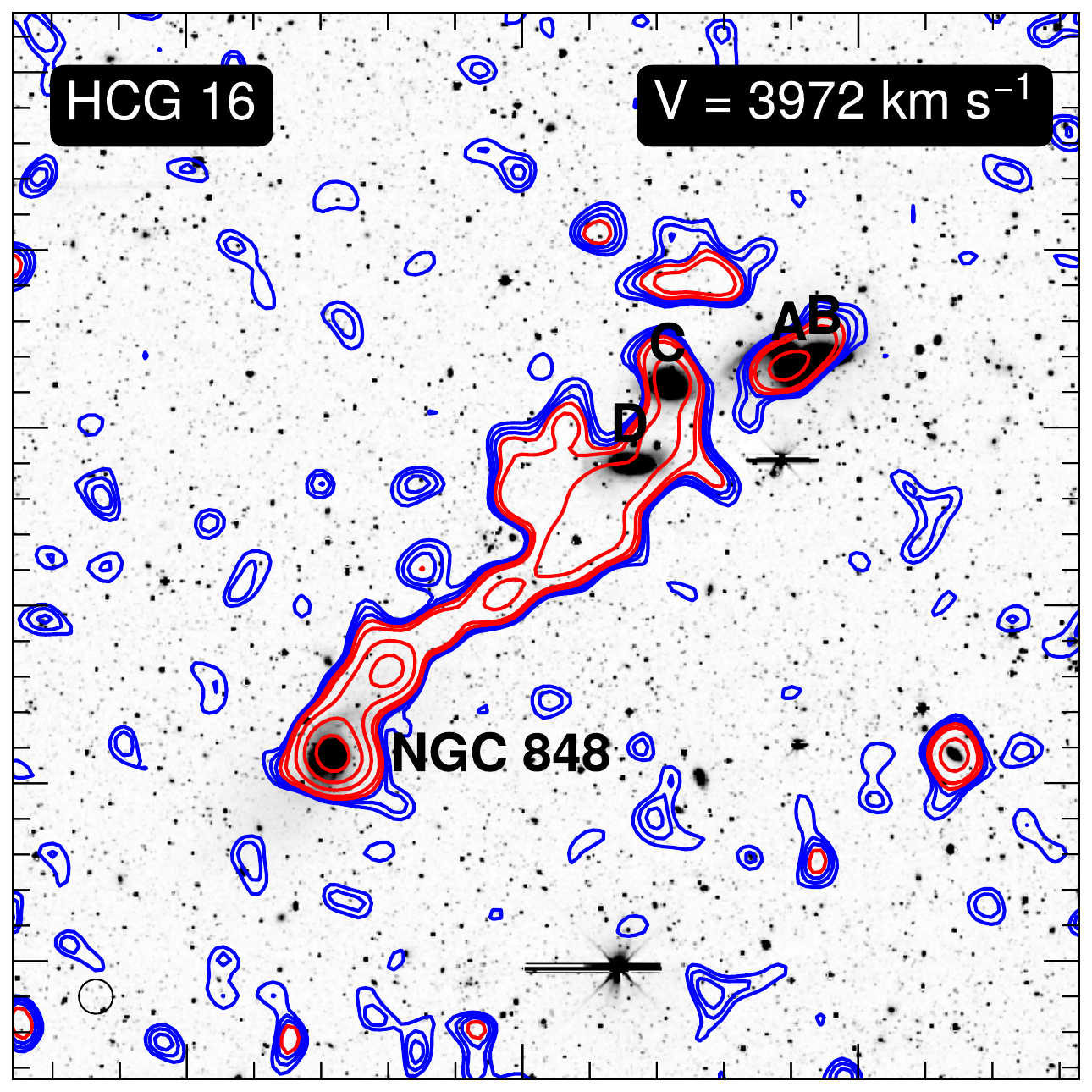} \\[-0.2cm]
     \includegraphics[scale=0.25]{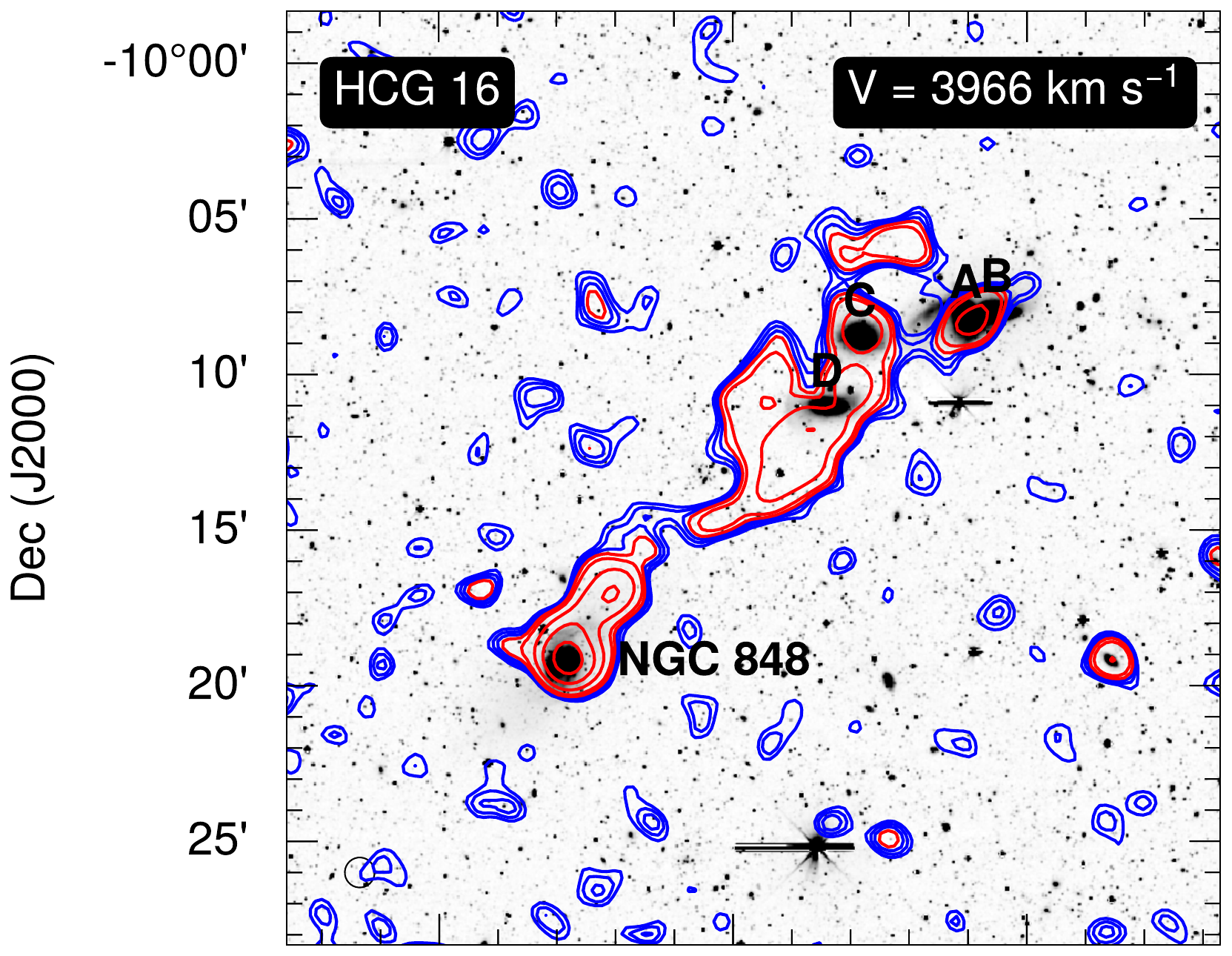} &
     \includegraphics[scale=0.25]{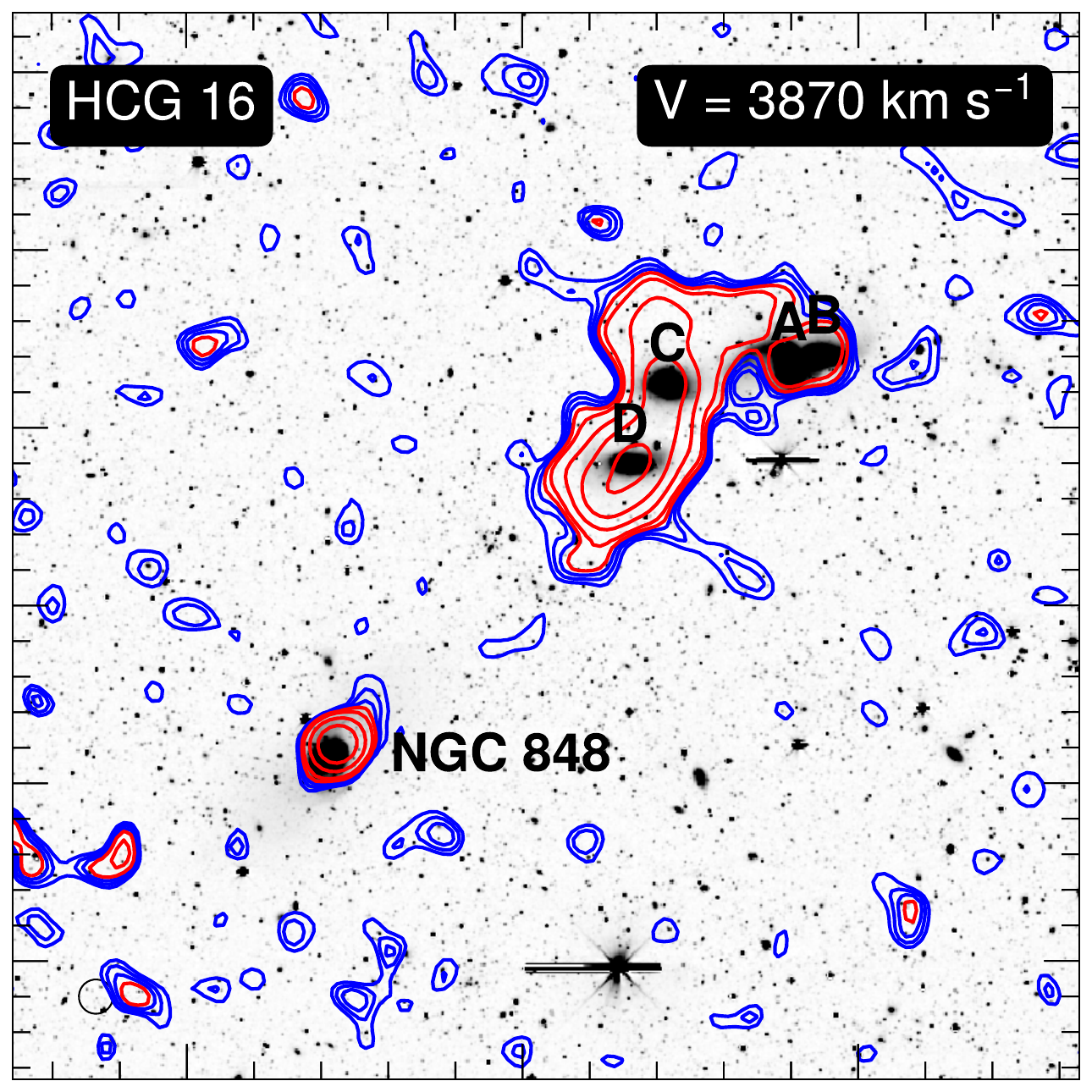} &
     \includegraphics[scale=0.25]{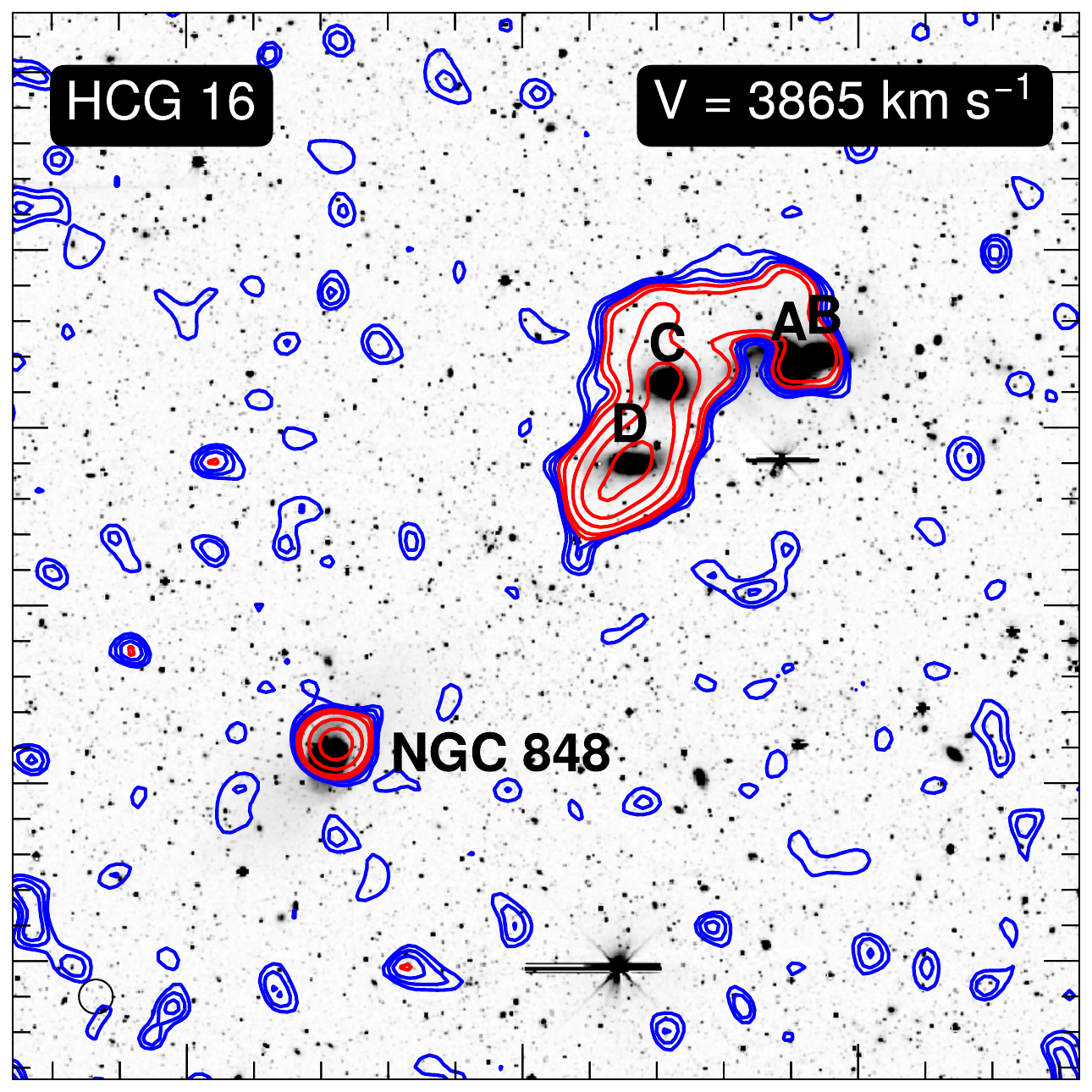} \\[-0.2cm] 
     \includegraphics[scale=0.25]{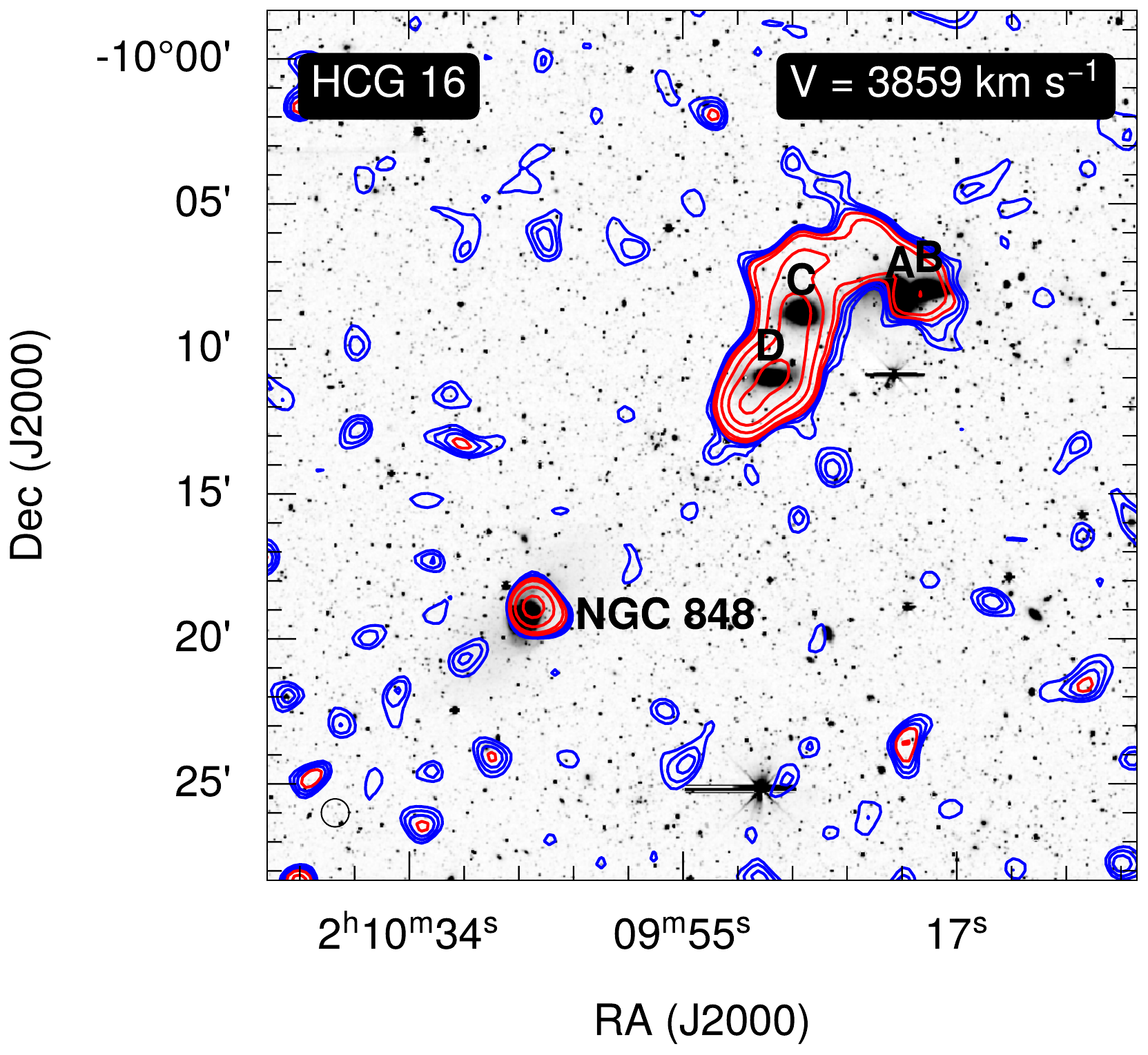}& 
     \includegraphics[scale=0.25]{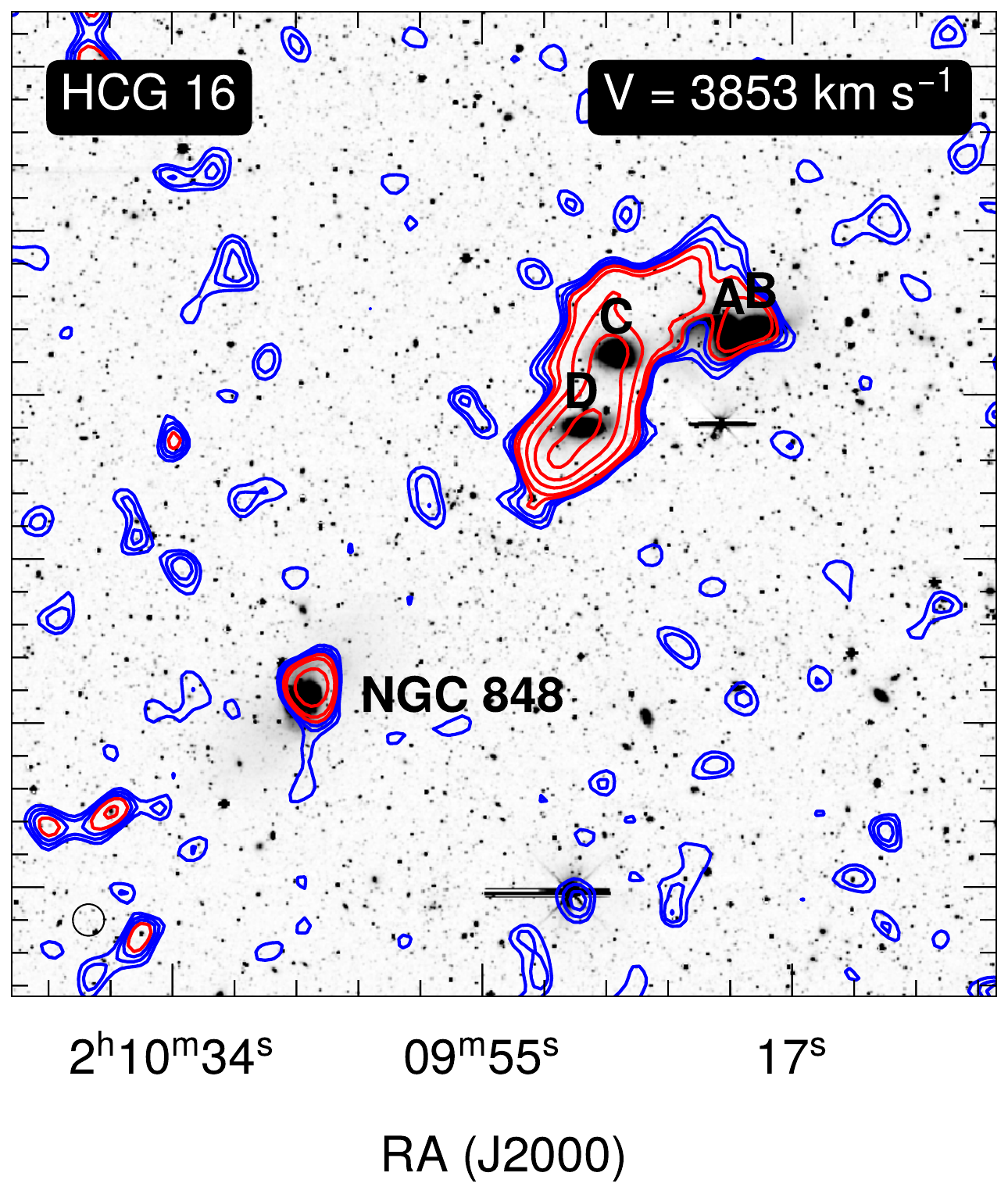}&
     \includegraphics[scale=0.25]{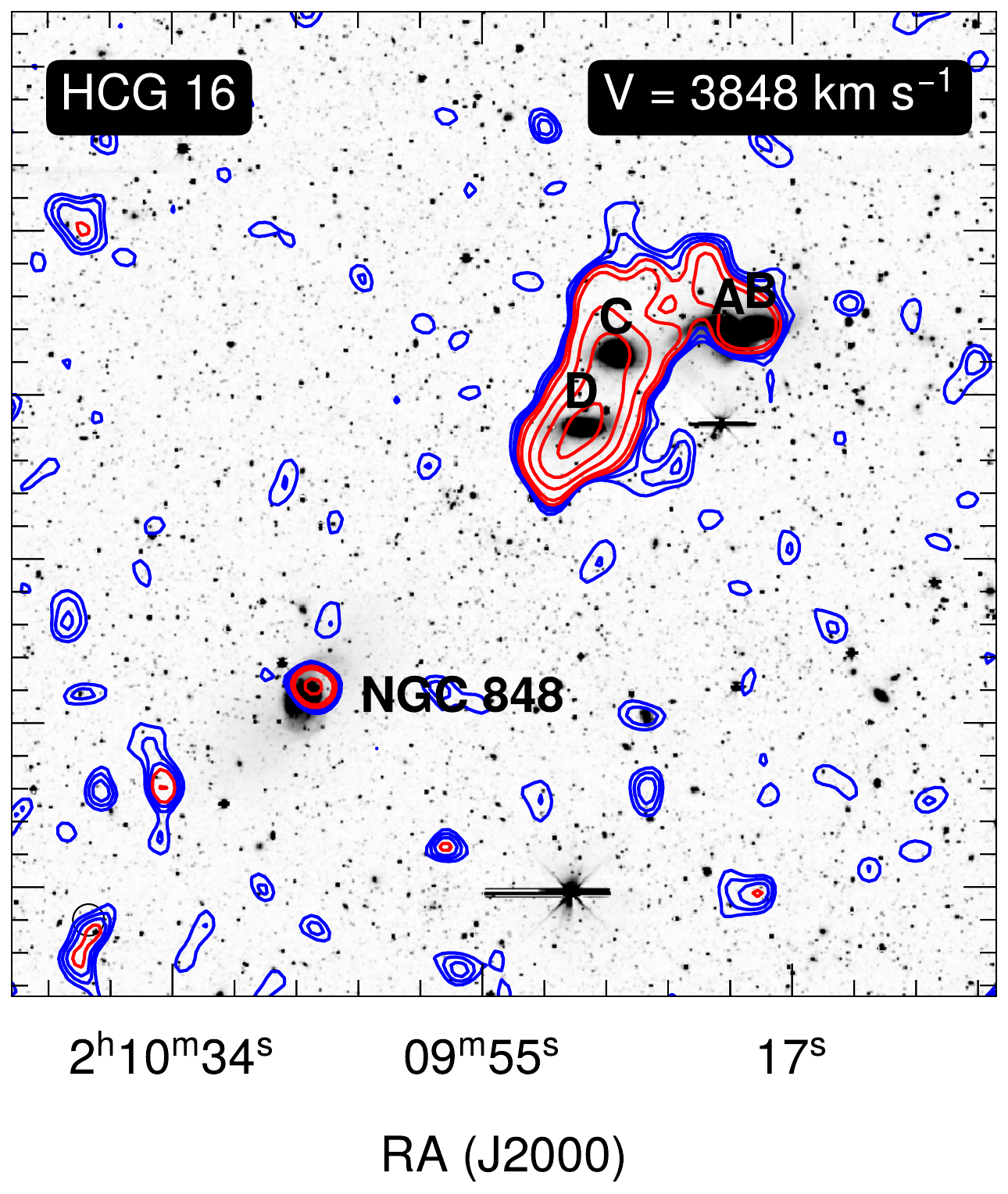}
   \end{tabular}
	 \caption{Example channel maps of the primary beam corrected cube of HCG 16 overlaid on DECaLS DR10 R-band optical images. Contour 
   levels are (1.5, 2, 2.5, 3,  4, 8, 16, 32) times the median noise level in the cube (0.58 $\mathrm{mJy~beam^{-1}}$). 
   The blue colours show contour levels below 3$\sigma$; the red colours represent contour levels at 3$\sigma$, or higher. More channel maps can be found \href{https://zenodo.org/records/14856489}{here}}.
   \label{fig:hcg16_chanmap}
  \end{figure*}

 We show example channel maps of the central part of HCG~16 in Figure~\ref{fig:hcg16_chanmap}. The rest is available \href{https://zenodo.org/records/14856489}{online}.  
 As also mentioned in \citet{2019A&A...632A..78J}, the NW tail, apparent in the moment maps, is not visible in the channel maps. In contrast, 
 the NE tail is clearly visible and appears to be dislodged gas from HCG 16c. 
 As can be seen from the channel map at $\mathrm{\sim 3966~km~s^{-1}}$, the SE tail is made of gas coming from both NGC 848 and HCG 16d, 
 extending across about seven channels. 
 The hook feature becomes more distinct at $\mathrm{\sim 4034~km~s^{-1}}$. The channel maps clearly show the double sided tails of 
 HCG 848, i.e., the hook-like feature and part of the gas that makes up the SE tail.  
 
%

\begin{figure*}
\begin{tabular}{l l}
    \includegraphics[scale=0.27]{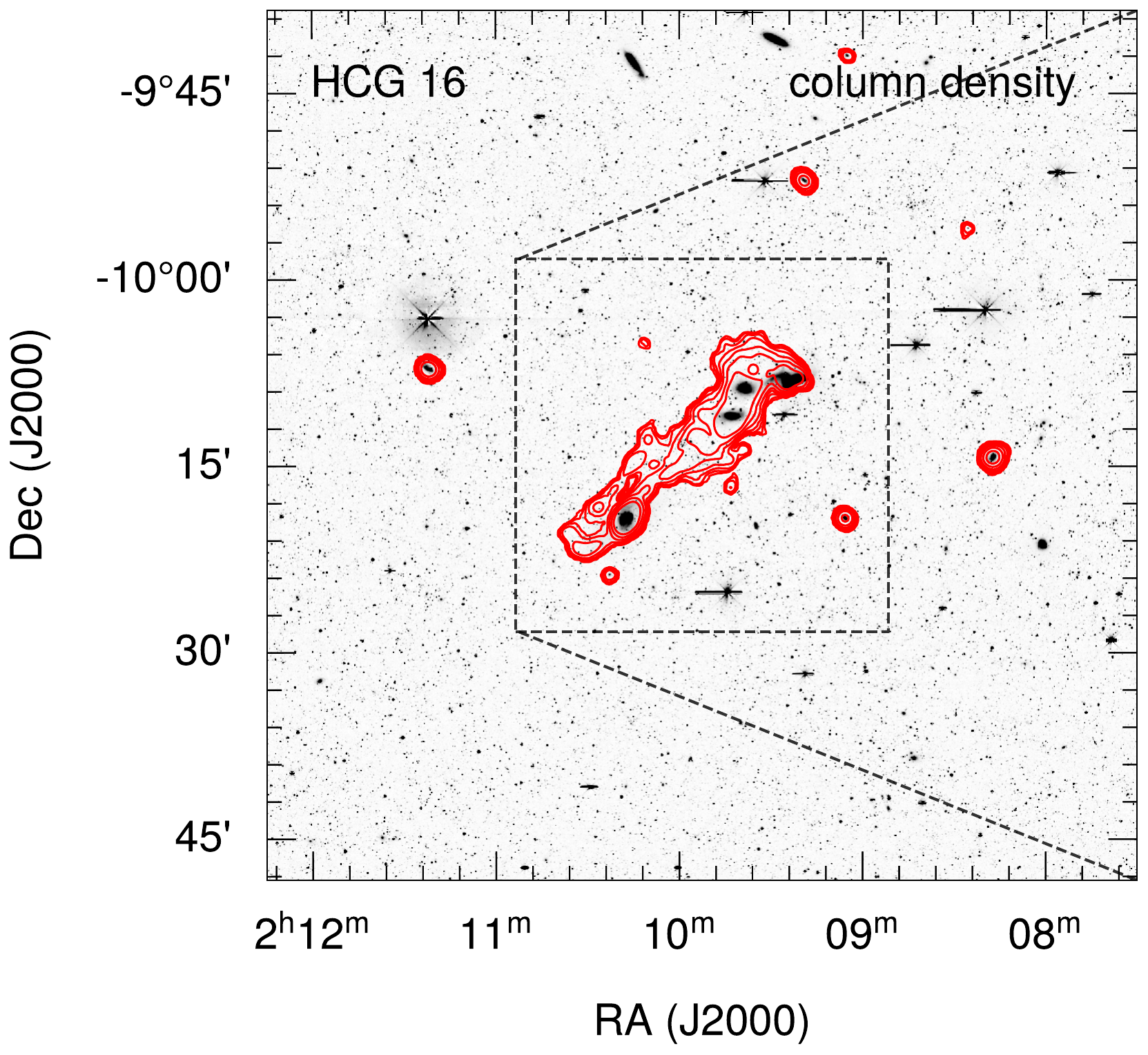}
    & \includegraphics[scale=0.27]{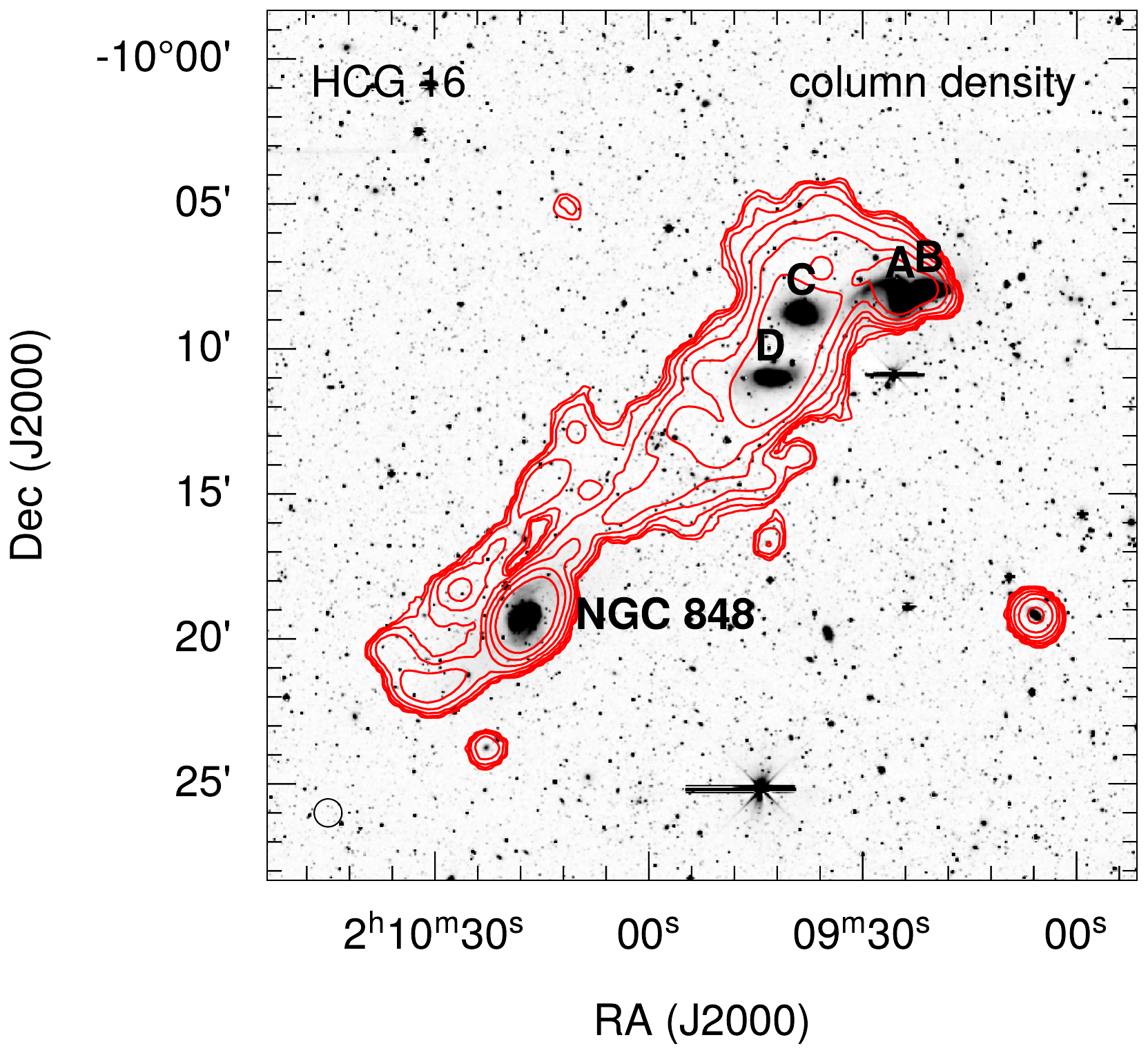}\\
    \includegraphics[scale=0.27]{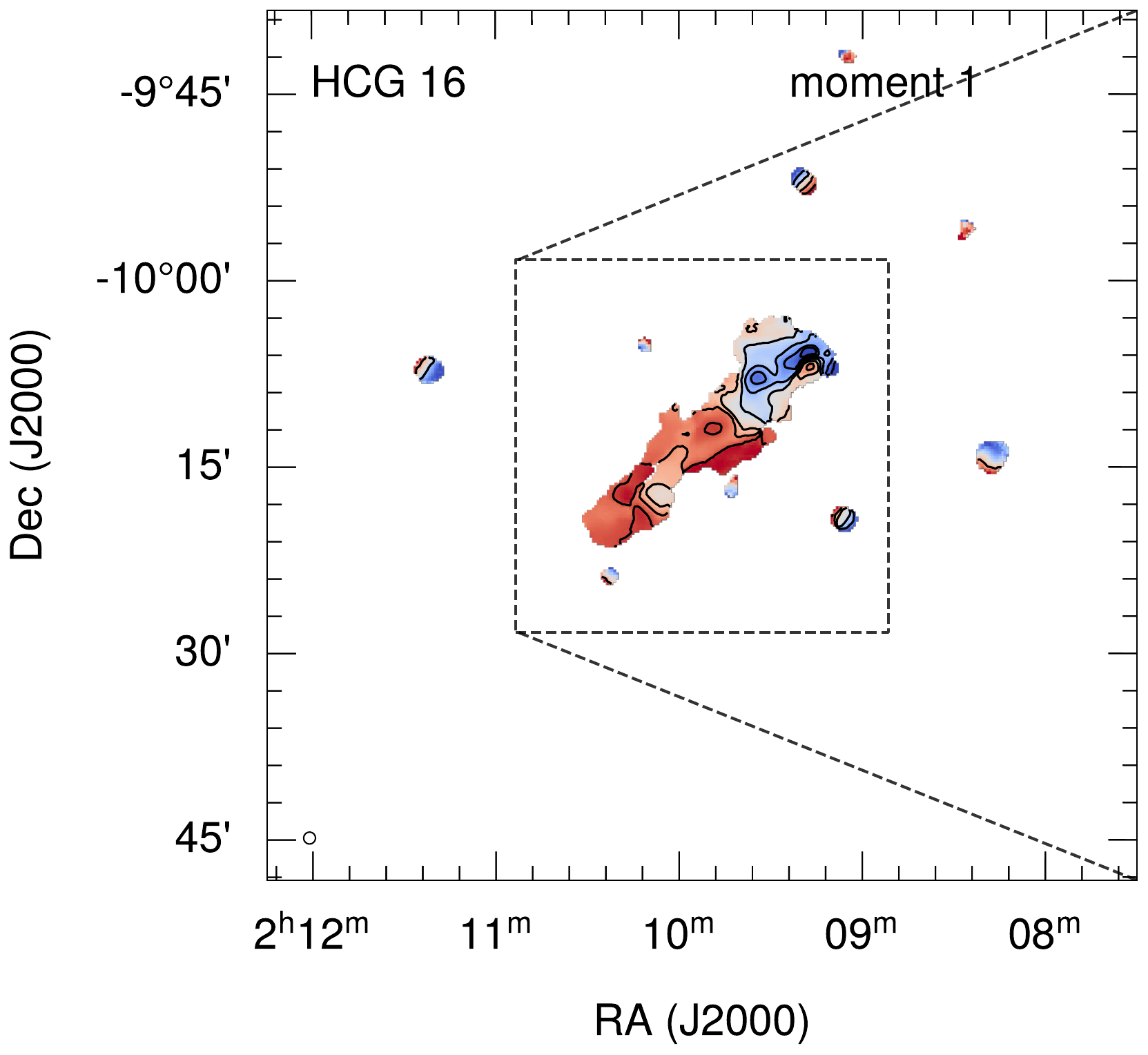} & 
    \includegraphics[scale=0.27]{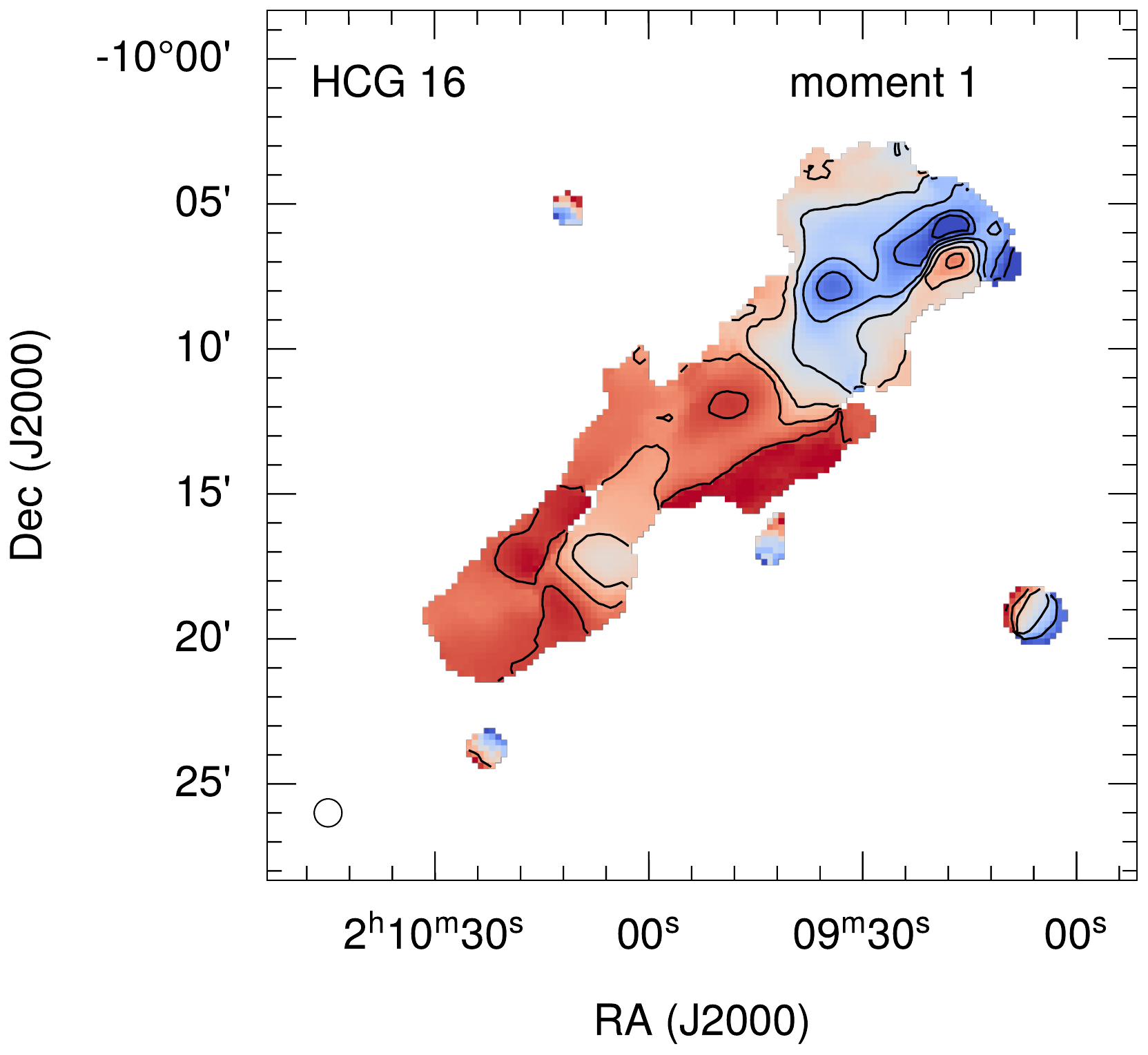}  
\end{tabular}
\caption{\HI\ Moment maps of HCG 16. Left panels show all sources detected by SoFiA. The right panels show sources within the rectangular box 
shown on the left to better show the central part of the group. The top panels show the column density maps with contour levels of 
	($\mathrm{3.1~\times~10^{18}}$, $\mathrm{6.2~\times~10^{18}}$, $\mathrm{1.2~\times~10^{18}}$, $\mathrm{2.5~\times~10^{19}}$, 
    $\mathrm{5.0~\times~10^{19}}$, $\mathrm{9.9~\times~10^{19}}$, $\mathrm{2.0~\times~10^{20}}$) $\mathrm{cm^{-2}}$. 
    The contours are overlaid on DECaLS DR10 R-band optical images. The bottom panels show the moment one map. Each individual 
    source has its own colour scaling and contour levels to highlight any rotational component.}
\label{fig:hcg16_mom}
\end{figure*}

\begin{figure*}
    \setlength{\tabcolsep}{-2pt}
\begin{tabular}{c c}
    \includegraphics[scale=0.37]{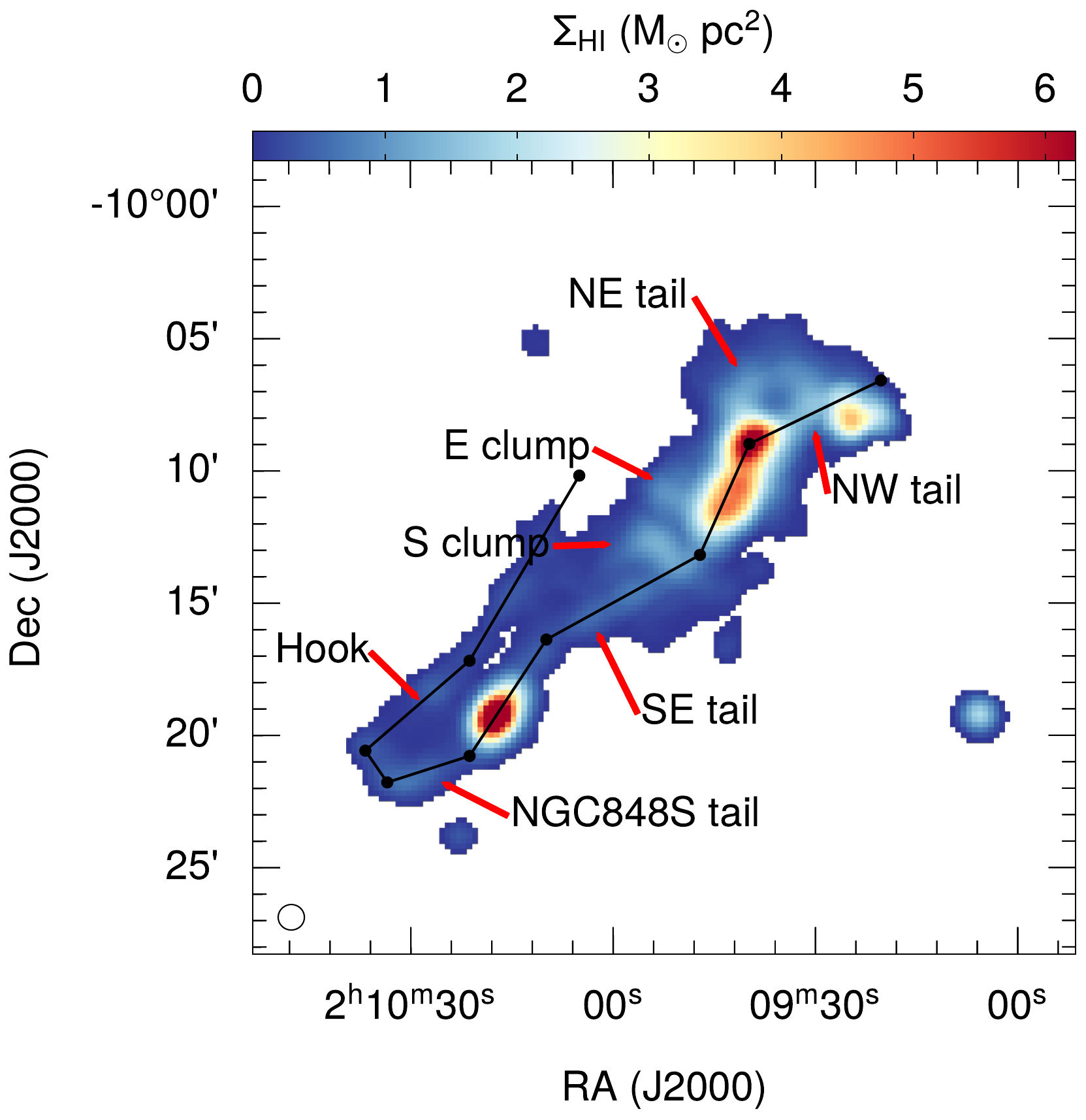} & 
    \settoheight{\imageheight}{\includegraphics[scale=0.37]{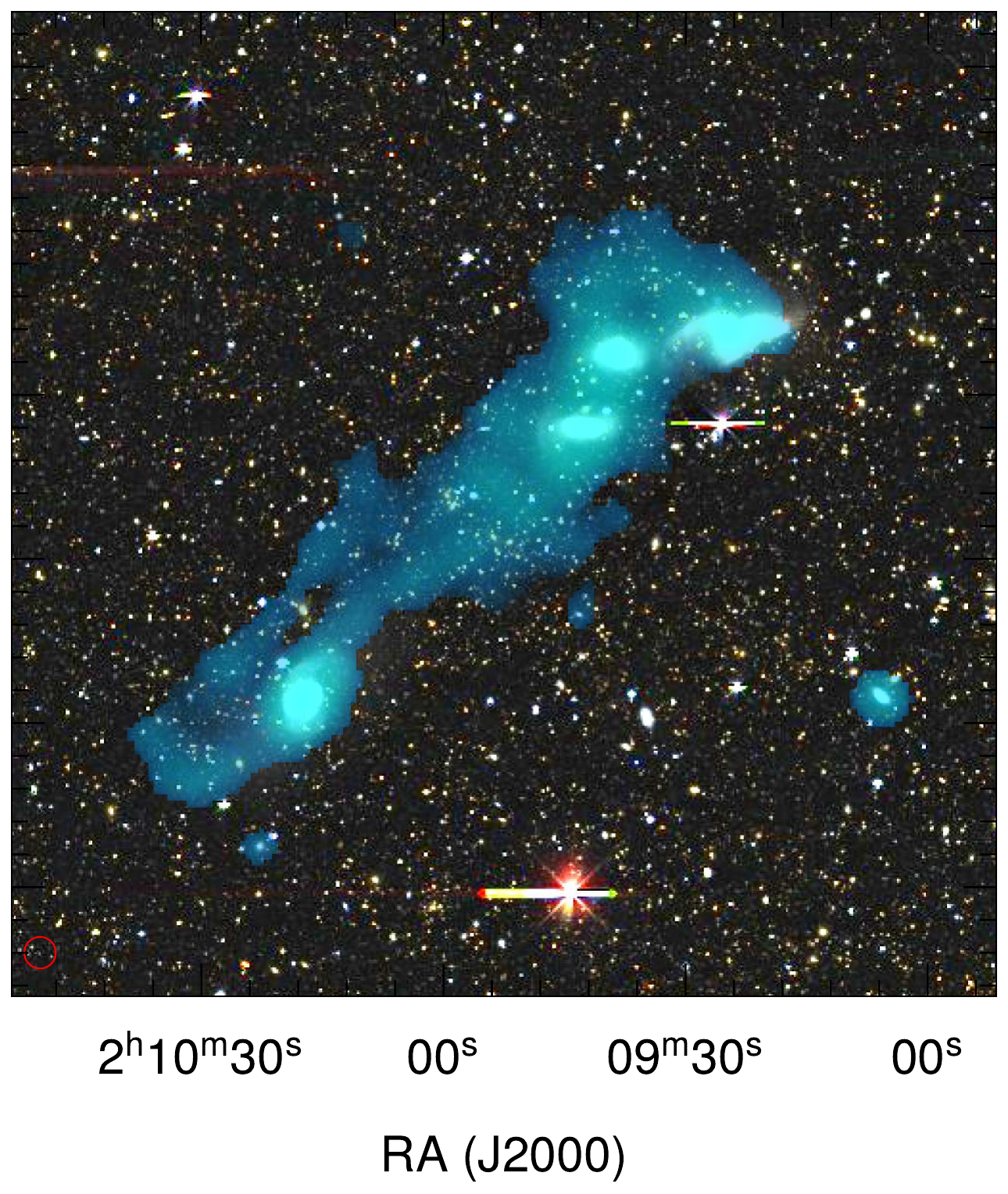}}
    \setlength{\imageheight}{0.974\imageheight}

    \includegraphics[height=\imageheight]{hcg16_optical_mom0.pdf}

  \end{tabular}
  \caption{Left panel: MeerKAT \HI\ surface density map of HCG 16 showing the paths (black lines) from which the segmented position-velocity 
  diagrams shown in Figure~\ref{fig:hcg16_pvd} are taken. 
  The black circles show the positions of the nodes that make up the different slices. Right panel: The same moment zero map overplotted on DECaLS optical images to highlight the core members.}
  \label{fig:hcg16_pvd_path}
 \end{figure*}

\begin{figure*}
\begin{tabular}{l l}
    \includegraphics[width=0.48\textwidth]{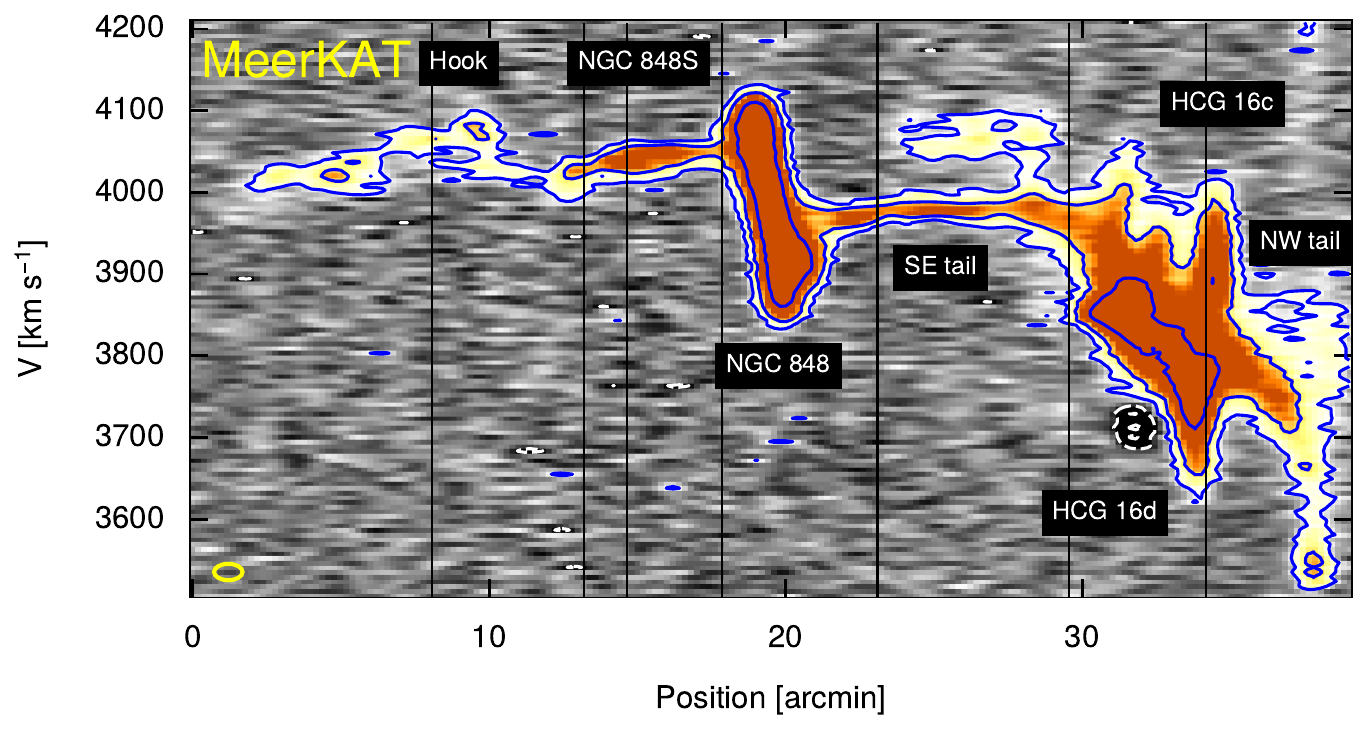} &
    \includegraphics[width=0.48\textwidth]{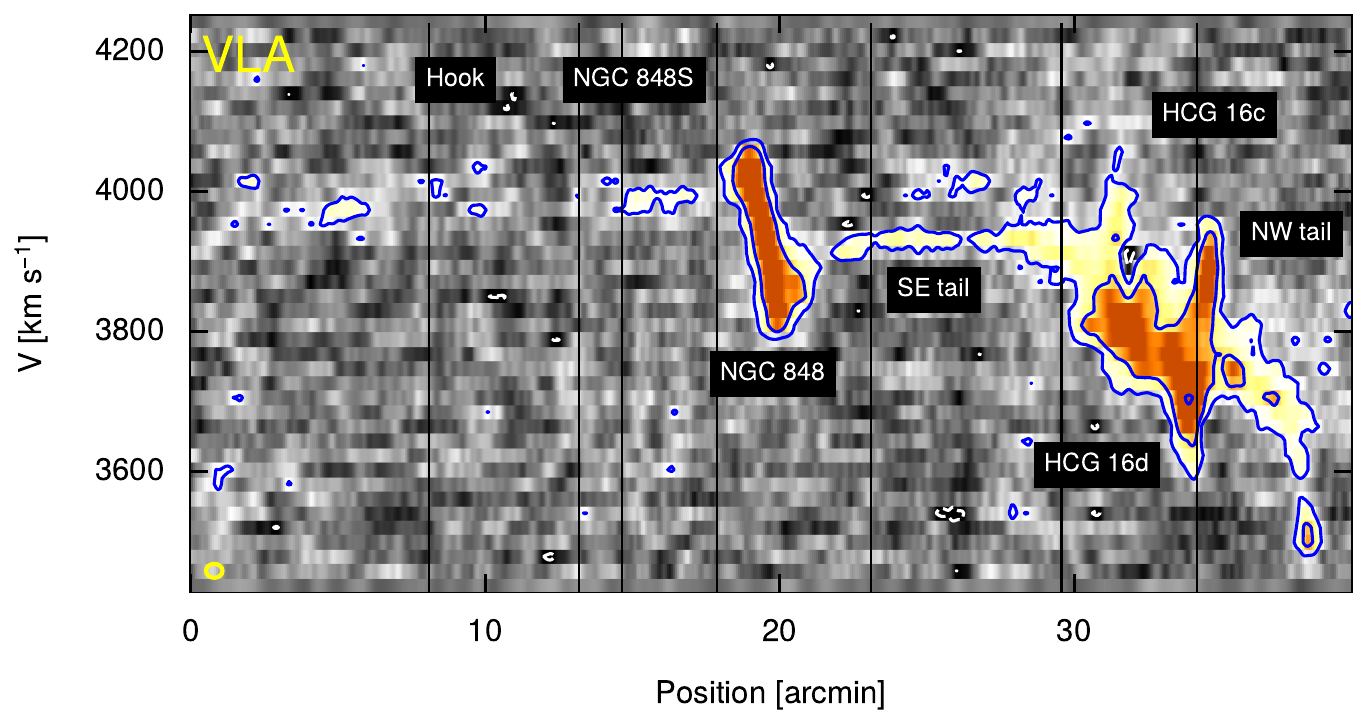} 
  \end{tabular}
  \caption{Segmented position-velocity diagrams of HCG 16 taken from the paths shown in Figure~\ref{fig:hcg16_pvd_path}. 
  Left panel: MeerKAT data from this paper, right panel: VLA data from \citet{2019A&A...632A..78J}. Blue contours show emission at 3~$\times$ the rms noise. 
  Dashed lines show negative contours. The vertical black lines indicate the positions of the nodes that make up the slices from which the position velocity 
  diagrams where taken. The yellow ellipses at the bottom left corner of each 
  panels show the half-power beam width $\mathrm{\sqrt{BMAJ * BMIN}}$ and 20~$\mathrm{km~s^{-1}}$ velocity width. }
  \label{fig:hcg16_pvd}
 \end{figure*}

In  Figure~\ref{fig:hcg16_mom}, we show maps of the \HI\ sources detected by SoFiA within the MeerKAT field of view. 
The left panels show all sources detected by SoFiA. The right panels highlight the central part of the group. In total, 
SoFiA detected nine surrounding members; their properties are discussed in an accompanying paper. In the top panels of the figure, we show the column density maps of HCG~16 
overplotted on top of DeCaLS R-band \citep{2019AJ....157..168D} optical images. The moment one map of the group is shown in the bottom panel of Figure~\ref{fig:hcg16_mom}, which is 
plotted in a way that each individual members are scaled by different colour-scaling factor to highlight any rotational 
patterns. The surrounding members show differential rotation with an apparently undisturbed \HI\ morphology. 
However, the integrated iso-velocity contours of the centre of the group as a whole are disturbed. Note though that the individual members may still have their rotation. 
The most remarkable finding here is the presence of numerous tidal features and clumps, as well as a long, continuous tidal tail connecting the core member of the group with NGC 848. 
In addition, a hook-like structure extends south-east of NGC 848 before curving to the north west in direction parallel to the main group. 
As shown in Figure~\ref{fig:hcg16_pvd_path}, all core members are linked by tails or high column density bridges. 
There is a continuity in velocity between NGC 848S and the curved tail. The hook appears to be a stretched gas from NGC 848 due to its encounter with the main group. 
The presence of numerous tails and clumps are typical of compact groups at the intermediate stage, indicating substantial gas loss into the IGM. 
To better visualise the tails, we show segmented position-velocity diagrams, taken from the paths shown in Figure~\ref{fig:hcg16_pvd_path}, 
in Figure~\ref{fig:hcg16_pvd} as previously done by \citet{2019A&A...632A..78J} using VLA data. 
Most of the tails were previously identified 
by the VLA map of \citet{2019A&A...632A..78J} and we adopt the author's nomenclature to name them. However, the tails are more pronounced in the MeerKAT map. 
Note also that the curved tail (hook) was not visible in the VLA map.   

\begin{figure*}
    \setlength{\tabcolsep}{0pt}
    \begin{tabular}{c c}
    \includegraphics[scale=0.32]{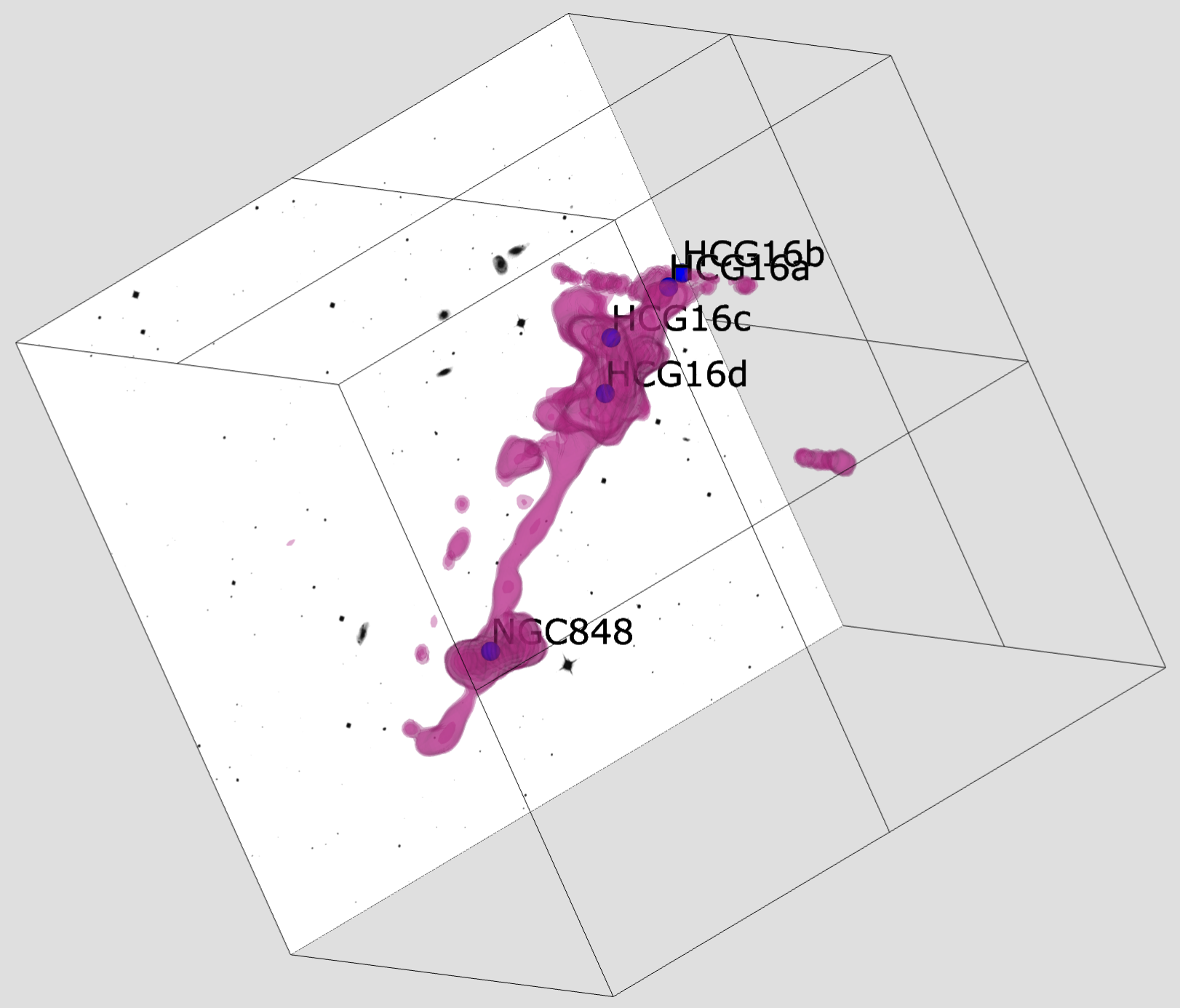} & 
    \includegraphics[scale=0.32]{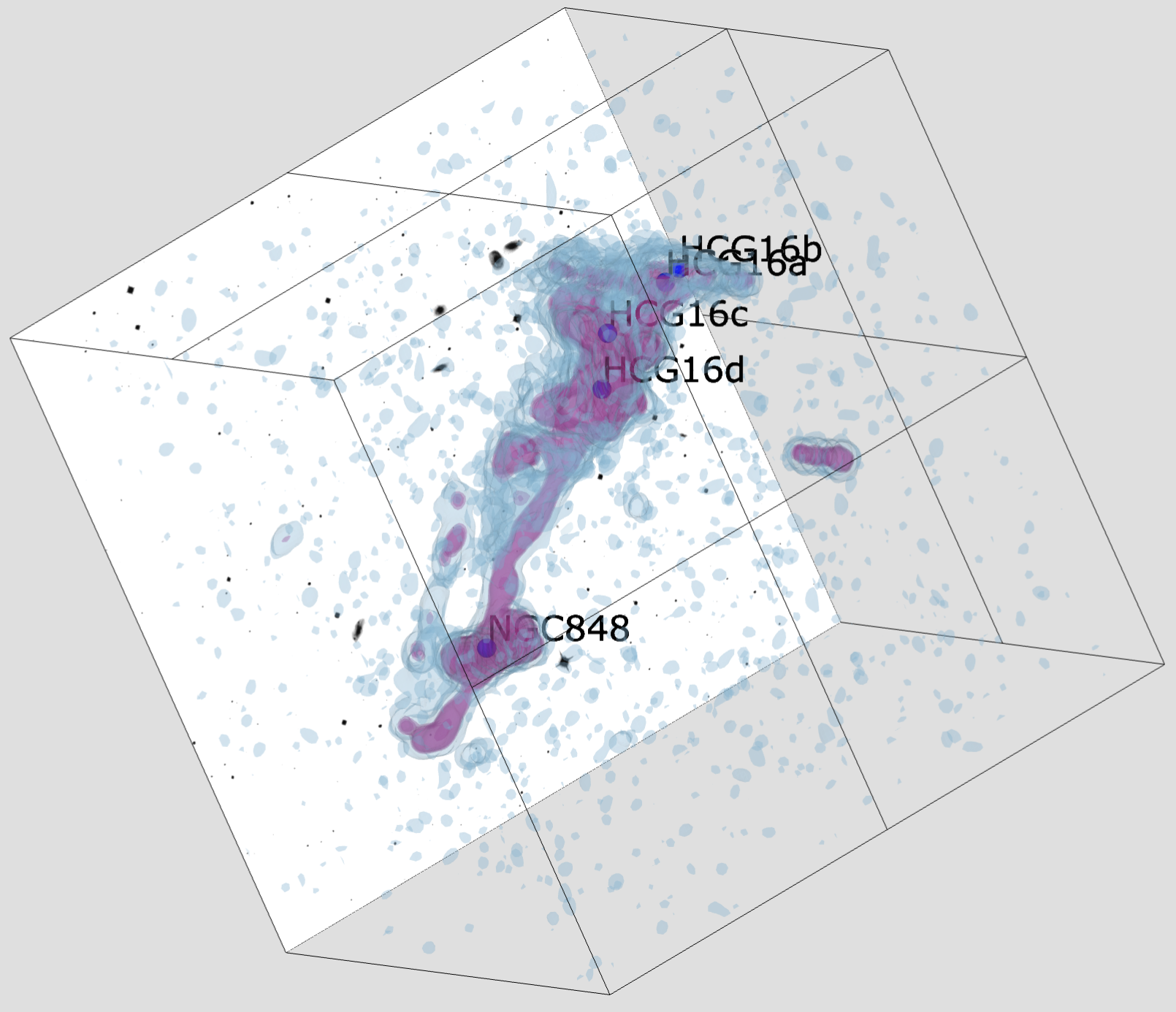}
    \end{tabular}
    \caption{3D visualisation of HCG 16. The left panel shows iso-surface level highlighting the high-column density gas. The right panel 
    showcases the low-column density \HI\ gas. The 2D grayscale image is a DeCaLS R-band optical image of the group. The blue circles indicate the position of the member galaxies. The 2D grayscale image is a DeCaLS R-band optical image of the group. 
    The online version of the cubes are available at \href{https://amiga.iaa.csic.es/x3d-menu/}{https://amiga.iaa.csic.es/x3d-menu/}.}
  \label{fig:hcg16_3dvis}
 \end{figure*}

 The 3D visualisation of the group is shown in Figure~\ref{fig:hcg16_3dvis}. The left panel highlights high column density HI, whereas the right 
 panel shows superposition of iso-surface corresponding to both low and high column density gas. This kind of visualisation is very crucial 
 to identify features not associated with the disk of the member galaxies. The connection between all member galaxies is evident. A high surface brightness bridge connects the core of the group and NGC 848. 
 The hook is composed of lower column density gas, stretching from NGC 848 before bending upward.
\subsection{HCG 31}\label{sec:hcg31}  

\begin{figure*}
\setlength{\tabcolsep}{0pt}
\begin{tabular}{c c c}
    \includegraphics[scale=0.215]{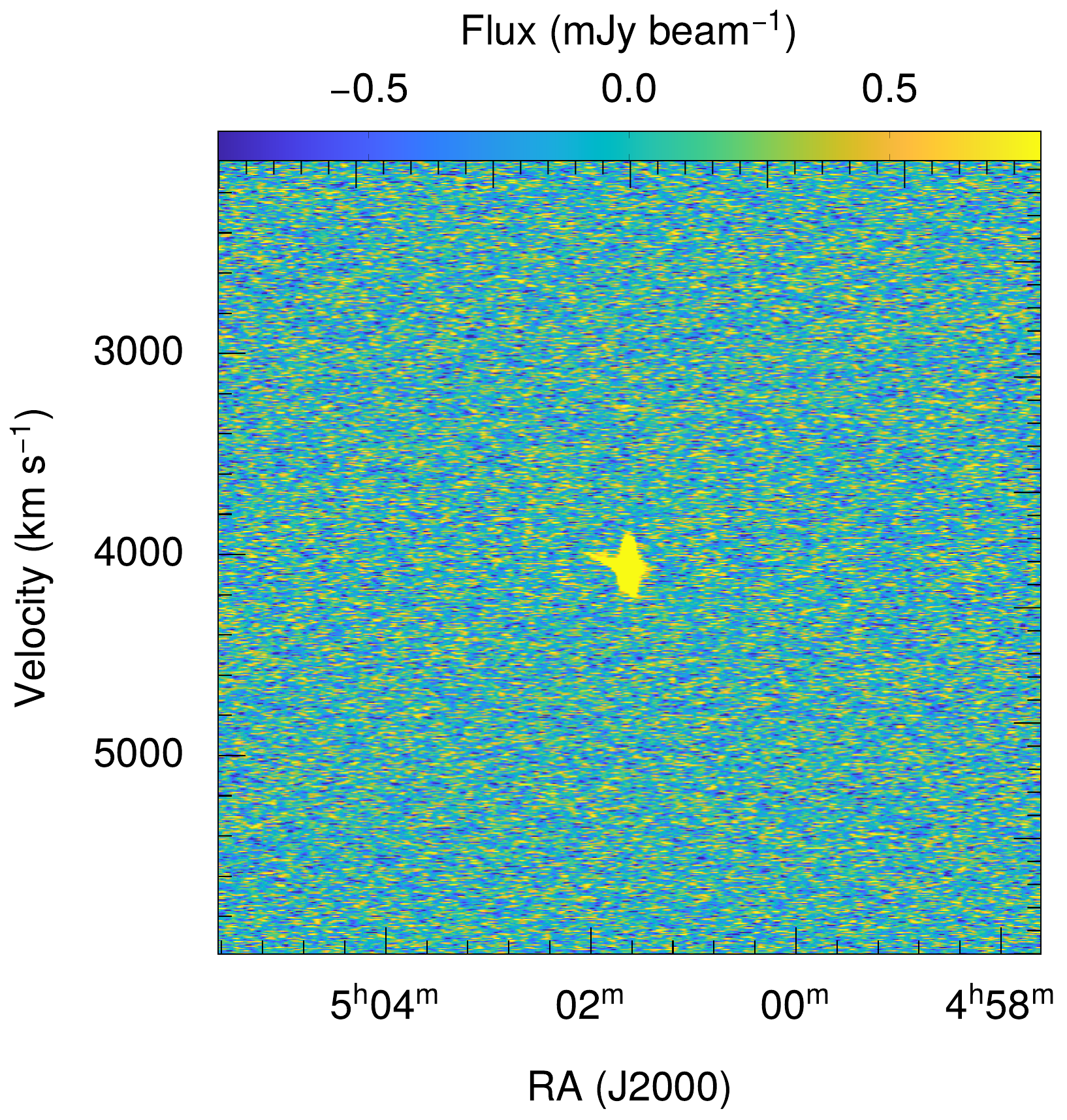} &
    \includegraphics[scale=0.215]{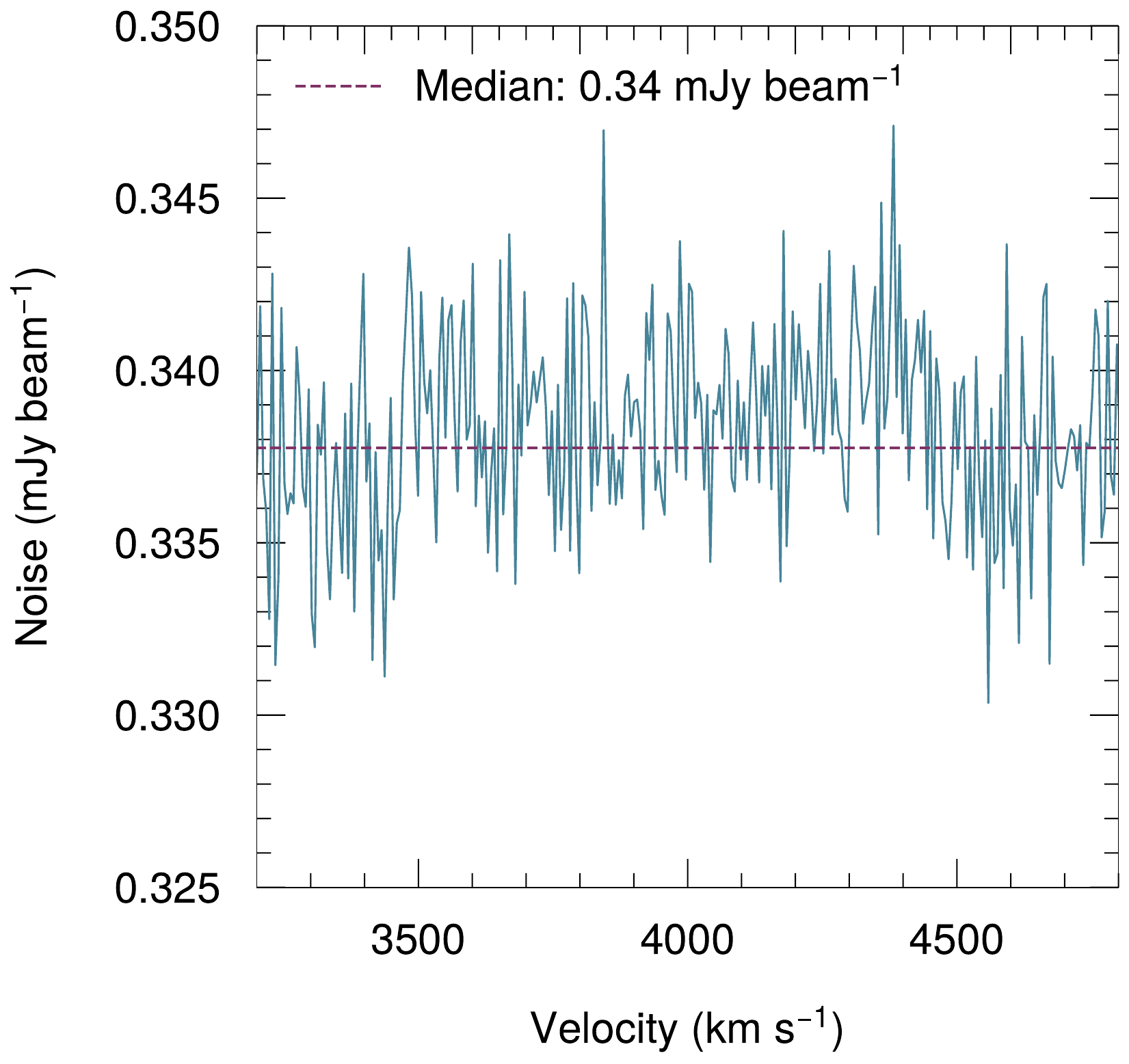} & 
    \includegraphics[scale=0.215]{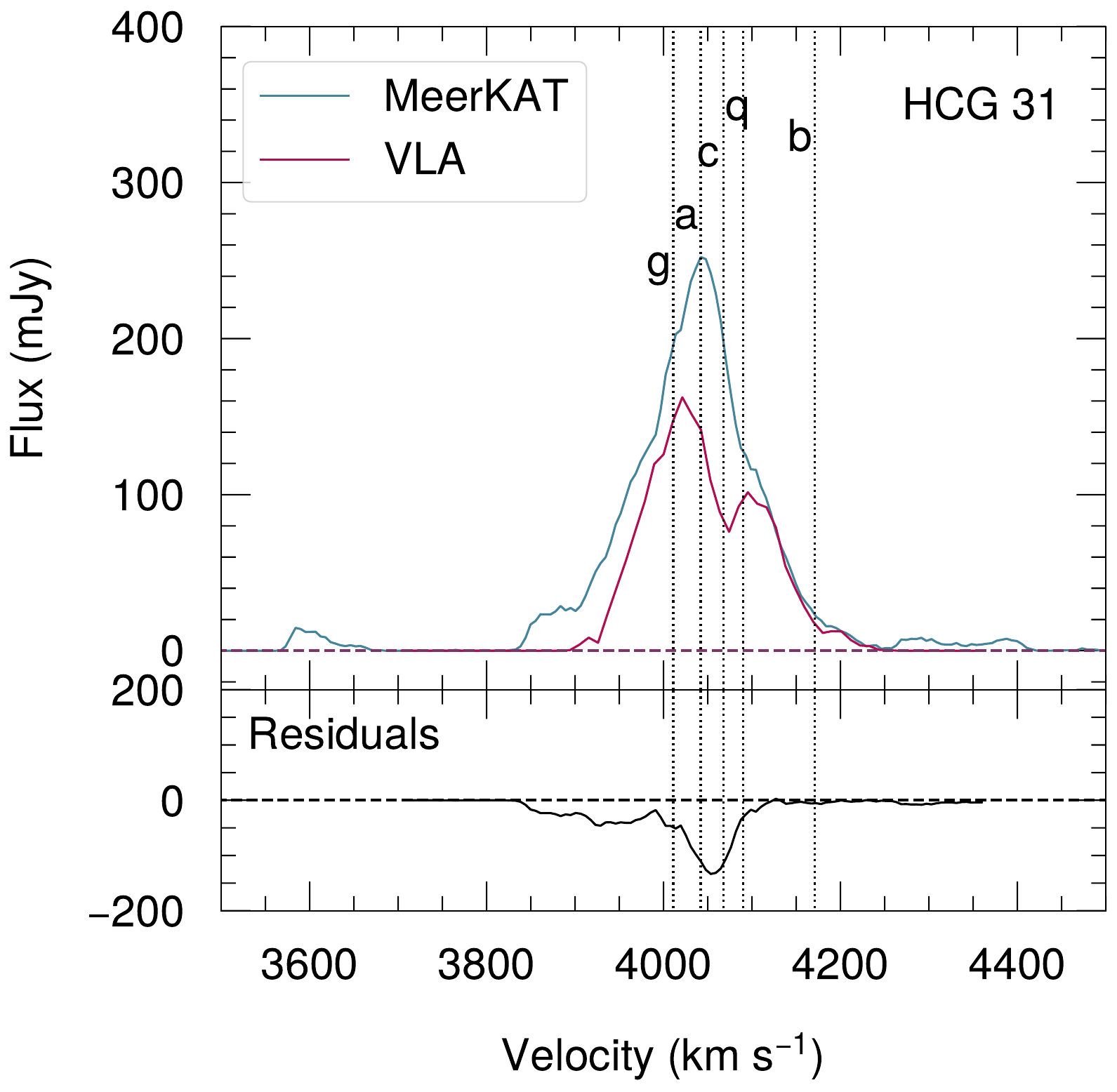}
  \end{tabular}
  \caption{Left panel: velocity vs right ascension of HCG~31. Middle panel: median noise values of each RA-DEC slice of the non-primary beam corrected 60\arcsec\ data cube of 
  HCG~31 as a function of velocity. The horizontal dashed line indicates the median of all the noise values from each slice. Right panel: the blue solid lines indicates the 
  MeerKAT integrated spectrum of HCG~31; the red solid line indicates VLA integrated spectrum of the group derived by \citep{2023A&A...670A..21J}. 
  The vertical dotted lines indicate the velocities of the galaxies in the core of the group. The spectra have been extracted from area containing only genuine \HI\ emission.}
  \label{fig:hcg31_noise}
 \end{figure*}

The RA-velocity slice shown in the left panel of Figure~\ref{fig:hcg31_noise} shows no apparent RFI or continuum residuals. In addition, the median noise values 
as a function of velocity, in the 
middle panel of the figure, do not change much. We compare the global spectrum of HCG~31 from the VLA with that from MeerkAT in 
the right panel of Figure~\ref{fig:hcg31_noise}. It's clear that MeerKAT recovers more \HI\ emission than VLA at virtually all the velocity range of the group. 
The largest difference in terms of recovered flux corresponds to the A+C complex. Therefore, most of the new tidal features we detected is expected to be coming from the A+C complex. 
 This is not surprising since HCG~31A and HCG~31C are known to be in the process of merging and their interactions are expected to produce many tidal features into the IGM. Note though 
 that separating features in HCG~31 is extremely challenging \citep{2005A&A...430..443V, 2023A&A...670A..21J} and may require complex kinematic analysis that we will investigate in a future paper.  
 
Example channel maps of HCG 31 are shown in Figure~\ref{fig:hcg31_chanmap}, and the rest is available \href{https://zenodo.org/records/14856489}{online}. Visual inspections indicate that the south-eastern and the north-western 
elongated \HI\ is most likely associated with the A+C complex. Part of the northern extension could be coming from HCG~31Q, though. The elongation of the south-eastern 
tail is likely a result of gas being stripped away during the interaction between HCG~31A and HCG~31C, which then subsequently extended by a fly-by encounter between 
member G and the A+C complex. There are no known optical counterparts associated with the extended \HI\ emission. Thus, they are not the results of interactions 
with members outside the central galaxies. The highest \HI\ peak is associated with HCG~31C, which also corresponds to sites of active star forming regions.   
 
We show the moment maps of HCG~31 in Figure~\ref{fig:hcg31_mom}. The central part of HCG~31 is detected as one source although 
it is composed of five late-type galaxies and four tidal dwarf candidates. The surrounding members seem to have differential rotation and present no obvious signs of interactions. 
However, the integrated iso-velocity contours of the central part of the group as a whole is clearly disturbed, although the \HI\ disk of the individual 
core members may still have rotation. The \HI\ moment map is elongated towards the south-east and the north-west, which was not visible in previous VLA maps. The DeCaLS image shows no optical counterparts within the elongated region.
We show the column density map of the central part of HCG~31 from a 15.47\arcsec $\times$ 11.86\arcsec datacube in 
Figure~\ref{fig:hcg31_optical_mom0}. We also show the previously 
identified tidal fragments and tidal dwarf candidates mentioned in \citet{2006AJ....132..570M}. 
The A+C complex, HCG~31B, and the tidal fragments/dwarf candidates are embedded in high column density features, with the highest contour corresponding to the overlap area between A and C and also to F1. All the tidal 
fragments and the tidal dwarf candidates are embedded in a tail connecting HCG~31~G and the other core members. 

 We show the 3D visualisation of HCG~31 in Figure~\ref{fig:hcg31_3dvis}. The left panel emphasises regions of high \HI\ column density, 
 while the right panel displays a superposition of iso-surfaces representing both low and high column density gas. Similar to HCG~16, significant tidal 
 features are observed in the group as a result of the ongoing interactions and mergers among the member galaxies.

\subsection{HCG 91}
The RA-velocity plot of HCG~91 is shown in the first panel of Figure~\ref{fig:hcg91_noise}. There are subtle vertical stripes at velocities corresponding to channels 
excluded during \texttt{UVLIN} fit. The effects can be seen in the noise-velocity plot in the middle panel of 
Figure~\ref{fig:hcg91_noise} where the noise increases at the channels where known emission is excluded while performing \texttt{UVLIN} fit. Unfortunately, we cannot correct for such effects. 
This is inherent to \texttt{UVLIN} as described in \citet{2023A&A...673A.146S}. We show the global profiles of HCG 91 from the VLA and MeerKAT in the right panel of Figure~\ref{fig:hcg91_noise}. 
MeerKAT detects much more \HI\ emission than the VLA due to its larger field of view of MeerKAT and better sensitivity. 
As shown in the global profiles, there are many sources beyond the velocity range of the central part of the group and some or many of them might still belong to the group. 
A more careful study of them will be done in an accompanying paper (Sorgho et al. in prep). 

We show example channel maps of HCG~91 in Figure~\ref{fig:hcg91_chanmap}, which contain the SE tail of HCG~91a. Additional channel maps are available \href{https://zenodo.org/records/14856489}{online}. Some of the 2$\sigma$ contours of HCG~91a bend toward the north, and appear 
to be aligned with its optical tails. In addition, the curved \HI\ emission that apparently connects HCG~91c and HCG~91d in projection appears as broken contours of low-column density 
gas coming from both HCG~91c and HCG~91b. Lastly, the channel maps show no \HI\ emission at the velocity and location of HCG~91d. This galaxy has never been detected in \HI\ emission 
before \citep{2023A&A...670A..21J}.

We show the moment maps of HCG~91 in Figure~\ref{fig:hcg91_mom}. We have detected many sources in HCG~91 but only those within $\mathrm{1000~km~s^{-1}}$ from the systemic velocity are shown here. 
On projection, the core members and a far-off members at the west side, LEDA 749936, are embedded 
in one common \HI\ envelop. Two \HI\ bridges connect HCG~91b and HCG~91c. The first was identified previously by \citet{2023A&A...670A..21J} which they separated as intra-group gas. 
The second one extends to the west from HCG~91c before curving toward the north to join another extended feature of HCG~91b. In addition, it seems to elongate towards LEDA 749936. 
Another higher S/N emission connects HCG~91c and HCG~91a. All the core members of HCG~91, including the detection slightly far west are detected as one source by SoFiA, highlighting 
the complex interactions between the members. We also show the velocity fields of the galaxies in the core of HCG~91 in Figure~\ref{fig:hcg91_mom_cores} from a data cube 
at 24.0\arcsec\ $\times$ 20.2\arcsec, as well as their DeCaLS optical image along with their \HI\ iso-velocity contours. Their optical major axis position angles are 
well aligned with their \HI\ major axis position angles. However, the outer disk of HCG~91b is warped. \\
HCG~91a has asymmetric double-horned profiles, typical of spiral galaxies. Its main disk still has a clear rotational pattern; however, 
its halo is disturbed. HCG~91c also has asymmetric double-horned profiles. In addition, its main disk has retained rotation despite the interaction with other members. The blue-shifted 
velocity field of HCG~91b is more or less regular, except the warp mentioned earlier. However, the red-shifted part is clearly disturbed due to the interaction with HCG~91c. The \HI\ emission in LEDA 749936 has 
very low S/N and its velocity field is disturbed. It has a single peaked global profile, typical of dwarf galaxies.  We detect many sources around the central part of HCG~91 that seem to 
have regular rotation. However, a group of four strongly interacting galaxies can be seen south-east of the core members. Their \HI\ velocity fields and morphologies are clearly disturbed. 
In addition, two interacting galaxies are found further east. Note though that their systemic velocities differ by about $\mathrm{1500~km~s^{-1}}$ compared to those of the core groups.   

We show the 3D visualisation of HCG~91 in Figure~\ref{fig:hcg91_3dvis}. Again, the left panel showcases regions of high \HI\ column density, 
while the right panel illustrates a composite view, overlaying both low and high column density gas distributions. This group appears to be less chaotic than 
HCG 16 and HCG 91, and has a lot less low surface brightness features. HCG~91b and HCG~91c are connected by a hook like high surface brightness 
features and a low column density bridge. However, there is no \HI\ connection between HCG~91c and HCG~91a. The overall morphology of the member galaxies 
are clearly disturbed, especially that of HCG~91a. It is also clear from the 3D view that no \HI\ has been detected in HCG~91d. 

\subsection{HCG 30}
The RA-velocity plot of HCG~30 is presented in the left panel of Figure~\ref{fig:hcg30_noise}, showing no apparent signs of continuum residuals or cleaning artefacts. 
The median noise of the data cube as a function of velocity is shown in the middle panel of the figure. 
The comparison of the MeerKAT and VLA global profile is shown in the right panel of Figure~\ref{fig:hcg30_noise}. The larger field of view of MeerKAT and the detection of 
new features and a member in the central of part of HCG~30 makes the global profile of MeerKAT much brighter than that of the VLA.    

We present the example channel maps of HCG~30 in Figure~\ref{fig:hcg30_chanmap}. These only show the channels corresponding to the receding side of HCG~30b, the rest is presented 
\href{https://zenodo.org/records/14856489}{online}. Overall, the emission is faint and is mostly below the 3$\sigma$ detection threshold.

HCG~30 is one of the most \HI\ deficient groups in our sample and previous observations failed to detect any \HI\ emission in the galaxies making up its core. However, 
we detect two clump-like \HI\ emission slightly offset from the optical centres of HCG30a and HCG30c as shown in Figure~\ref{fig:hcg30_mom}. 
One clump is located at 0.86\arcmin\ (15 kpc) from HCG30a, while the other one is at 0.33\arcmin\ (6 kpc) from HCG30c. They could be 
the remains of stripped \HI\ from HCG30a and HCG30c. The \HI\ detection in HCG~30a remains spurious and unresolved despite being a large spiral galaxy. 
We show overview plots of HCG~30c in \ref{fig:hcg30c}. These plots were generated 
using the SoFiA Image Pipeline (SIP) by Hess K.M.\footnote{\url{https://github.com/kmhess/SoFiA-image-pipeline}}\citep{SIP}. 
As shown in the figure, the \HI\ emission in HCG30c is faint and is barely resolved.   

The \HI\ emission corresponding to HCG~30b was split as three sources by SoFiA. We combined the three 
individual cubelettes from SoFiA using the MIRIAD task \texttt{IMCOMB} and rederived the moment maps, which we show in Figure~\ref{fig:hcg30b}. The moment one 
map presents apparent signs of rotation but the iso-velocity contours do not show the spider patterns expected from Sa galaxies. Instead, they show systematically convex 
curvatures. Note though that the emission along the minor axis is barely resolved. The global \HI\ profile exhibits an enhanced intensity at either side of a more subdued 
central peak. However, the overall S/N of the profile is low.  The systemic velocity we got from SIP, $\mathrm{4519~km~s^{-1}}$, corresponds well to the one quoted 
by \citet{2023A&A...670A..21J}, $\mathrm{4508~km~s^{-1}}$, 
from its optical redshift. 

 We show the 3D visualisation of HCG~30 in Figure~\ref{fig:hcg30_3dvis}. The 3D plot does not show a hint of detection at the 
 location of the core members. The tentative detection we observed in the moment maps needs to be confirmed by future more sensitive 
 observations. In addition, the surrounding galaxies do not show any signs of morphological disturbance.

\subsection{HCG 90}
\subsubsection{Data cube}
The RA-velocity plot of HCG~90 is shown in the left panel of Figure~\ref{fig:hcg90_noise}, indicating no obvious continuum residuals. The median noise values of the non-primary 
beam corrected cube as a function of velocity is presented in the middle panel of the figure. We compare the VLA and the MeerKAT global profile of HCG~90 in the right panel of 
Figure~\ref{fig:hcg90_noise}. Since the VLA observations only detected \HI\ emission in HCG~90a, the MeerKAT global profile appears much brighter than that of the VLA as MeerKAT 
detects emission both at the central part and in the vicinity of HCG~90. 

We show example channel maps of the central part of HCG~90 in Figure~\ref{fig:hcg90_chanmap}. More channel maps, covering a wider velocity range, can be downloaded \href{https://zenodo.org/records/14856489}{here}. 
The tail described previously appears as broken contours of low-column density \HI, with the 
brightest contour found at 2271 to 2293 $\mathrm{km~s^{-1}}$, at the southern optical tail. No 3-$\sigma$ contour is found at the location of HCG~90b, HCG~90c, and HCG~90d.  

The moment maps of HCG~90 are shown in Figure~\ref{fig:hcg90_mom}. The maps indicate a previously undetected \HI\ tail in the central region of HCG~90. When plotted against a deep DECaLS image, 
the \HI\ tail seems to be aligned with the optical tail. It has a total mass of $\mathrm{2.81~\times~10^{8} ~ M_{\sun}}$, and is about 128 kpc long. On projection, it is difficult to assess which members the tails are associated with. However, 
part of the tail coincides with the location of HCG90b and HCG90d. Previous H$\alpha$ kinematics by \citet{1998AJ....116.2123P} shows evidence of ongoing interaction 
between HCG~90b and HCG~90d, with HCG~90d acting as a gas provider. Spatially, the northern tail coincides more with the location of HCG~90d than that of HCG~90b. 
We also show in Figure~\ref{fig:hcg90_mom} a position-velocity map of the tail taken from a thick slice shown at the top right panel of the figure. A continuity in velocity 
is clearly visible, suggesting that this is not a projection of multiple features but rather a single, coherent structure. A more 
compact \HI\ structure is found further north of HCG~90c, at the south western side of HCG~90a. 
We therefore hypothesise that the observed tail is part of a more extended structure connecting HCG~90c with the other members, which may have escaped our detection, 
or have already been dispersed by other processes such as ionisation or star formation.
HCG~90a has been detected before by the VLA, and we also clearly detect it with MeerKAT. 

Figure~\ref{fig:hcg90_3dvis} presents the 3D view of HCG 90. The tail we see in the moment map is also visible in the 3D plot and 
appears to be a tidal remnant of HCG~91c. This group might be a good candidate to look for diffuse \HI\ components as the detection of this 
tidal fragment suggests that it is at a less advanced evolutionary stage than HCG 30 and HCG 97. 
\subsection{HCG 97}
The RA-velocity plot of HCG~97 is shown in the left panel of Figure~\ref{fig:hcg97_noise}, presenting some continuum residual 
emission that appears as positive and negative vertical stripes. These artefacts also manifest as increased noise levels starting 
around $\mathrm{7500~km~s^{-1}}$ and above as shown in the middle panel of the figure. 
The comparison between the VLA global profile and that of MeerKAT is presented in the right panel of Figure~\ref{fig:hcg97_noise}.  
Another \HI\ emission peak beyond the velocity covered by the VLA is detected by MeerKAT. 
The moment maps are shown in Figure~\ref{fig:hcg97_mom}. Many galaxies are detected beyond the central part of HCG~97. 
However, no \HI\ emission is detected in HCG~97a, 
HCG~97c, HCG~97d, and HCG~97e. Only HCG~97b is detected, whose approaching side was also detected by the VLA \citep{2023A&A...670A..21J}. 
\citet{2023A&A...670A..21J} suggested that HCG~97b might be disturbed since they only detected one side of the galaxy. However, we do not 
see any \HI\ extension in our map despite the fact that MeerKAT detected the two sides of the galaxies. The receding part of this galaxy is indeed fainter 
as previously mentioned by \citet{2023A&A...670A..21J}. HCG~97b is an edge-on spiral galaxy, but its velocity field is typical of a dwarf with solid body 
rotation curve. This is further corroborated by the position-velocity diagram shown in Figure~\ref{fig:hcg97_mom}, which shows that HCG~97b has a 
very steep inner rotation curve, which most spirals with flat rotation curve also have \citep{1978PhDT.......195B}. 
If this galaxy has a flat rotation curve, then we are only probing its inner disk, corresponding to the rising part of the rotation curve. 
Its outer disk might be too faint to be detected. Thus, we do not exclude the possibility that this galaxy is interacting. 

Figure~\ref{fig:hcg97_3dvis} presents the 3D view of HCG 97. No signs of detection are observed at the location of the core members. Only HCG 97b 
is detected. The surrounding members are not morphologically disturbed. Our observations confirm that this group is at an advanced evolutionary stage.
\section{Summary and conclusion}\label{summary}
We have presented MeerKAT \HI\ observations of six HCGs (3 phase 2 and 3 in phase 3), providing essential data to understand the 
transition between the two most evolved phases in the evolutionary sequence of these compact galaxy aggregations. 
We aimed to detect diffuse \HI\ gas that was apparent in previous 
GBT observations but missed by the VLA telescope. Our observations have revealed significantly more extended 
tidal features in phase 2 groups compared to those detected by the VLA. In addition, we have detected new high 
surface brightness features in phase 3 groups. The presence of tidal features, such as \HI\ tails, bridges, 
and clumps, suggests substantial gas loss into the IGrM due to gravitational interactions. When derived within the field of view of GBT, our measured \HI\ 
flux is similar to that of the GBT. However, we have not detected the diffuse \HI\ component apparent in previous GBT spectra for phase 3 groups despite MeerKAT's 
superb sensitivity. This indicates that part of the missing \HI\ could be too diffuse to be detected by MeerKAT and the rest might be ionised.  
Numerous surrounding galaxies have been detected for both phase 2 and phase 3 groups, most of which are normal disk galaxies. 
This suggests that these groups might be embedded in larger structures. 
MeerKAT detected significantly more extended \HI\ gas in Phase 2 groups than the VLA, 
thanks to its superior short baseline coverage. Thus, a single-dish telescope such as FAST is expected to detect \HI\ at even larger 
angular scales beyond MeerKAT's coverage. 
Apart from data cubes, source catalogues, and moment maps, we have released 3D visualisation of our data. 
This offers a more advanced approach to visually inspect different \HI\ features, 
which can help in separating the complex substructures present in phase 2 groups. 

\section{Data availability}
All the necessary scripts and data inputs are released along with this paper and can be found at \url{https://github.com/ianjarog/hcg-data-paper}; 
the instructions to download and process the data are also given there. 
\begin{acknowledgements}
Authors RI, LVM, AS, IL, CC, TW, BN, JM, SSE, JG acknowledge financial support from the grant PID2021-123930OB-C21 funded by MICIU/AEI/ 10.13039/501100011033 and by ERDF/EU, and the grant CEX2021-001131-S funded by MICIU/AEI/ 10.13039/501100011033, and the grant TED2021-130231B-I00 funded by MICIU/AEI/ 10.13039/501100011033 and by the European Union NextGenerationEU/PRTR, and acknowledge the Spanish Prototype of an SRC (espSRC) service and support funded by the Ministerio de Ciencia, Innovación y Universidades (MICIU), by the Junta de Andalucía, by the European Regional Development Funds (ERDF) and by the European Union NextGenerationEU/PRTR. The espSRC acknowledges financial support from the Agencia Estatal de Investigación (AEI) through the "Center of Excellence Severo Ochoa" award to the Instituto de Astrofísica de Andalucía (IAA-CSIC) (SEV-2017-0709) and from the grant CEX2021-001131-S funded by MICIU/AEI/ 10.13039/501100011033. Part of BN's work was supported by the grant PTA2023-023268-I funded by MICIU/AEI/ 10.13039/501100011033 and by ESF+. IL also acknowledges financial support from 
PRE2021-100660 funded by MICIU/AEI /10.13039/501100011033 and by ESF+. JMS acknowledge financial support from the Spanish state agency MCIN/AEI/10.13039/501100011033 and by 'ERDF A way of making Europe' funds through research grant PID2022-140871NB-C22. MCIN/AEI/10.13039/501100011033 has also provided additional support through the Centre of Excellence Mar\'\i a de Maeztu's award for the Institut de Ci\`encies del Cosmos at the Universitat de Barcelona under contract CEX2019–000918–M. JR acknowledges financial support from the Spanish Ministry of Science and Innovation through the project PID2022-138896NB-C55. MEC acknowledges the support of an Australian Research Council Future Fellowship (Project No. FT170100273) funded by the Australian Government. TW acknowledges financial support from the grant CEX2021-001131-S funded by MICIU/AEI/ 10.13039/501100011033, from the coordination of the participation in SKA-SPAIN, funded by the Ministry of Science, Innovation and Universities (MICIU). A. del Olmo and J. Perea acknowledge financial support from the Spanish MCIU through project PID2022-140871NB-C21 by ‘ERDF A way of making Europe’, and the Severo Ochoa grant CEX2021- 515001131-S funded by MCIN/AEI/10.13039/501100011033. JMS acknowledge financial support from the Spanish state agency MCIN/AEI/10.13039/501100011033 and by 'ERDF A way of making Europe' funds through research grant PID2022-140871NB-C22. MCIN/AEI/10.13039/501100011033 has also provided additional support through the Centre of Excellence Mar\'\i a de Maeztu's award for the Institut de Ci\`encies del Cosmos at the Universitat de Barcelona under contract CEX2019-000918-M. EA and AB gratefully acknowledge support from the Centre National d’Etudes Spatiales (CNES), France. RGB acknowledges financial support from the Severo Ochoa grant CEX2021-001131- S funded by MCIN/AEI/ 10.13039/501100011033 and PID2022-141755NB-I00. JM acknowledges financial support from  grant PID2023-147883NB-C21, funded by MCIU/AEI/ 10.13039/501100011033

OMS's research is supported by the South African Research Chairs Initiative of the Department of Science and Technology and National Research Foundation (grant No. 81737).

The MeerKAT telescope is operated by the South African Radio Astronomy Observatory, which is a facility of the National Research Foundation, an agency of the Department of Science and Innovation.
\end{acknowledgements}

\begin{appendix}
\section{Additional figures of HCG~31} 
\subsection{Channel maps}
\begin{figure*}
    \setlength{\tabcolsep}{0pt}
    \begin{tabular}{l l l}
        \includegraphics[scale=0.25]{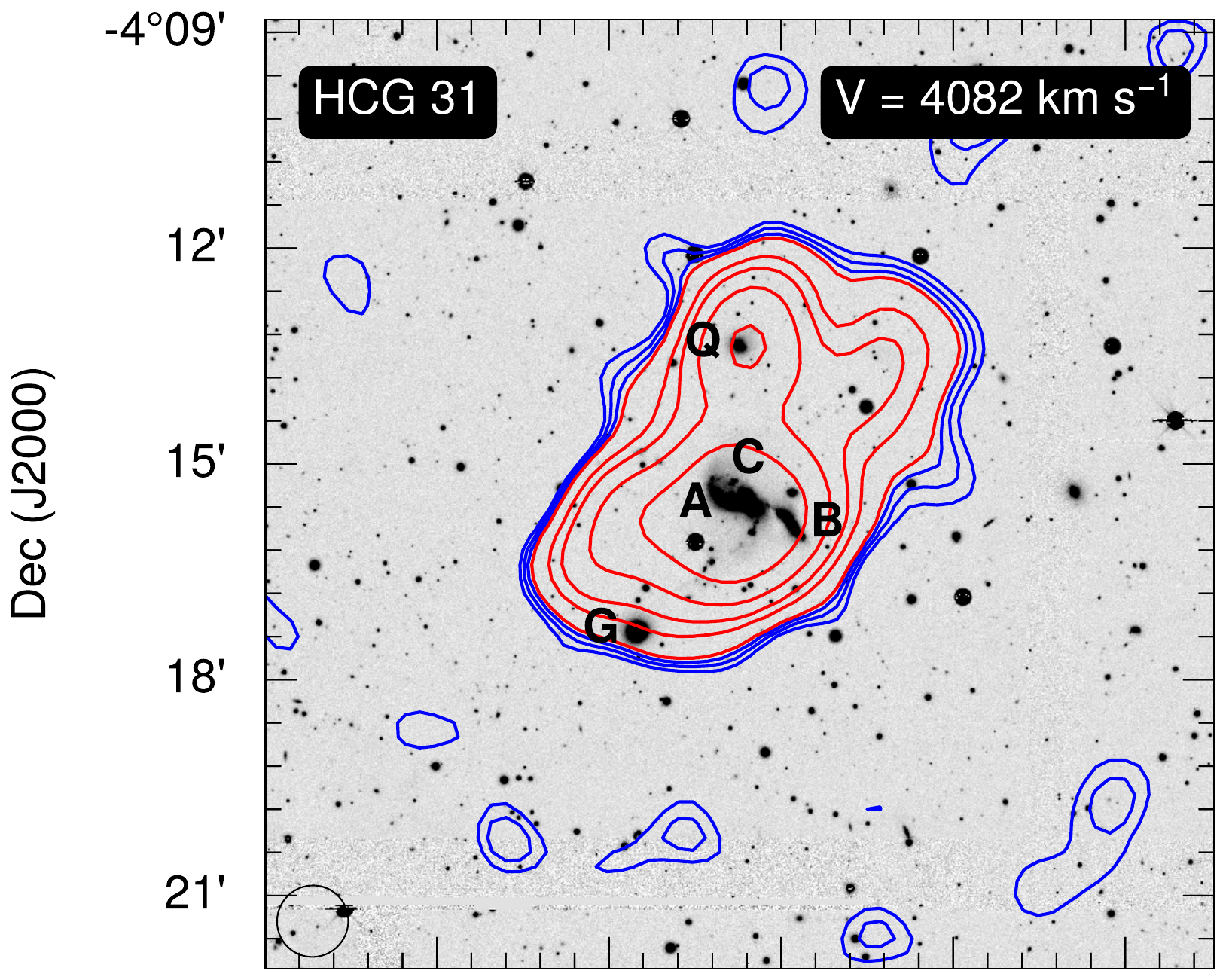} &
        \includegraphics[scale=0.25]{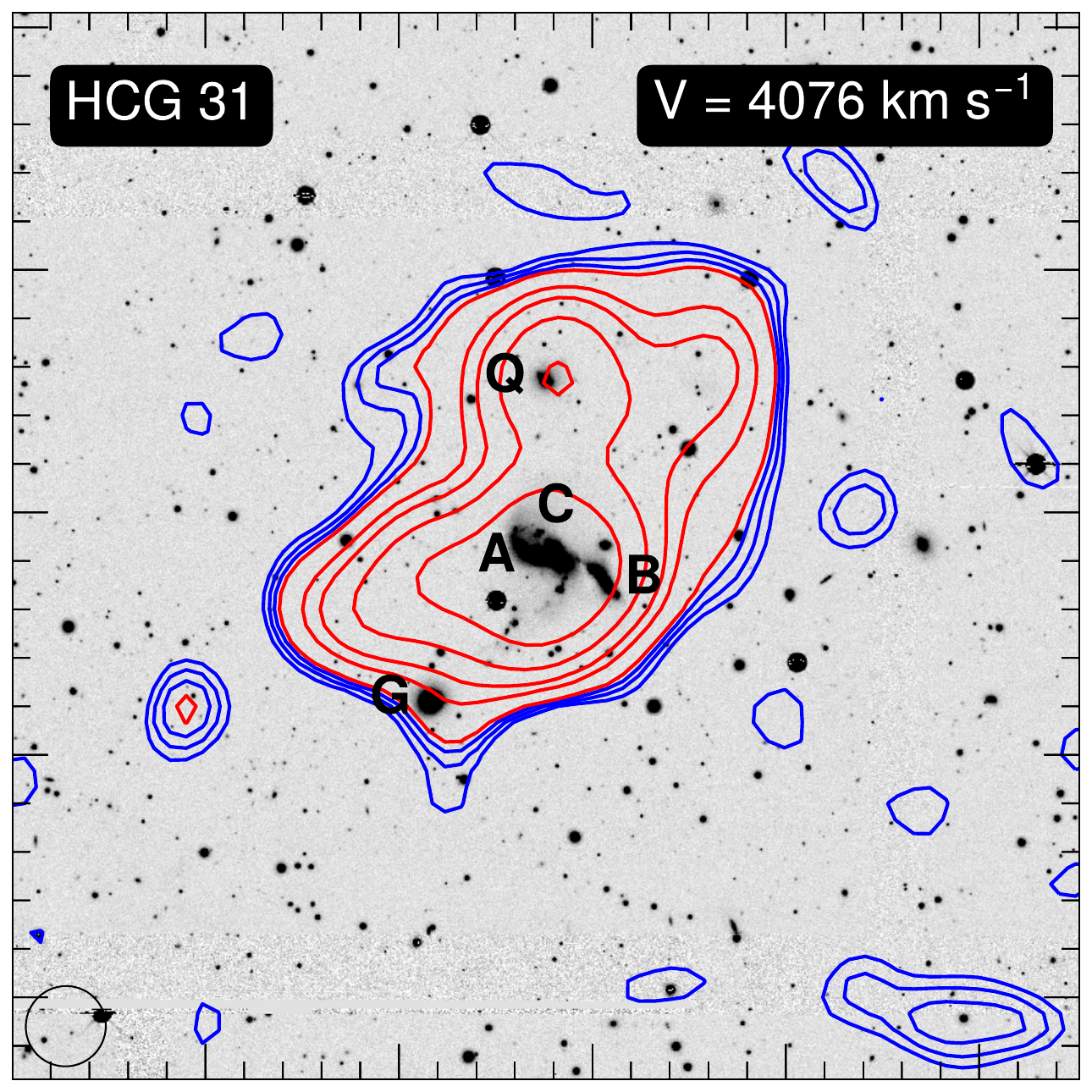} &
        \includegraphics[scale=0.25]{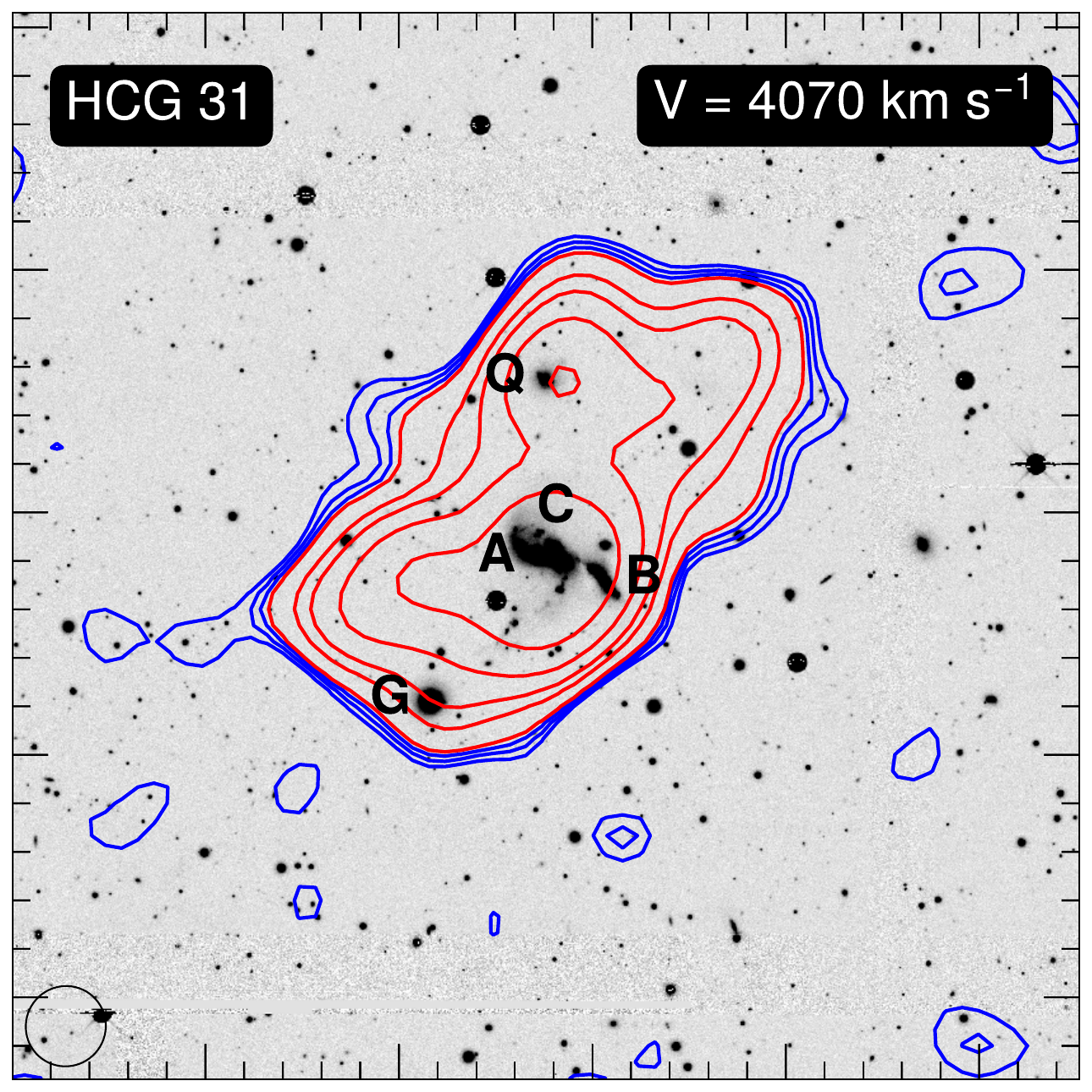} \\[-0.2cm]
        \includegraphics[scale=0.25]{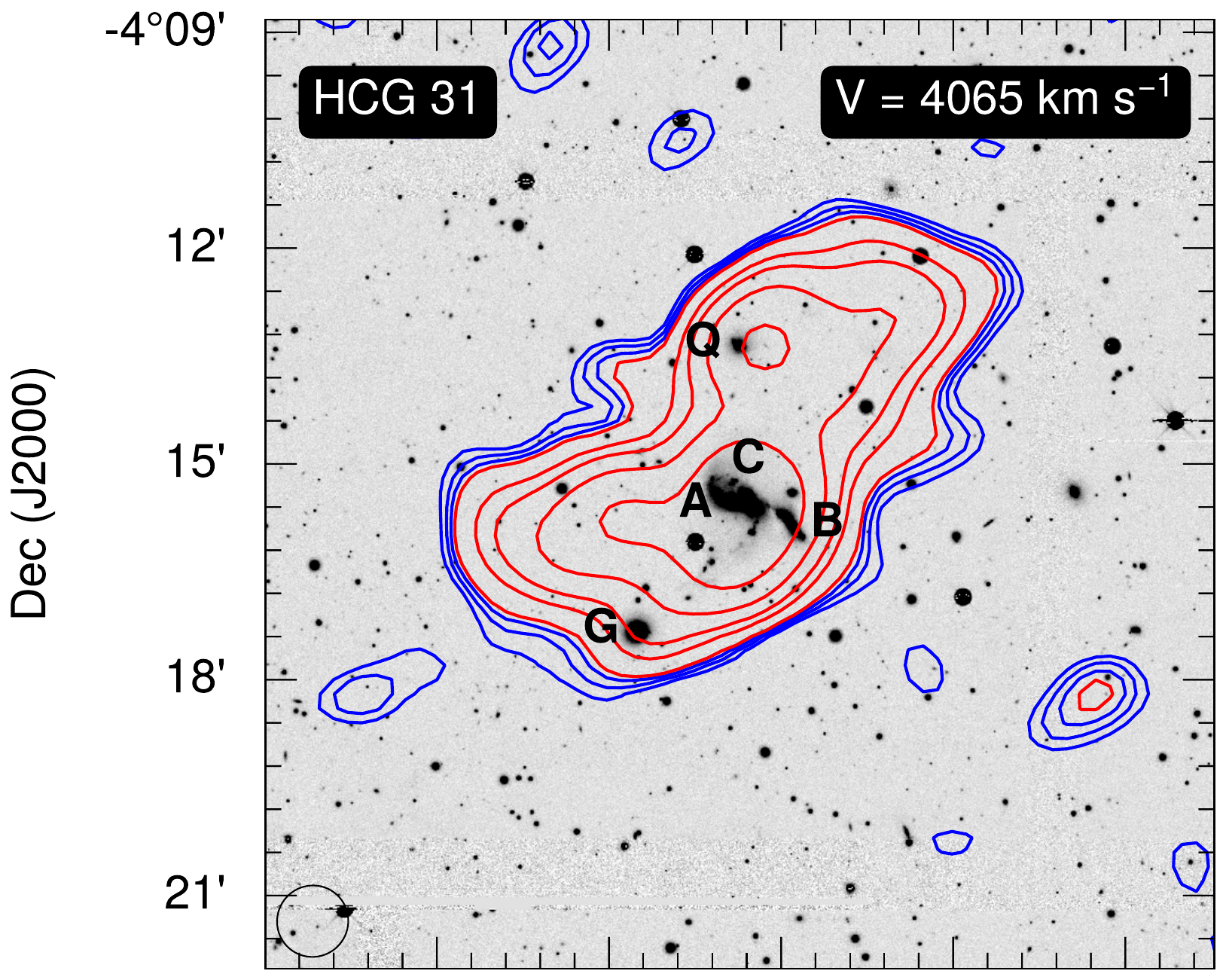} &
        \includegraphics[scale=0.25]{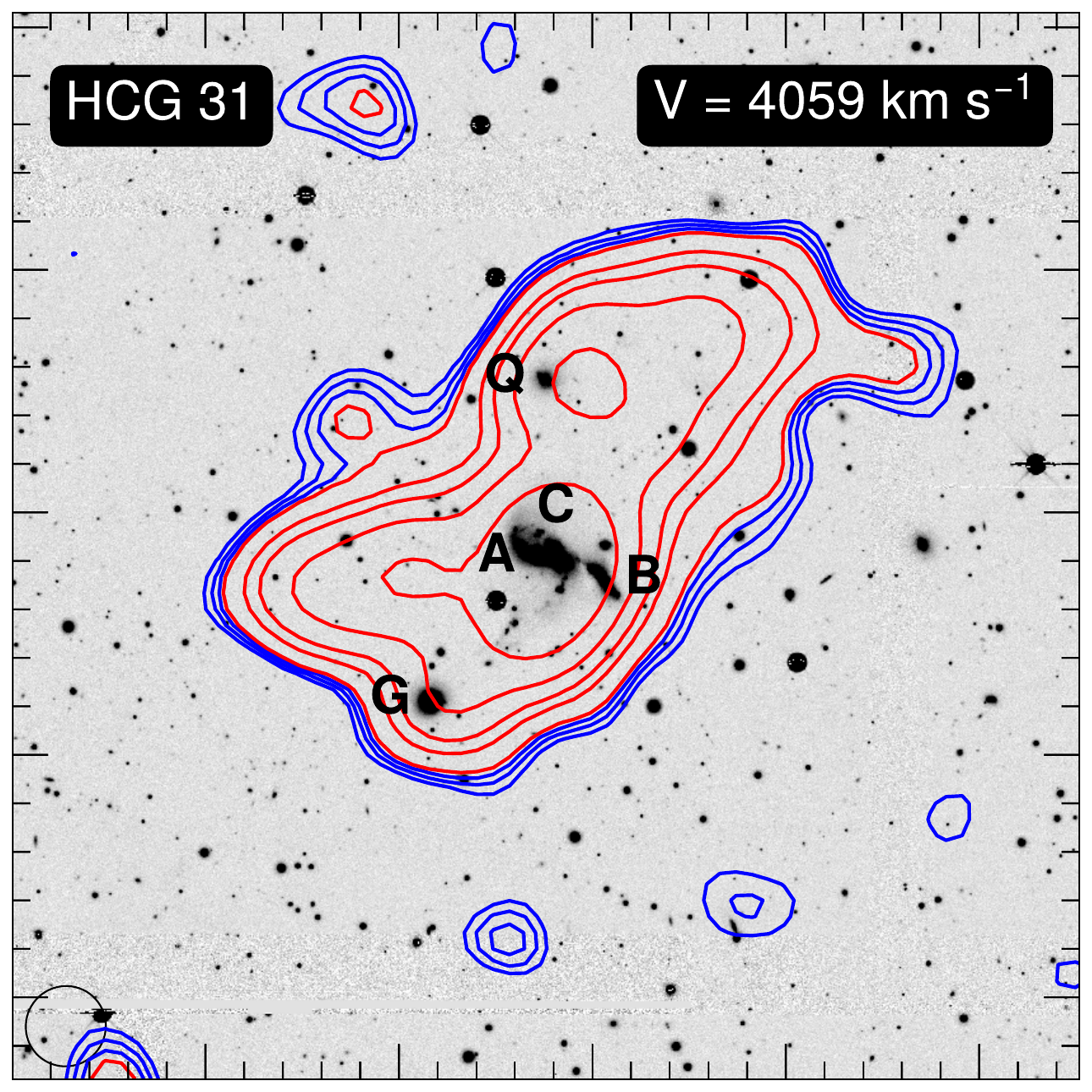} &
        \includegraphics[scale=0.25]{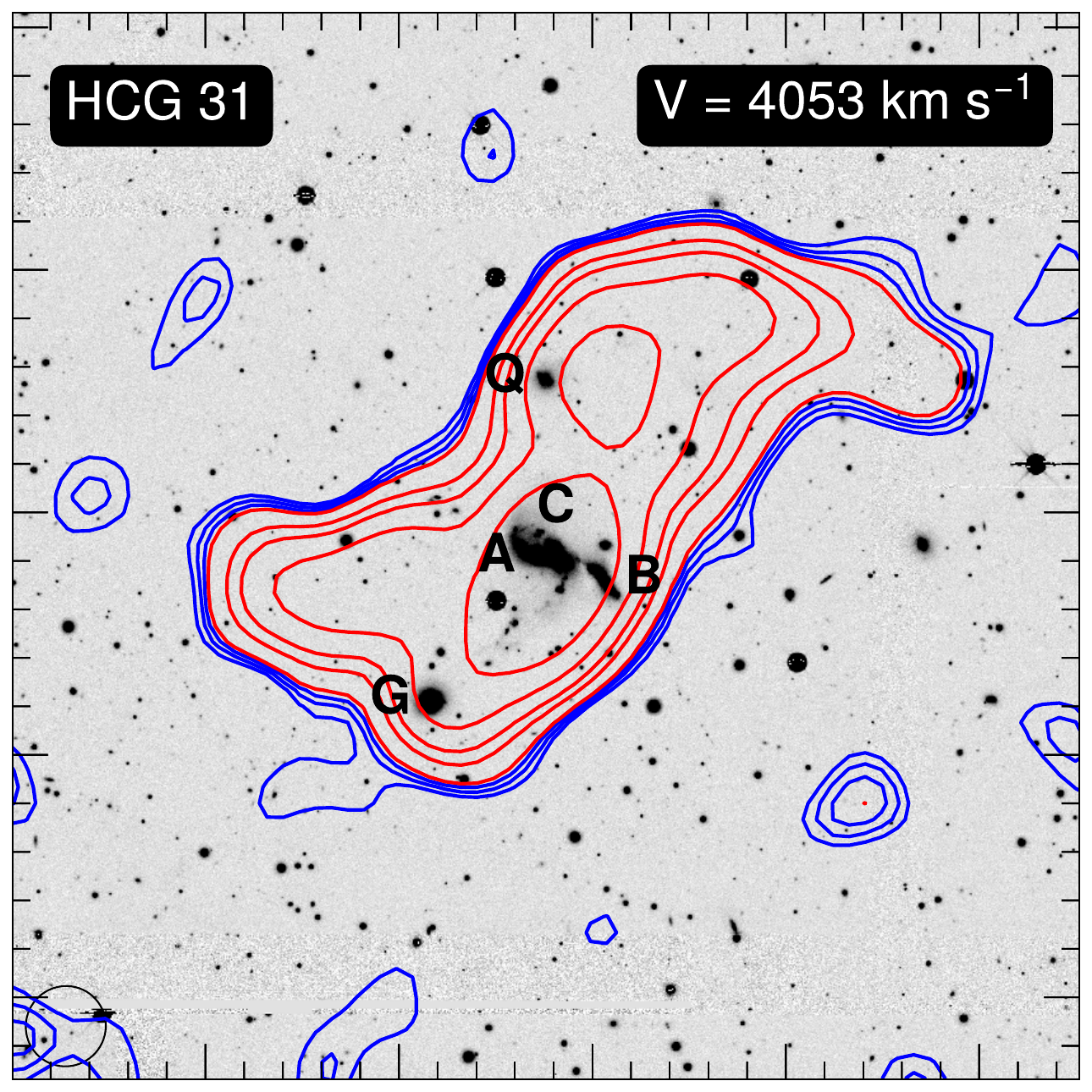} \\[-0.2cm]
        \includegraphics[scale=0.25]{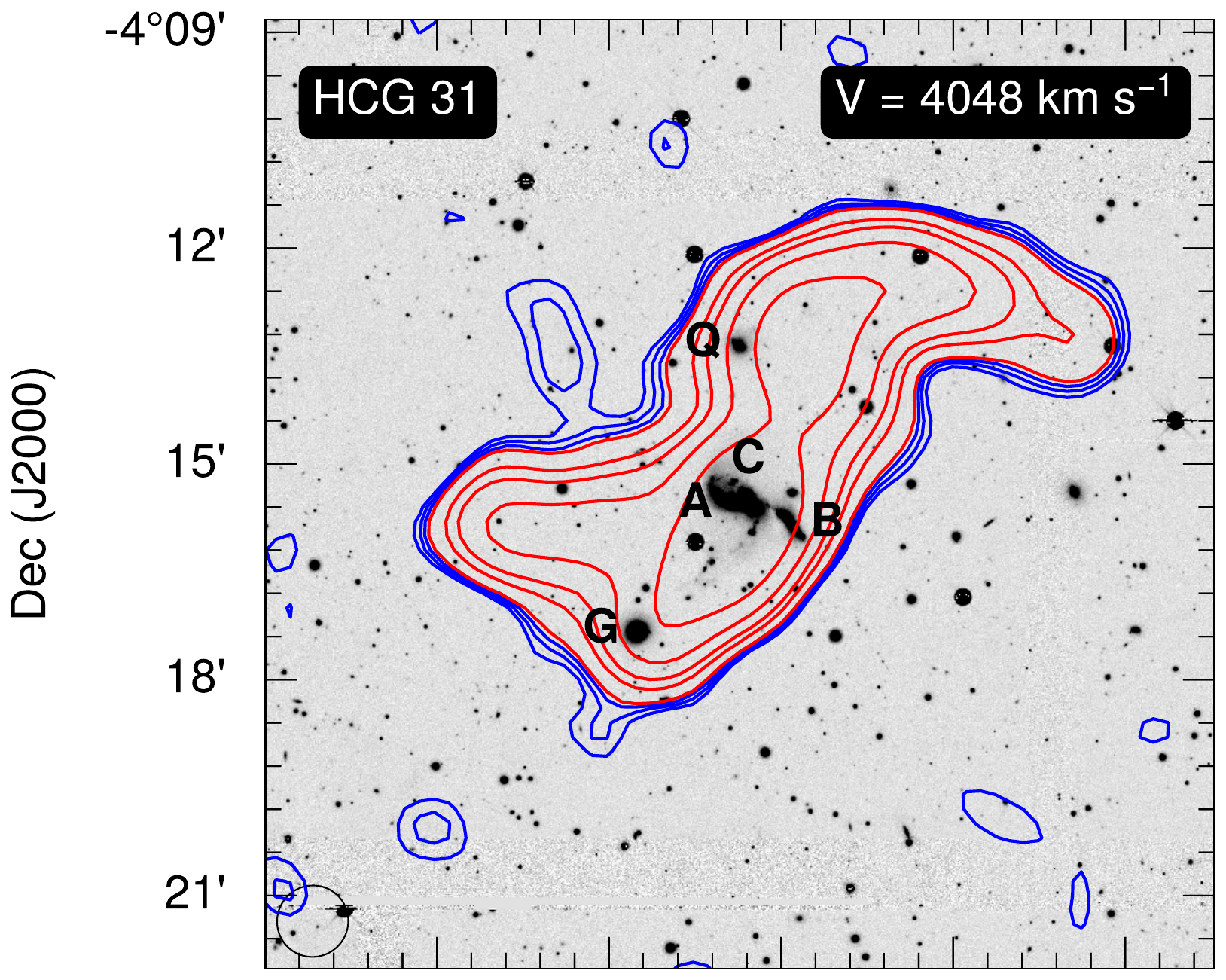} &
        \includegraphics[scale=0.25]{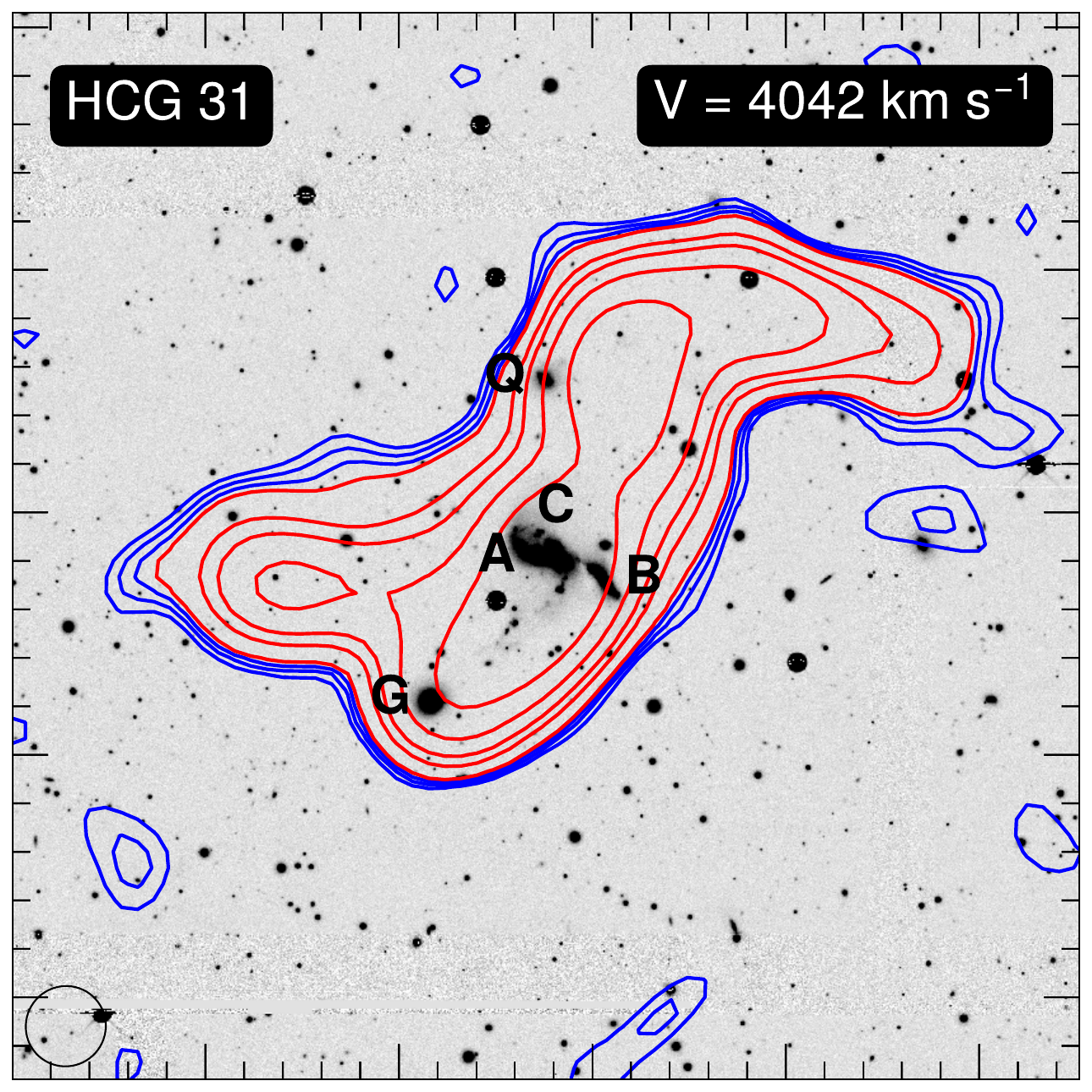} & 
        \includegraphics[scale=0.25]{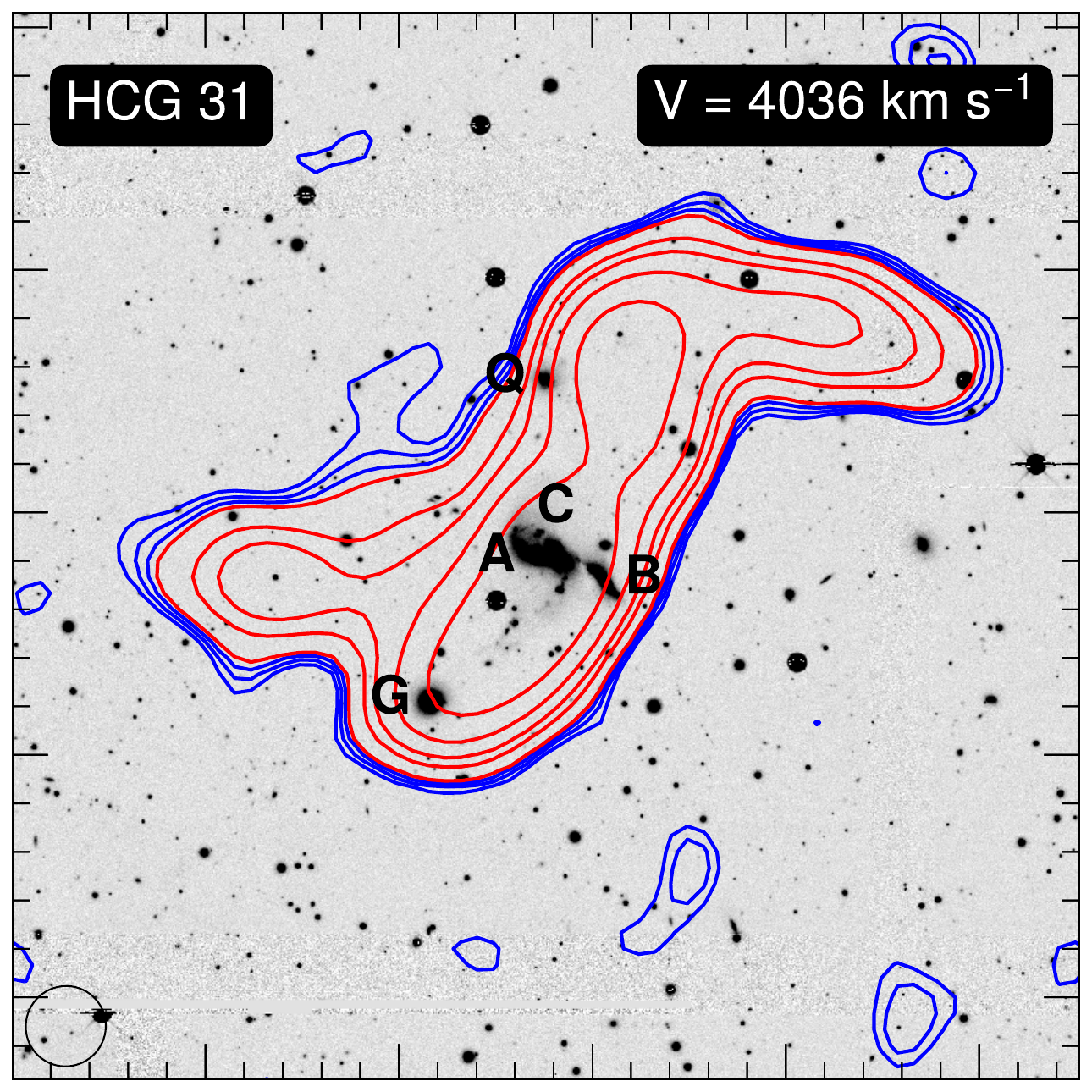} \\[-0.2cm] 
        \includegraphics[scale=0.25]{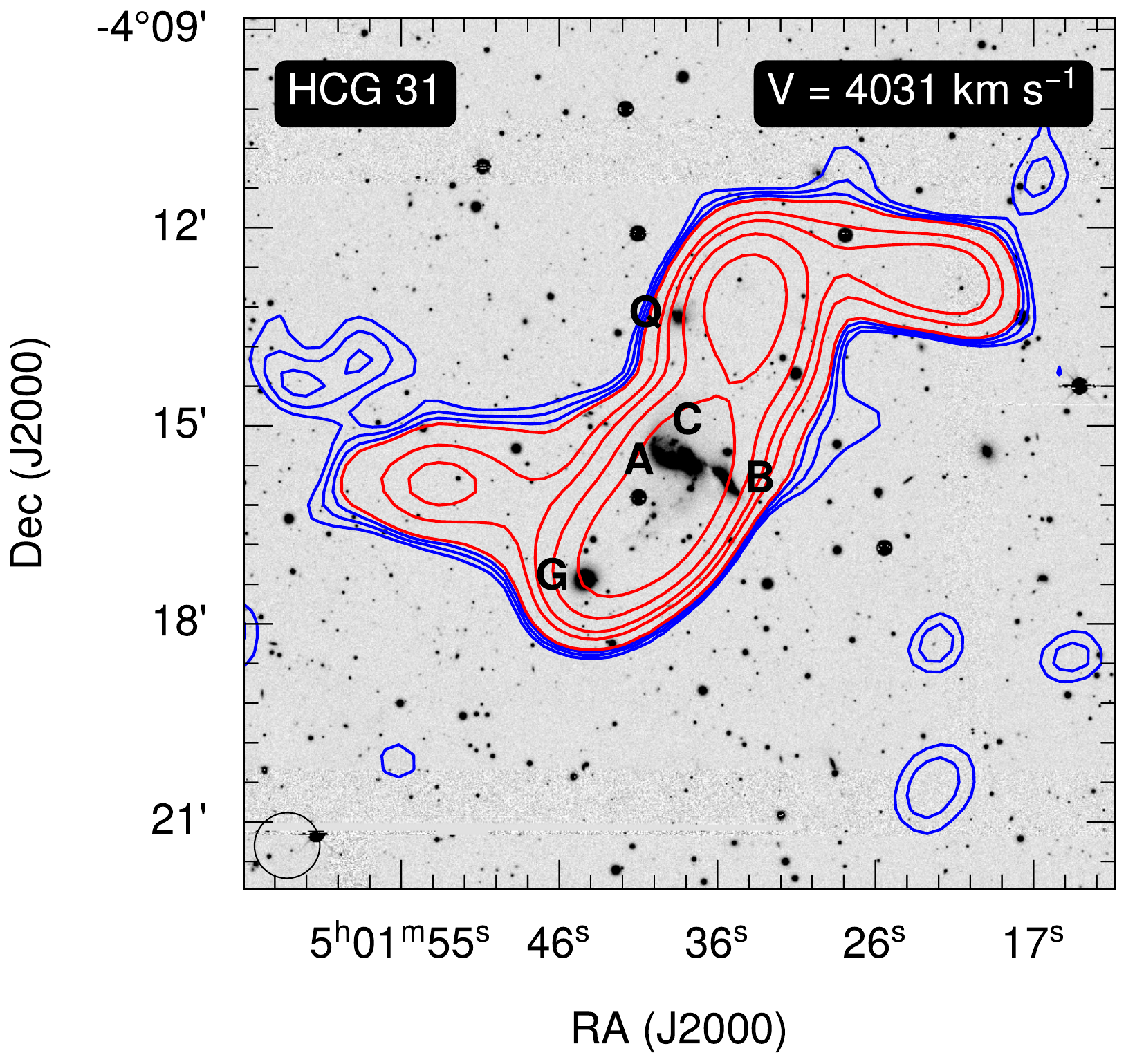} & 
        \includegraphics[scale=0.25]{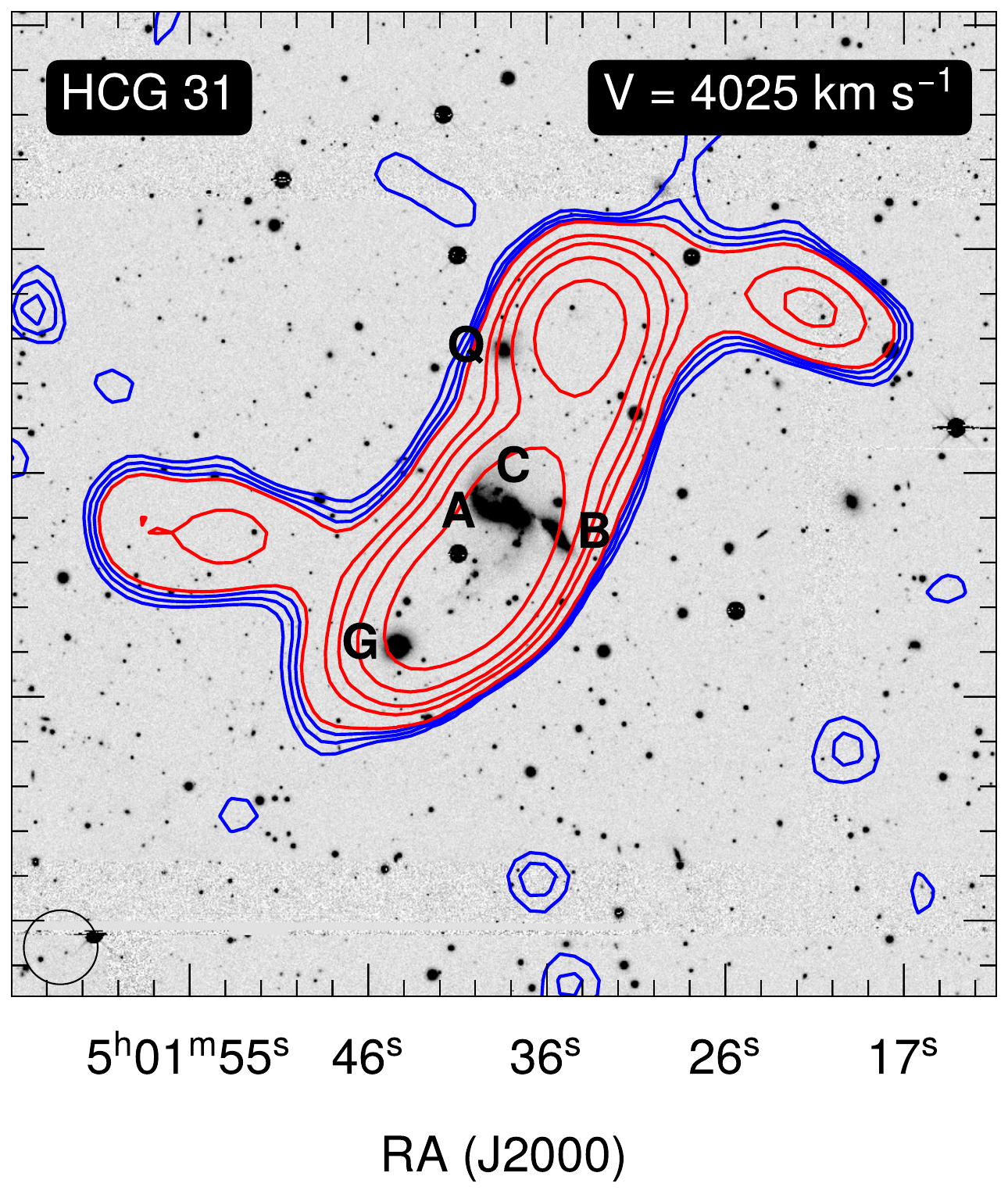} & 
        \includegraphics[scale=0.25]{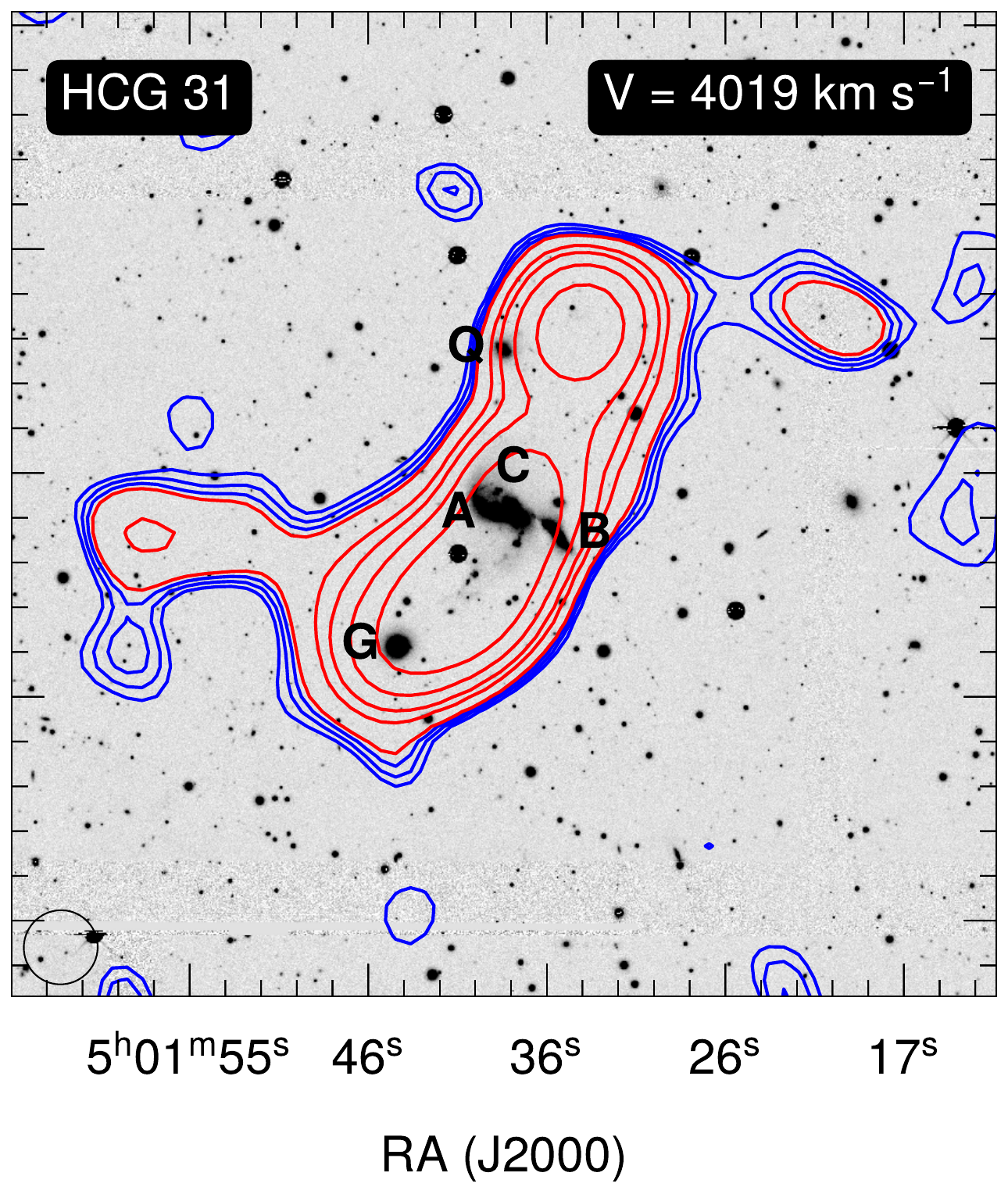}  
      \end{tabular}
      \caption{Example channel maps of the primary beam corrected cube of HCG 31 overlaid on DECaLS DR10 R-band optical images. Contour levels are (1.5, 2, 2.5, 3, 6, 9, 16, 32) 
      times the median noise level in the cube (0.71 $\mathrm{mJy~beam{-1}}$). 
      The blue colours show contour levels below 3$\sigma$; the red colours represent contour levels at 3$\sigma$, or higher. More channel maps are available \href{https://zenodo.org/records/14856489}{online}.}
      \label{fig:hcg31_chanmap}
     \end{figure*}

Figure~\ref{fig:hcg31_chanmap} shows example channel maps from the primary beam corrected data cube of HCG 31, overlaid on DECaLS DR10 R-band optical images.
\subsection{Moment maps}  
\begin{figure*}
  \begin{tabular}{l l}
    \includegraphics[scale=0.27]{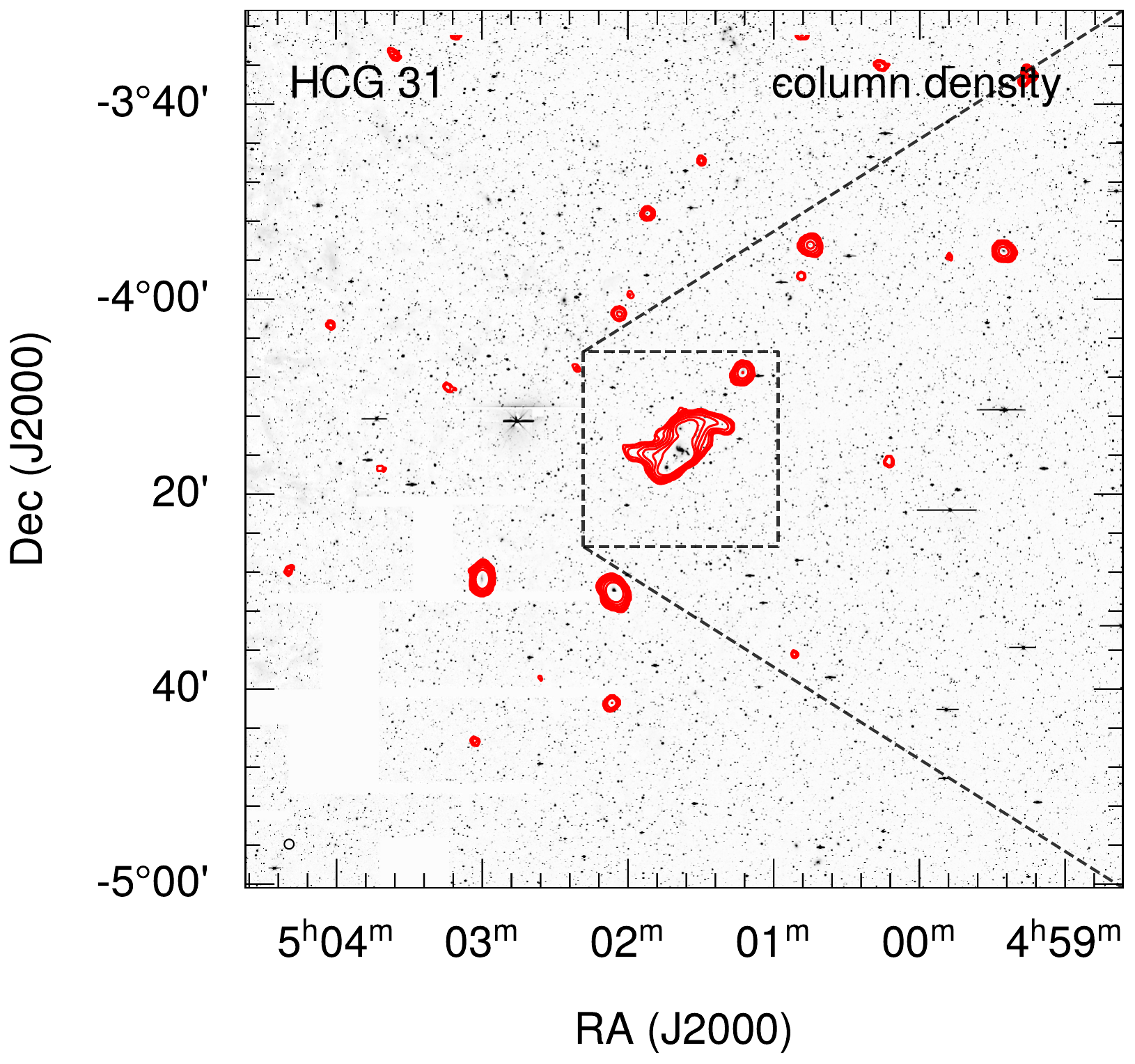}
    & \includegraphics[scale=0.27]{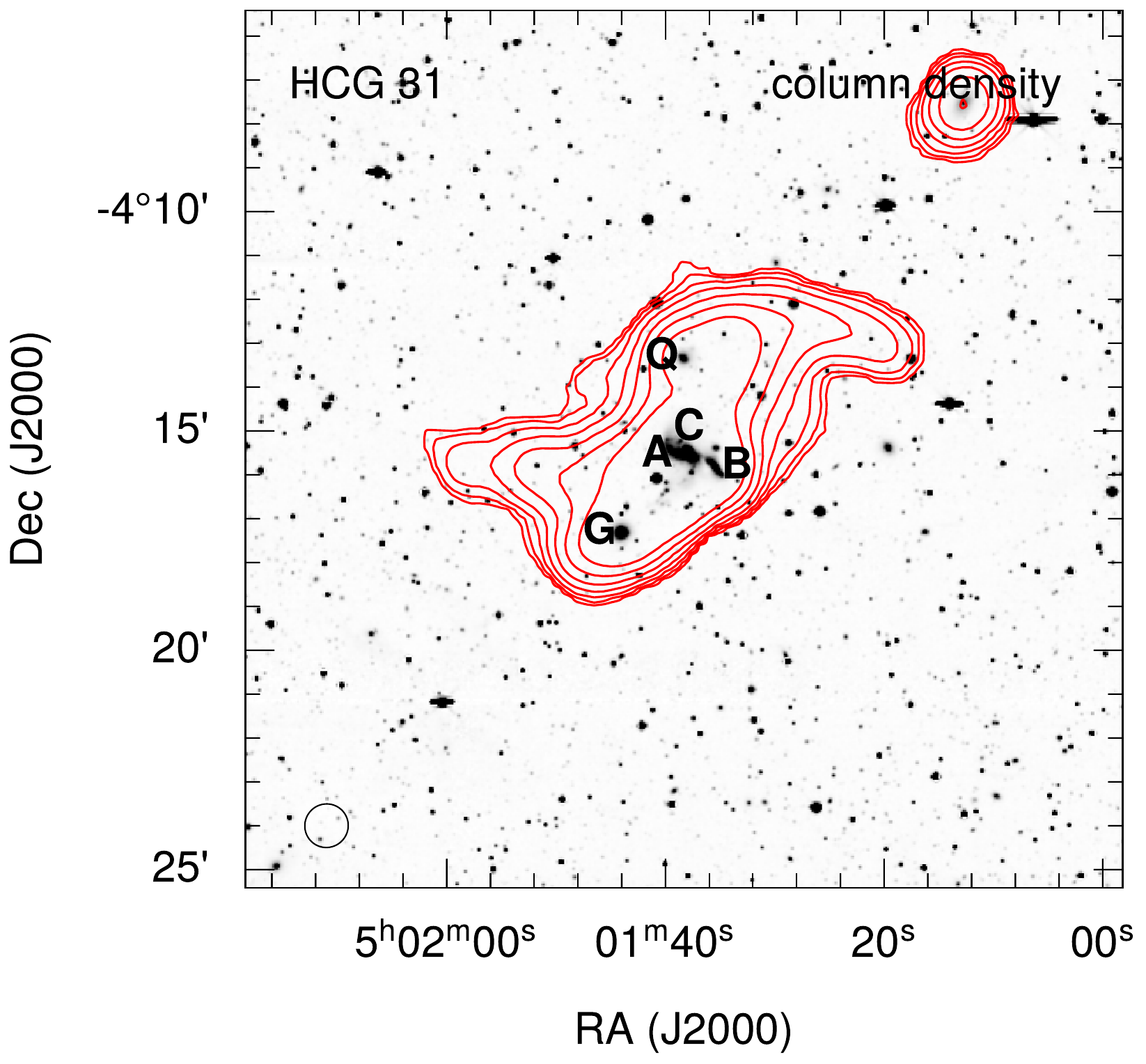}\\
    \includegraphics[scale=0.27]{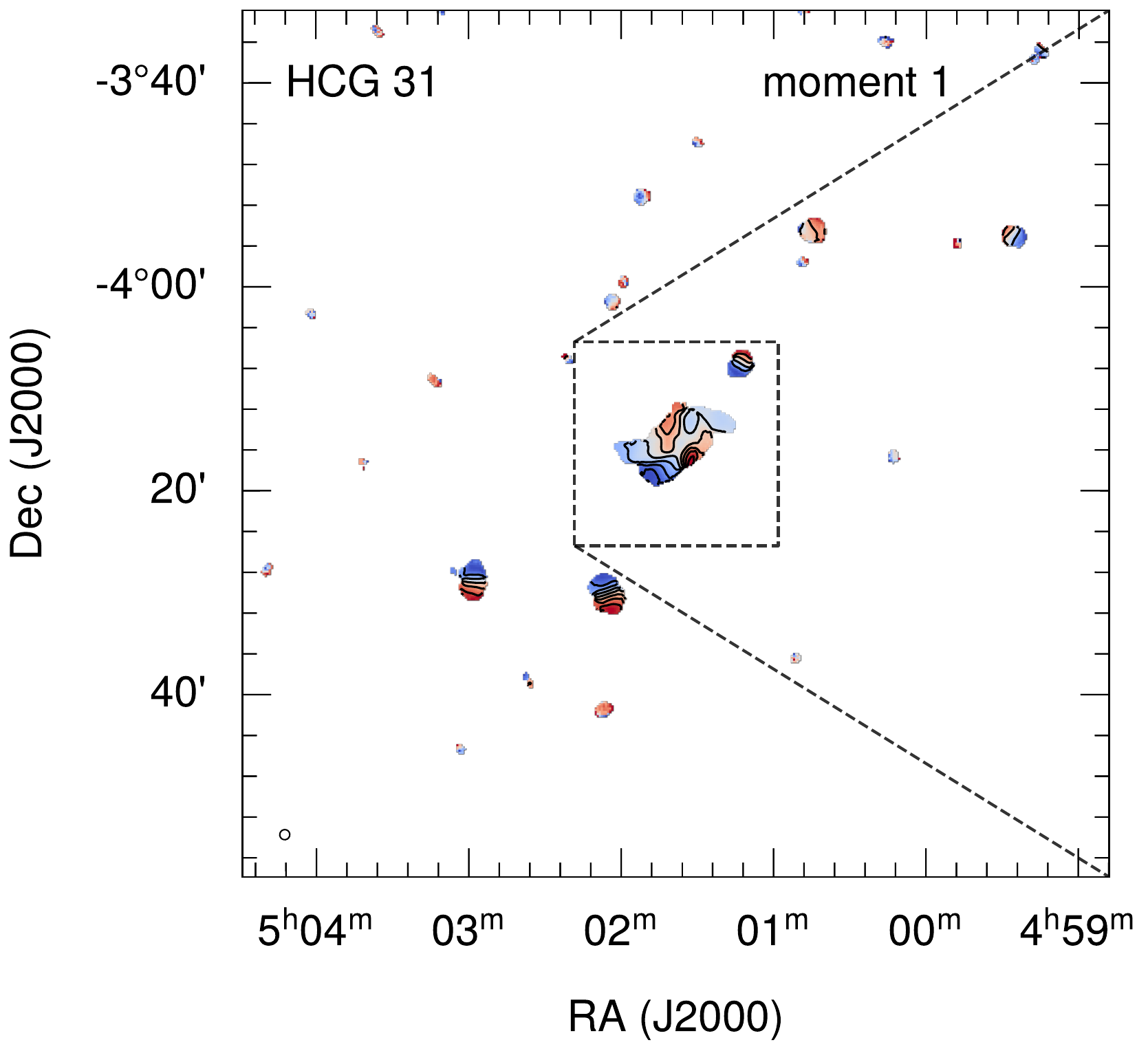} &
    \includegraphics[scale=0.27]{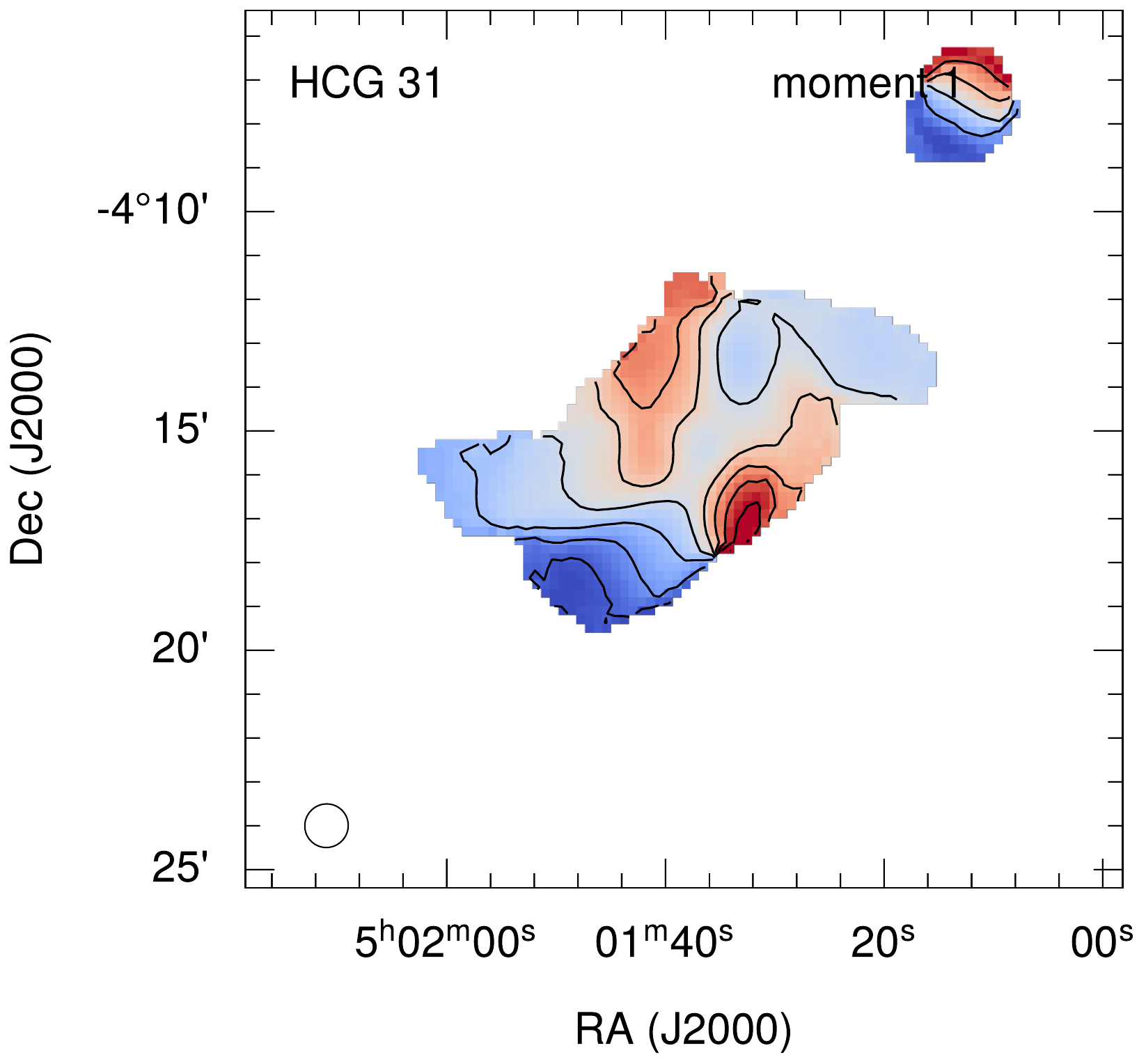}
  \end{tabular}
  \caption{\HI\ Moment maps of HCG 31. Left panels show all sources detected by SoFiA. The right panels show sources within the rectangular box shown 
  on the left to better show the central part of the group. The top panels show the column density maps with contour levels of 
  ($\mathrm{4.0~\times~10^{18}}$, $\mathrm{8.0~\times~10^{18}}$, $\mathrm{1.6~\times~10^{19}}$, $\mathrm{3.2~\times~10^{19}}$, 
  $\mathrm{6.4~\times~10^{19}}$, $\mathrm{1.3~\times~10^{20}}$, $\mathrm{2.6~\times~10^{20}}$) $\mathrm{cm^{-2}}$. 
  The contours are overlaid on DECaLS DR10 R-band optical images. The bottom panels show the moment one map. Each individual source has its own colour scaling and contour levels to highlight any rotational component.}
  \label{fig:hcg31_mom}
  \end{figure*}

\begin{figure*}
  \begin{tabular}{c}
    \includegraphics[scale=0.45]{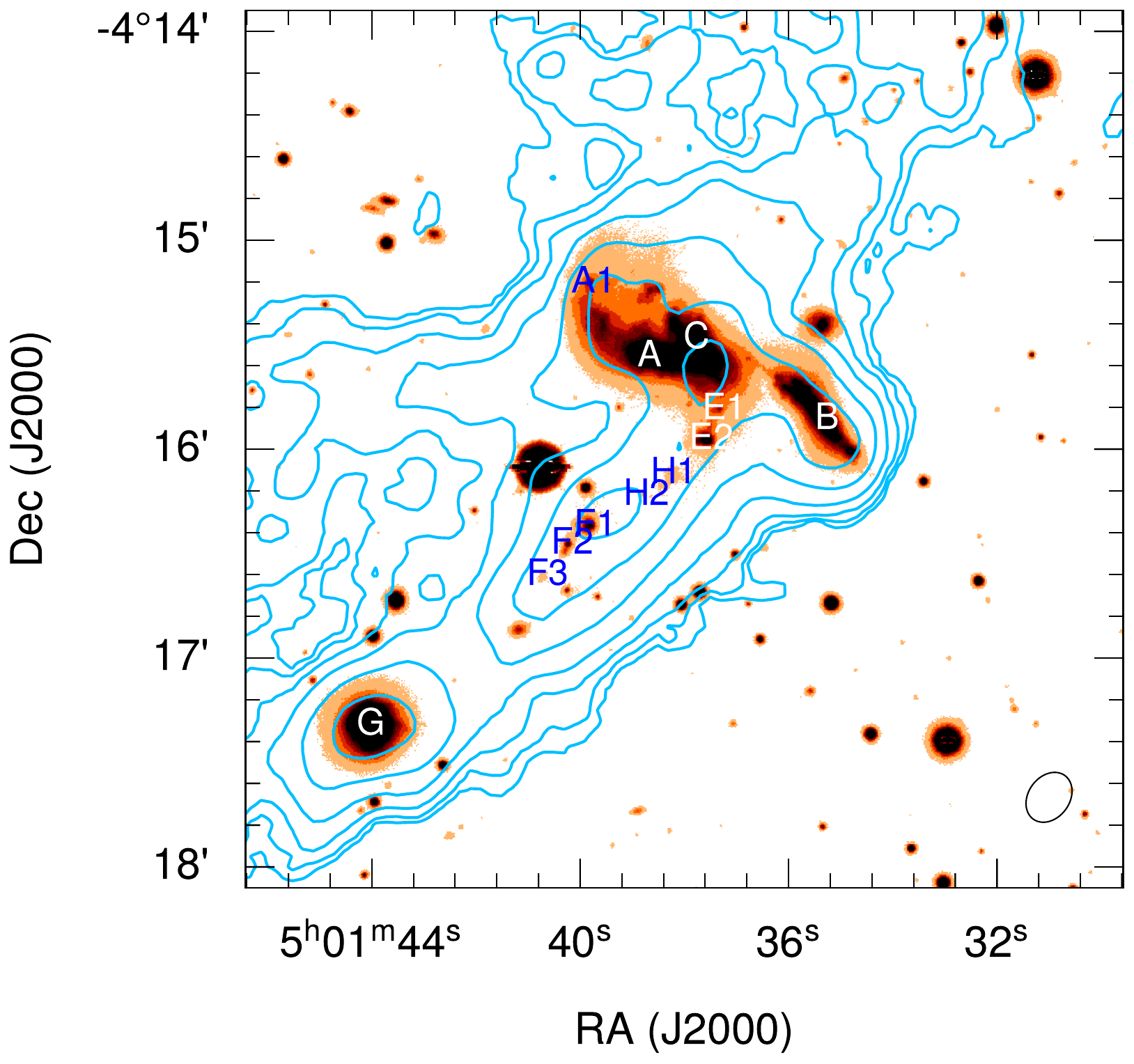} 
  \end{tabular}
  \caption{Column density map of the central part of HCG~31 from a 15.47\arcsec $\times$ 11.86\arcsec datacube, overlaid on DECaLS filter optical images to highlight the core members. 
  The contour levels are ($\mathrm{5.96~\times~10^{19}}$, $\mathrm{1.19~\times~10^{20}}$, $\mathrm{2.38~\times~10^{20}}$, $\mathrm{4.77~\times~10^{20}}$, 
  $\mathrm{9.53~\times~10^{20}}$, $\mathrm{1.91~\times~10^{21}}$, $\mathrm{3.20~\times~10^{21}}$) $\mathrm{cm^{-2}}$.} 
  \label{fig:hcg31_optical_mom0}
  \end{figure*}

Figure~\ref{fig:hcg31_mom} shows the column density maps and moment one (velocity field) of HCG~31. The left panel highlights the large-scale structure of the group, whereas the right panel 
shows its central part. To better illustrate the previously identified tidal dwarf galaxies within HCG~31 (see section~\ref{sec:hcg31}), 
we present in Figure~\ref{fig:hcg31_optical_mom0} a higher-resolution view of the group's central region, derived from a data cube at 15.47\arcsec $\times$ 11.86\arcsec. 
  \subsection{3D visualisation}
\begin{figure*}
      \setlength{\tabcolsep}{0pt}
      \begin{tabular}{c c}
      \includegraphics[scale=0.345]{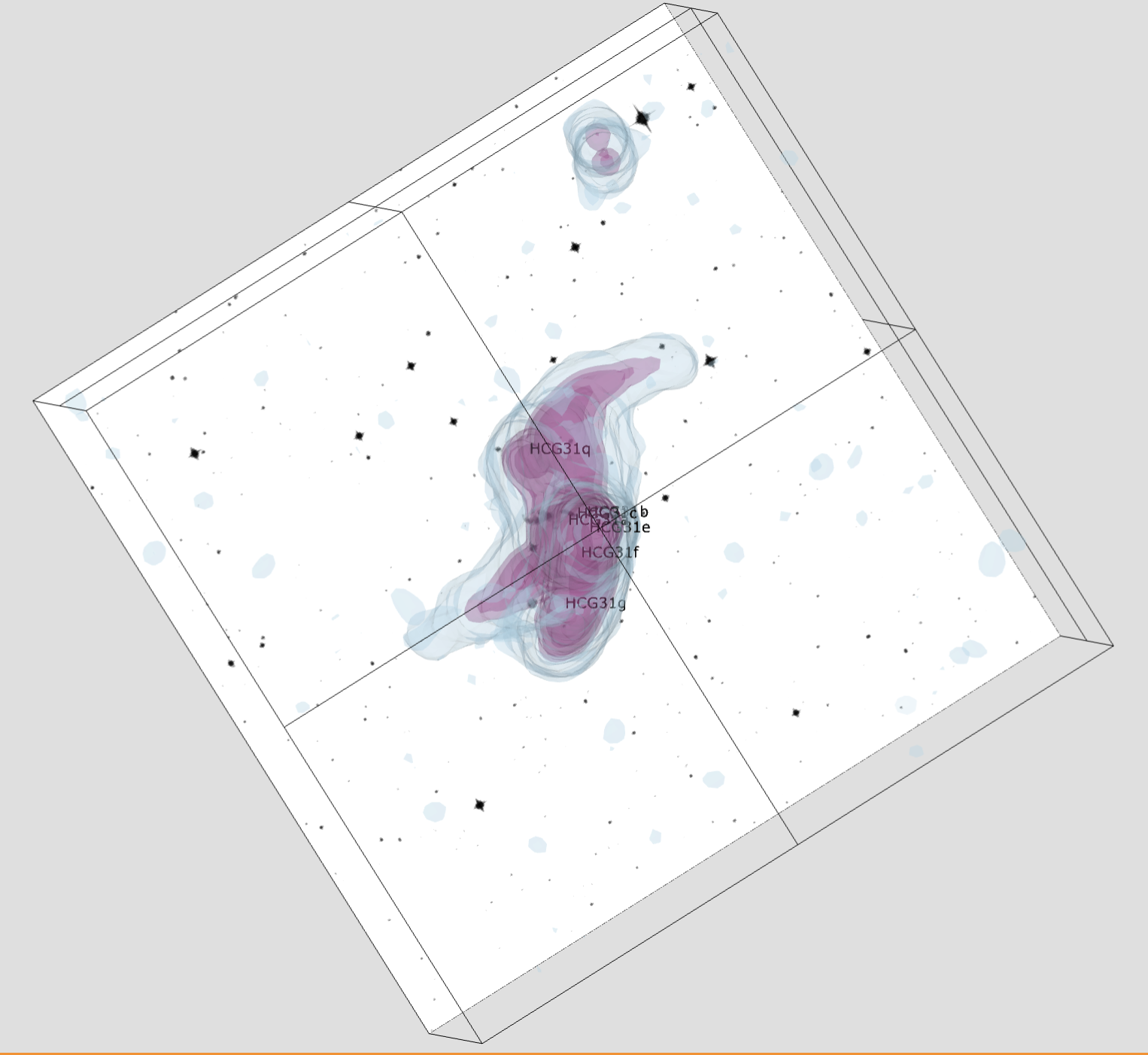} & 
      \includegraphics[scale=0.345]{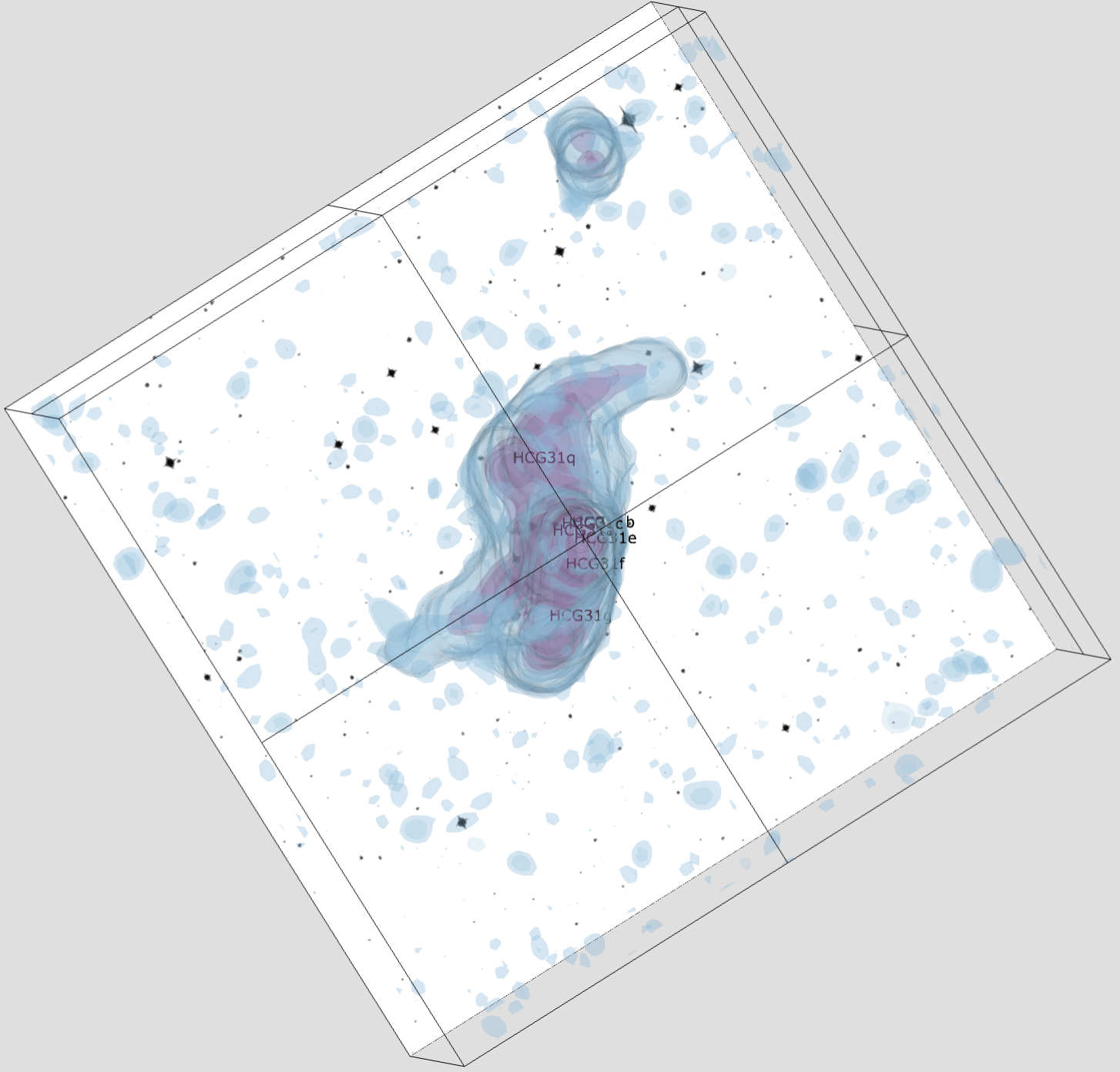}
      \end{tabular}
      \caption{3D visualisation of HCG 31. The left panel shows iso-surface level highlighting the high-column density gas. The right panel 
      showcases the low-column density \HI\ gas. The blue circles indicate the position of the member galaxies. 
      The 2D grayscale image is a DeCaLS R-band optical image of the group. 
      The online version of the cubes are available at \href{https://amiga.iaa.csic.es/x3d-menu/}{https://amiga.iaa.csic.es/x3d-menu/}.}
    \label{fig:hcg31_3dvis}
   \end{figure*}

We present in Figure~\ref{fig:hcg31_3dvis} a three-dimensional visualisation of HCG 31. The left panel displays iso-surfaces corresponding to high-column-density gas, emphasising the group’s densest structures, 
while the right panel highlights the diffuse low-column-density \HI\ gas. Blue circles mark the positions of the member galaxies. The 2D grayscale background is a 
DeCaLS R-band optical image of the group. An interactive version of these 3D cubes is publicly accessible at \href{https://amiga.iaa.csic.es/x3d-menu/}{https://amiga.iaa.csic.es/x3d-menu/}.
\section{Additional figures of HCG~91}  
\subsection{Noise properties and global profiles}
\begin{figure*}
  \begin{tabular}{l l l}
      \includegraphics[scale=0.215]{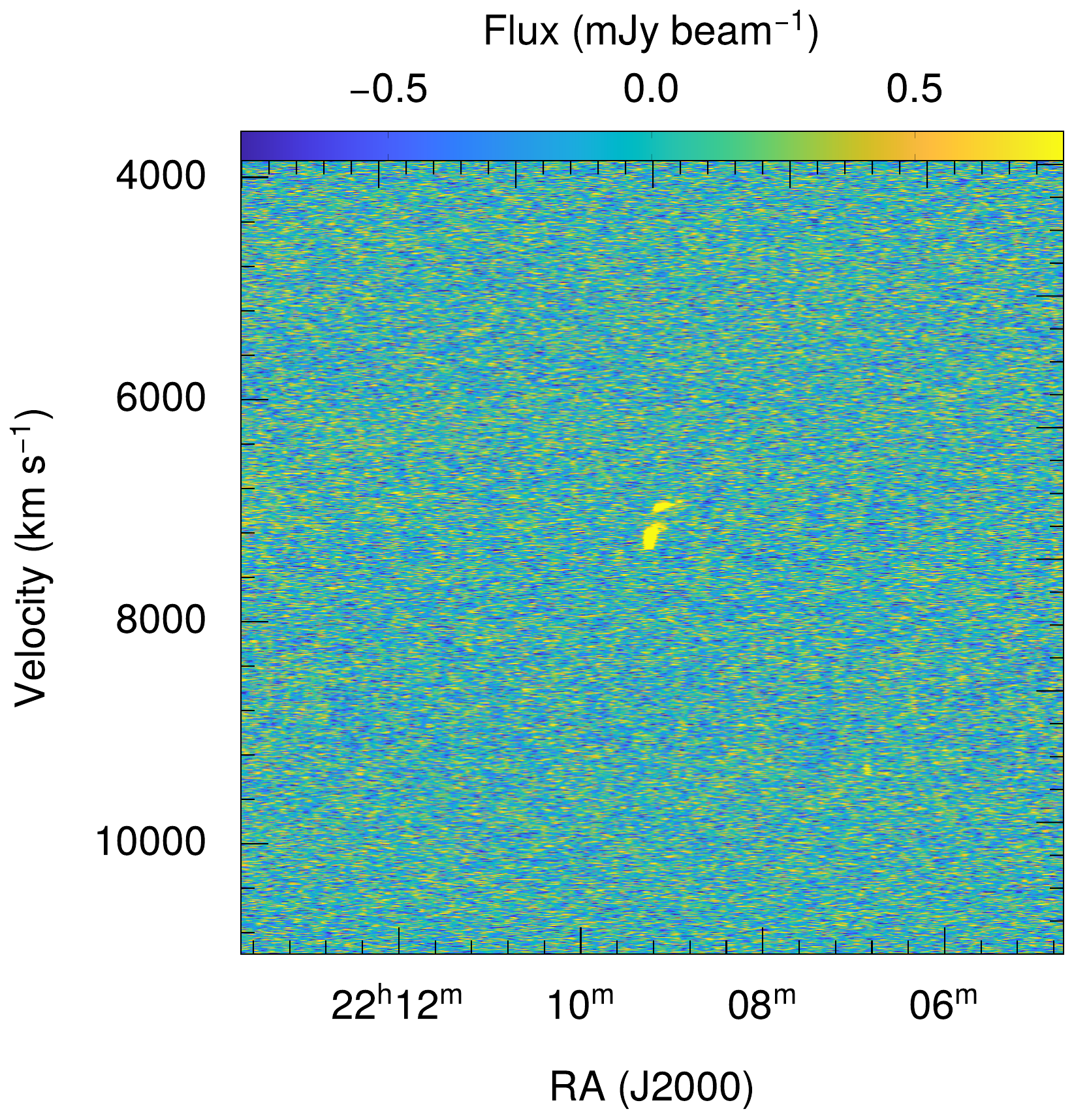} &
      \includegraphics[scale=0.215]{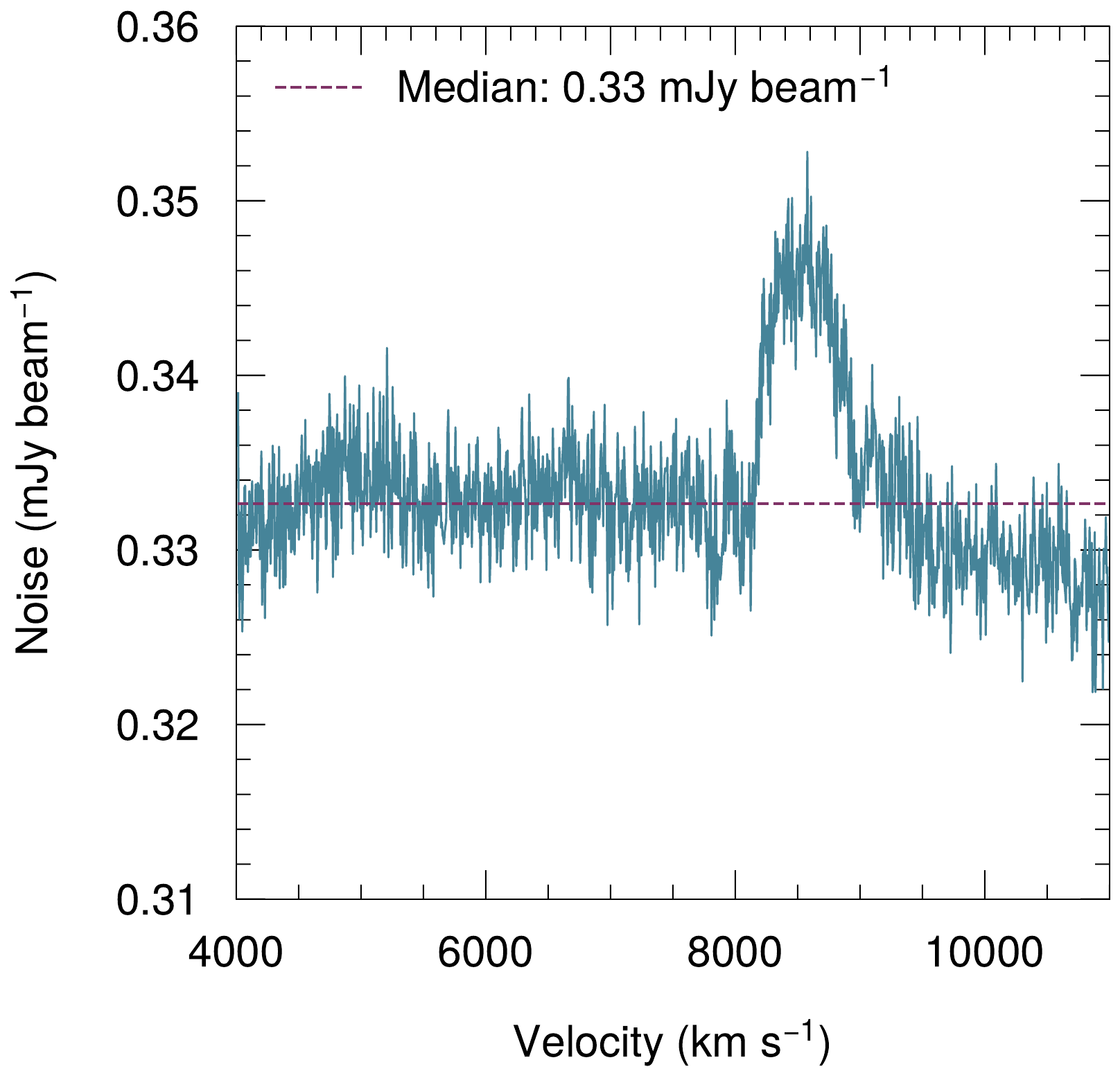}&
      \includegraphics[scale=0.215]{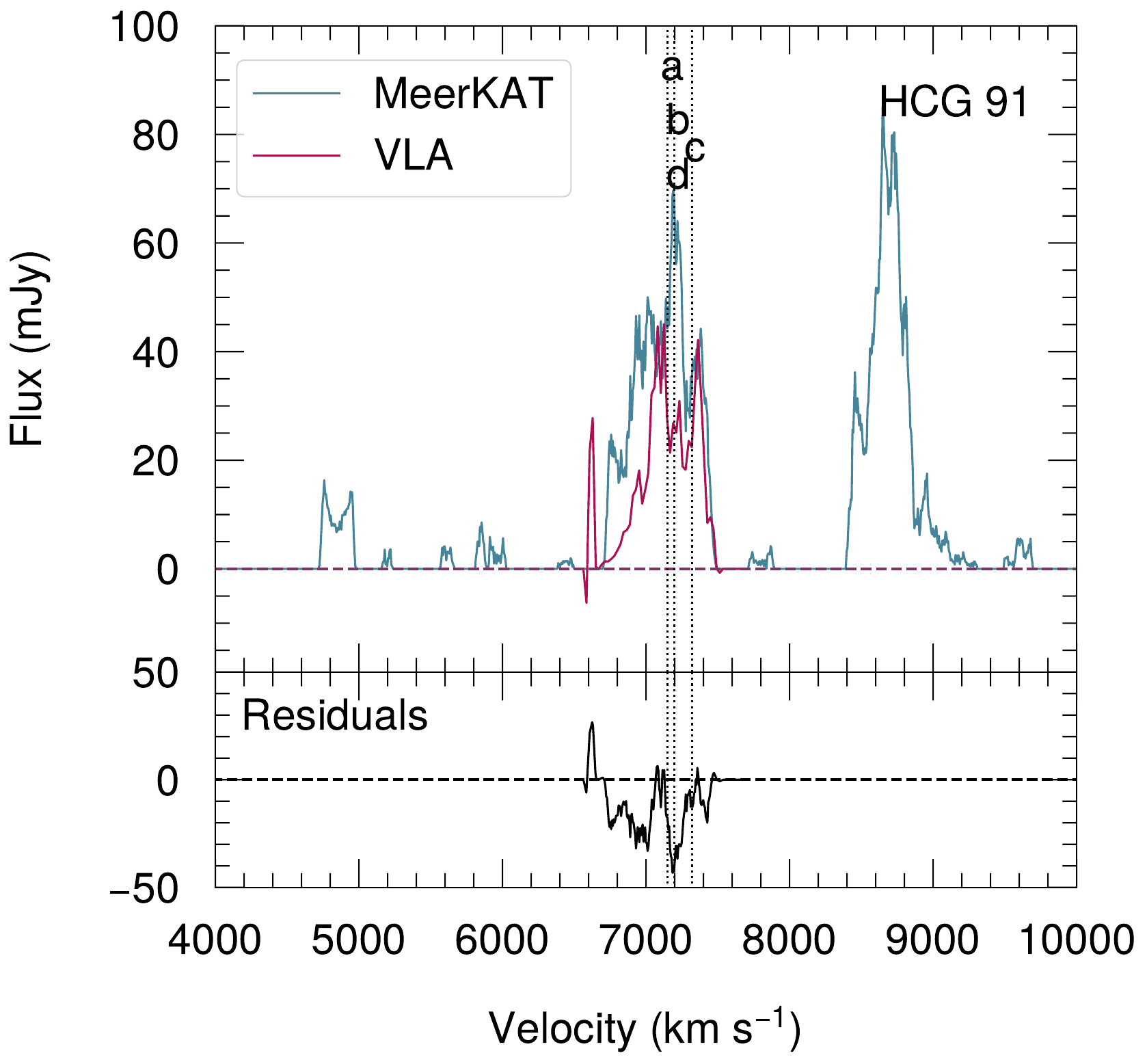}  
    \end{tabular}
    \caption{Left panel: velocity vs right ascension of HCG~91. Middle panel: median noise values of each RA-DEC slice of the non-primary beam corrected 60\arcsec\ data cube of 
    HCG~91 as a function of velocity. The horizontal dashed line indicates the median of all the noise values from each slice. Right panel: the blue solid lines indicates the 
    MeerKAT integrated spectrum of HCG~91; the red solid line indicates VLA integrated spectrum of the group derived by \citep{2023A&A...670A..21J}. The vertical dotted lines indicate the velocities of the galaxies in the core of the group. 
    The spectra have been extracted from areas containing only genuine \HI\ emission. }
    \label{fig:hcg91_noise}
   \end{figure*}

The RA–velocity plot of HCG~91 is shown in the first panel of Figure~\ref{fig:hcg91_noise}. 
The middle panel shows the median noise levels in each RA–DEC slice of the non primary beam corrected data cube as a function of velocity. The horizontal 
dashed line indicates the global median noise. The right panel compares the MeerKAT \HI\ integrated spectra with archival VLA measurements 
from \citep{2023A&A...670A..21J}, where the vertical dotted lines denote the systemic velocities of the galaxies in the group’s core. 
All spectra were extracted from regions containing only genuine \HI\ emission.
  %
  
  \subsection{Channel maps}  
\begin{figure*}
      \setlength{\tabcolsep}{0pt}
      \begin{tabular}{l l l}
          \includegraphics[scale=0.25]{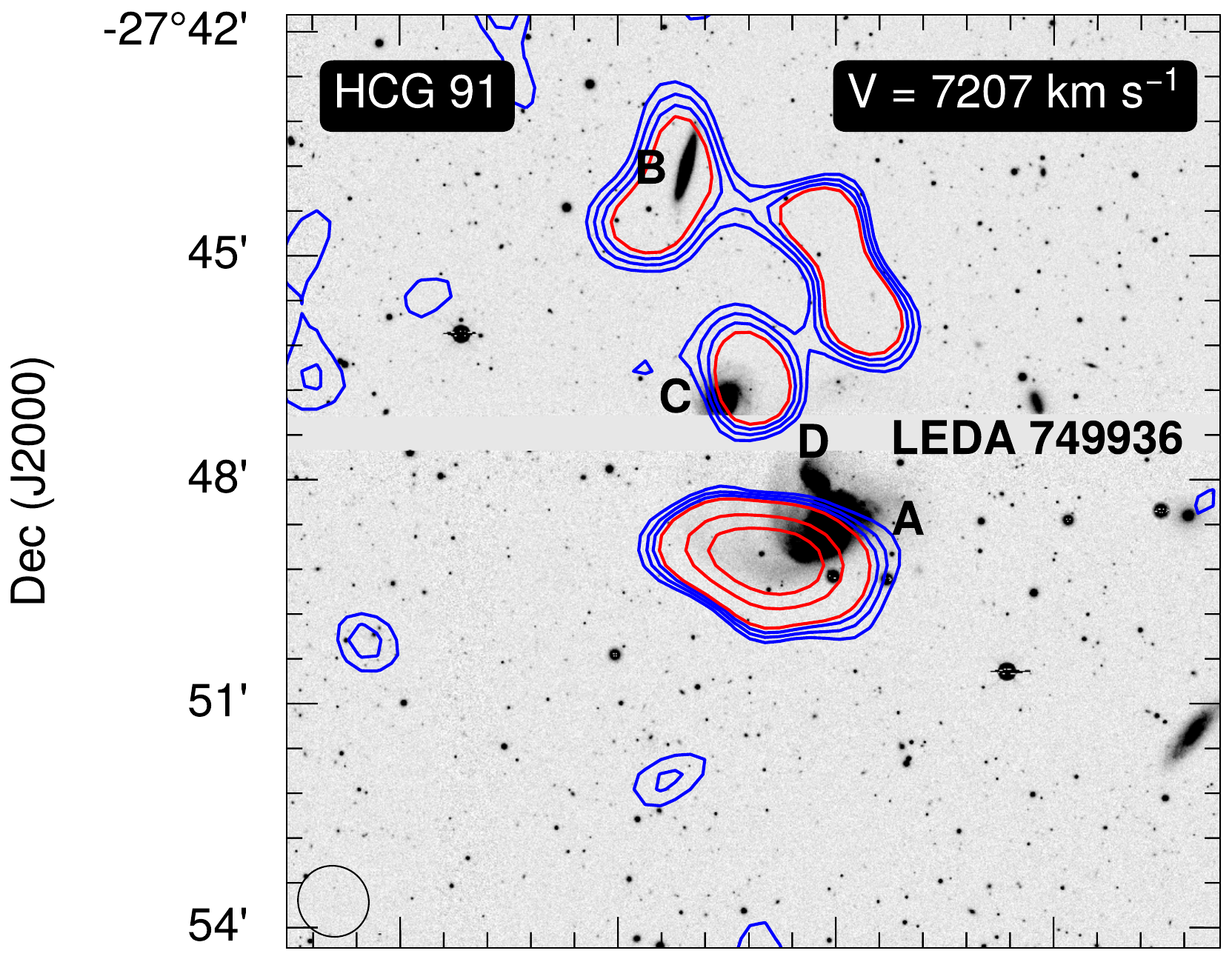} &
          \includegraphics[scale=0.25]{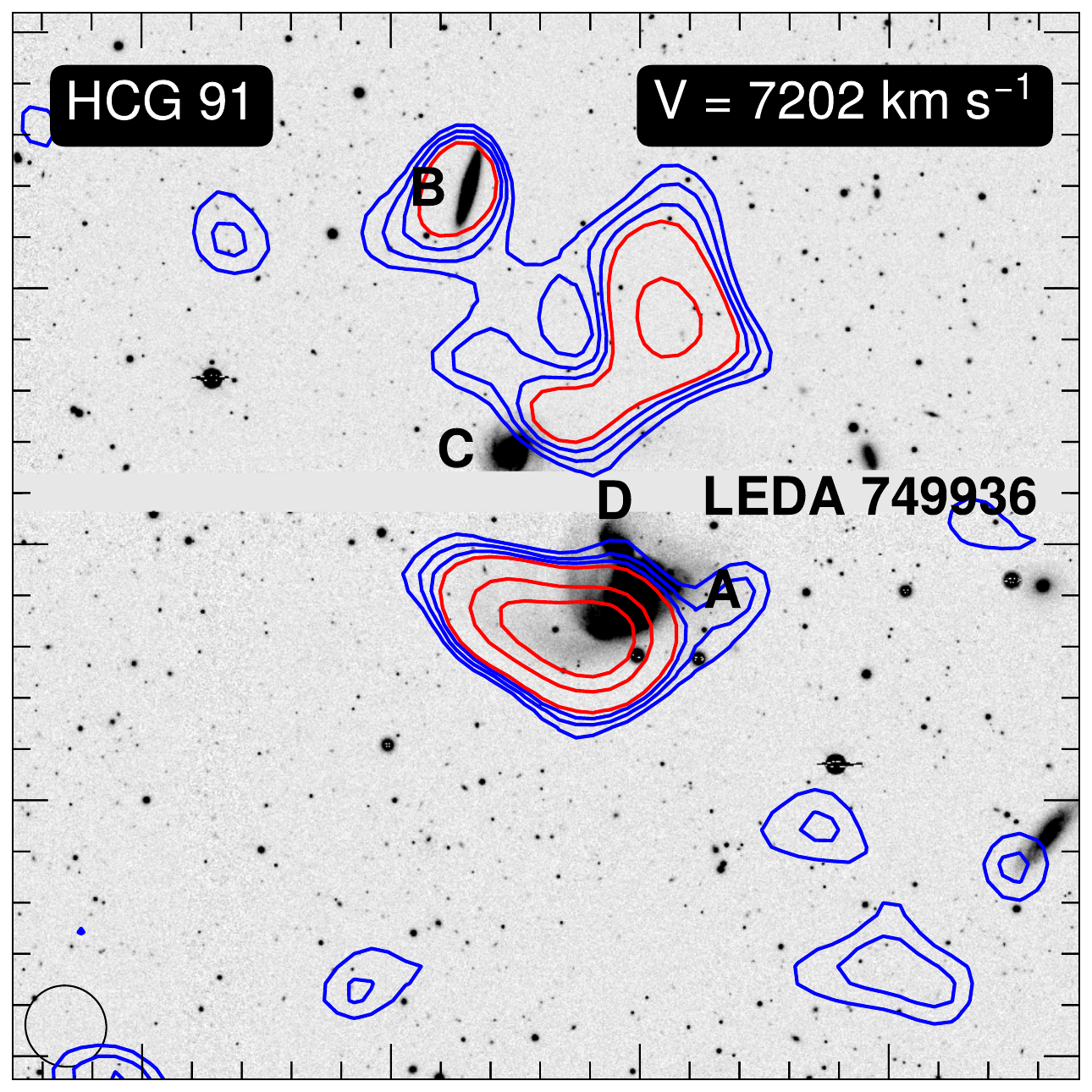} &
          \includegraphics[scale=0.25]{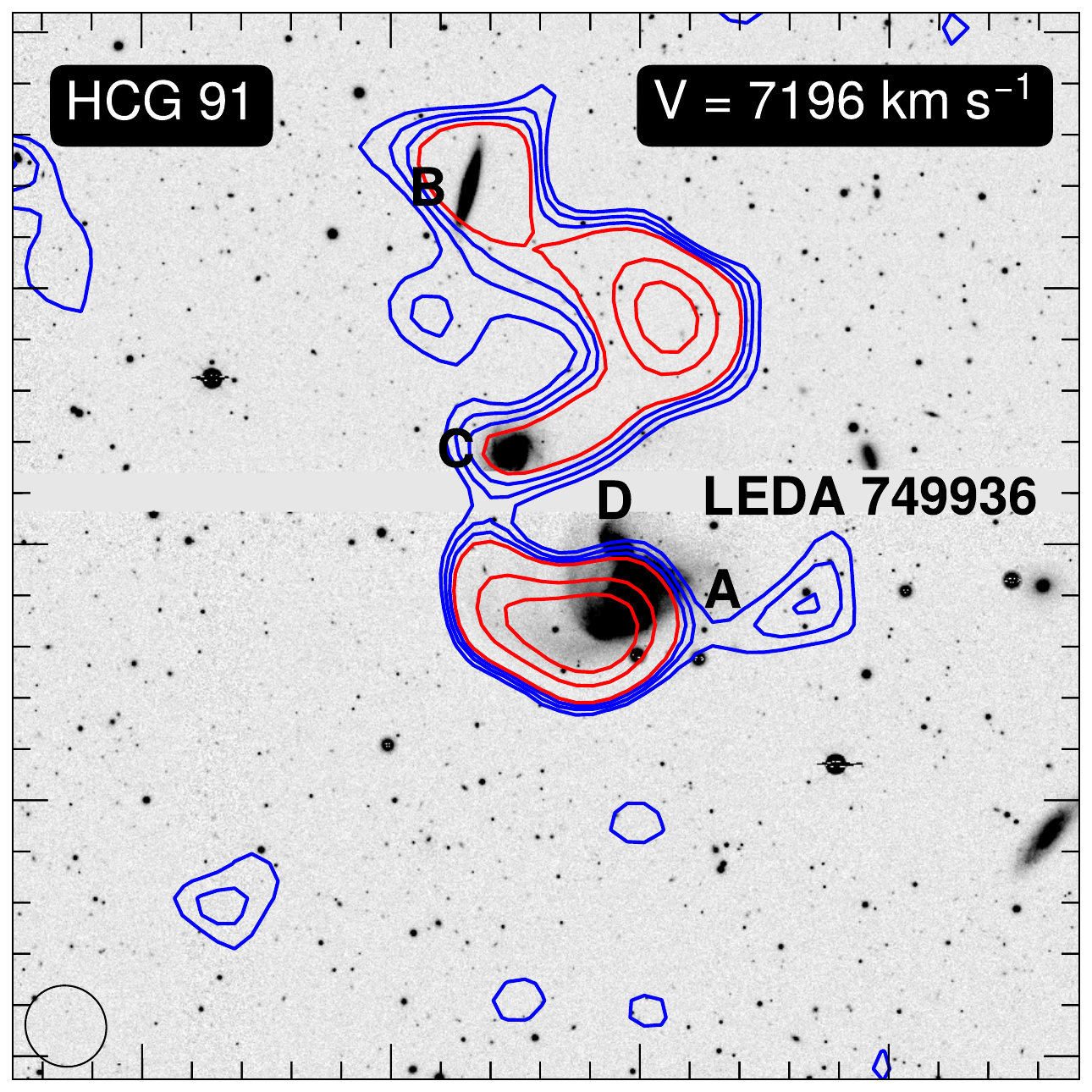} \\[-0.2cm]
          \includegraphics[scale=0.25]{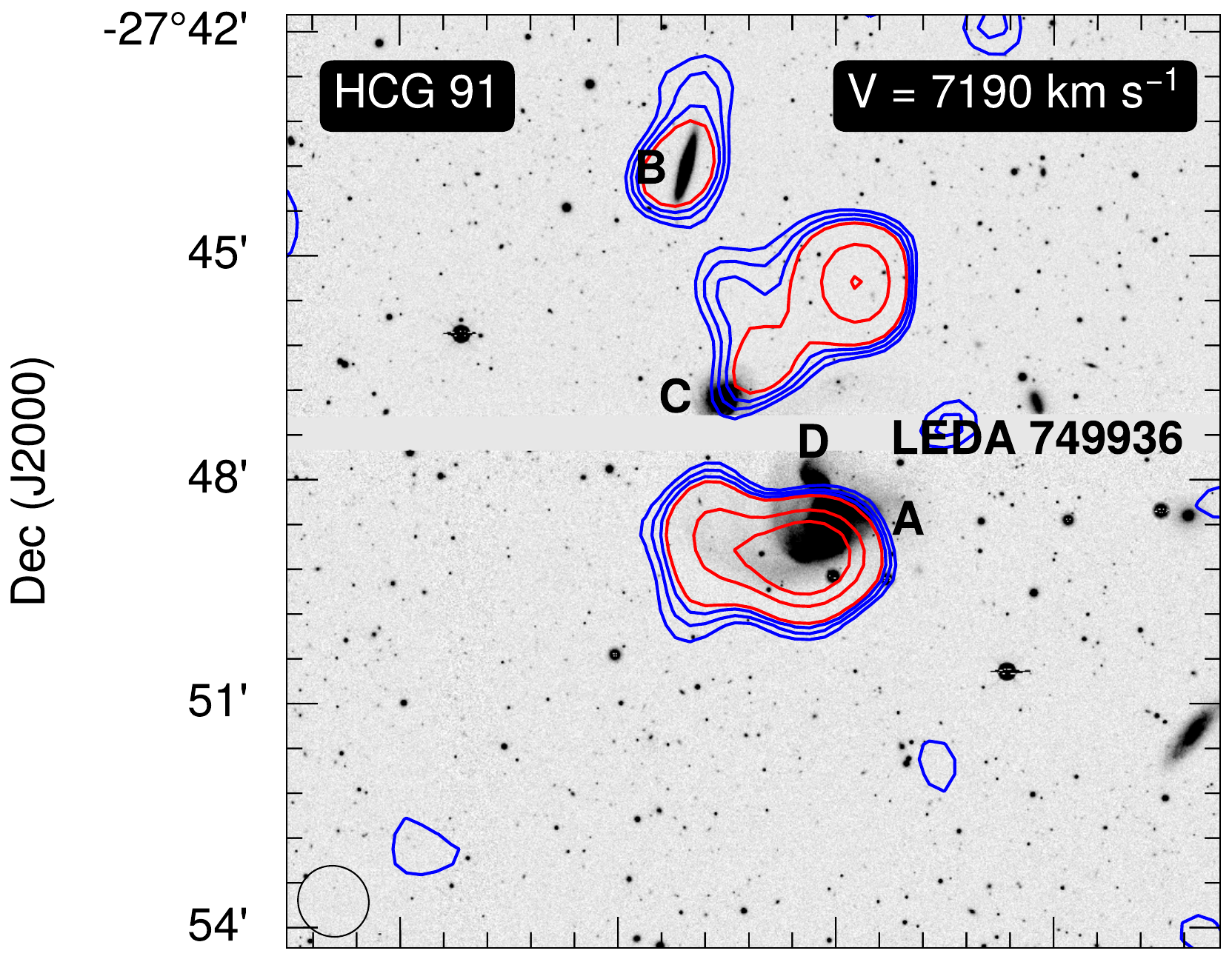} &
          \includegraphics[scale=0.25]{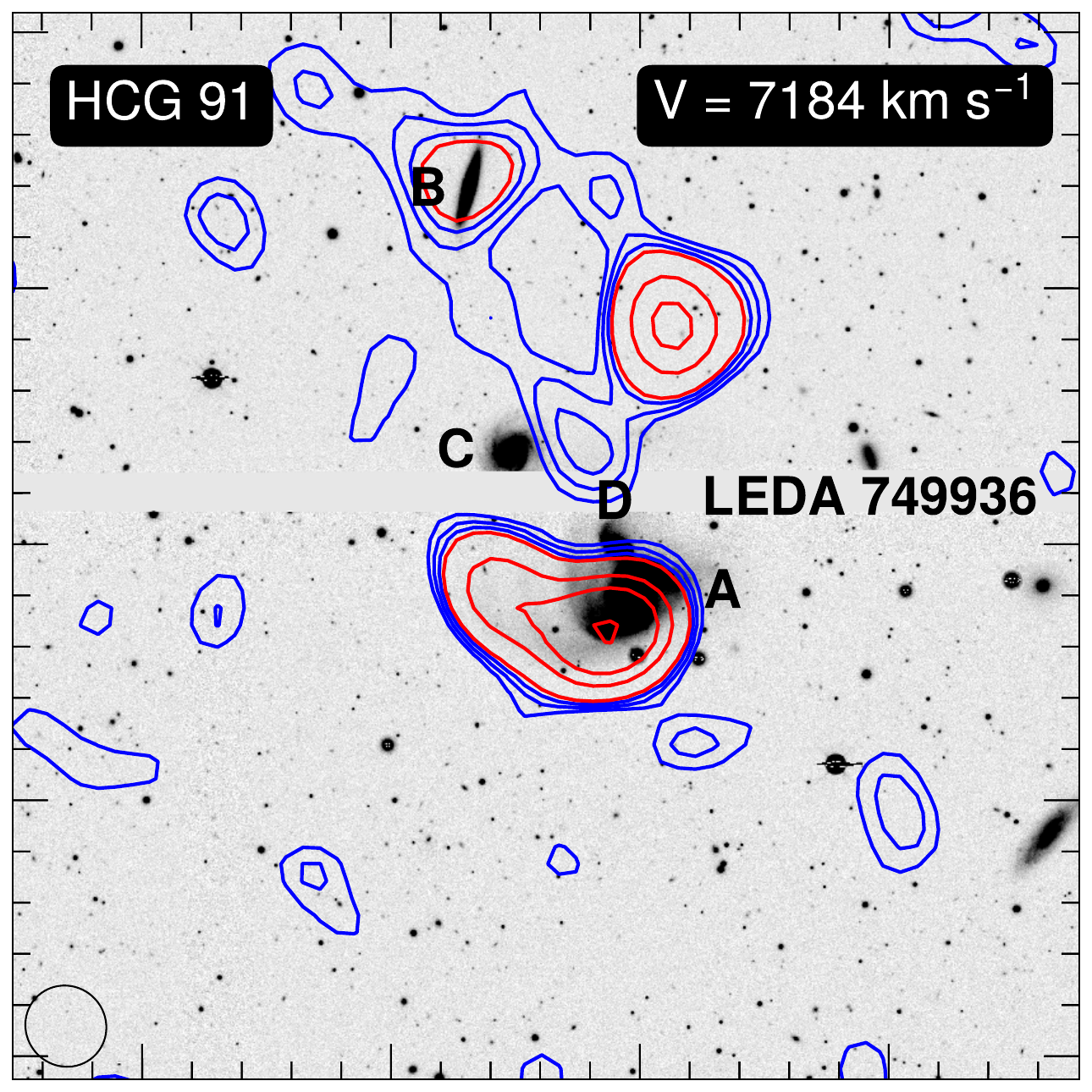} &
          \includegraphics[scale=0.25]{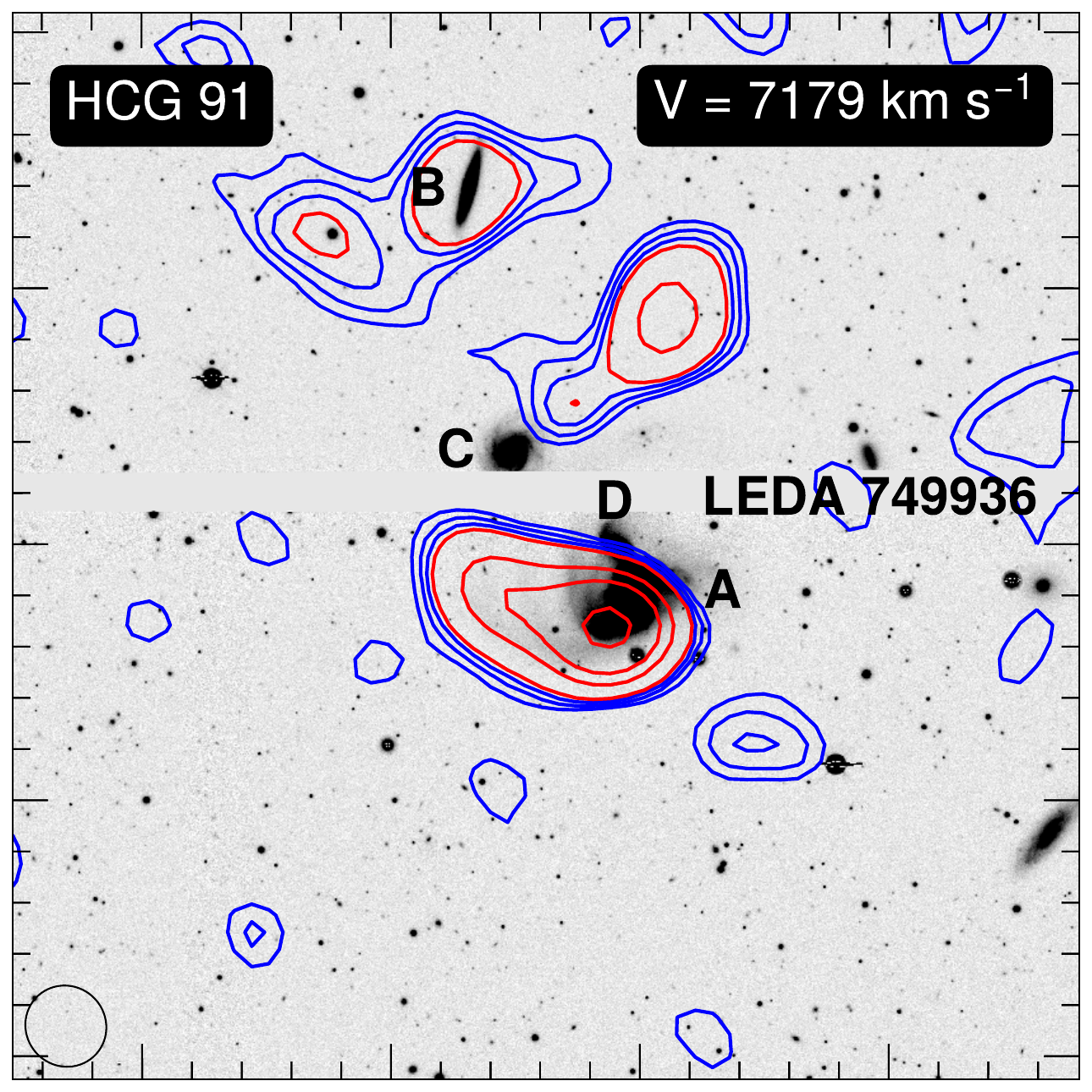} \\[-0.2cm]
          \includegraphics[scale=0.25]{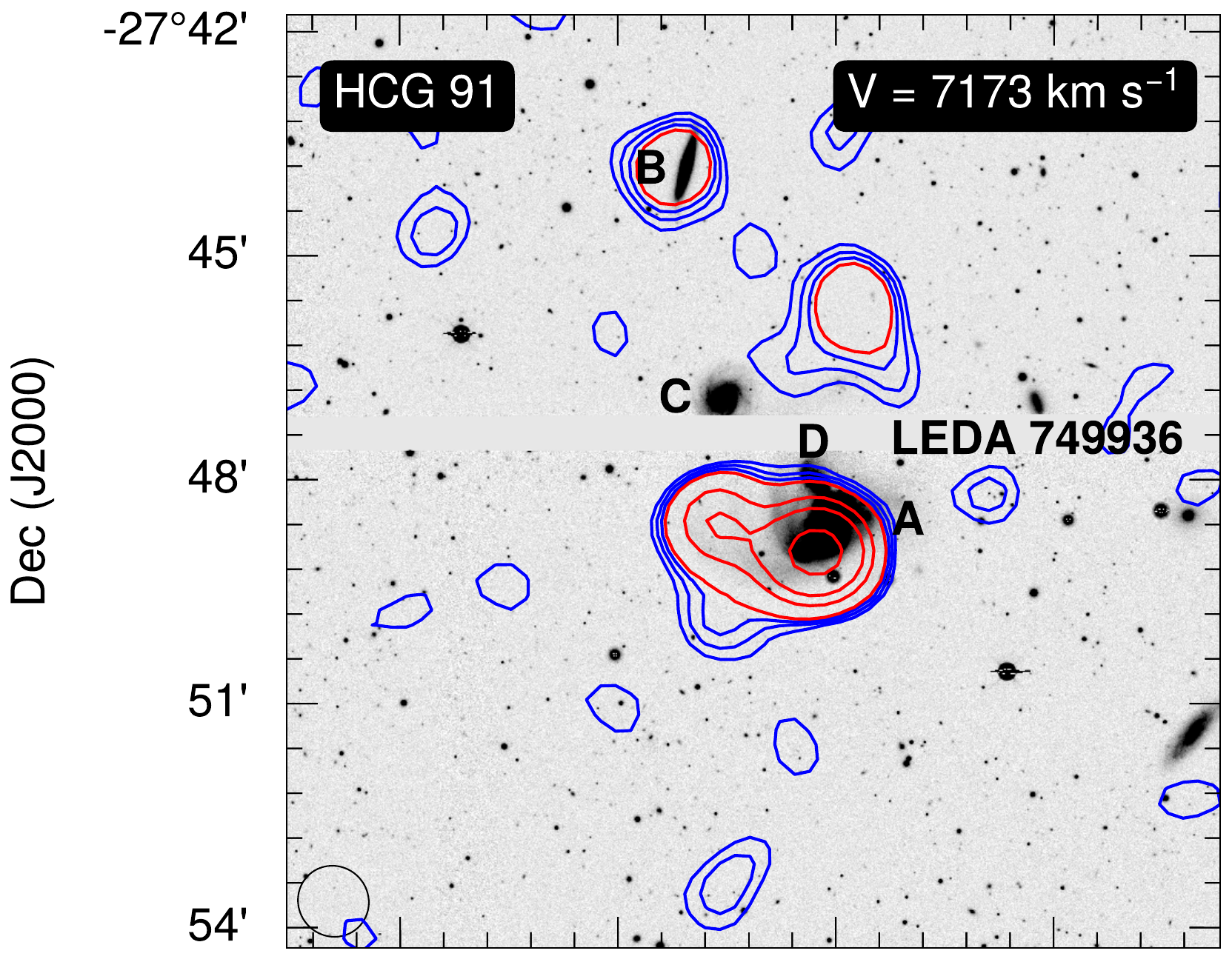} &
          \includegraphics[scale=0.25]{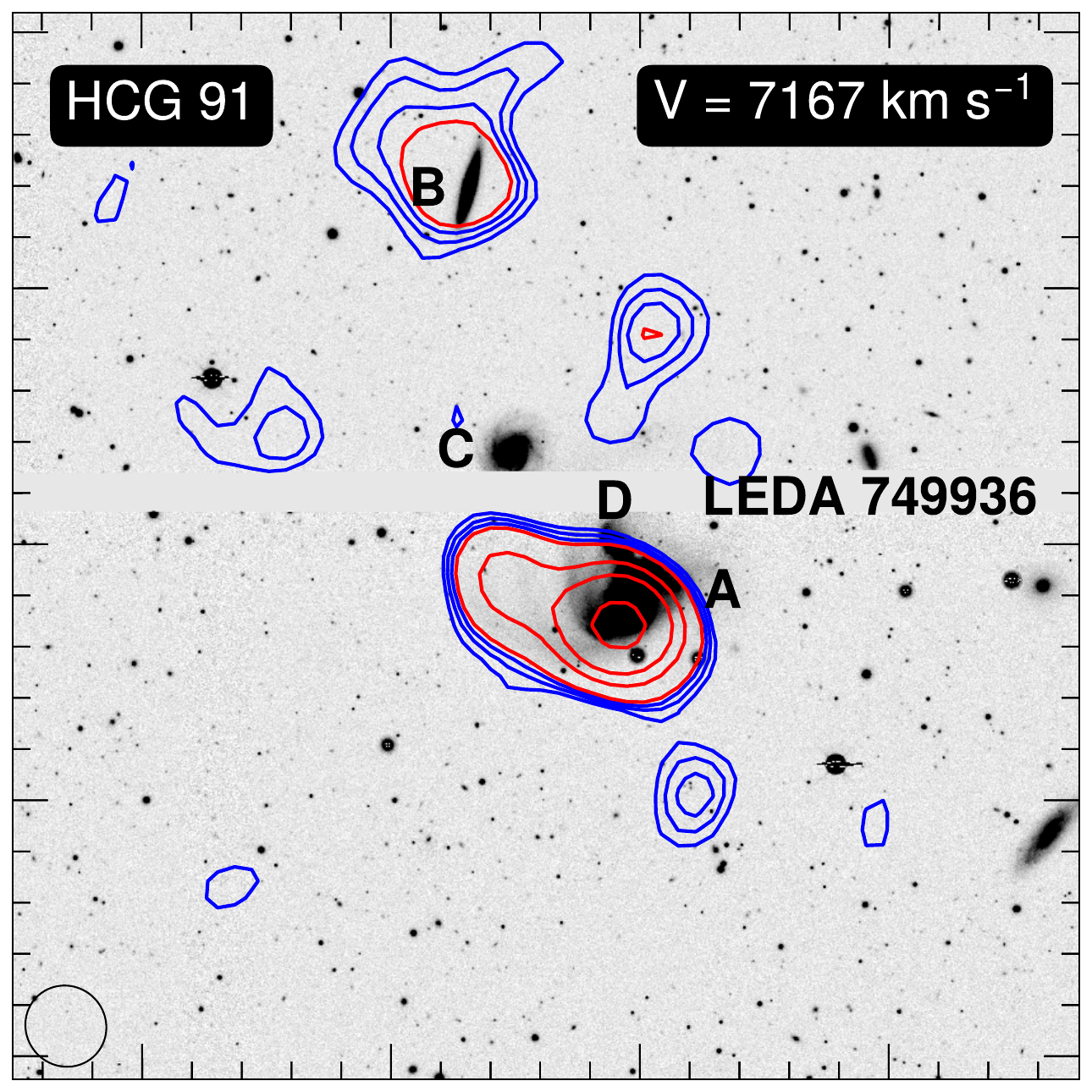} &
          \includegraphics[scale=0.25]{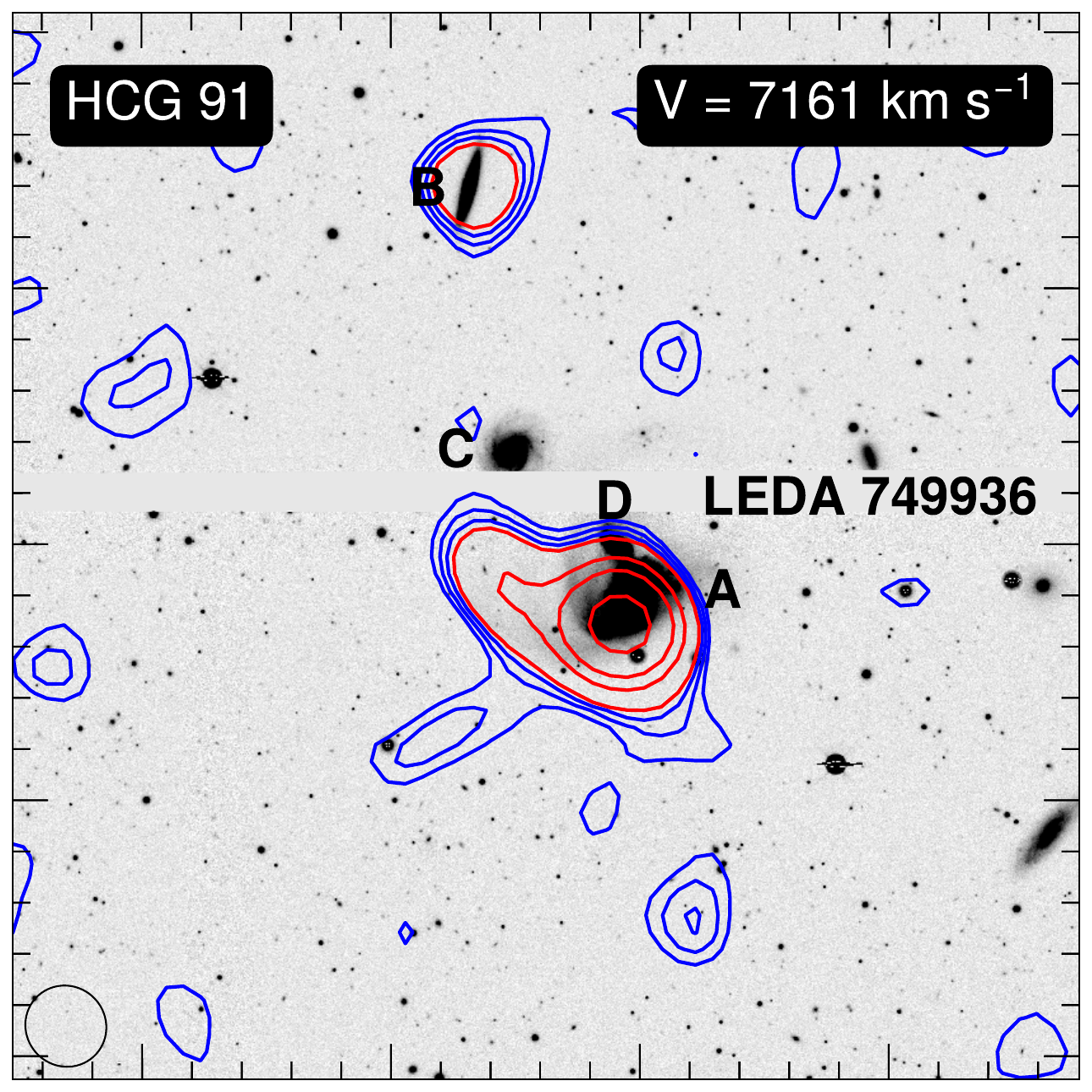} \\[-0.2cm]
          \includegraphics[scale=0.25]{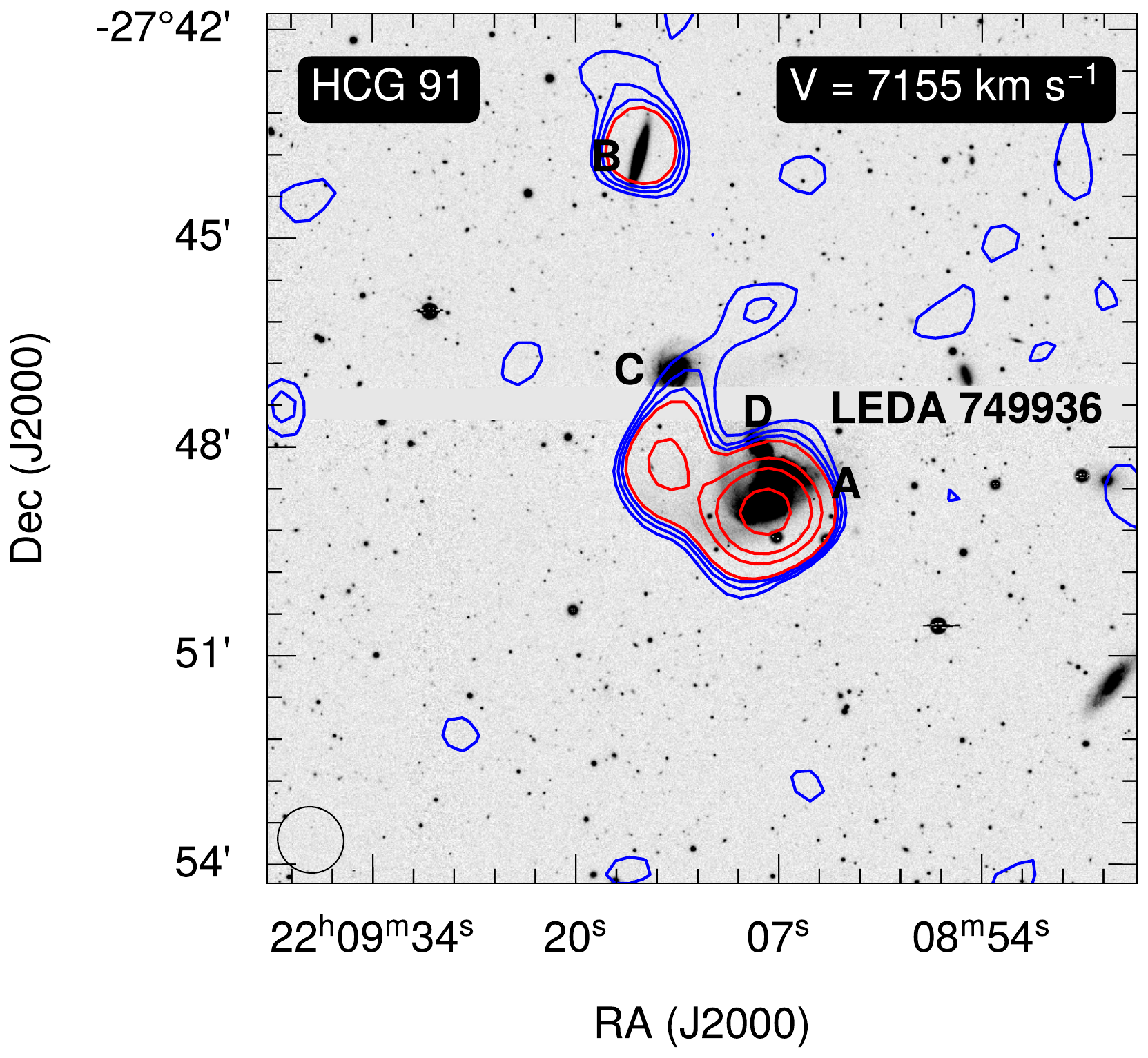} &
          \includegraphics[scale=0.25]{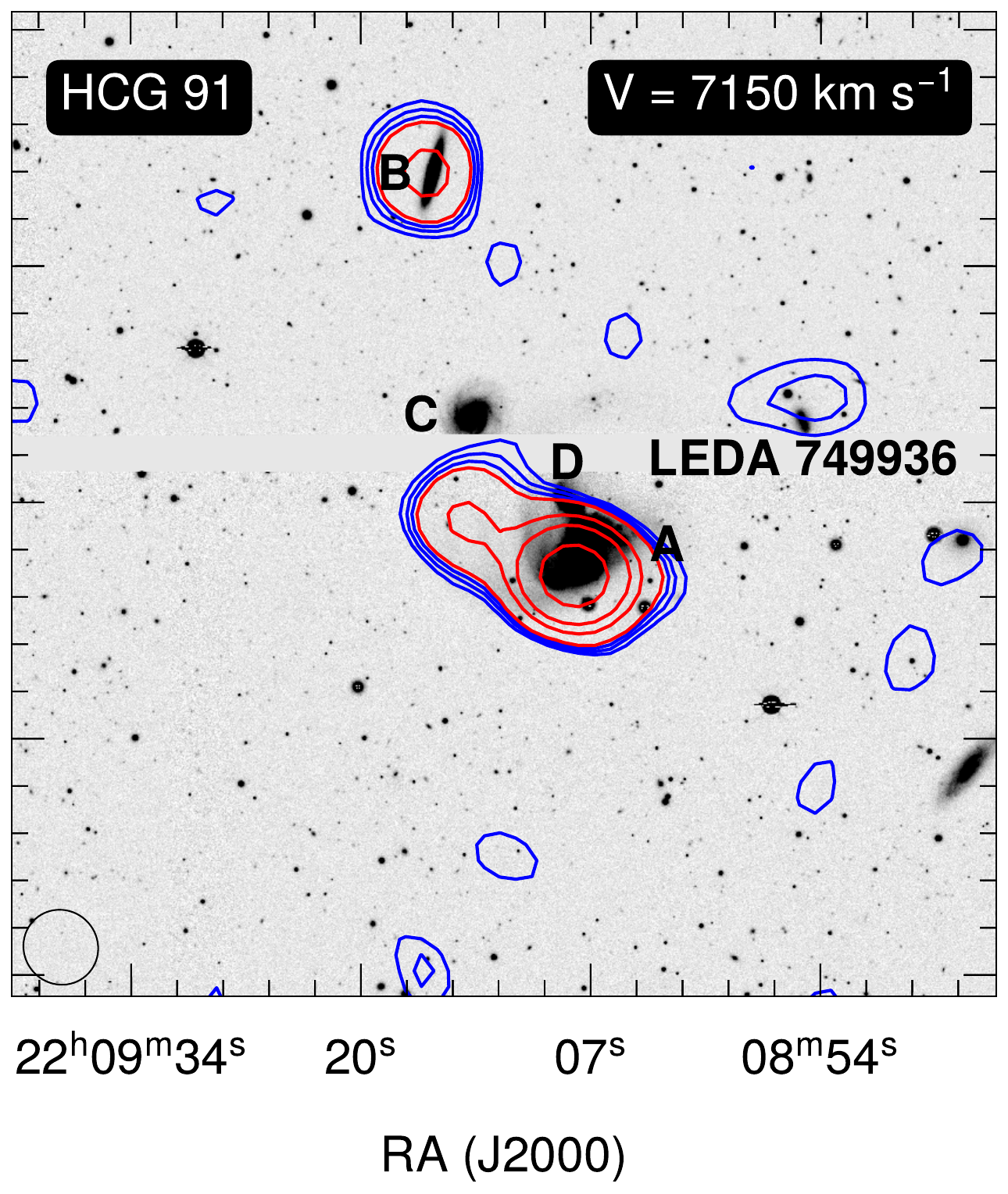} &
          \includegraphics[scale=0.25]{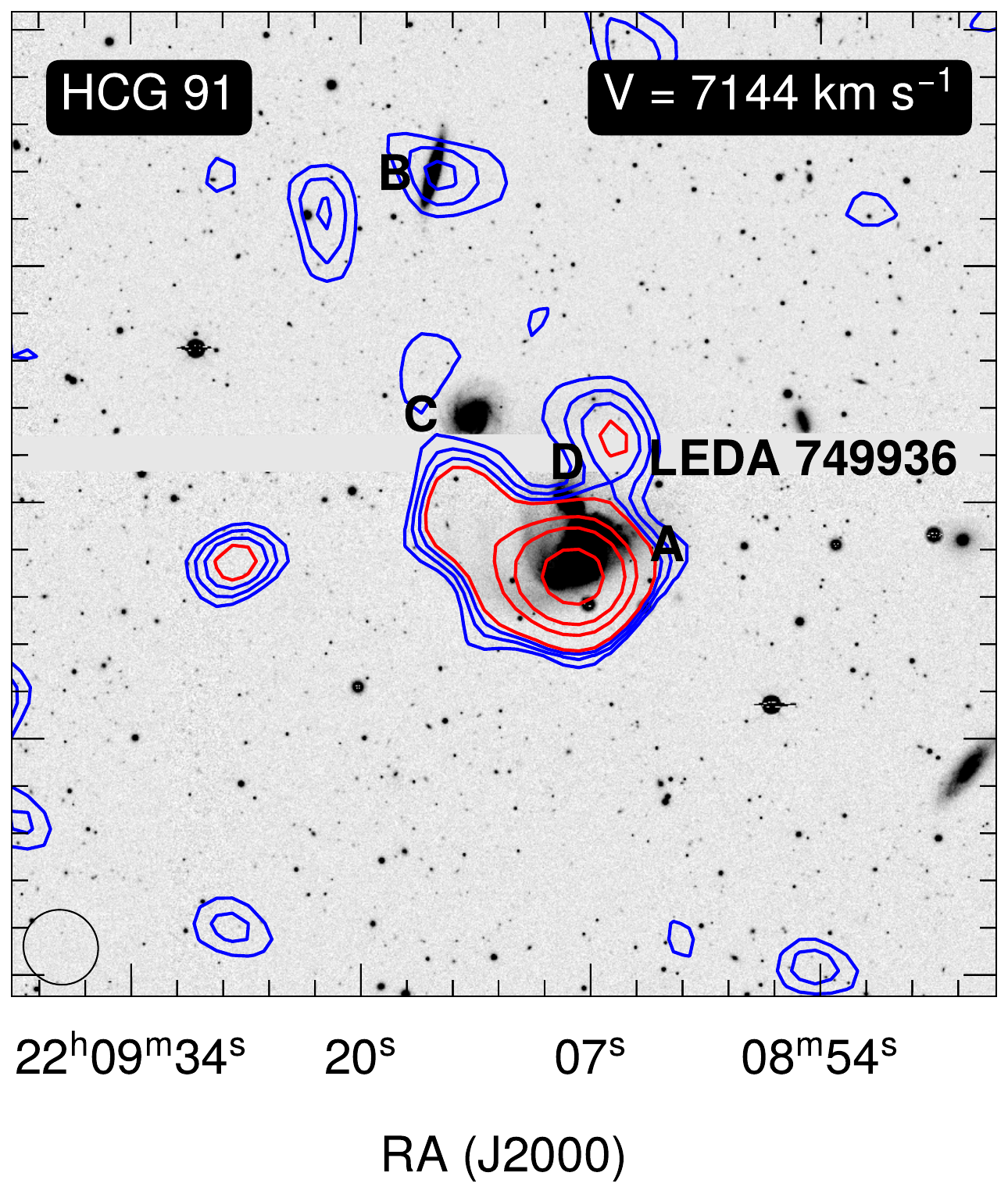} 
        \end{tabular}
        \caption{Example channel maps of the primary beam corrected cube of HCG~91 overlaid on DECaLS DR10 I-band optical image. Contour levels are (1.5, 2,  2.5, 3, 6, 9, 16, 32) 
        times the median noise level in the cube (0.69 $\mathrm{mJy~beam{-1}}$). The blue colours show contour levels below 3$\sigma$; the red colours represent contour levels at 3$\sigma$, or higher. 
        Additional channel maps can be downloaded \href{https://zenodo.org/records/14856489}{here}.}
        \label{fig:hcg91_chanmap}
   \end{figure*}

  Figure~\ref{fig:hcg91_chanmap} presents the channel maps of HCG~91 overlaid on DECaLS DR10 I-band optical image. More channel maps, covering a larger velocity range, can be found 
  \href{https://zenodo.org/records/14856489}{here}. 
\subsection{Moment maps} 
\begin{figure*}
  \begin{tabular}{l l}
      \includegraphics[scale=0.27]{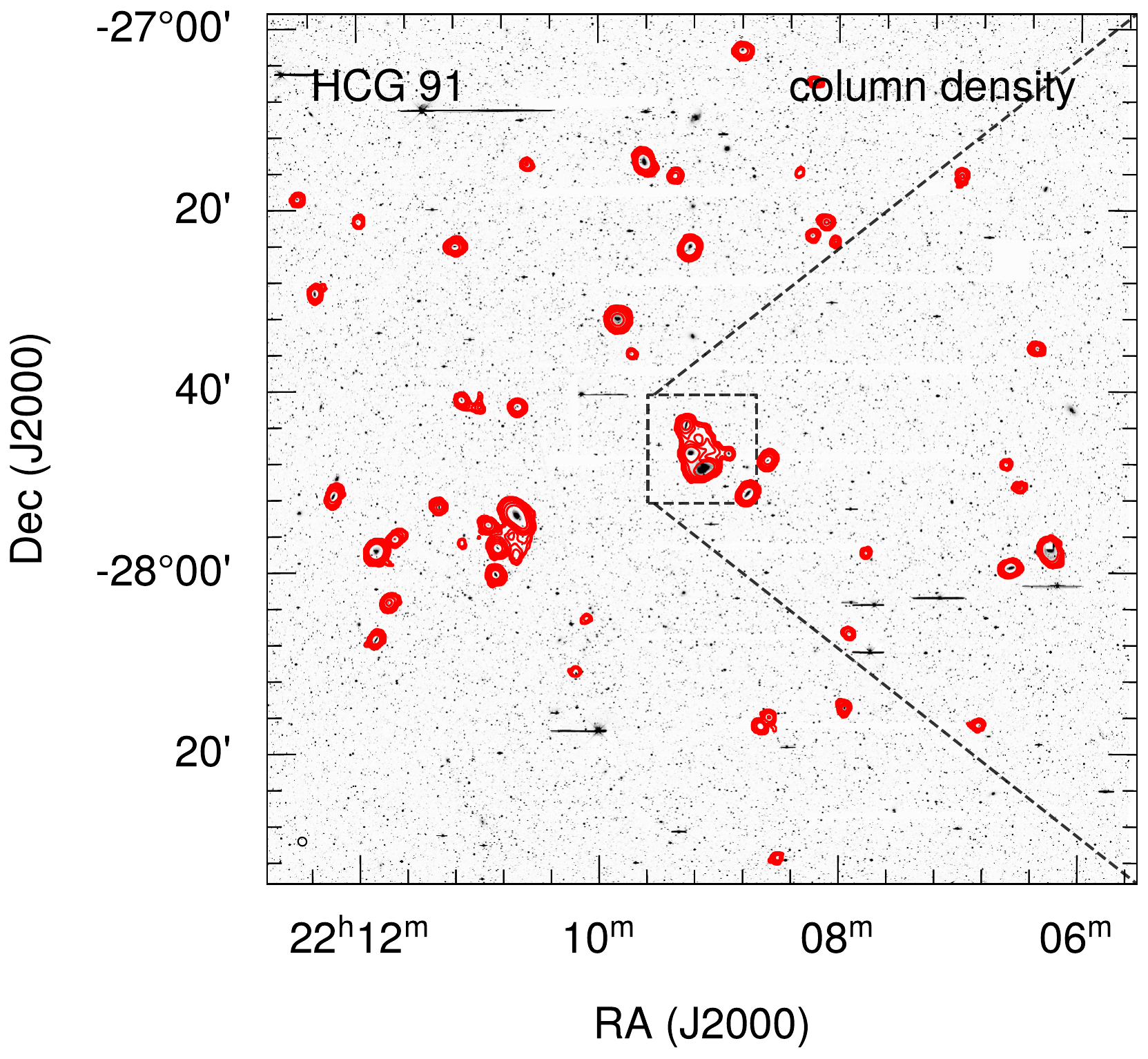}
      & \includegraphics[scale=0.27]{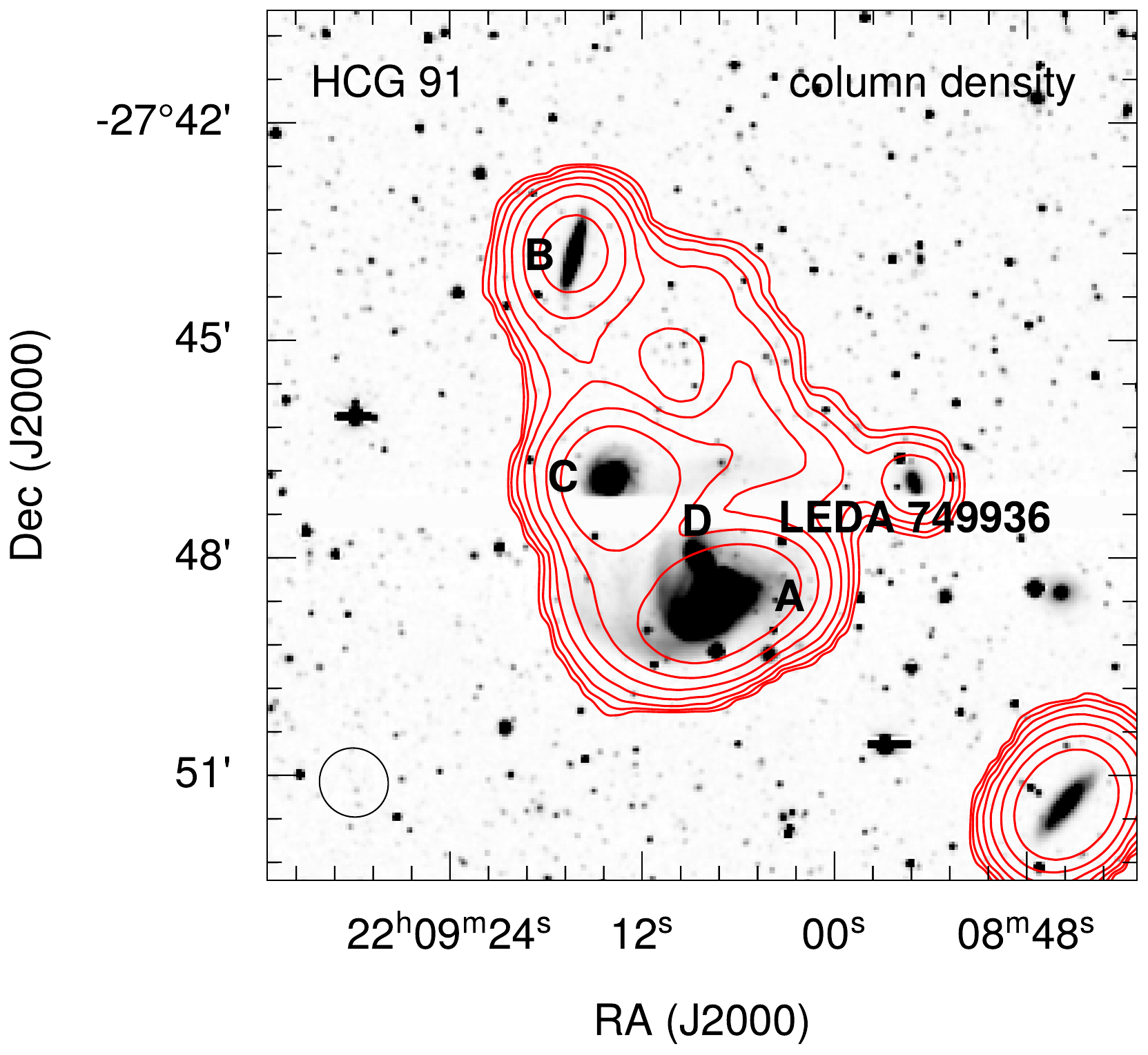}\\
      \includegraphics[scale=0.27]{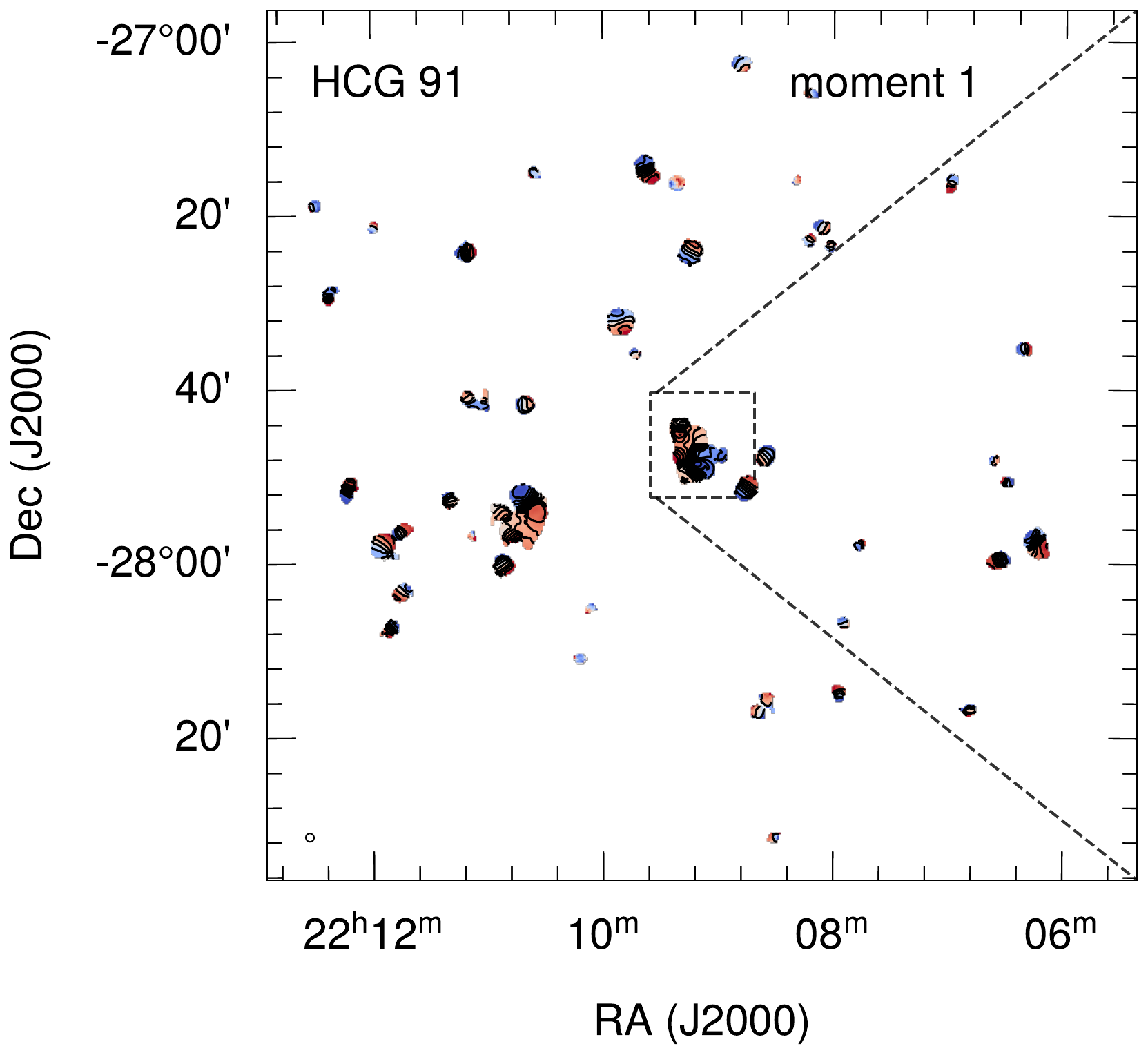} &
      \includegraphics[scale=0.27]{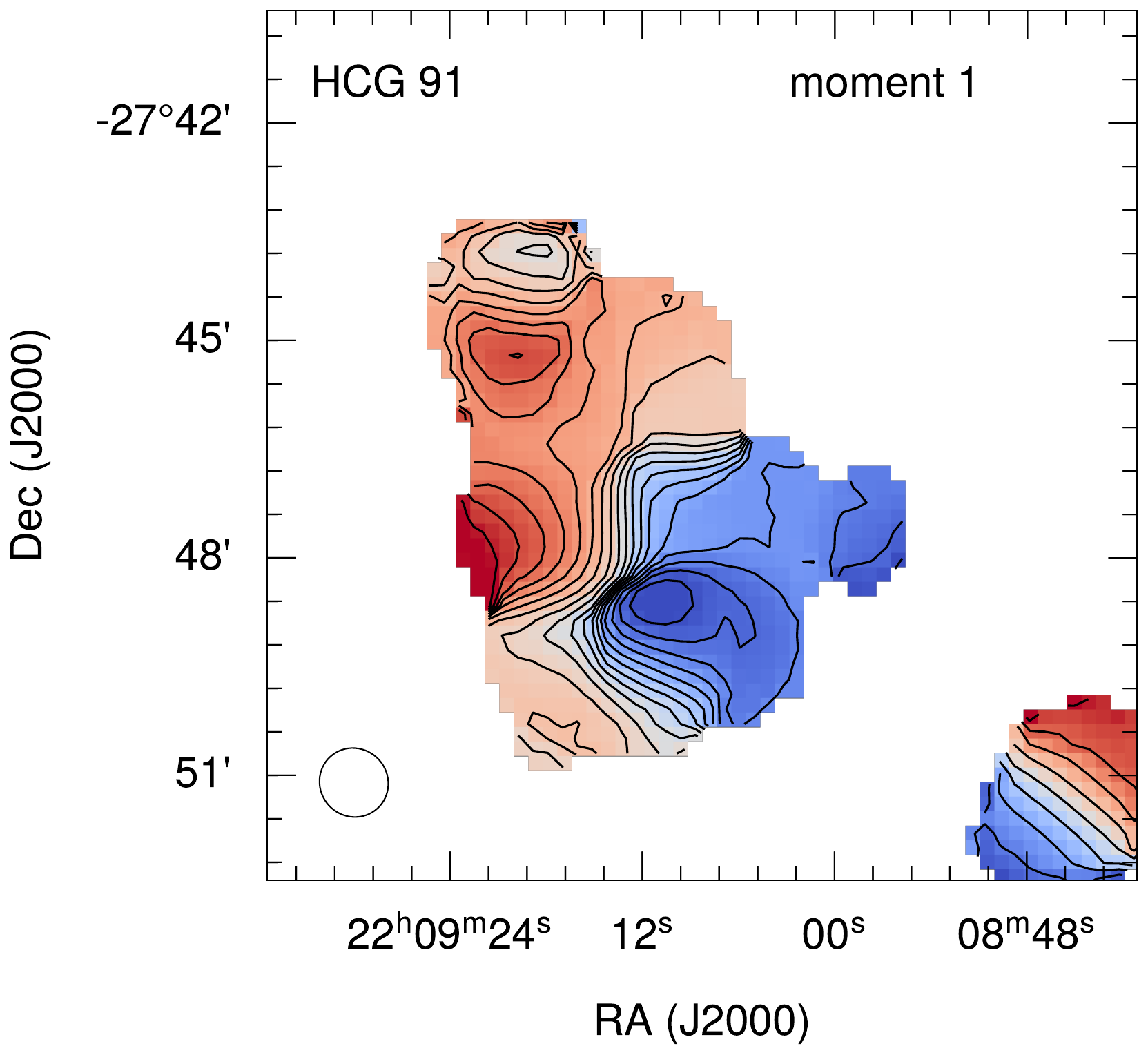}
  \end{tabular}
  \caption{\HI\ Moment maps of HCG 91. Left panels show all sources detected by SoFiA. The right panels show sources within the rectangular box shown on the 
  left to better show the central part of the group. The top panels show the column density maps with contour levels of
          ($\mathrm{4.6~\times~10^{18}}$, $\mathrm{9.2~\times~10^{18}}$, $\mathrm{1.8~\times~10^{19}}$, $\mathrm{3.7~\times~10^{19}}$, 
          $\mathrm{7.4~\times~10^{19}}$, $\mathrm{1.5~\times~10^{20}}$, $\mathrm{3.0~\times~10^{20}}$) $\mathrm{cm^{-2}}$. The contours 
          are overlaid on DECaLS DR10 I-band optical images. The bottom panels show the moment one map. Each individual source has its own colour 
          scaling and contour levels to highlight any rotational component.}
  \label{fig:hcg91_mom}
  \end{figure*}

The column density maps of HCG~91 is shown in Figure~\ref{fig:hcg91_mom}. The left panels show all detected sources within the field of view of MeerKAT. The right panels 
present the central part of the group. The \HI\ maps are overlaid on DECaLS DR10 I-band optical images.    
\subsection{Velocity fields of the core members of HCG~91}  
\begin{figure*}
  \begin{tabular}{c c}
      \includegraphics[scale=0.44]{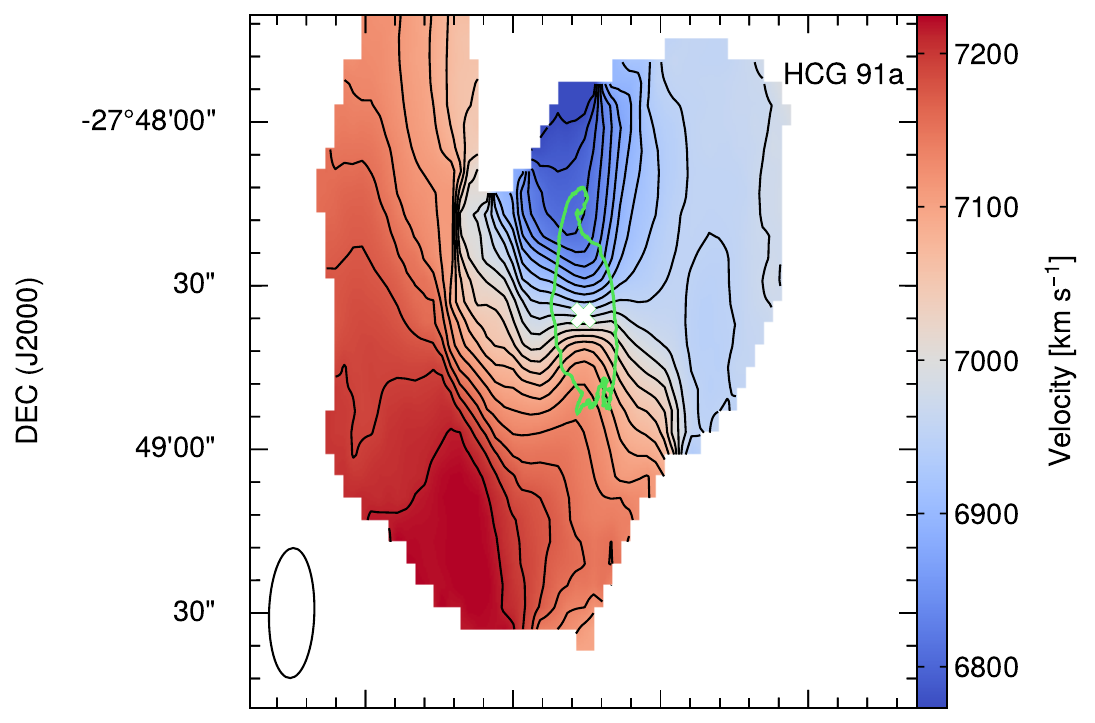} &
      \includegraphics[scale=0.44]{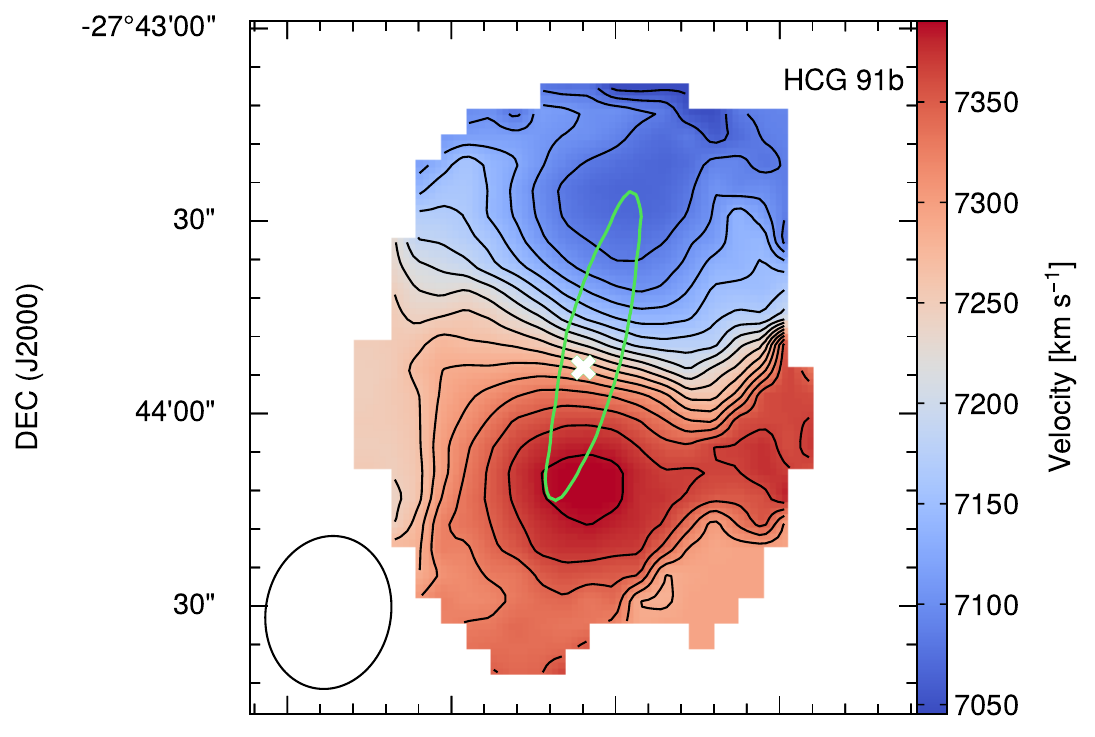} \\[-0.2cm]
      \includegraphics[scale=0.44]{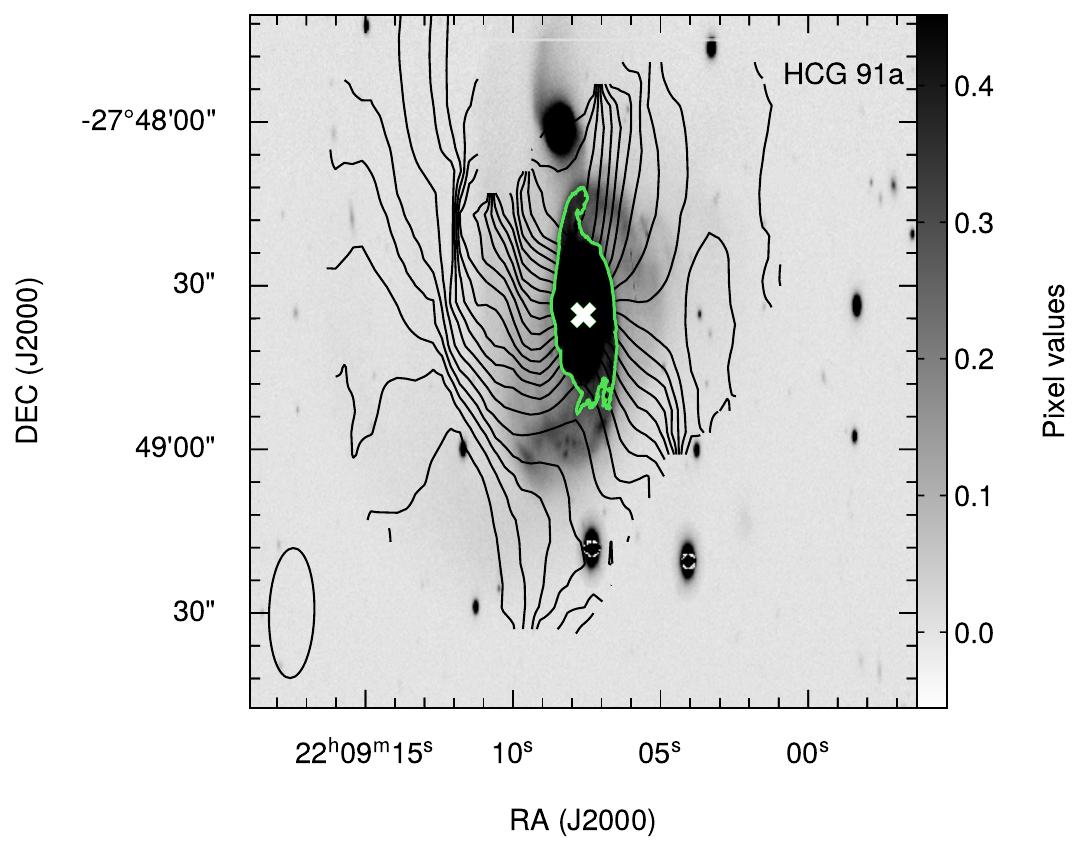} &
      \includegraphics[scale=0.44]{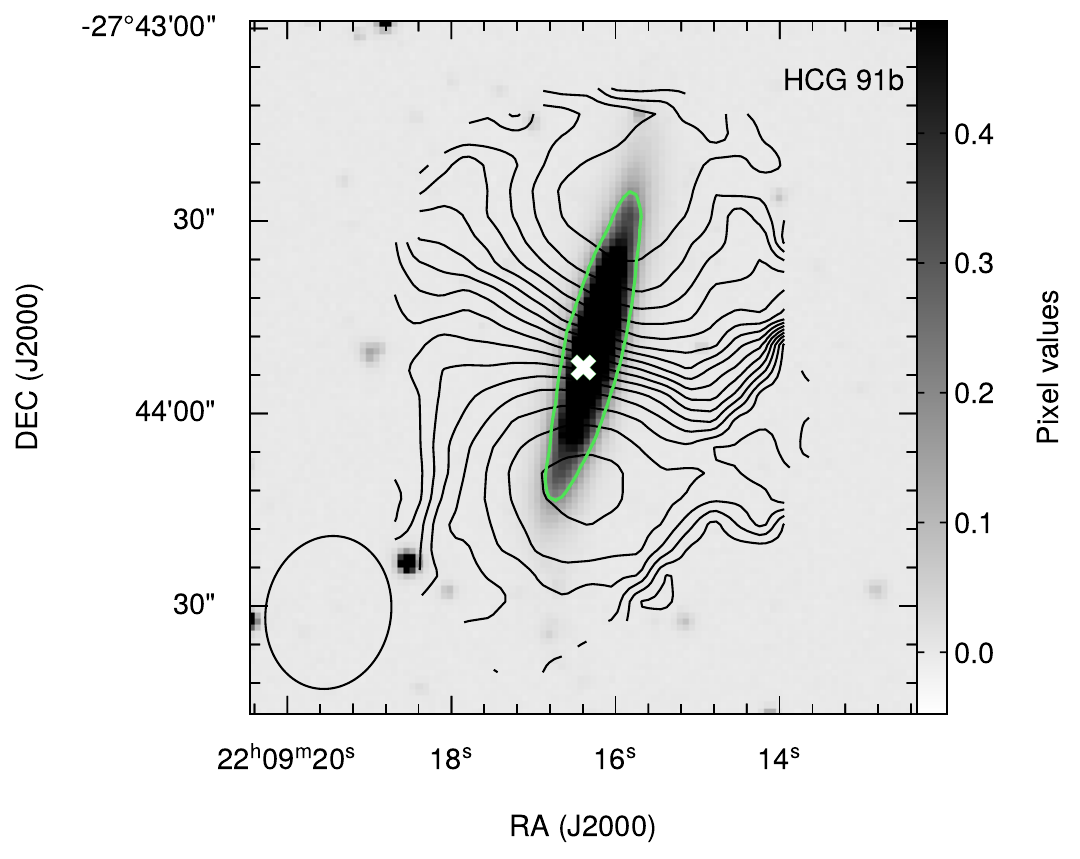} \\[-0.2cm]
      \includegraphics[scale=0.44]{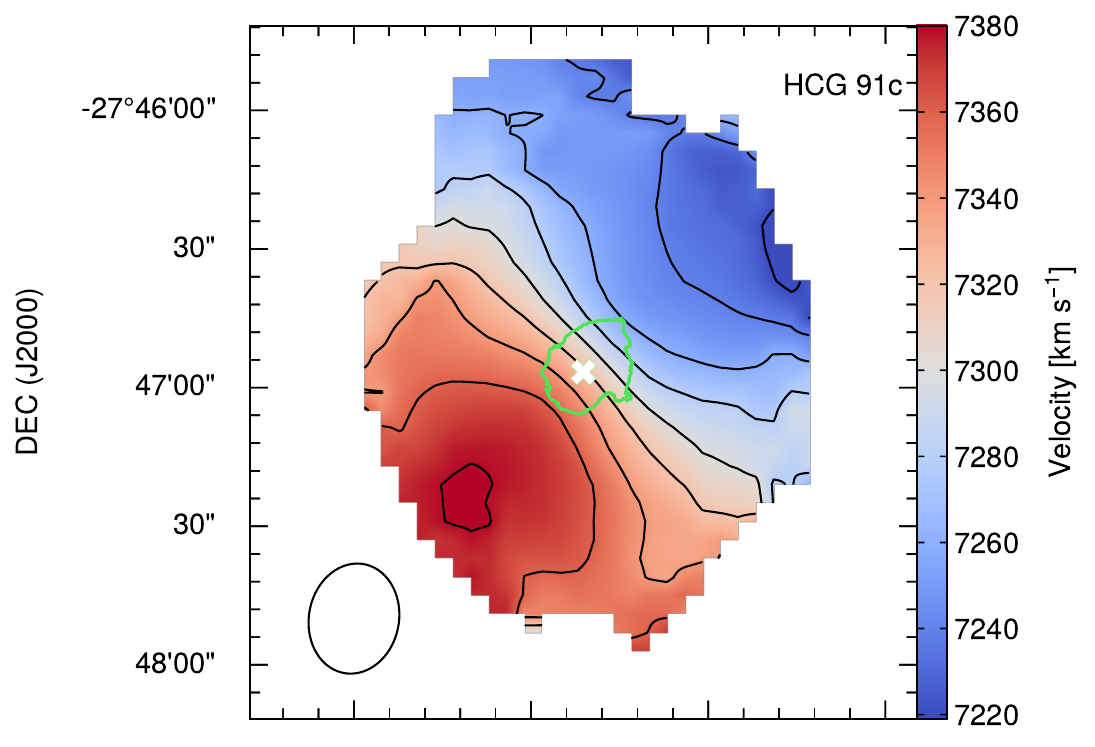} & 
      \includegraphics[scale=0.44]{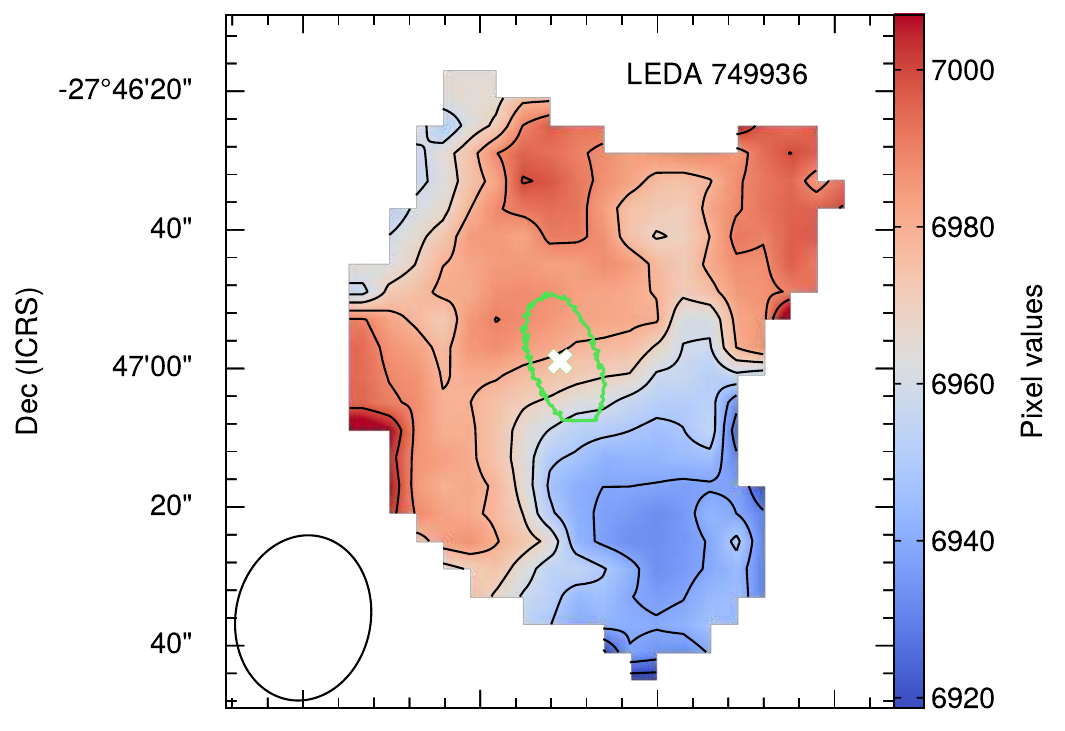} \\[-0.2cm]
      \includegraphics[scale=0.44]{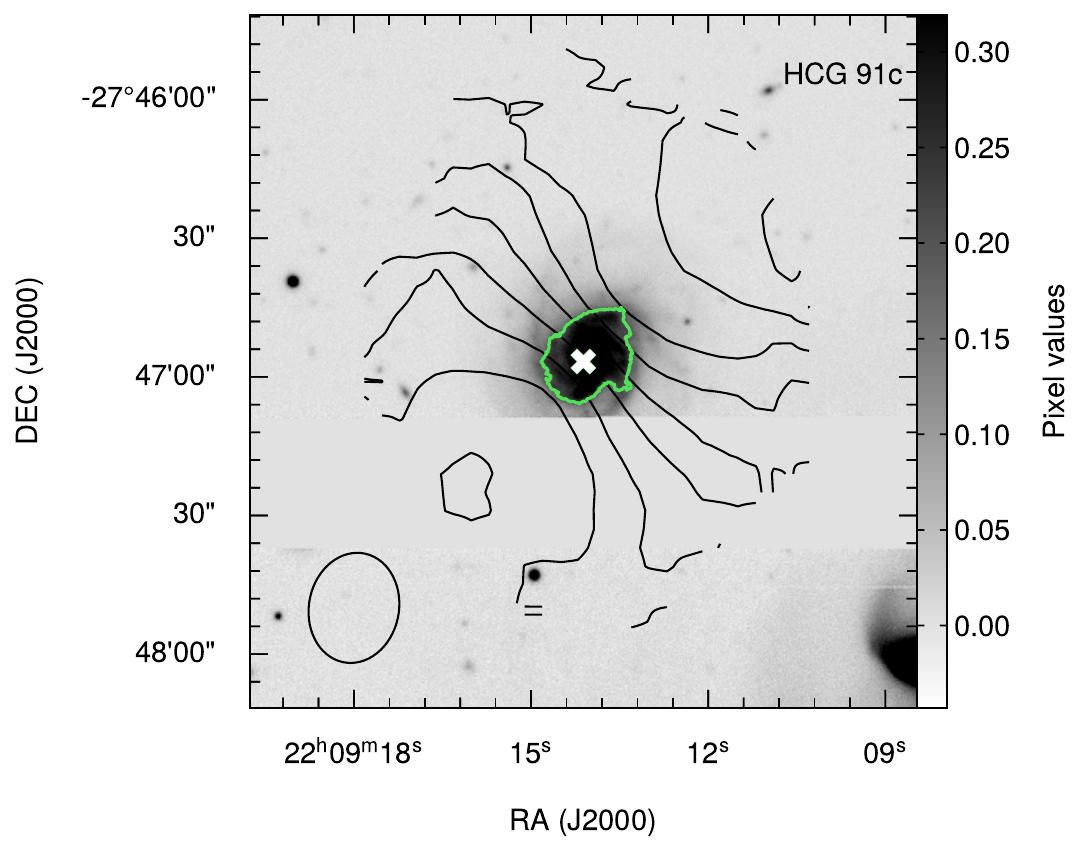} & 
      \includegraphics[scale=0.44]{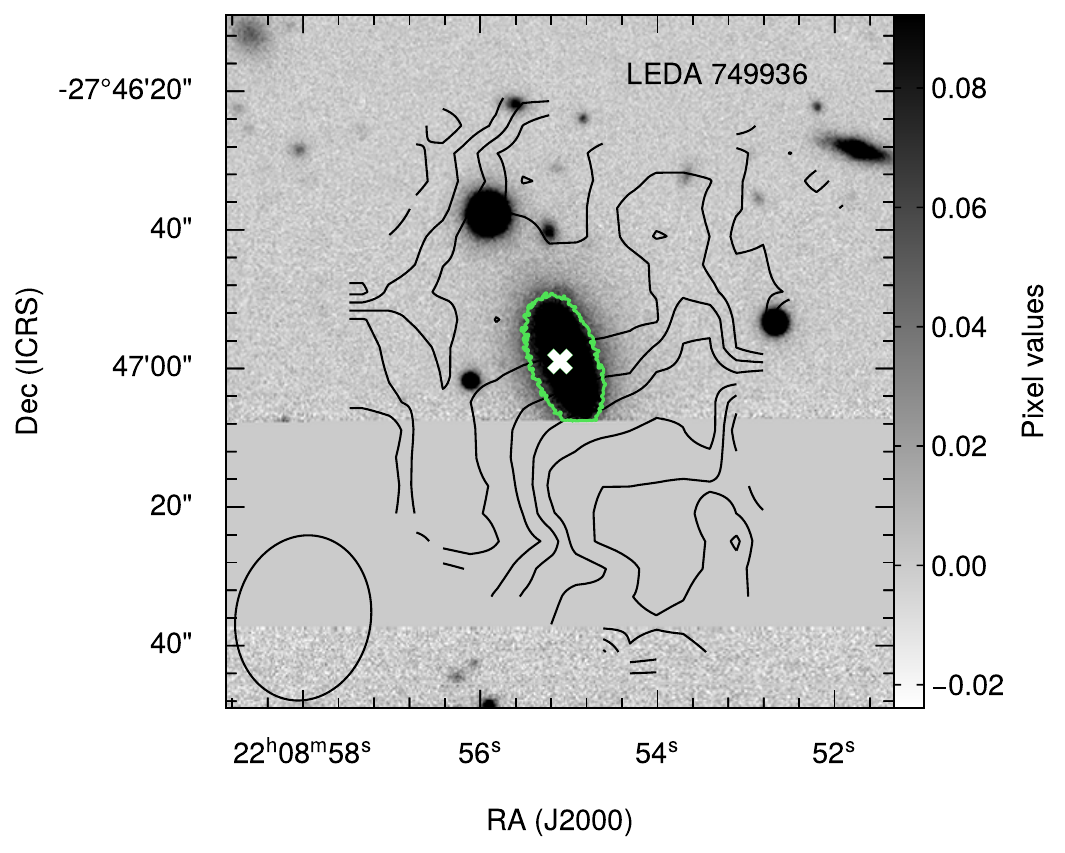}
    \end{tabular}
    \caption{\HI\ Velocity fields of the galaxies in the core of HCG~91. The grayscale images show DeCaLS DR10 I-band optical data. 
    The crosses indicate the optical centre, whereas the green contours highlight the optical disk.}
    \label{fig:hcg91_mom_cores}
   \end{figure*}

The velocity fields of the galaxies in the core of HCG~91 are shown in Figure~\ref{fig:hcg91_mom_cores}. The grayscale images show optical I-band DeCALS DR10 data.  
  
  \subsection{3D visualisation}  
\begin{figure*}
     \setlength{\tabcolsep}{0pt}
     \begin{tabular}{c c}
     \includegraphics[scale=0.285]{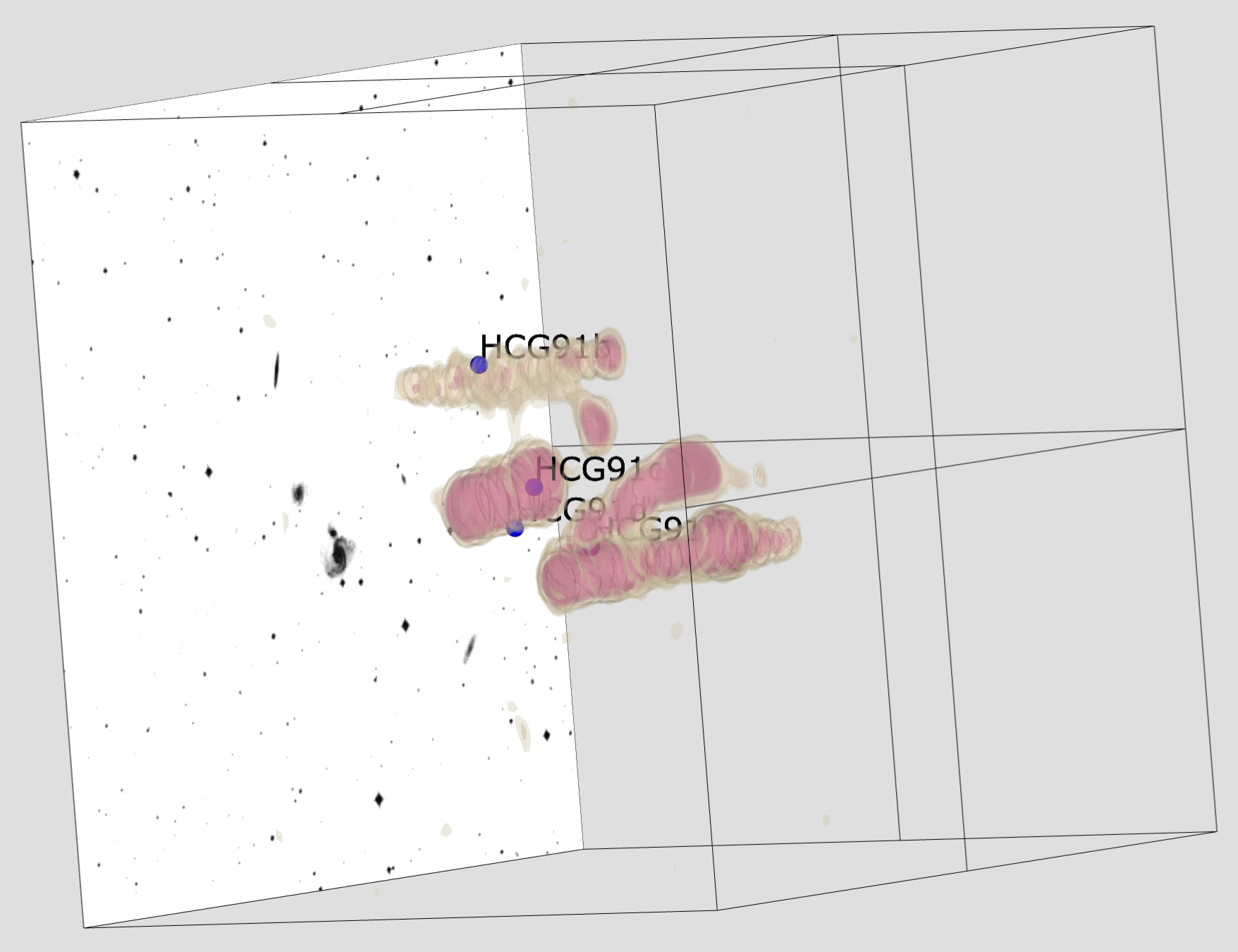} & 
     \includegraphics[scale=0.285]{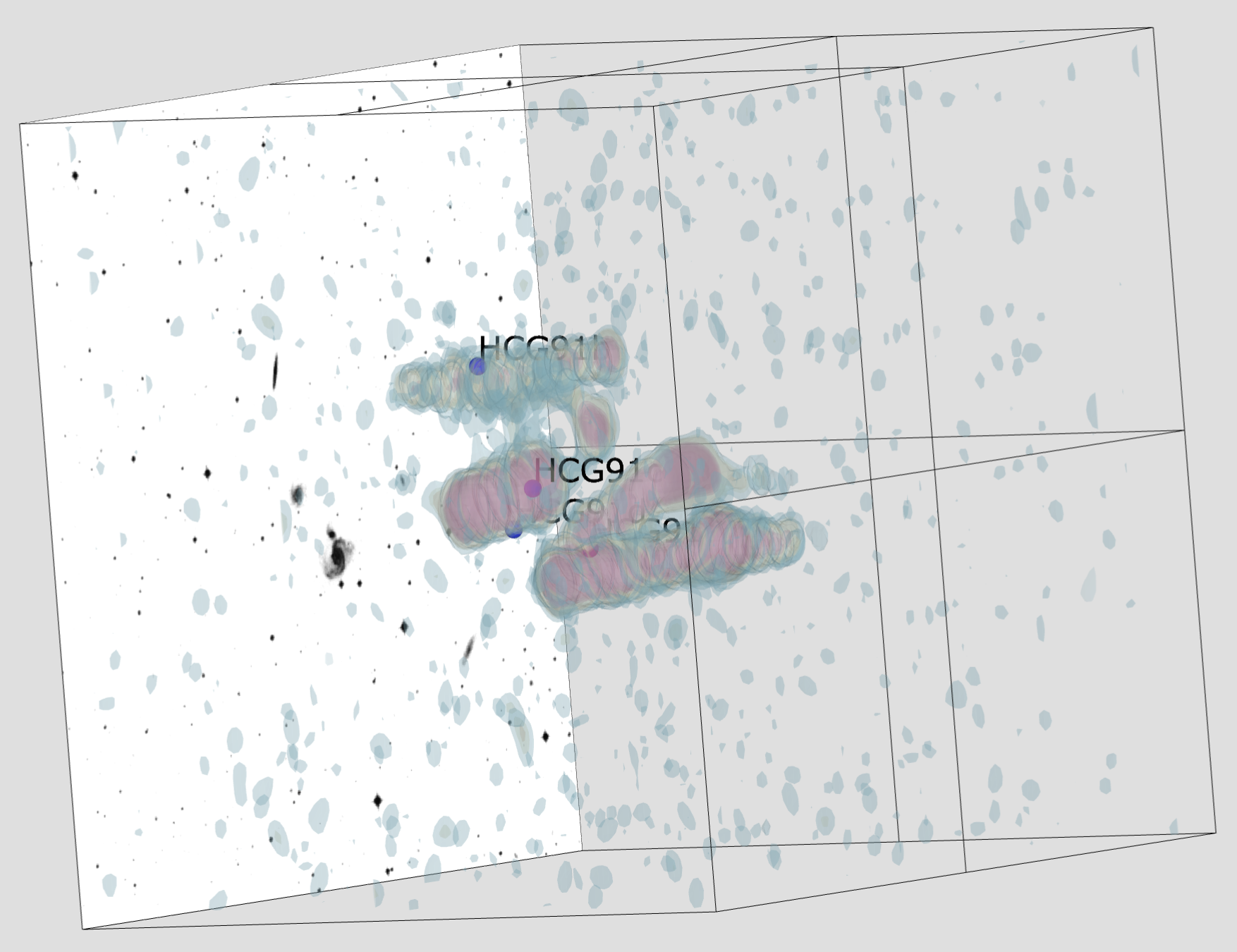}
     \end{tabular}
     \caption{3D visualisation of HCG~91. The left panel shows iso-surface level highlighting the high-column density gas. The right panel 
     showcases the low-column density \HI\ gas. The blue circles indicate the position of the member galaxies. The 2D grayscale image is a DeCaLS R-band optical image of the group. The online version of the cubes are available at \href{https://amiga.iaa.csic.es/x3d-menu/}{https://amiga.iaa.csic.es/x3d-menu/}.}
   \label{fig:hcg91_3dvis}
  \end{figure*}

The 3D visualisation of HCG~91 is presented in Figure~\ref{fig:hcg91_3dvis}. 
The left panel displays iso-surfaces that highlight regions of high \HI\ column density, while the right panel overlays areas of low and high column density. 
The blue circles mark the positions of the group’s member galaxies, and the 2D grayscale background is a DECaLS R-band image. 
An interactive version of these cubes can be found at \href{https://amiga.iaa.csic.es/x3d-menu/}{https://amiga.iaa.csic.es/x3d-menu/}.
  \section{Extra figures of HCG~30}
  \subsection{Noise properties and global profiles}
\begin{figure*}
  \setlength{\tabcolsep}{0pt}
  \begin{tabular}{l l l}
      \includegraphics[scale=0.22]{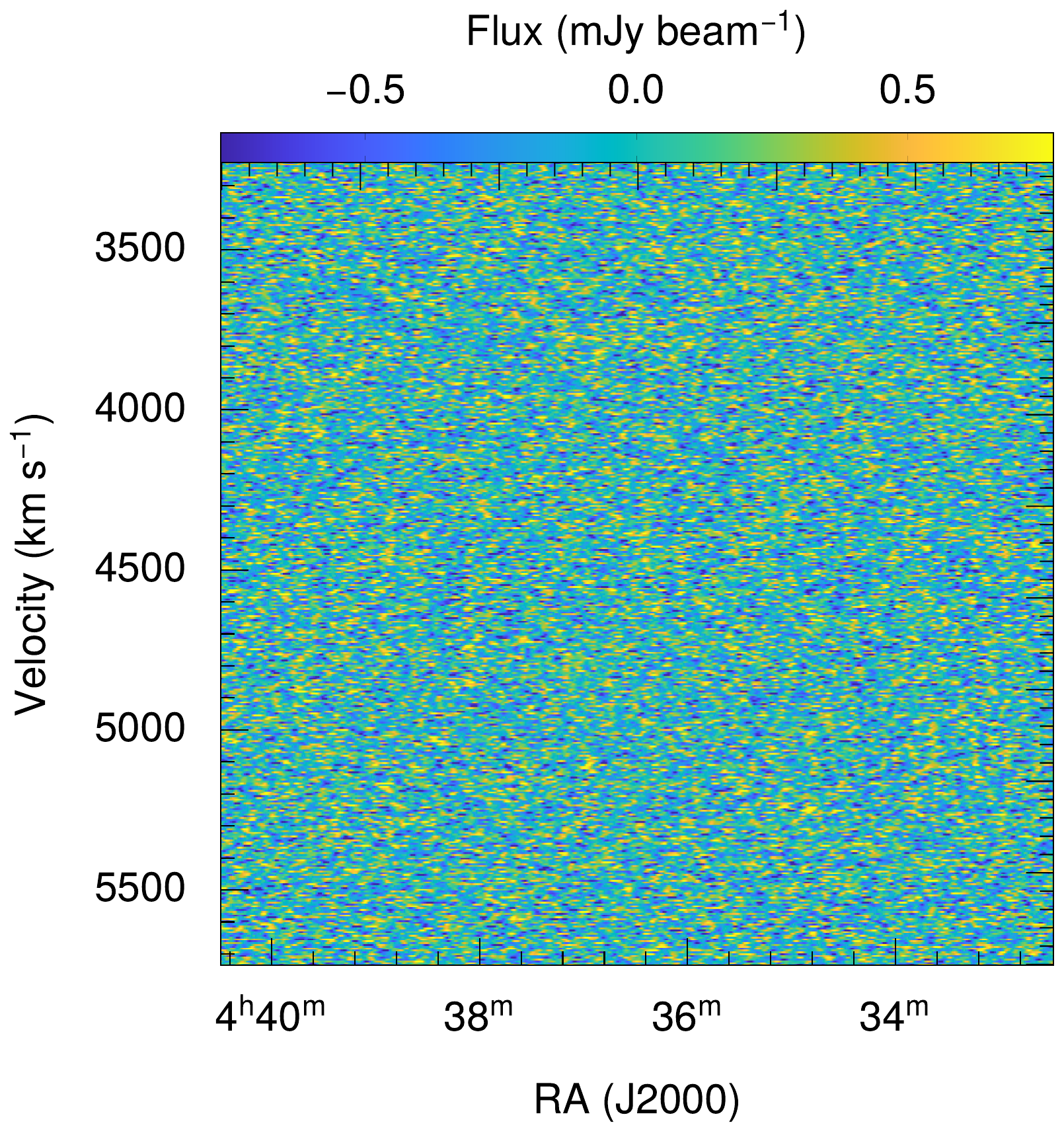} &
      \includegraphics[scale=0.22]{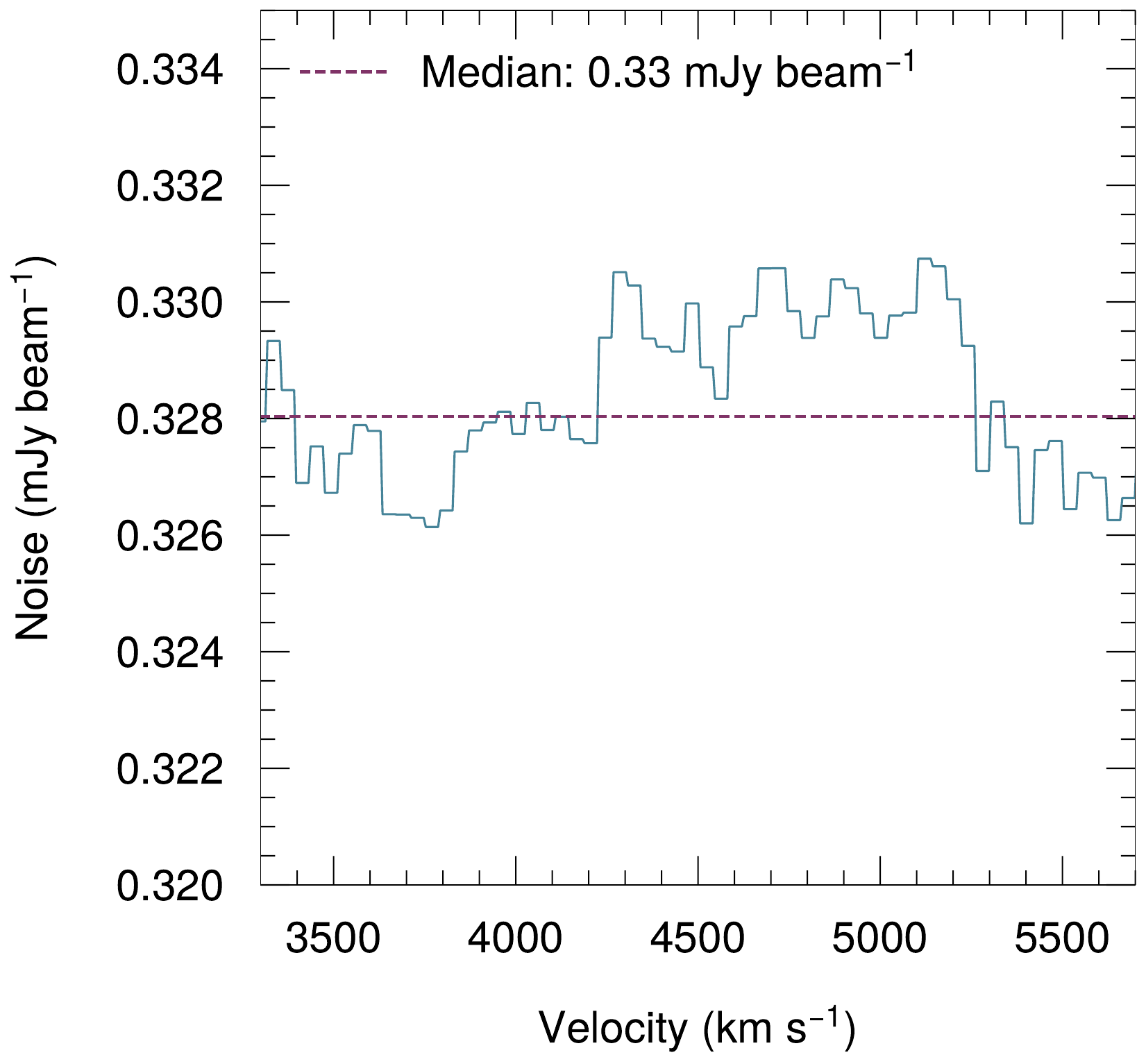} &
      \includegraphics[scale=0.22]{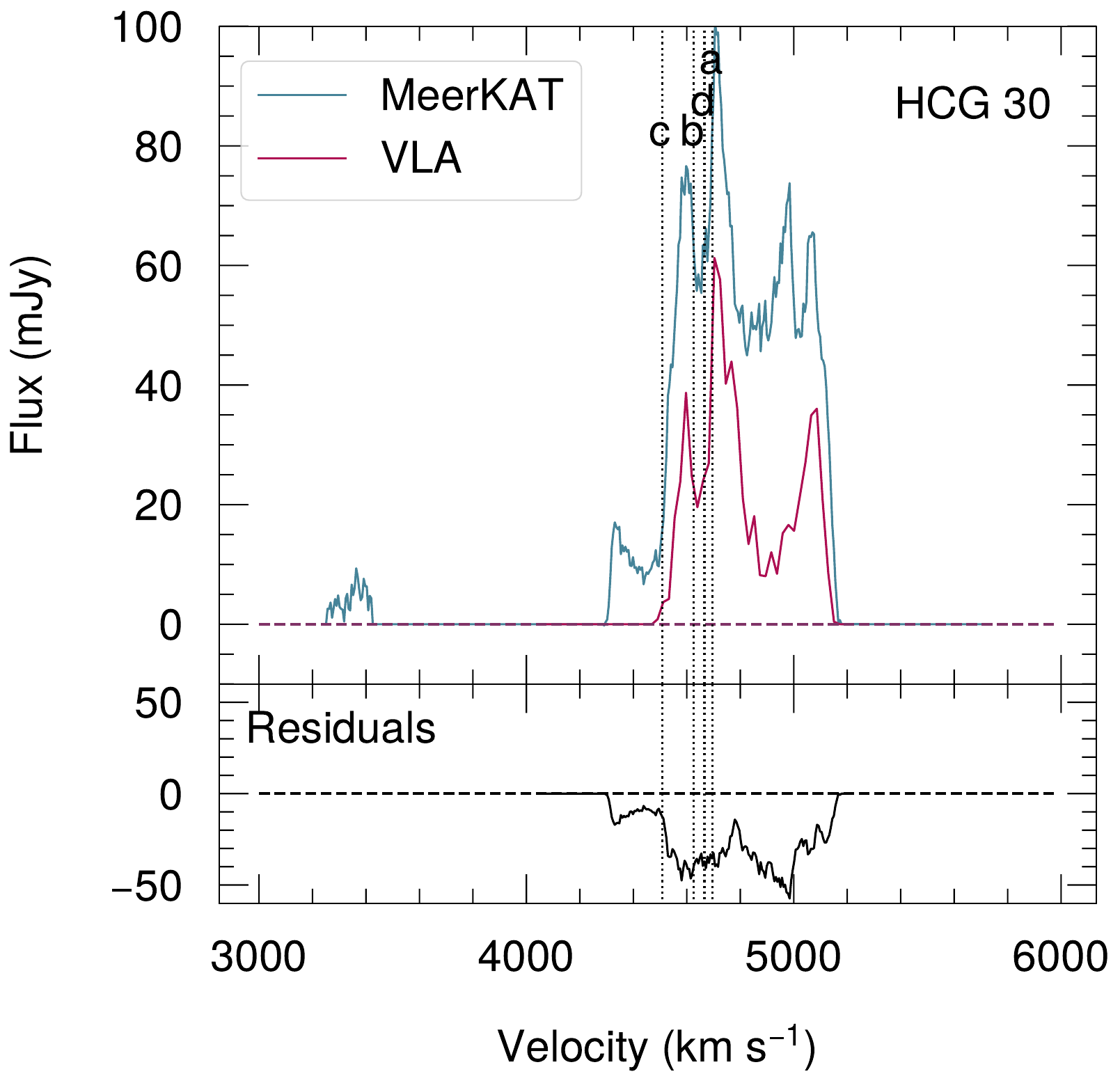}
    \end{tabular}
    \caption{Left panel: velocity vs right ascension of HCG~30. Middle panel: median noise values of each RA-DEC slice of the non-primary beam corrected 60\arcsec\ data cube of 
    HCG~30 as a function of velocity. The horizontal dashed line indicates the median of all the noise values from each slice. Right panel: the blue solid lines indicates the 
    MeerKAT integrated spectrum of HCG~30; the red solid line indicates VLA integrated spectrum of the group derived by \citep{2023A&A...670A..21J}. 
    The vertical dotted lines indicate the velocities of the galaxies in the core of the group. The spectra have been extracted from areas containing only genuine \HI\ emission. }
    \label{fig:hcg30_noise}
  \end{figure*}

  Figure~\ref{fig:hcg30_noise} illustrates the noise properties in HCG~30 and compares the MeerKAT and VLA integrated spectra of the group. 
  The left panel displays the RA-velocity plot. The middle panel shows the median noise levels across individual RA–DEC slices of the non primary beam corrected data cube, 
  with a horizontal dashed line indicating the global median noise. The right panel compares the \HI\ integrated spectrum derived from the MeerKAT data (blue solid line) 
  with the VLA measurement from \citet{2023A&A...670A..21J} (red solid line). The vertical dotted lines mark the systemic velocities of the core galaxies. 
  The spectra were extracted from regions containing only genuine \HI\ emission.
  \subsection{Channel maps}
\begin{figure*}
      \setlength{\tabcolsep}{0pt}
      \begin{tabular}{l l l}
          \includegraphics[scale=0.255]{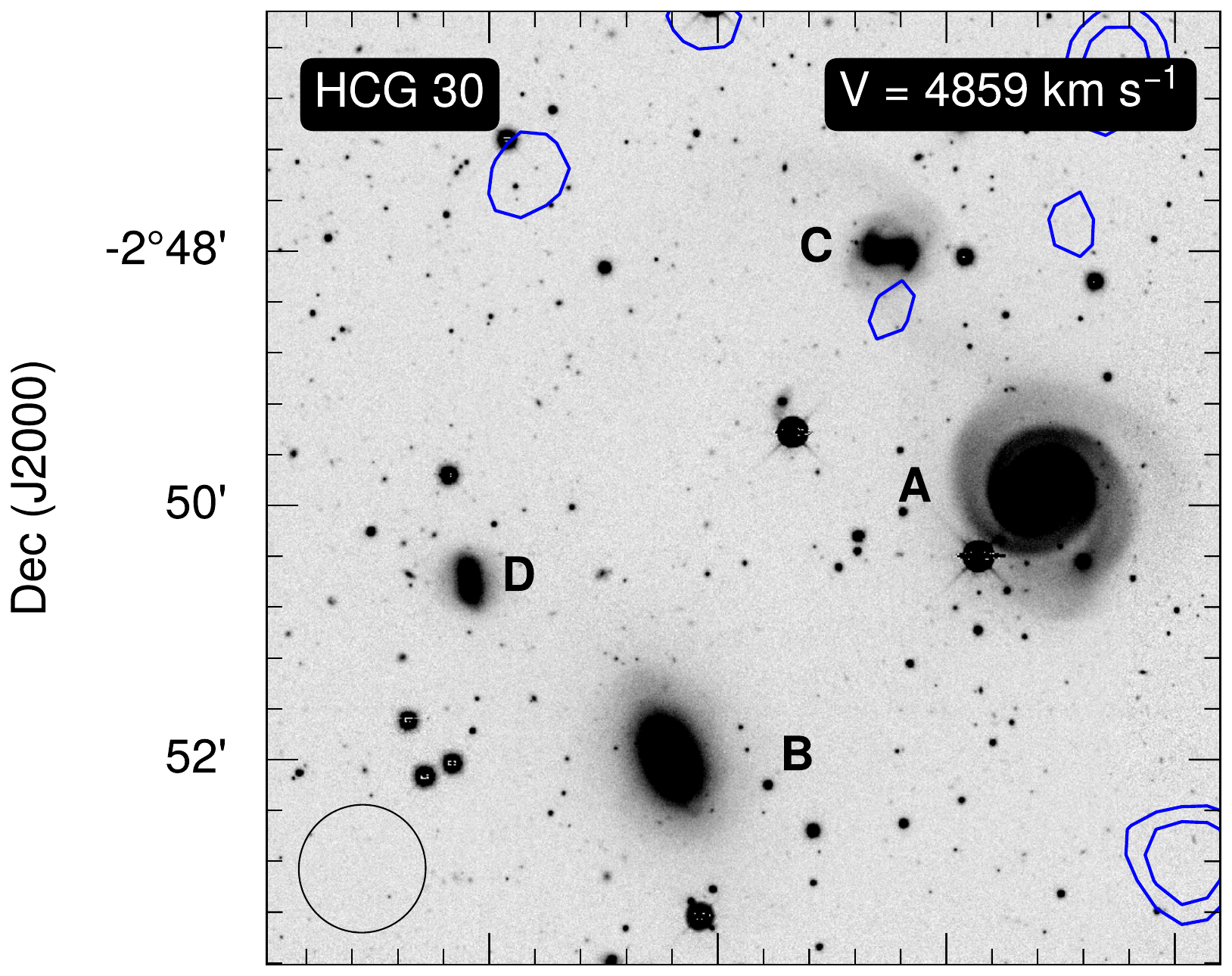} &
          \includegraphics[scale=0.255]{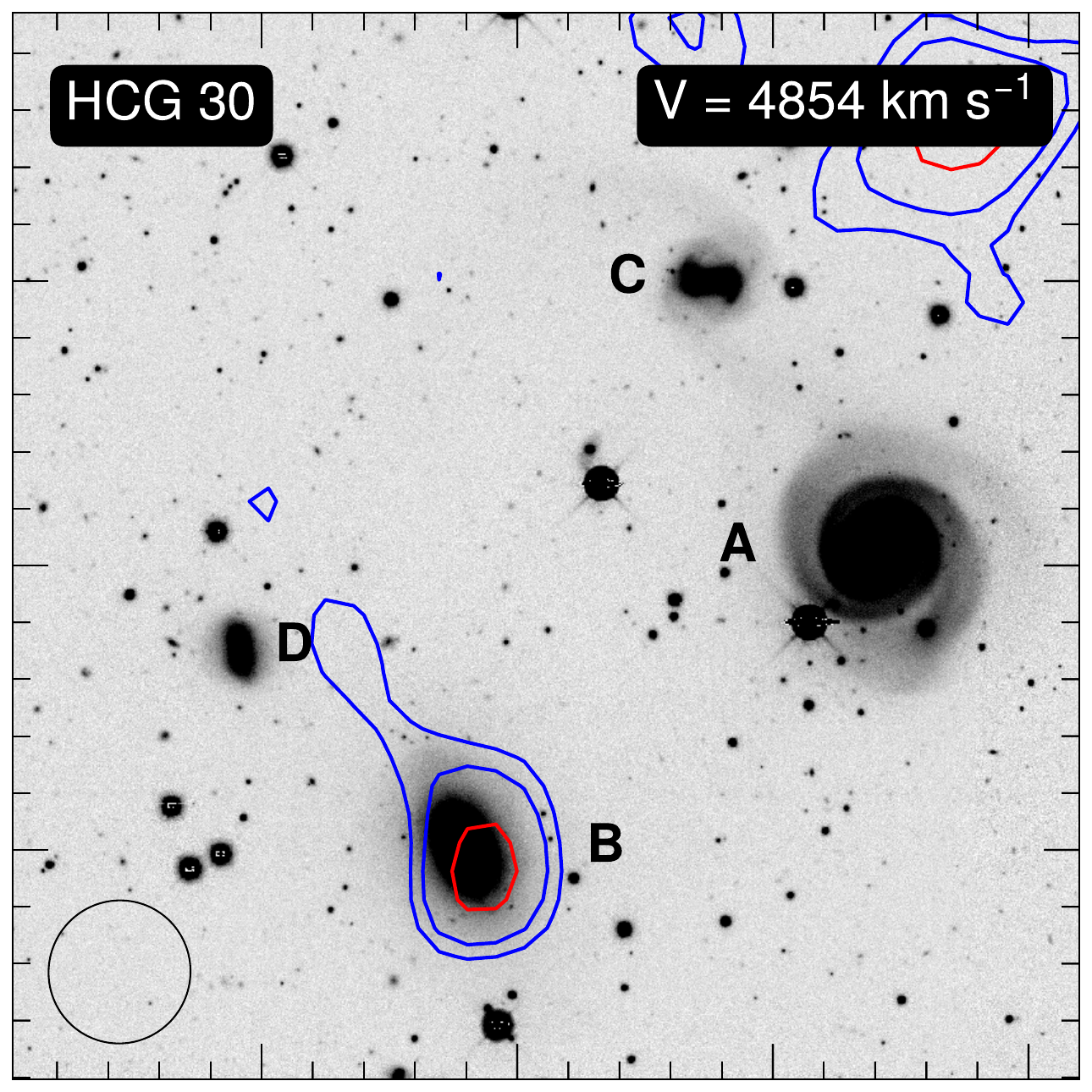} &
          \includegraphics[scale=0.255]{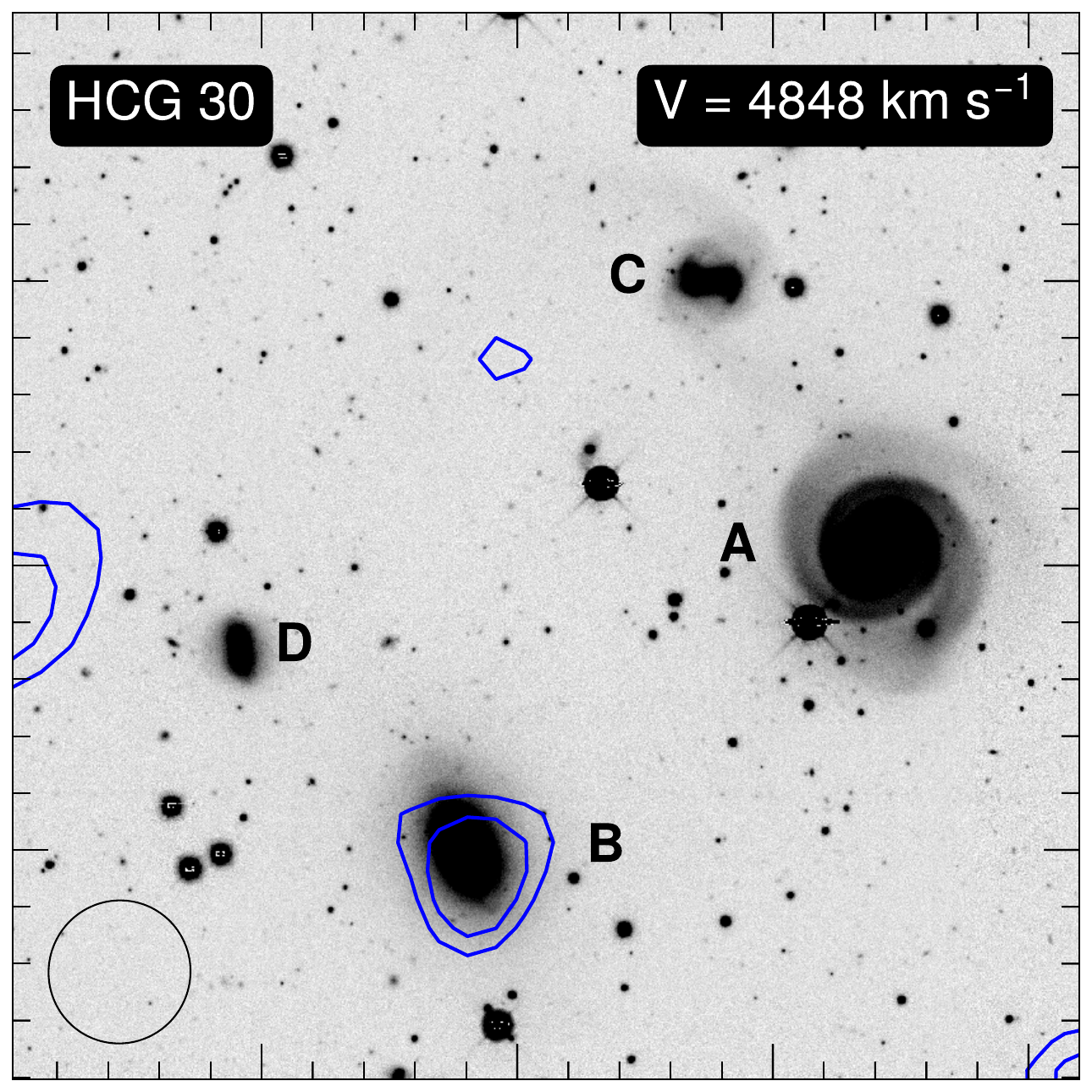} \\[-0.2cm]
          \includegraphics[scale=0.255]{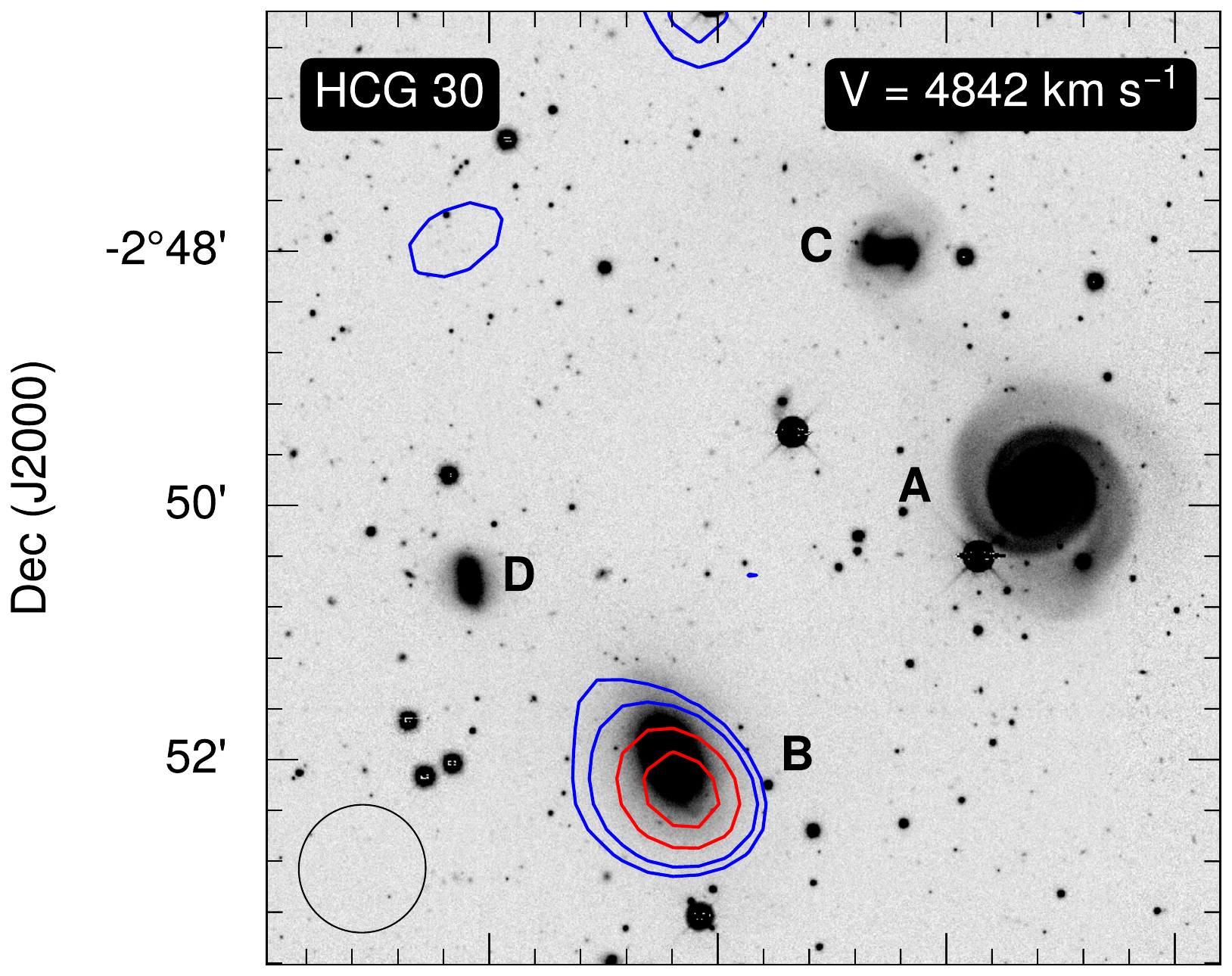} &
          \includegraphics[scale=0.255]{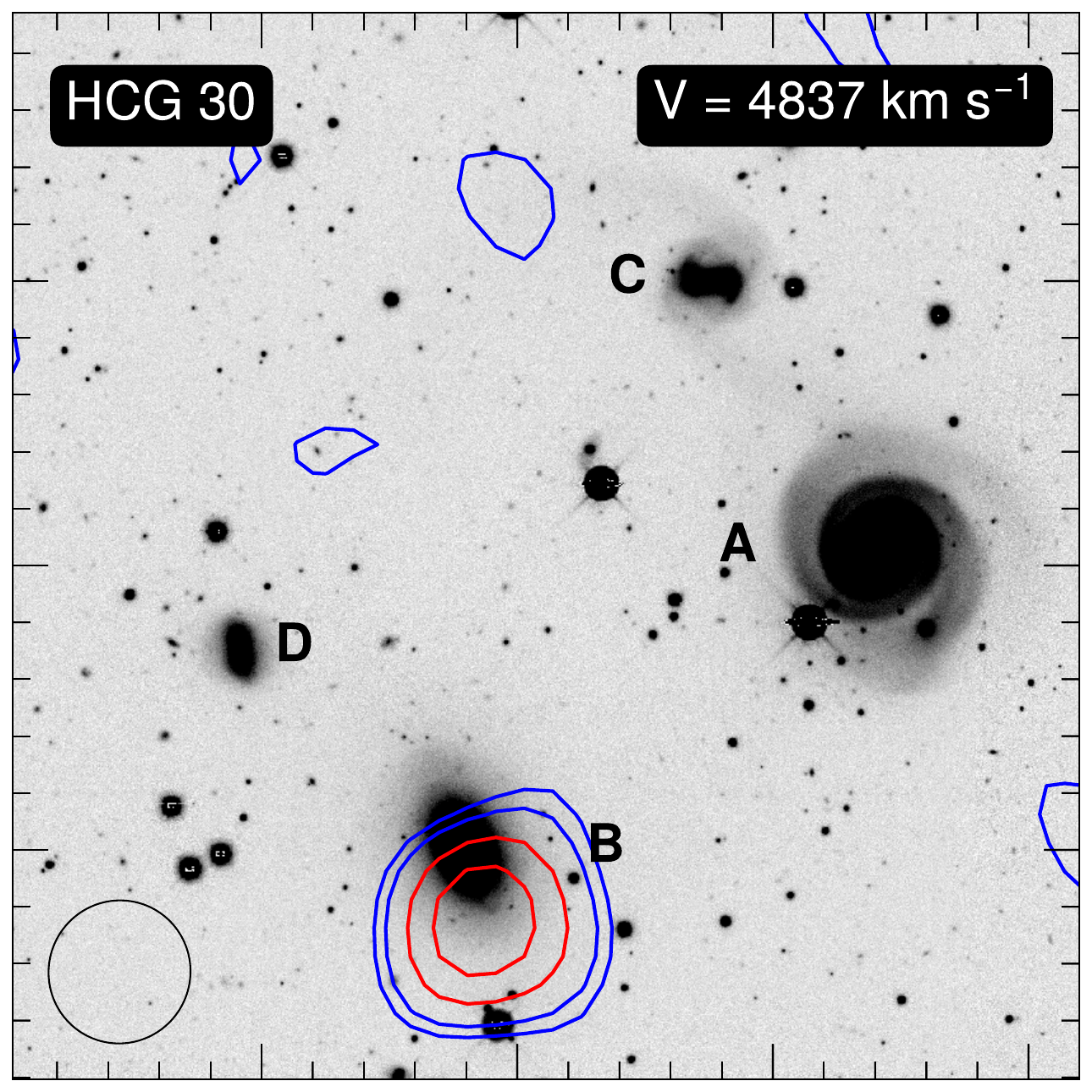} &
          \includegraphics[scale=0.255]{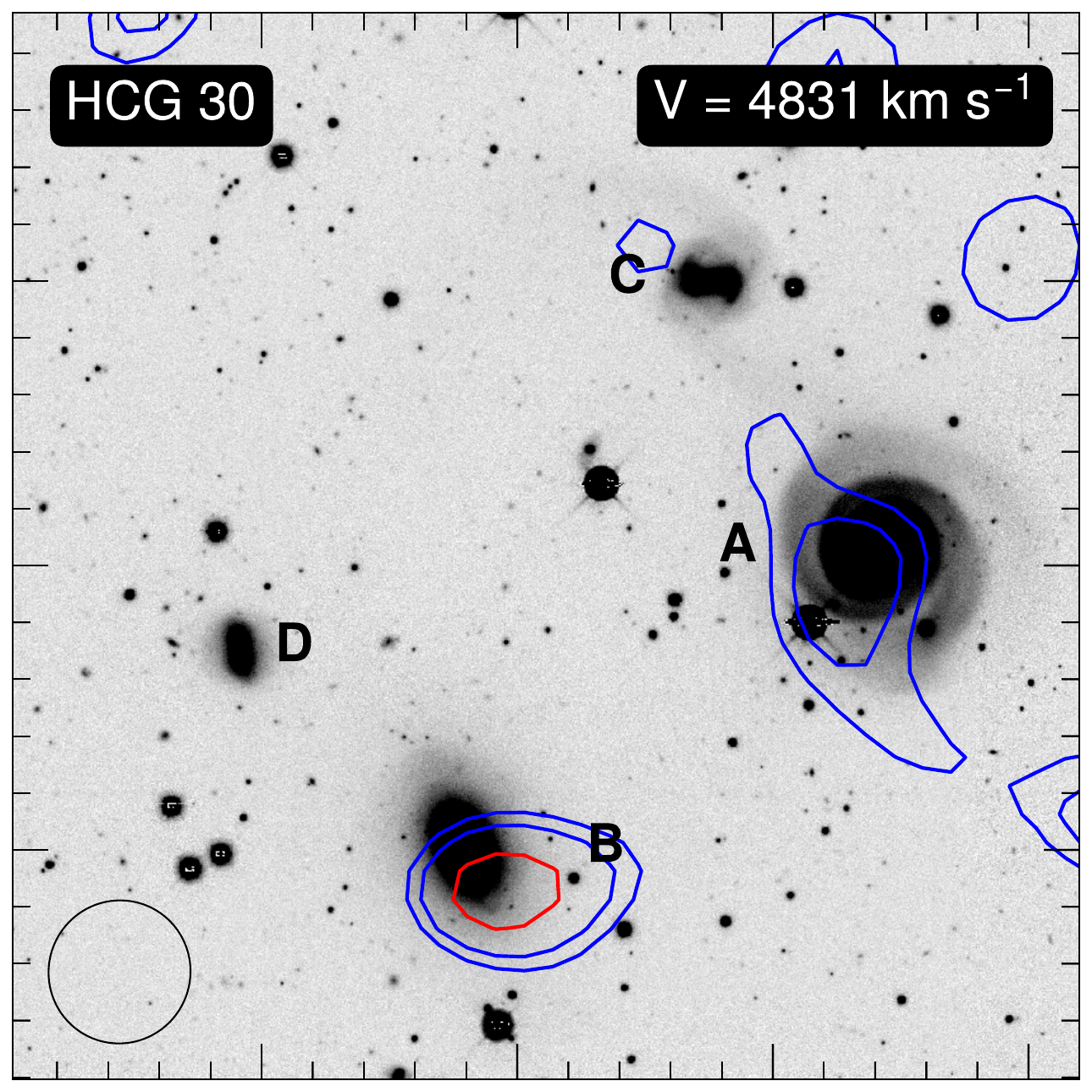}\\[-0.2cm]
          \includegraphics[scale=0.255]{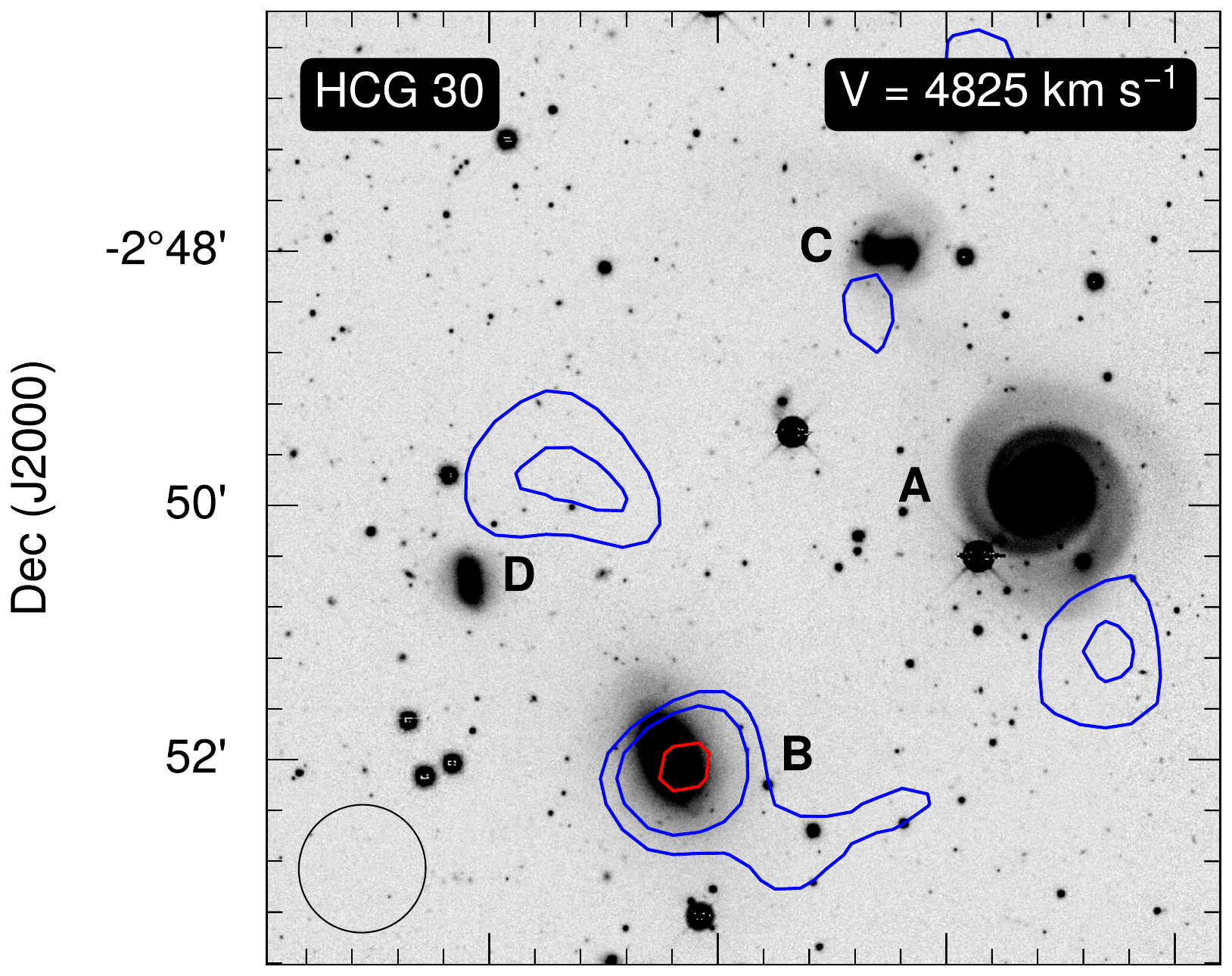} &
          \includegraphics[scale=0.255]{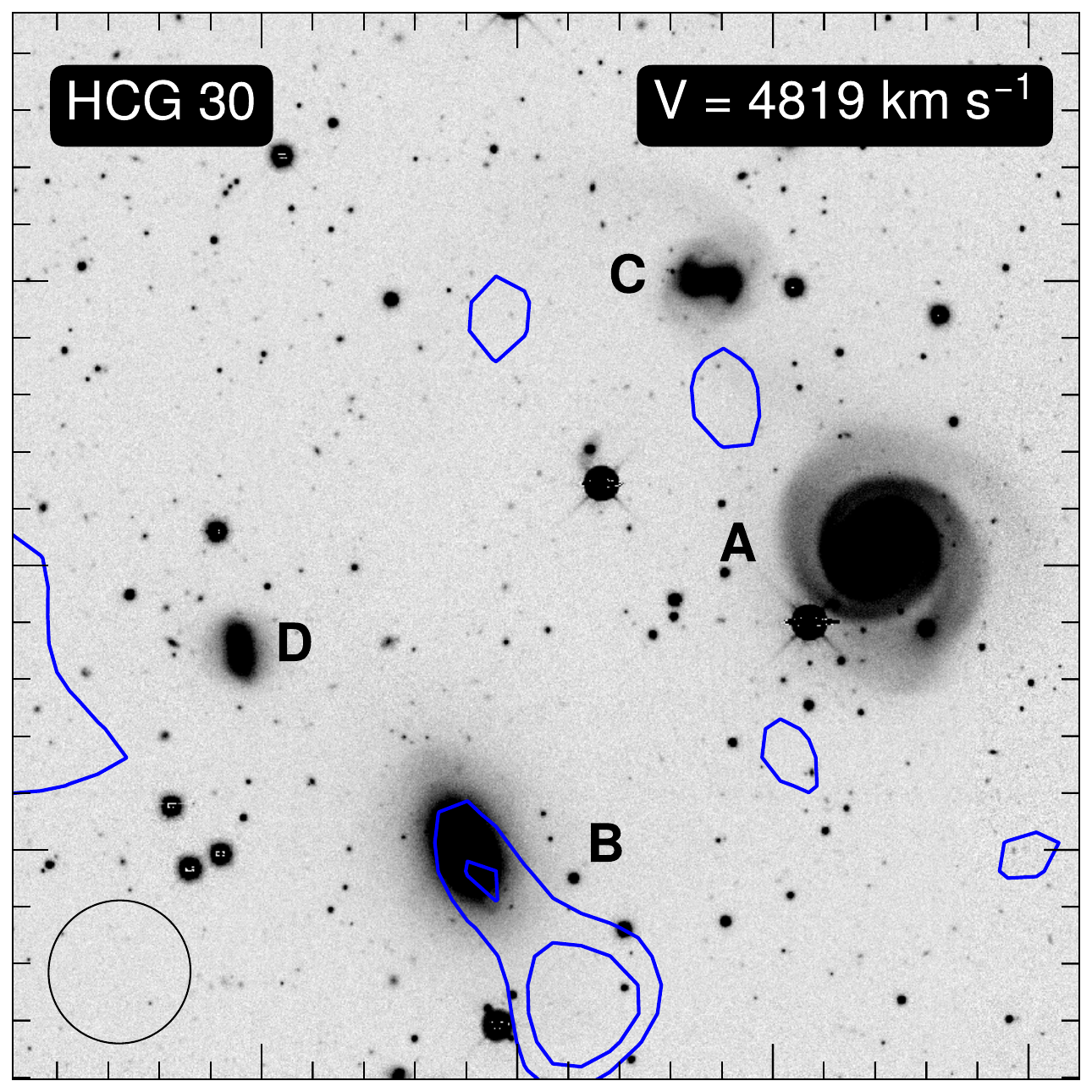} & 
         \includegraphics[scale=0.255]{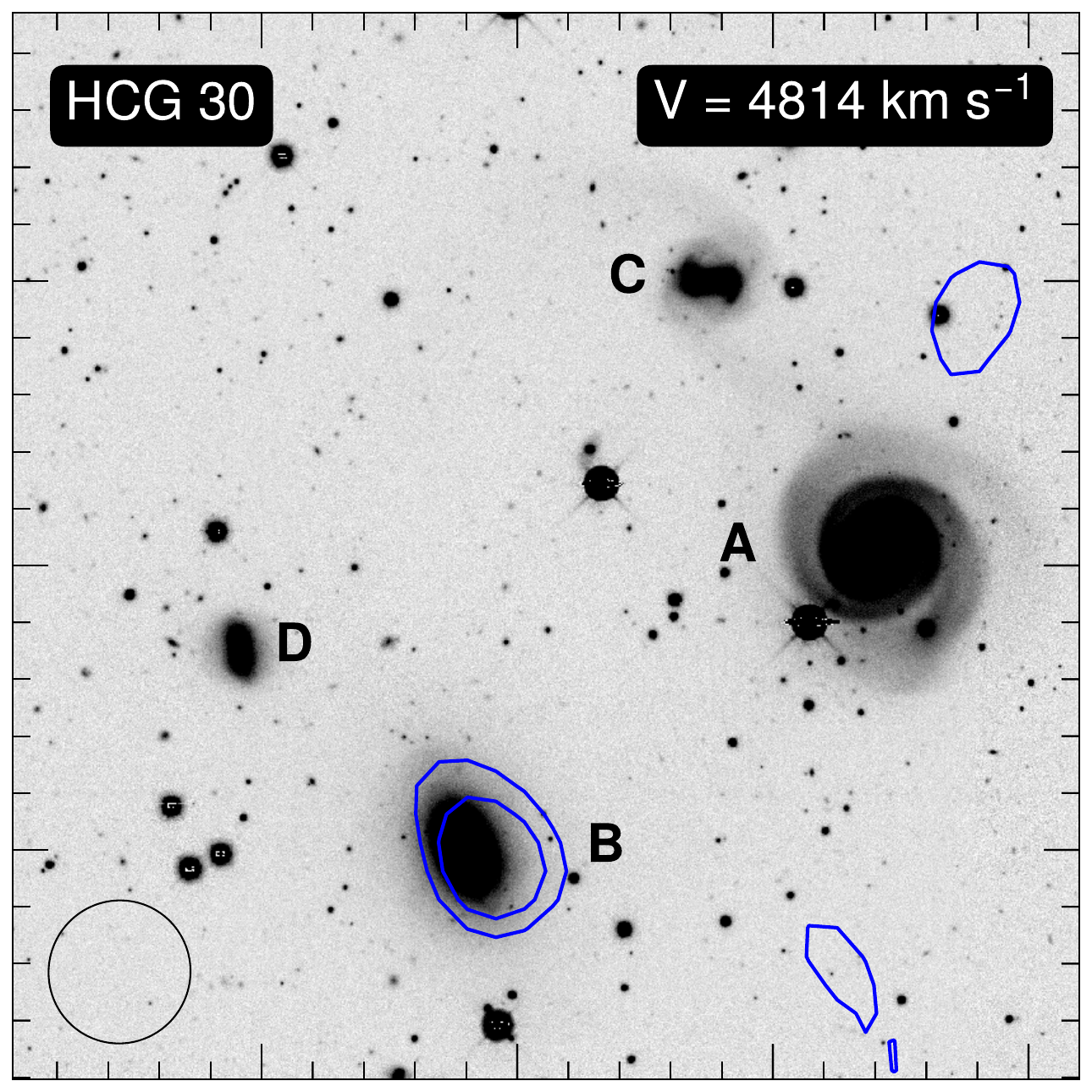}\\[-0.2cm]
          \includegraphics[scale=0.255]{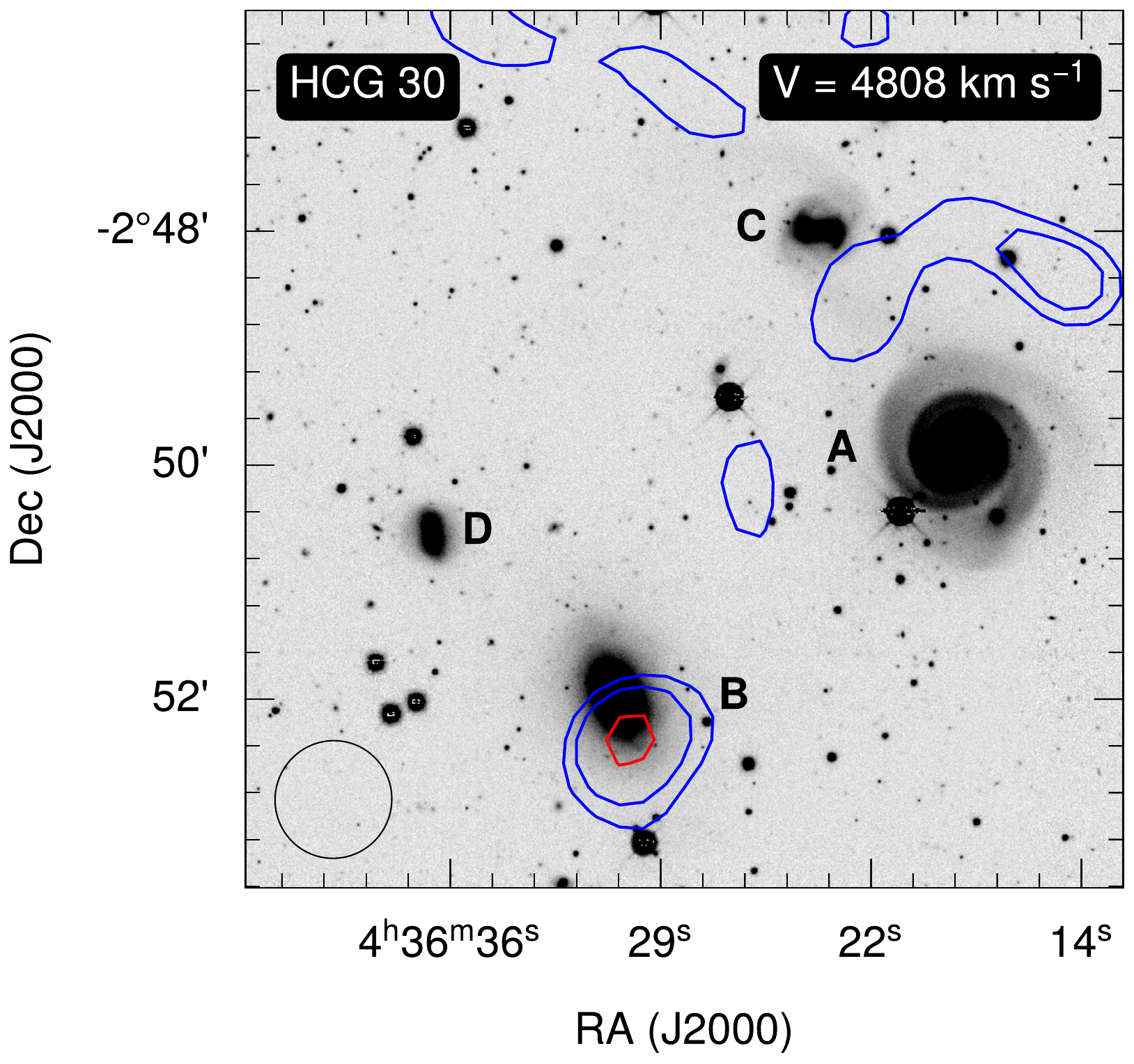} &
          \includegraphics[scale=0.255]{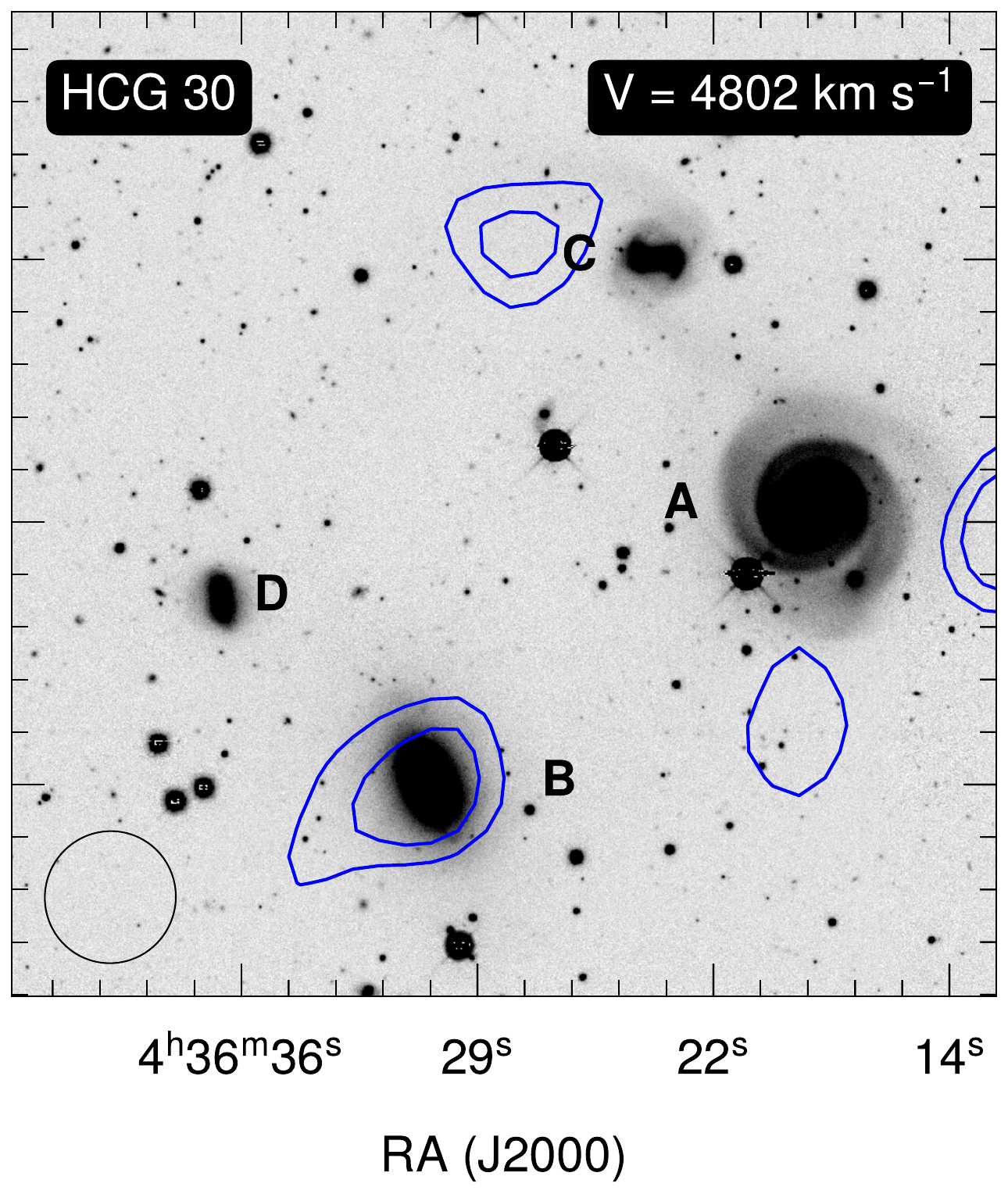} &
         \includegraphics[scale=0.255]{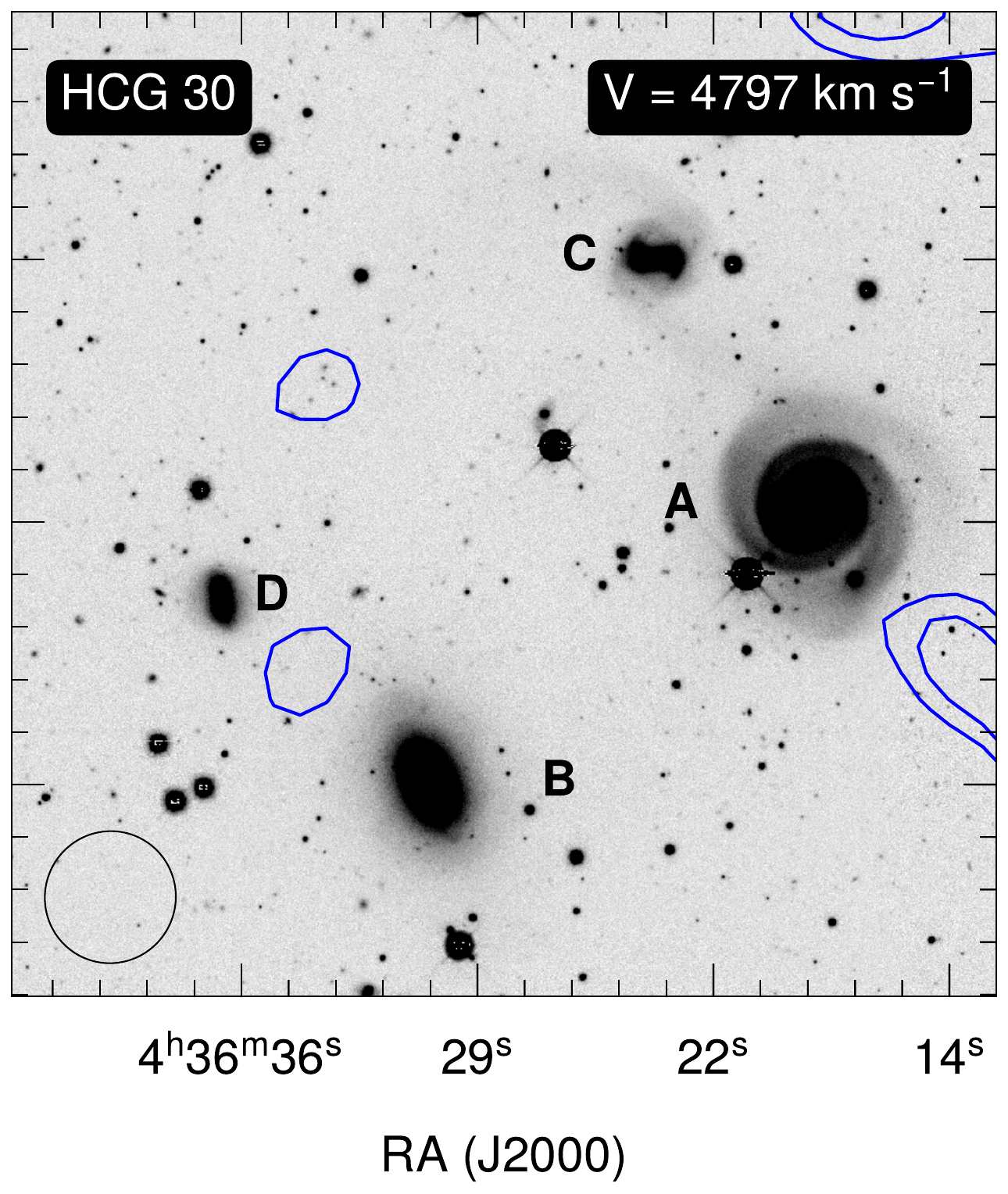}
        \end{tabular}
        \caption{Example channel maps of the primary beam corrected cube of HCG 30 overlaid on DECaLS DR10 R-band optical image. Contour levels are (1.5, 2, 3, 4, 5, 6) times 
        the median noise level in the cube (0.59 $\mathrm{mJy~beam{-1}}$). The blue colours show contour levels below 3$\sigma$; the red colours represent contour levels at 3$\sigma$, or higher. 
        Additional channel maps can be accessed \href{https://zenodo.org/records/14856489}{online}.}
        \label{fig:hcg30_chanmap}
  \end{figure*}

  Figure~\ref{fig:hcg30_chanmap} shows example channel maps from the primary beam corrected data cube of HCG~30. Additional channel maps can be retrieved from 
  \href{https://zenodo.org/records/14856489}{https://zenodo.org/records/14856489}.
  
  \subsection{Moment maps}
\begin{figure*}
  \begin{tabular}{c c}
      \includegraphics[scale=0.26]{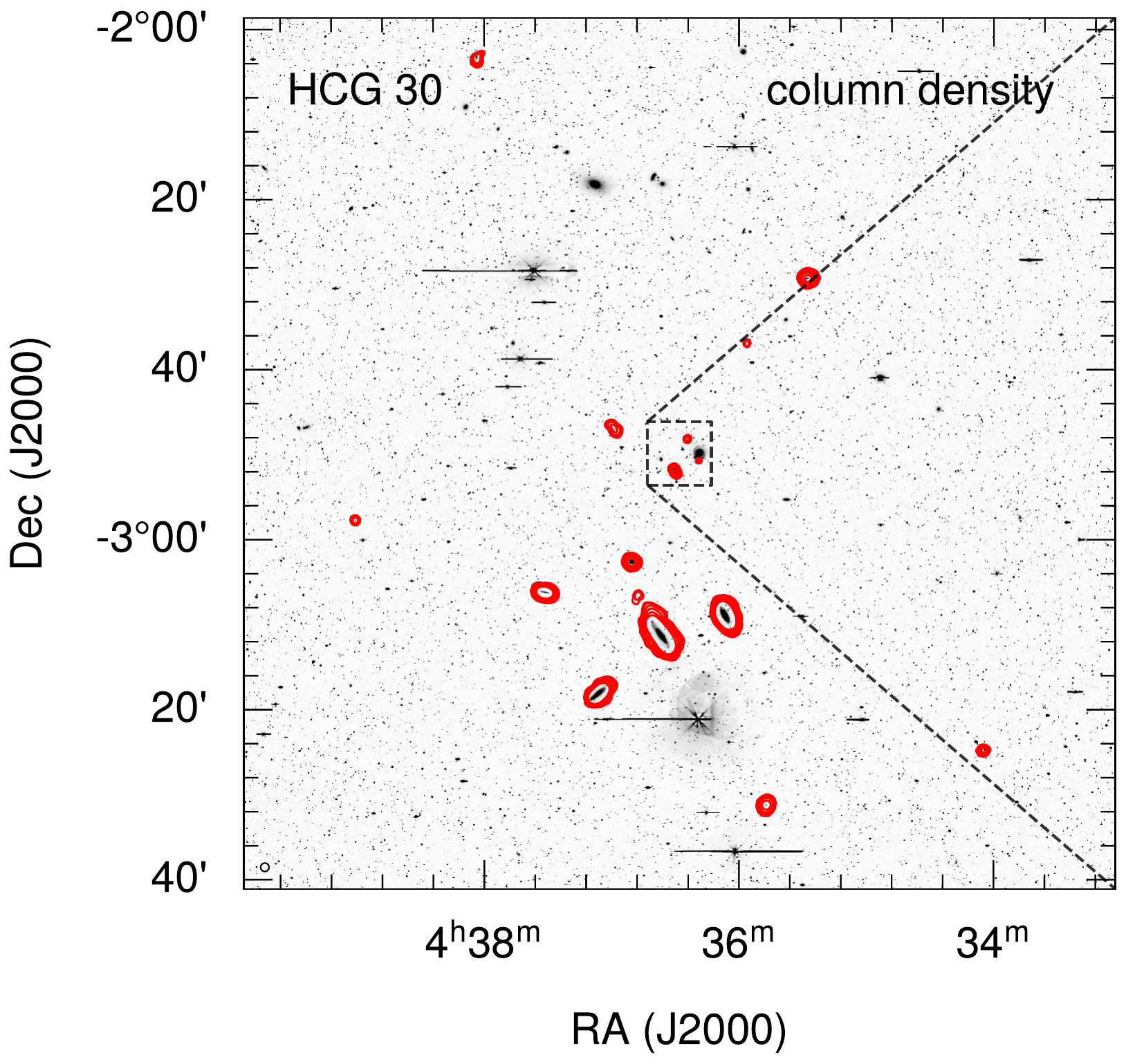}
      & \includegraphics[scale=0.26]{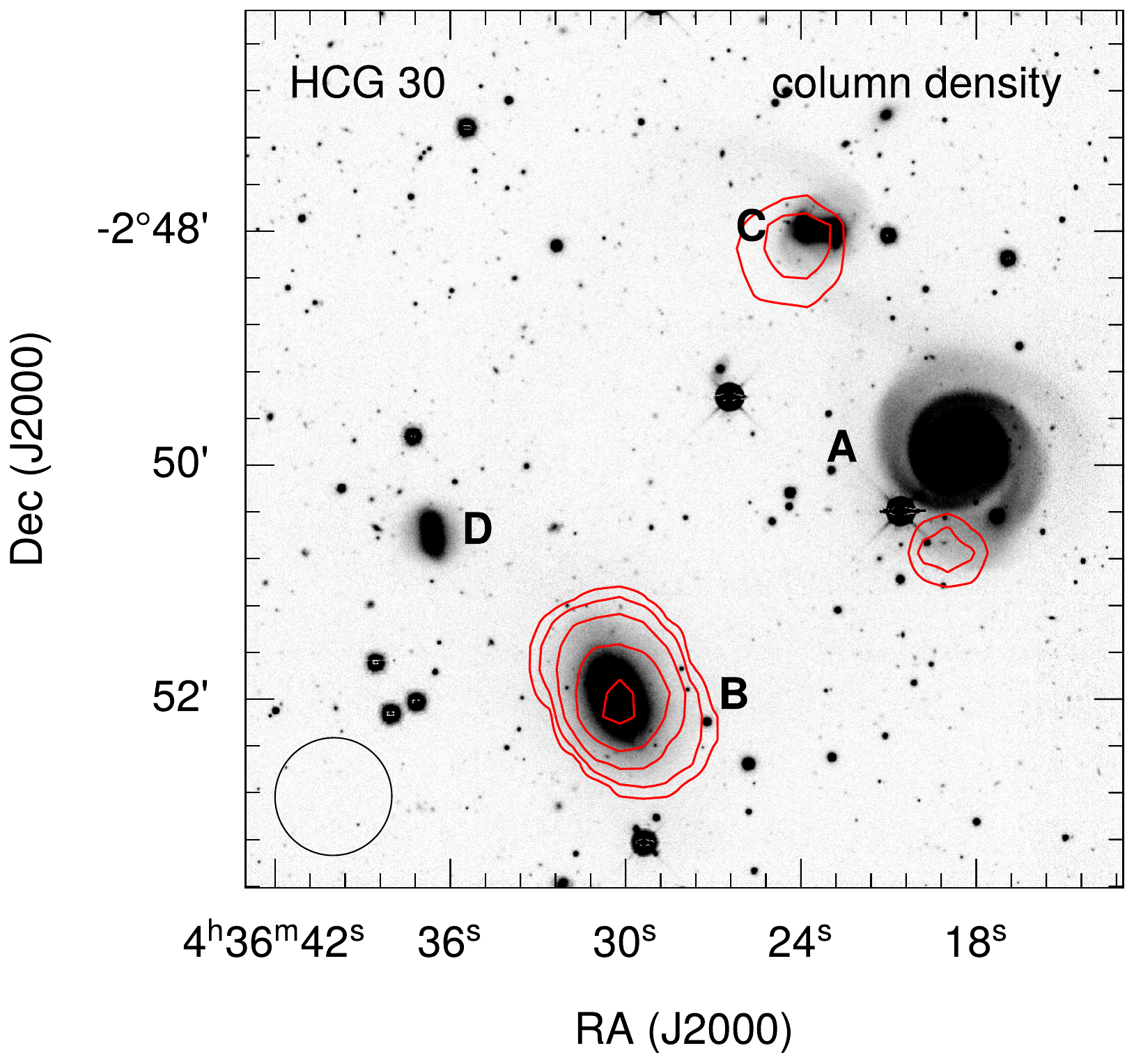}\\
      \includegraphics[scale=0.26]{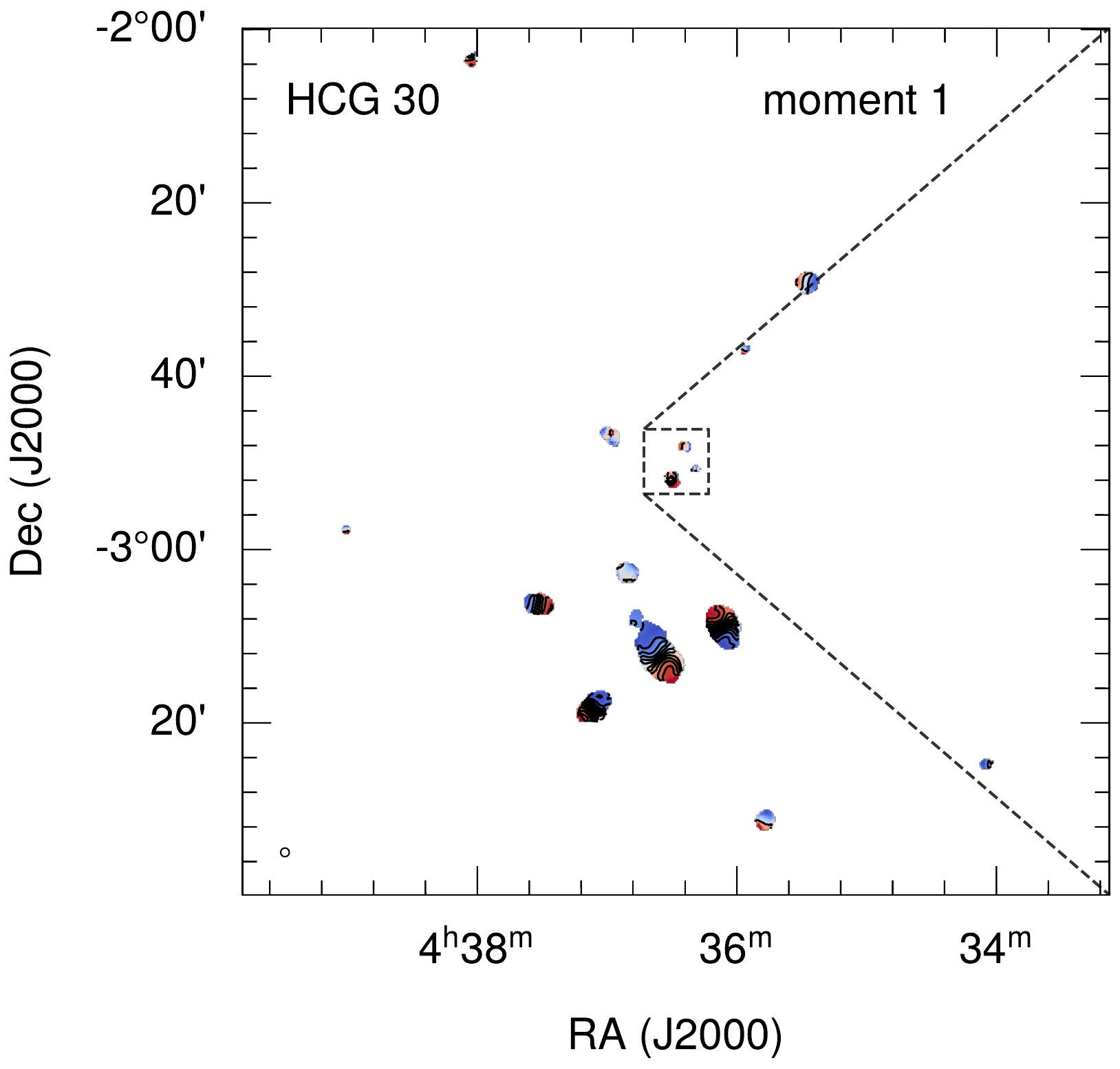} &
      \includegraphics[scale=0.26]{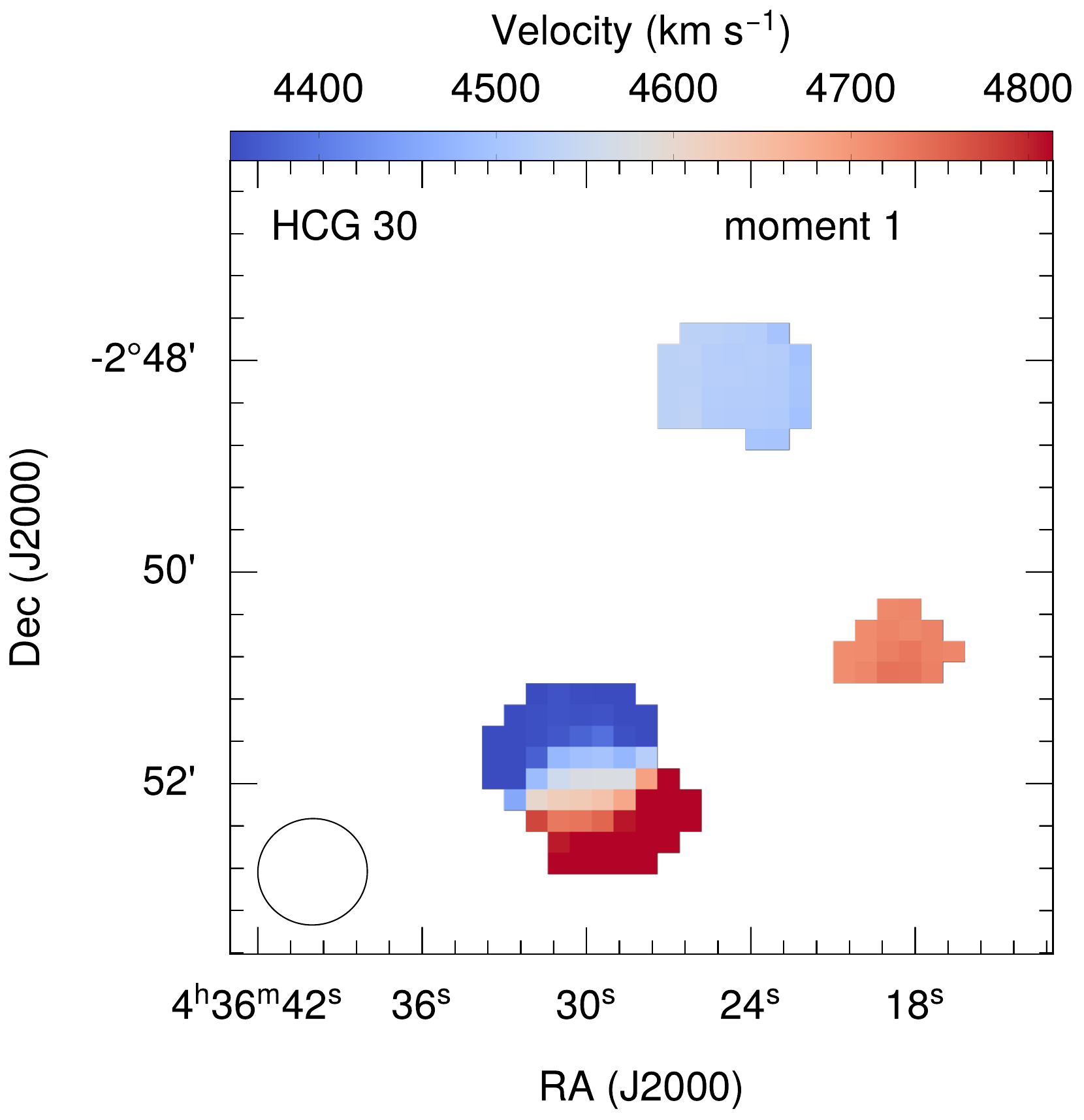}
  \end{tabular}
  \caption{\HI\ Moment maps of HCG 30. Left panels show all sources detected by SoFiA. The right panels show sources within the rectangular box shown on 
  the left to better show the central part of the group. The top panels show the column density maps with contour levels of
          ($\mathrm{6.0~\times~10^{18}}$, $\mathrm{1.2~\times~10^{19}}$, $\mathrm{2.4~\times~10^{19}}$, $\mathrm{4.8~\times~10^{19}}$, 
          $\mathrm{9.6~\times~10^{19}}$, $\mathrm{1.9~\times~10^{20}}$) $\mathrm{cm^{-2}}$. 
          The contours are overlaid on DECaLS DR10 R-band optical images. The bottom panels show the moment one map. Each individual source has its own colour scaling and contour levels to highlight any rotational component.}
  \label{fig:hcg30_mom}
  \end{figure*}

Figure~\ref{fig:hcg30_mom} presents the \HI\ moment maps of HCG~30. The left panels show all \HI\ sources detected by SoFiA within the MeerKA field of view, 
while the right panels display zoomed-in views of the group’s core. 
The top panels show column density maps overlaid on DECaLS DR10 R-band optical images. The bottom panels display the moment-one (velocity field) map for each detected source, 
highlighting the rotational signatures of the surrounding galaxies.
  
\subsection{The core members of HCG~30}
\begin{figure*}
      \setlength{\tabcolsep}{1.2pt}
      \begin{tabular}{c c c}
          \includegraphics[scale=0.29]{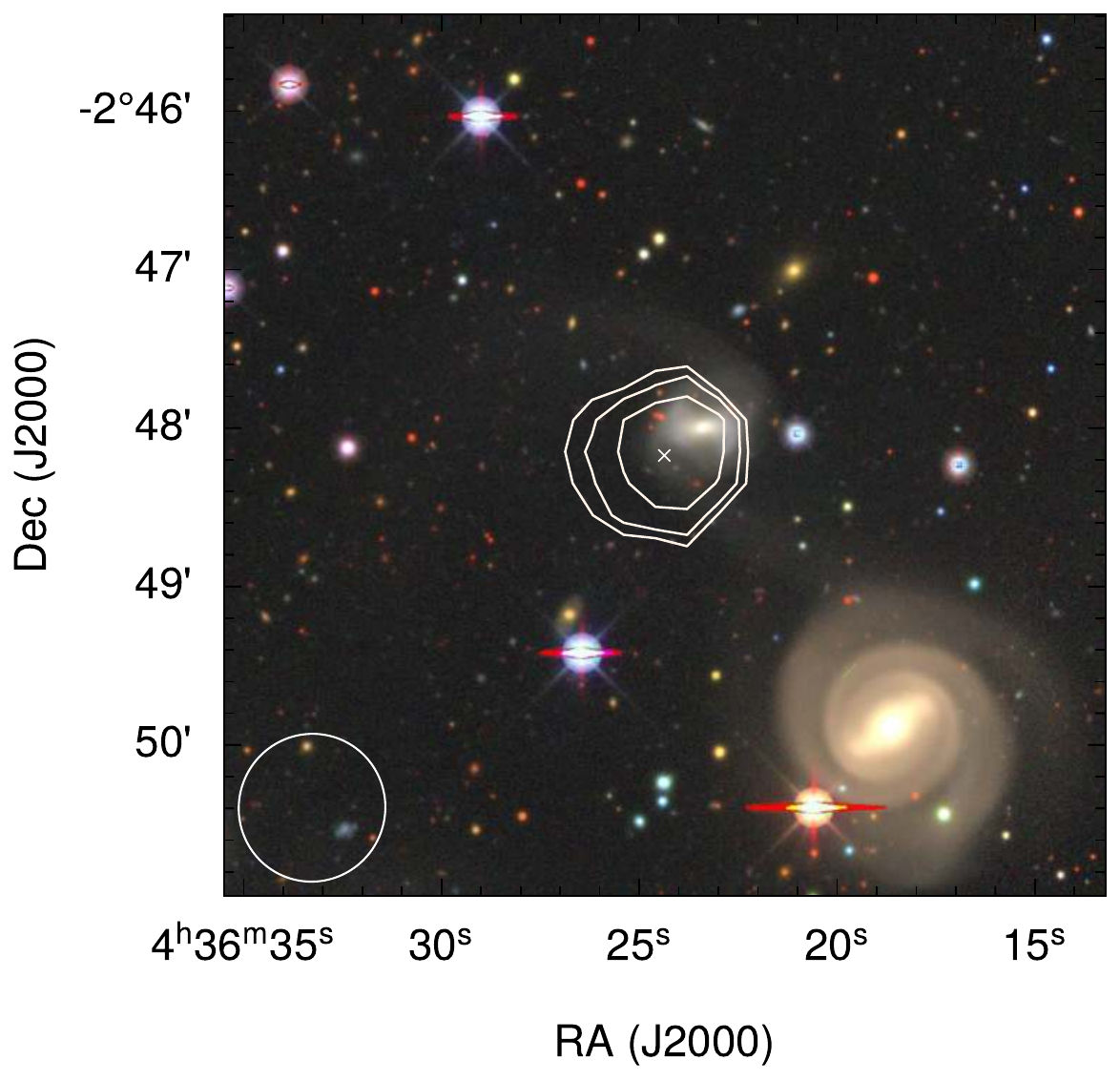} &
          \includegraphics[scale=0.29]{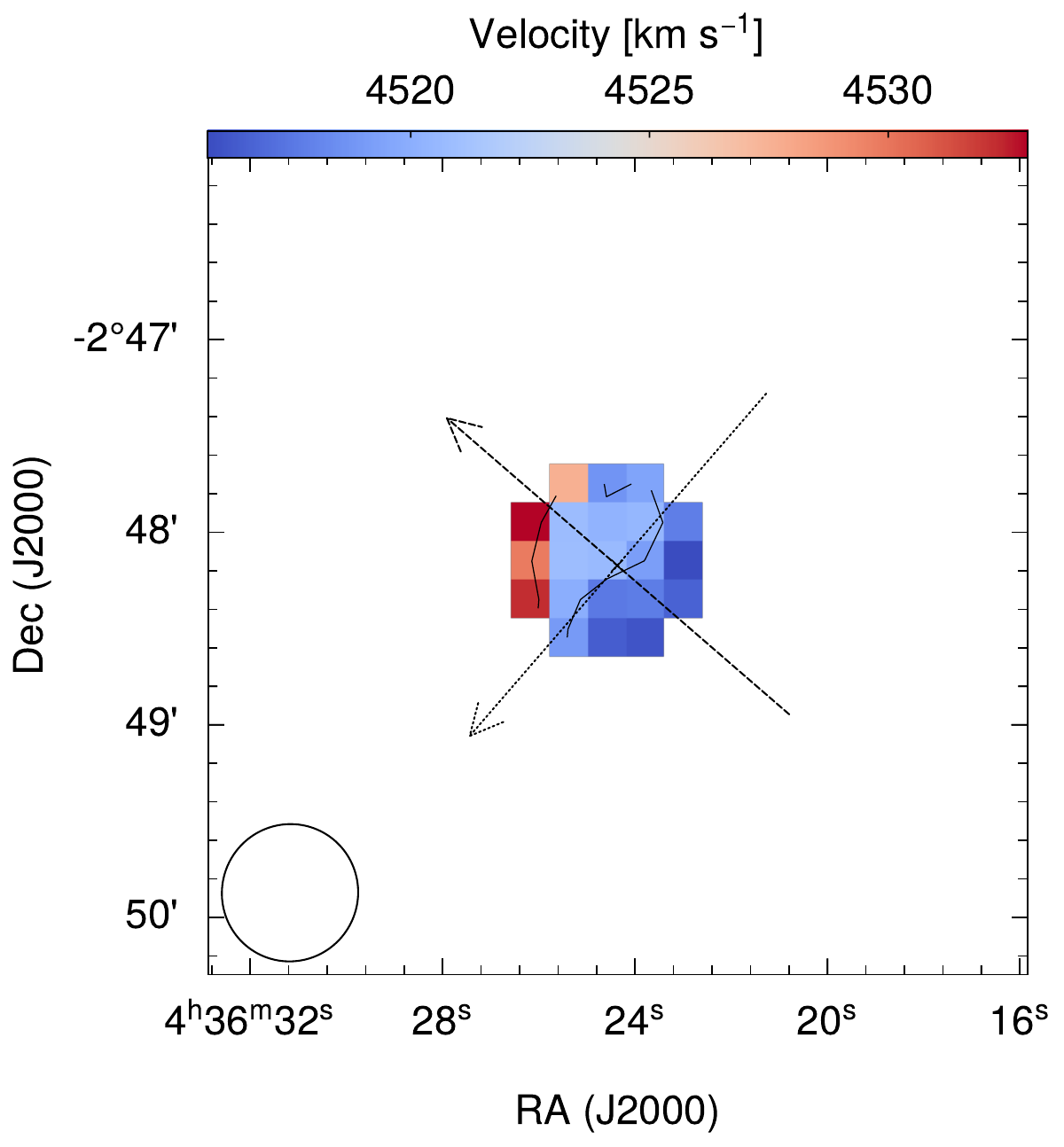} &
          \includegraphics[scale=0.29]{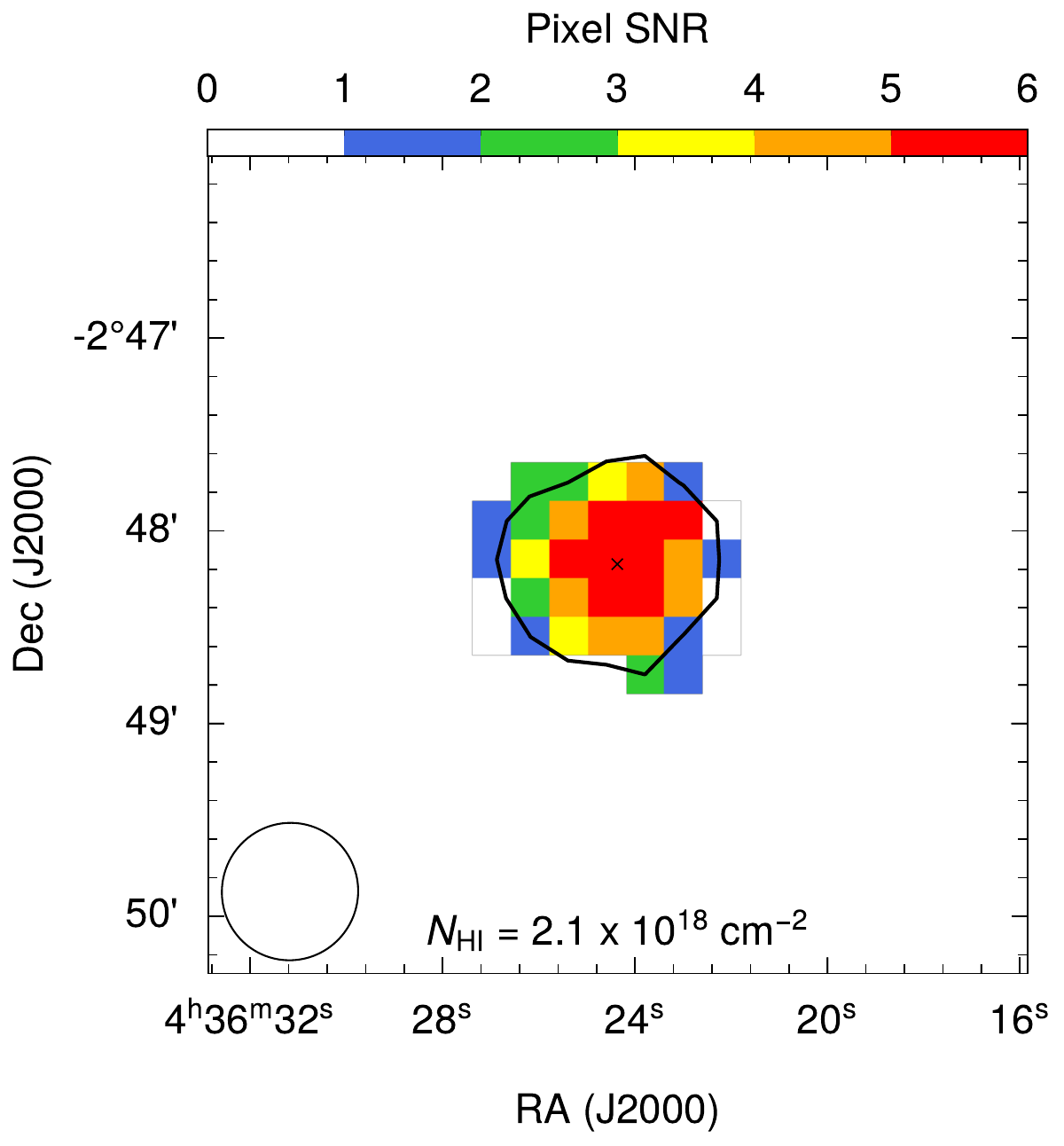}\\ 
          \includegraphics[scale=0.29]{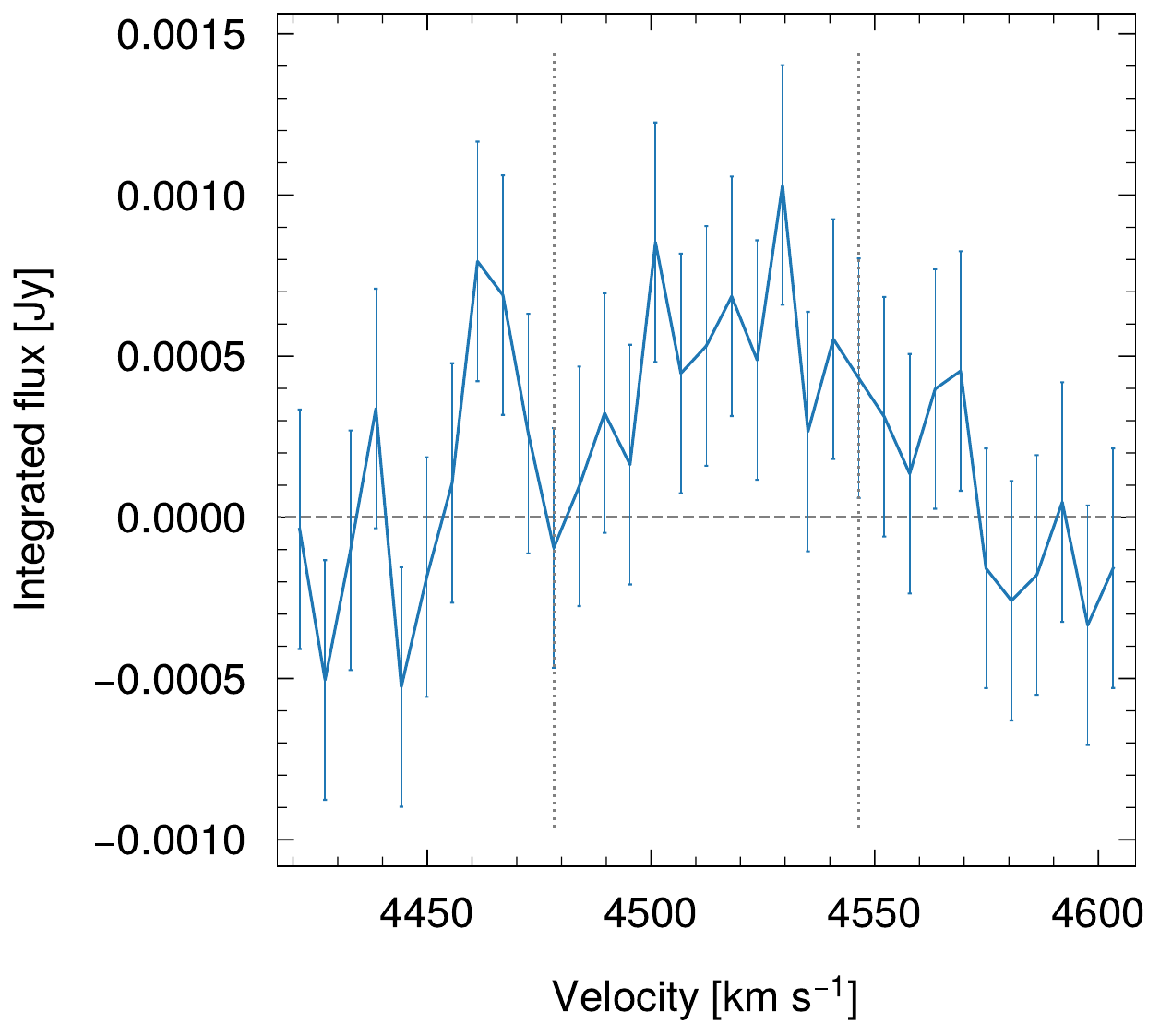} &
          \includegraphics[scale=0.29]{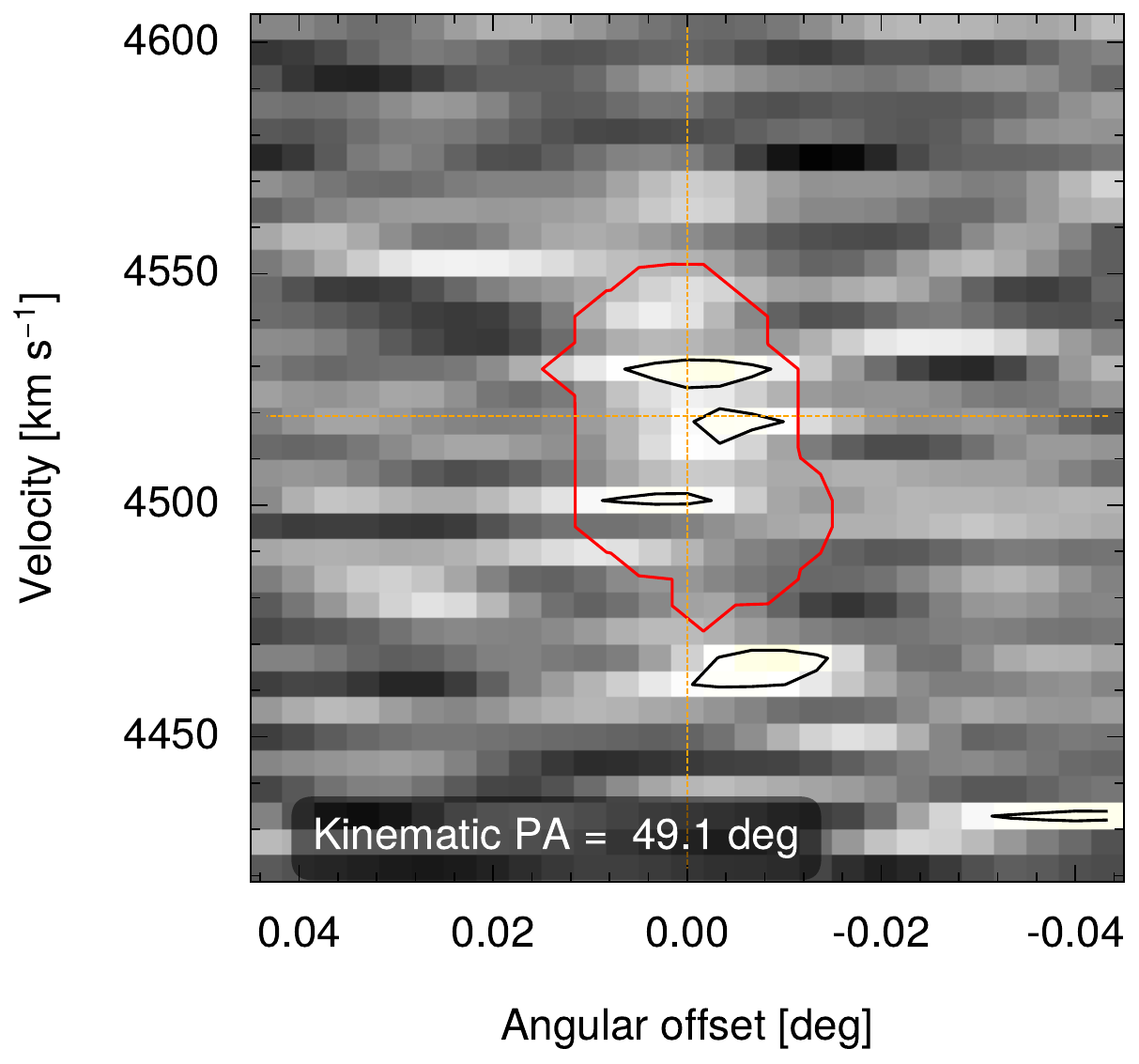}&
          \includegraphics[scale=0.29]{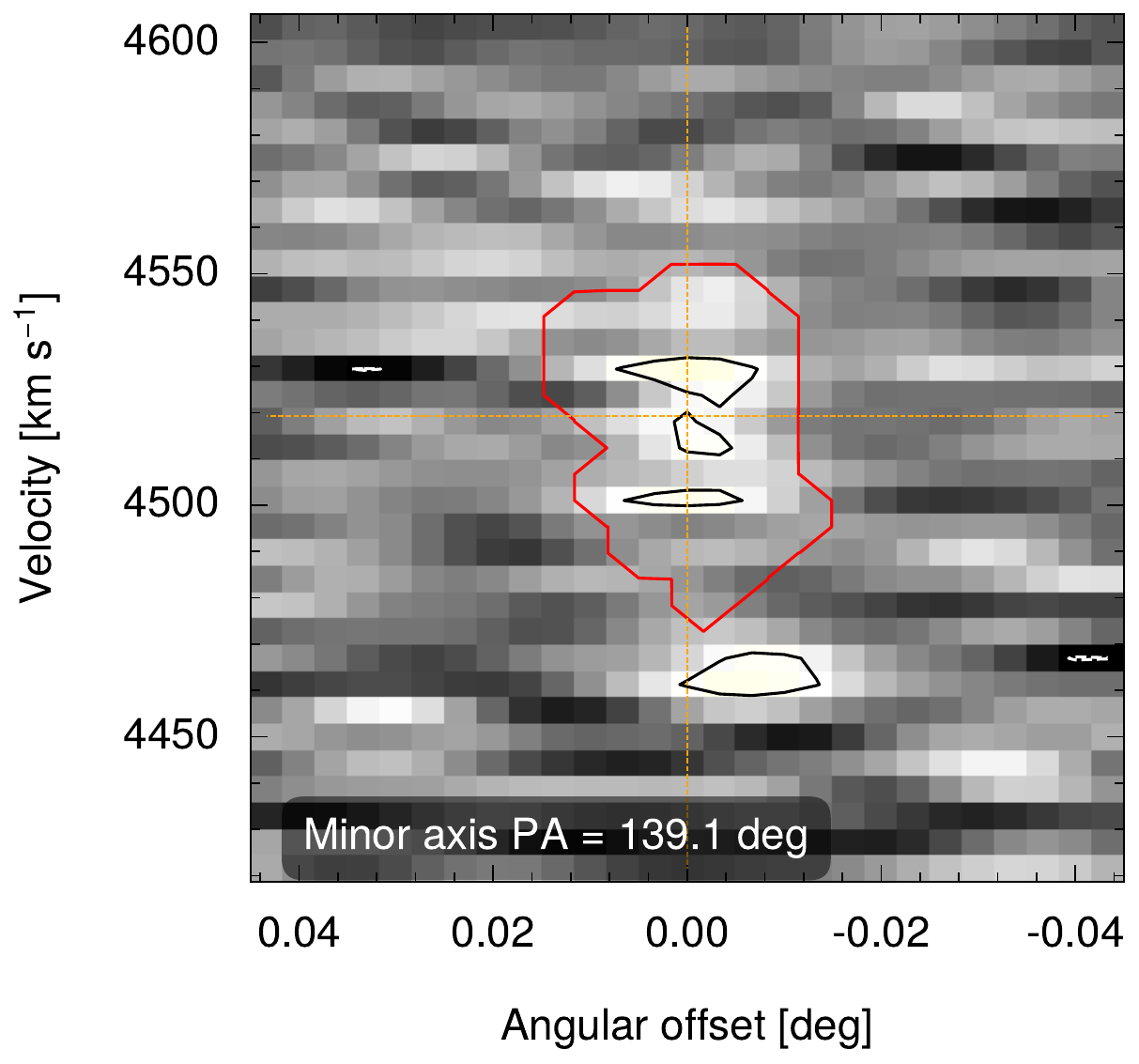}
        \end{tabular}
        \caption{Top left: \HI\ column density map overlaid on DECaLS optical image of HCG~30c. The contour levels are (2.10, 4.20, 8.41)~$\times~\mathrm{10^{18}~cm^{-2}}$. 
        Top centre: moment-1 map. The arrows indicate the slices from which the position-velocity diagrams shown at the bottom panels were derived. Top right: signal-to-noise ratio map. 
        Bottom left: global \HI\ profile. Bottom centre: major axis position-velocity diagram. Bottom right: minor axis position-velocity diagram.}
        \label{fig:hcg30c}
       \end{figure*}

\begin{figure*}
  \setlength{\tabcolsep}{1.2pt}
  \begin{tabular}{c c c}
      \includegraphics[scale=0.29]{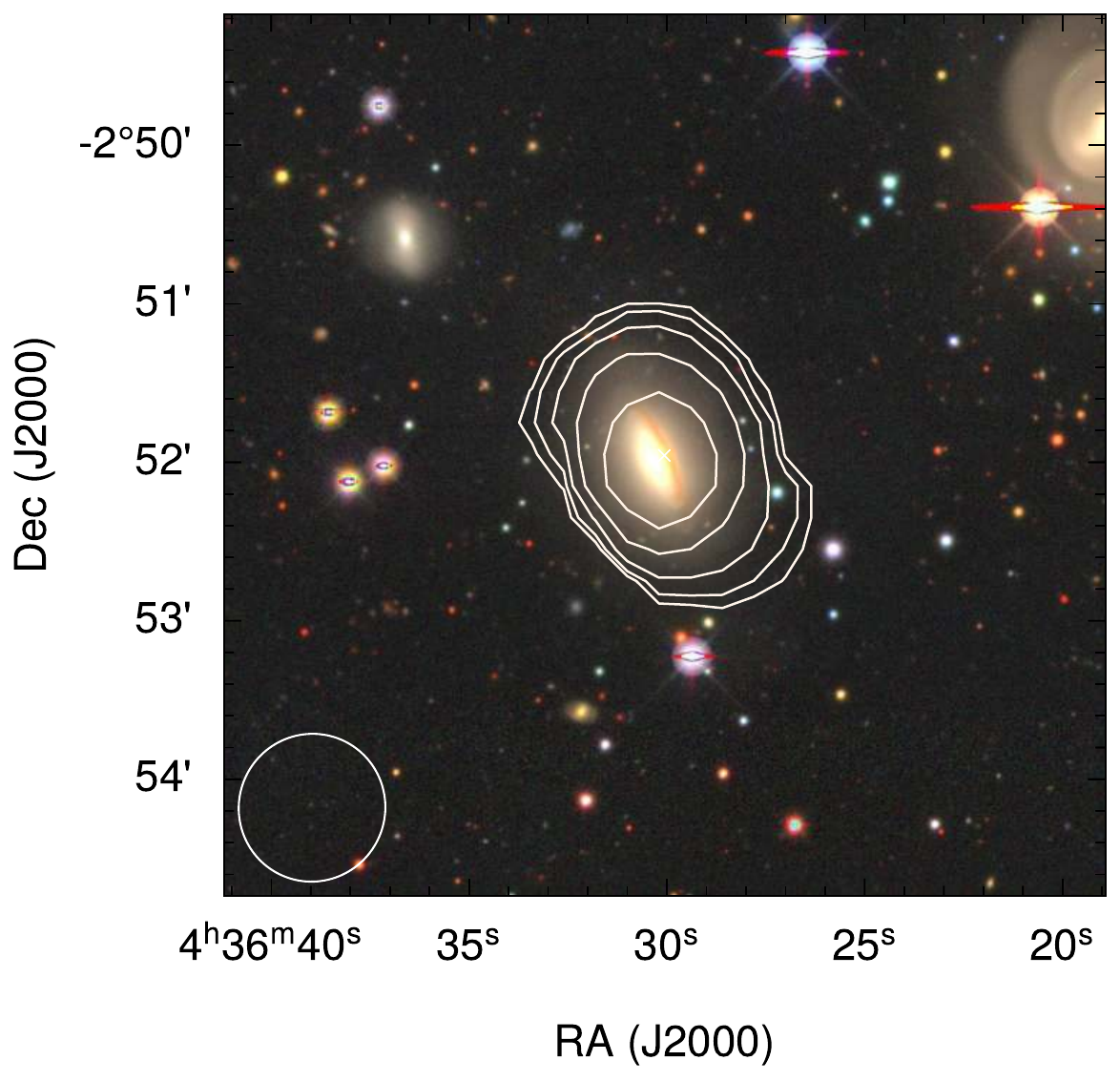} &
      \includegraphics[scale=0.29]{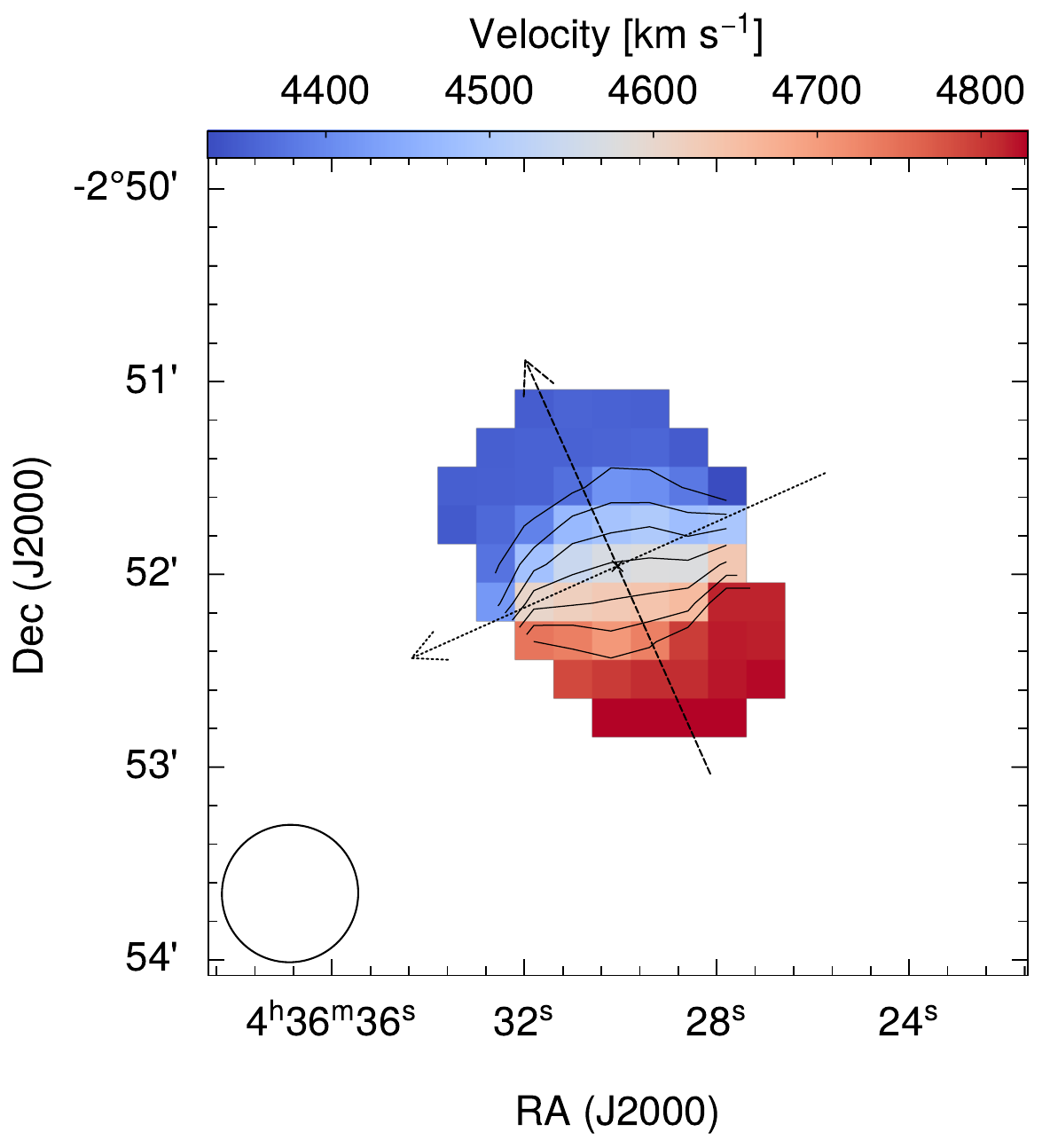} &
      \includegraphics[scale=0.29]{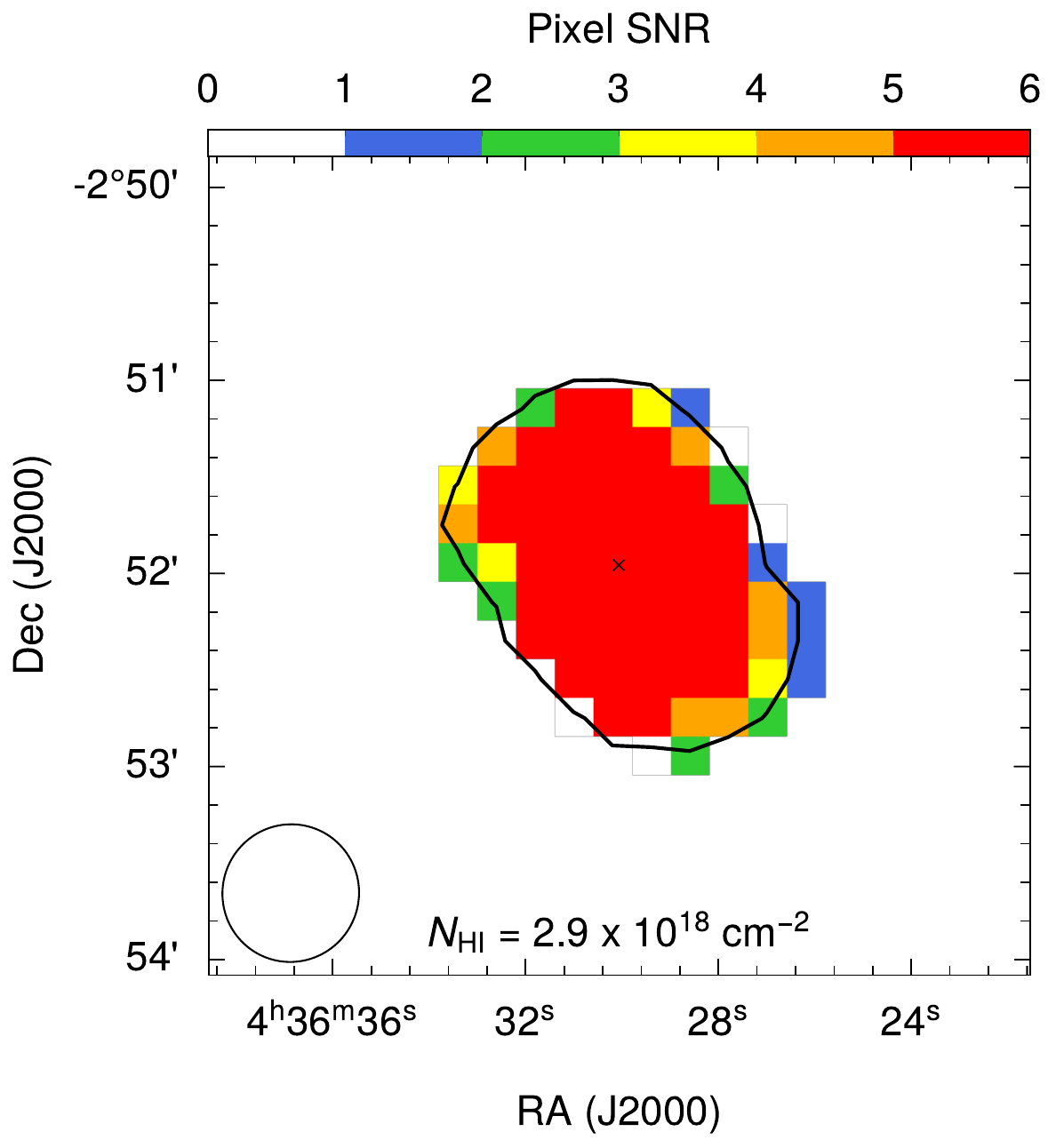}\\ 
      \includegraphics[scale=0.29]{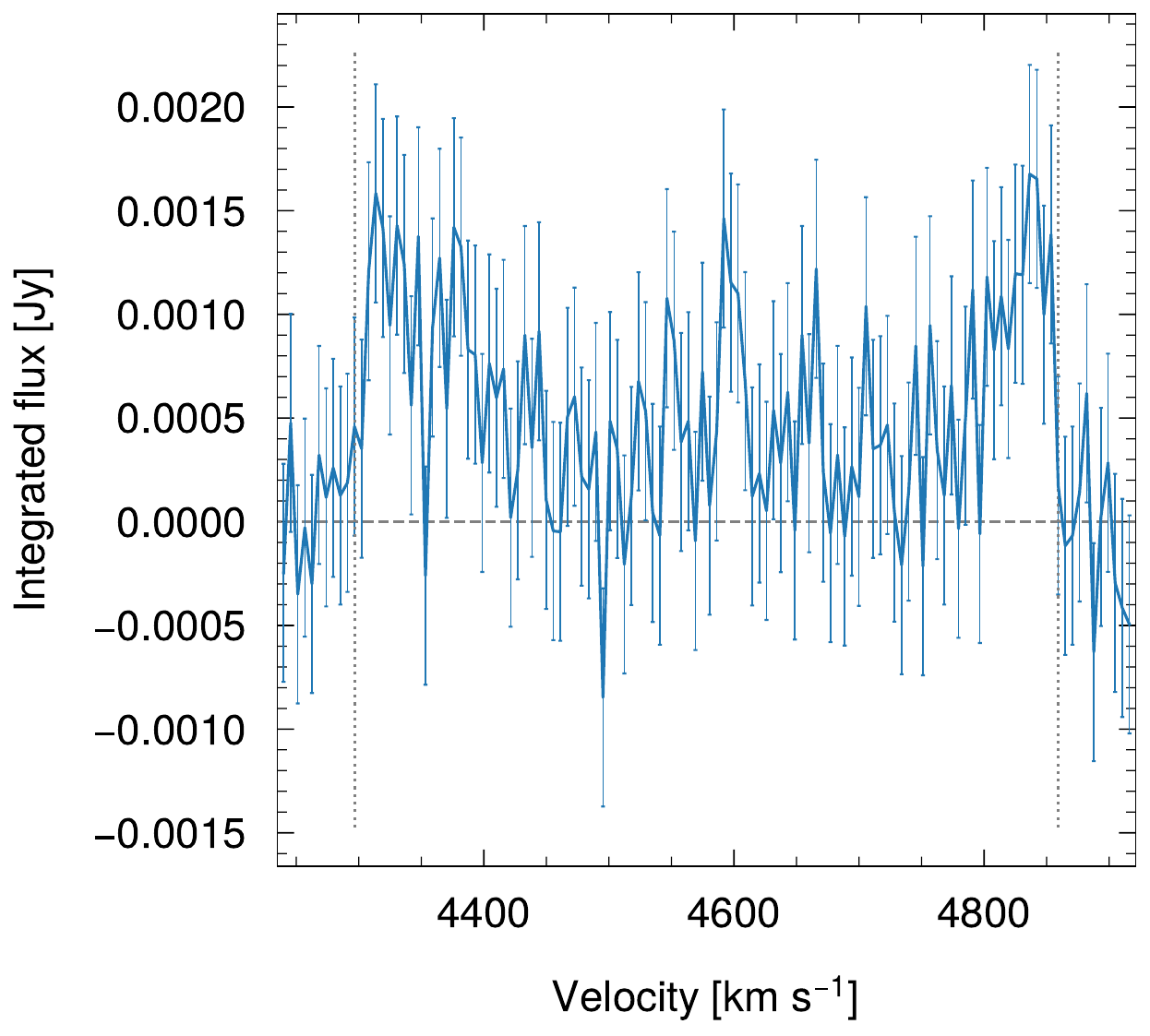} &
      \includegraphics[scale=0.29]{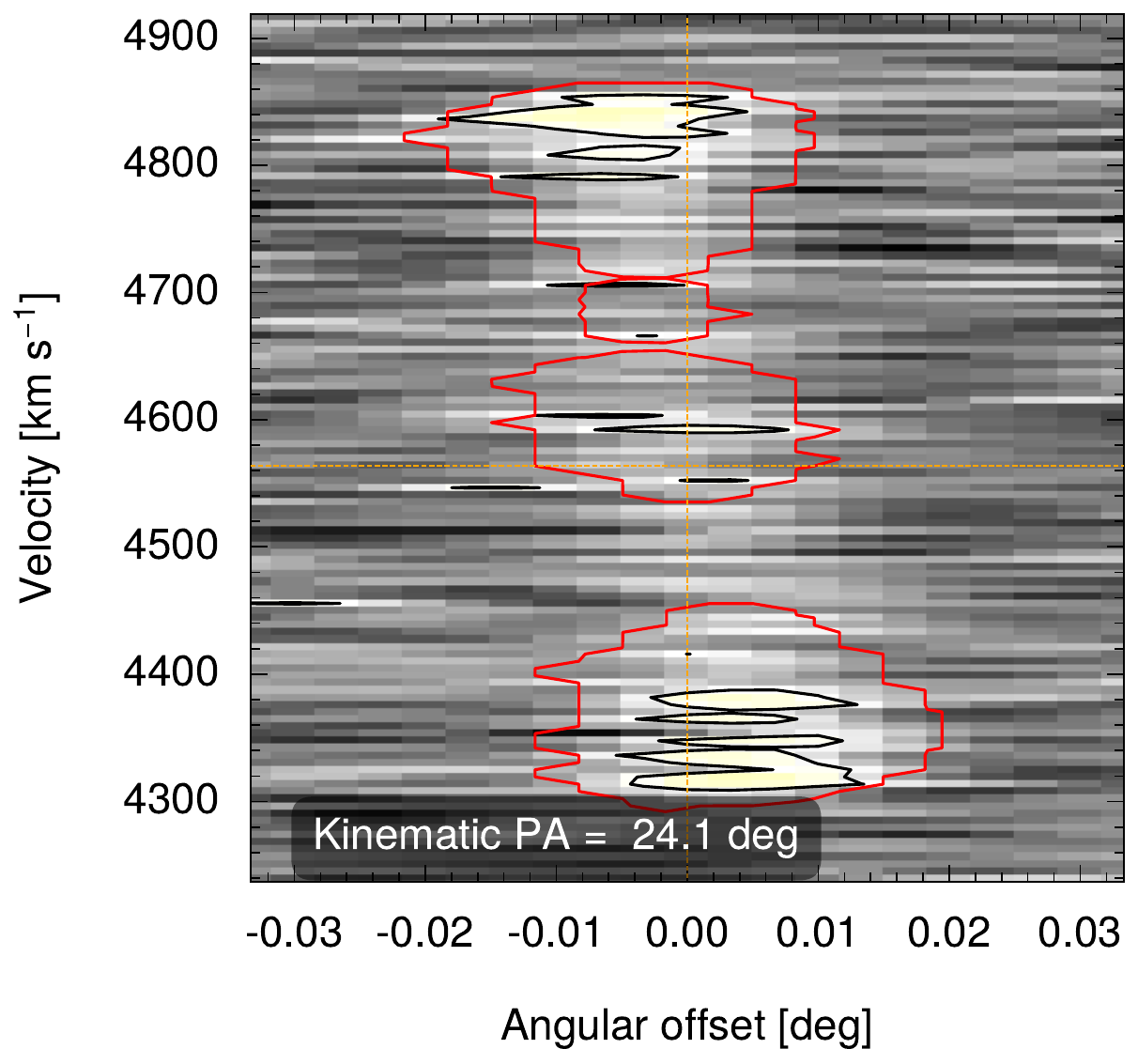}&
      \includegraphics[scale=0.29]{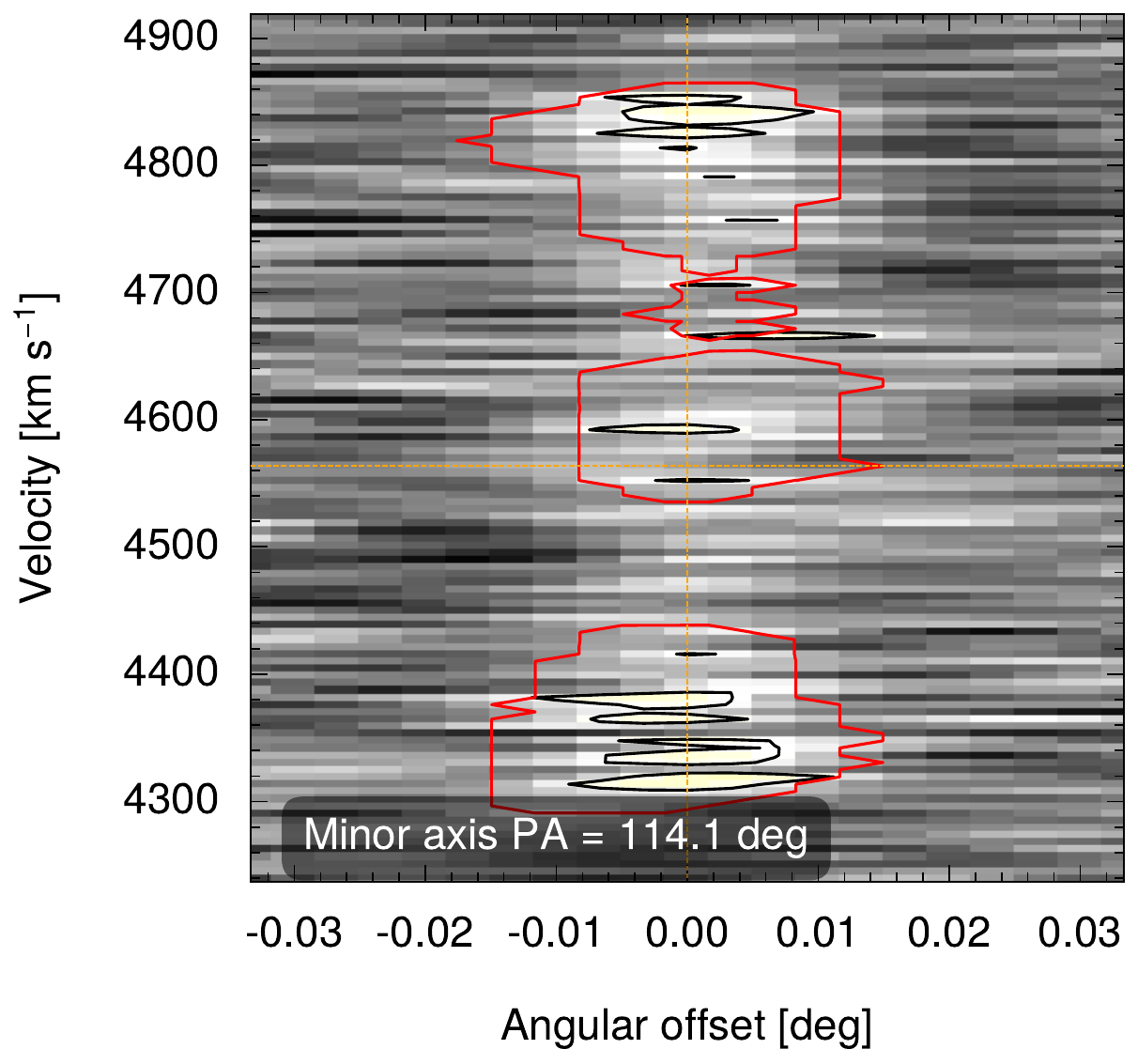}
    \end{tabular}
    \caption{Top left: \HI\ column density map overlaid on DECaLS optical image of HCG~30b. The contour levels are (0.29, 0.57, 1.14, 2.28, 4.57)~$\times~\mathrm{10^{19}~cm^{-2}}$. 
    Top centre: moment-1 map, the contour levels are (4384, 4444, 4504, 4564, 4624, 4684, 4744)~$\mathrm{km~s^{-1}}$. The arrows indicate the slices from which the position-velocity 
    diagrams shown at the bottom panels were derived. Top right: signal-to-noise ratio map. Bottom left: global \HI\ profile. Bottom centre: major axis position-velocity diagram. 
    Bottom right: minor axis position-velocity diagram.}
    \label{fig:hcg30b}
   \end{figure*}

Our tentative detection of \HI\ in the core members of HCG~30 is illustrated in Figure~\ref{fig:hcg30c} and Figure~\ref{fig:hcg30b}. These plots were generated 
by the SoFiA Image Pipeline\footnote{\url{https://github.com/kmhess/SoFiA-image-pipeline}}\citep{SIP}, an automated tool that takes as an input a SoFiA-generated source catalogue 
and produces overview plots of the \HI\ properties of the sources. Figure~\ref{fig:hcg30c} and Figure~\ref{fig:hcg30b} show \HI\ contours overlaid on DeCALS false-colour images, 
\HI\ moment maps, pixel-by-pixel signal-to-noise ratio maps, position–velocity diagrams, and global \HI\ spectra.  

\subsection{3D visualisation} 
\begin{figure*}
      \setlength{\tabcolsep}{0pt}
      \begin{tabular}{c c}
      \includegraphics[scale=0.345]{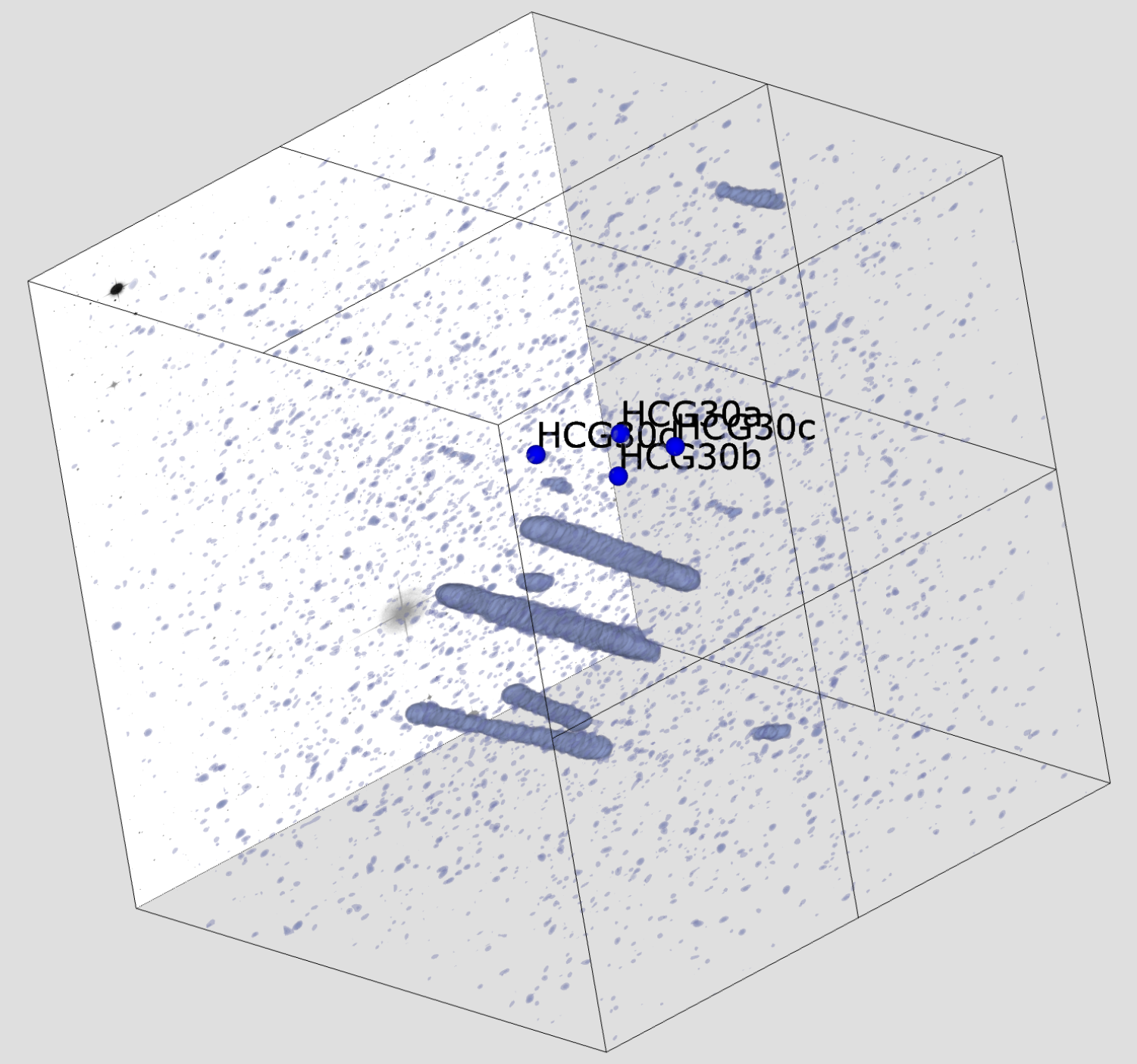}
      \end{tabular}
      \caption{3D visualisation of HCG 30. The blue circles indicate the position of the member galaxies. The 2D grayscale image is a DeCaLS R-band optical image of the group. 
      The online version of the cubes are available at \href{https://amiga.iaa.csic.es/x3d-menu/}{https://amiga.iaa.csic.es/x3d-menu/}.}
    \label{fig:hcg30_3dvis}
   \end{figure*}

Figure~\ref{fig:hcg30_3dvis} presents a 3D visualisation of HCG~30. The member galaxies are indicated by the blue circles, and the grayscale background is a DECaLS R-band optical image. 
An interactive 3D version of these cubes can be found at \href{https://amiga.iaa.csic.es/x3d-menu/}{https://amiga.iaa.csic.es/x3d-menu/}. 
\section{Supplementary figures of HCG~90}
\subsection{Noise properties and global profiles}
\begin{figure*}
  \setlength{\tabcolsep}{0pt}
  \begin{tabular}{l l l}
      \includegraphics[scale=0.215]{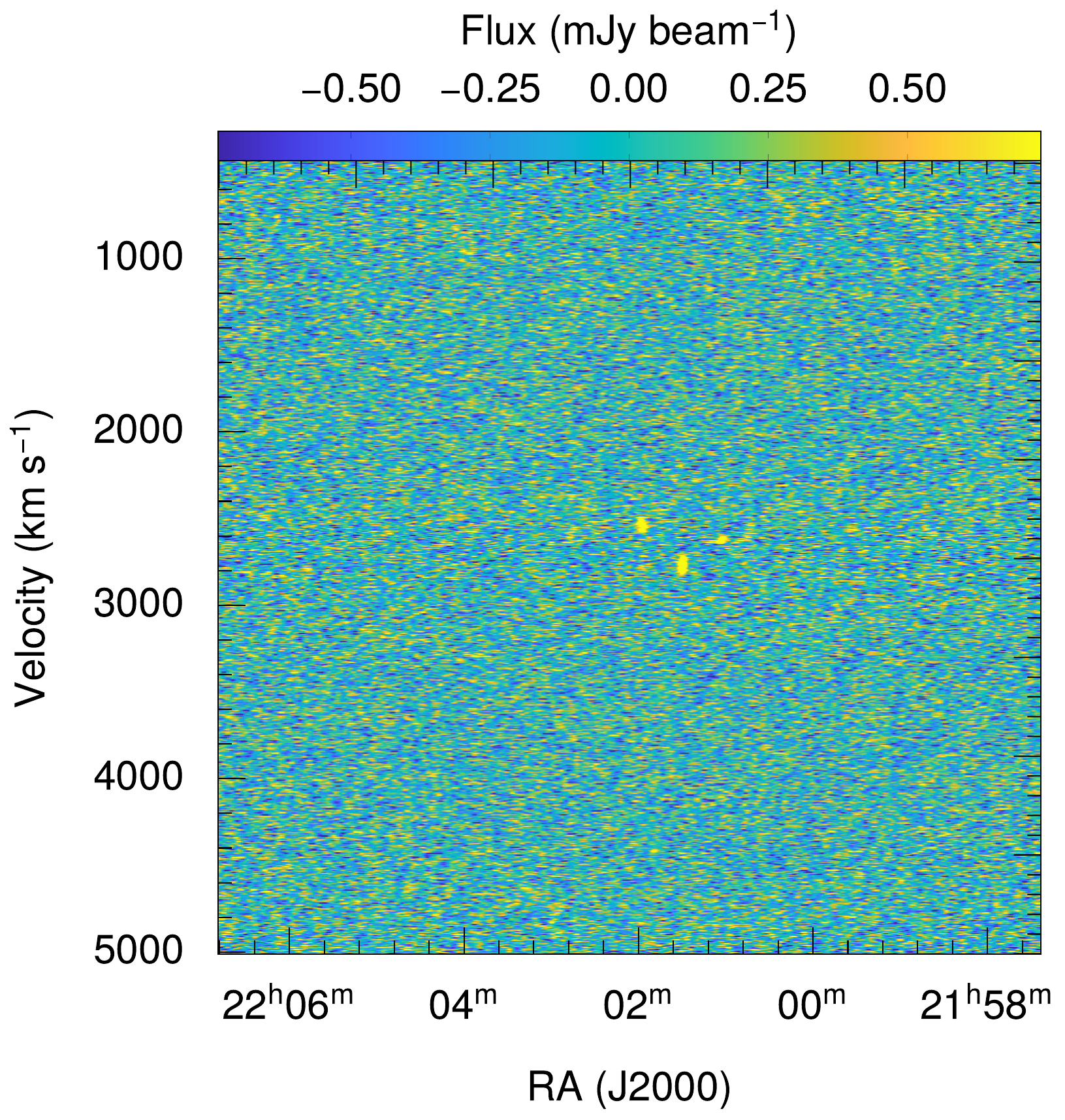} &
      \includegraphics[scale=0.215]{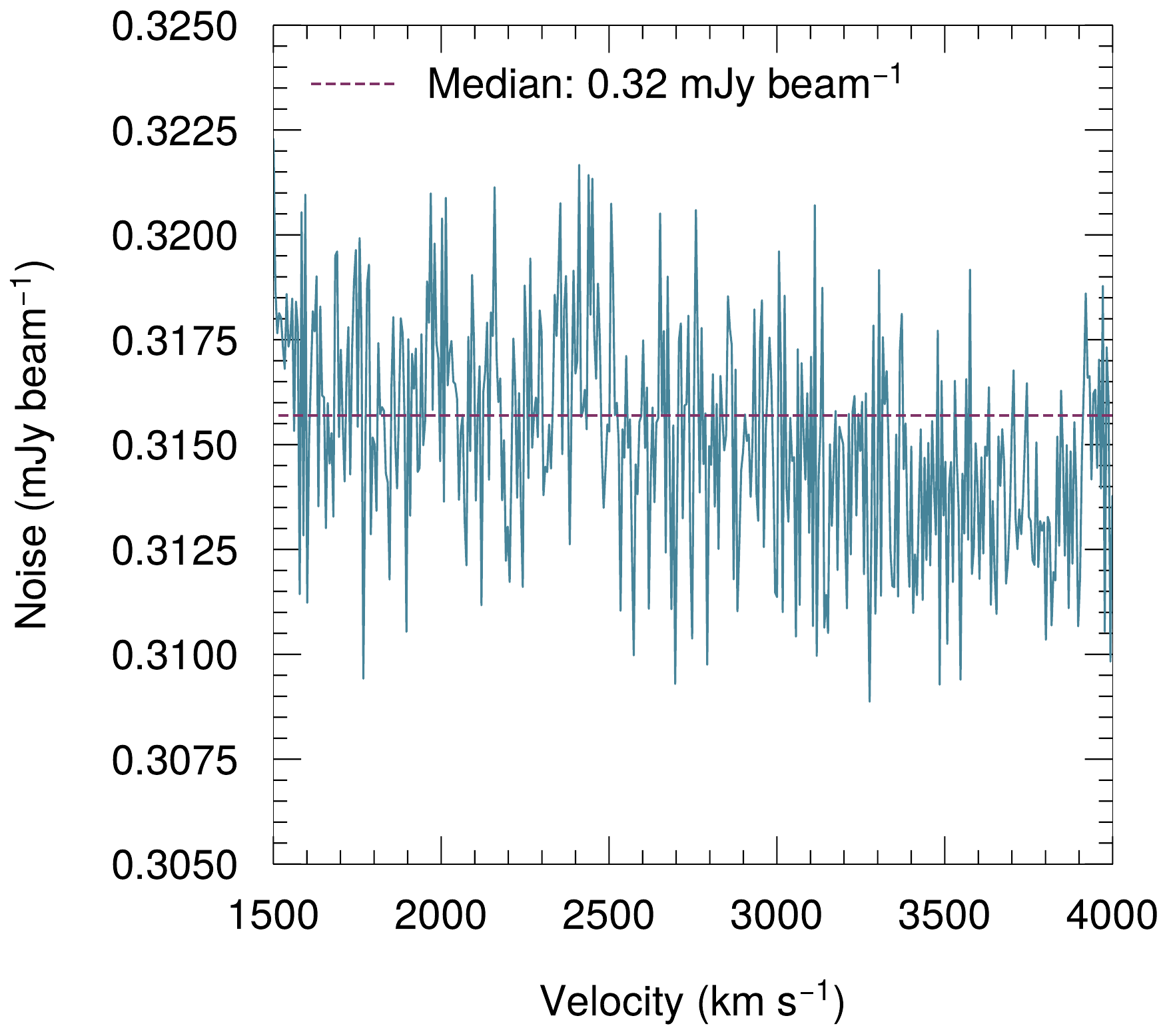}&
      \includegraphics[scale=0.215]{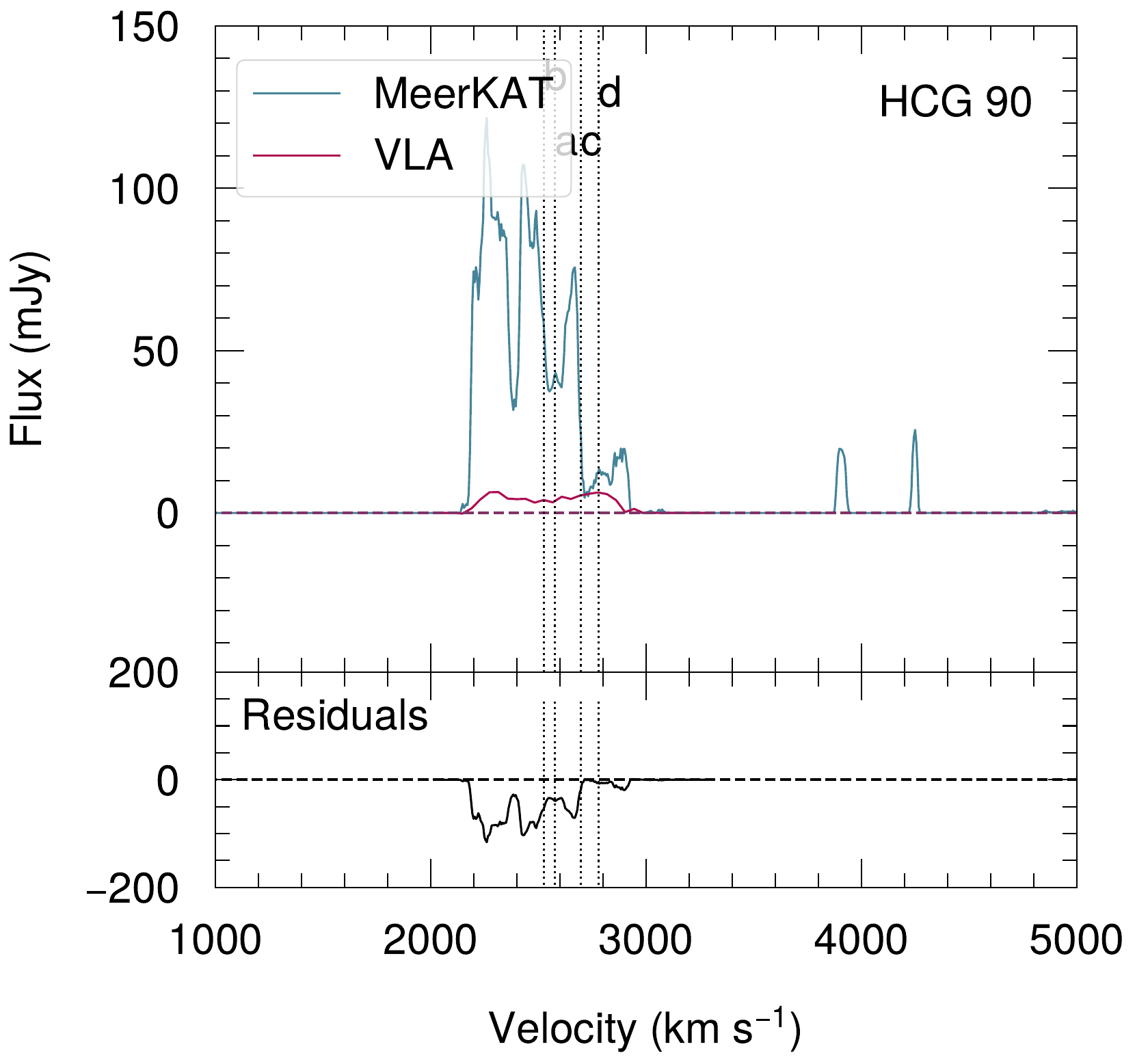}
    \end{tabular}
    \caption{Left panel: velocity vs right ascension of HCG~90. Middle panel: median noise values of each RA-DEC slice of the non-primary beam corrected 60\arcsec\ data cube of 
    HCG 90 as a function of velocity. The horizontal dashed line indicates the median of all the noise values from each slice. Right panel: the blue solid lines indicates the 
    MeerKAT integrated spectrum of HCG~90; the red solid line indicates VLA integrated spectrum of the group derived by \citep{2023A&A...670A..21J}. 
    The vertical dotted lines indicate the velocities of the galaxies in the core of the group. The spectra have been extracted from areas containing only genuine \HI\ emission.}
    \label{fig:hcg90_noise}
   \end{figure*}

Figure~\ref{fig:hcg90_noise} shows the noise variations and global profiles of HCG~90. 
The left panel shows a RA-velocity slice. The middle panel plots the median noise levels per RA–DEC slice. The horizontal dashed line marks the overall median noise level. 
The right panel presents the MeerKAT integrated spectrum (blue solid line) compared with the VLA data (red solid line) originally presented by \citet{2023A&A...670A..21J}. 
The vertical dotted lines indicate the systemic velocities of the galaxies in the core of the group.
  %
  
\subsection{Channel maps}  
\begin{figure*}
      \setlength{\tabcolsep}{0pt}
      \begin{tabular}{l l l}
          \includegraphics[scale=0.25]{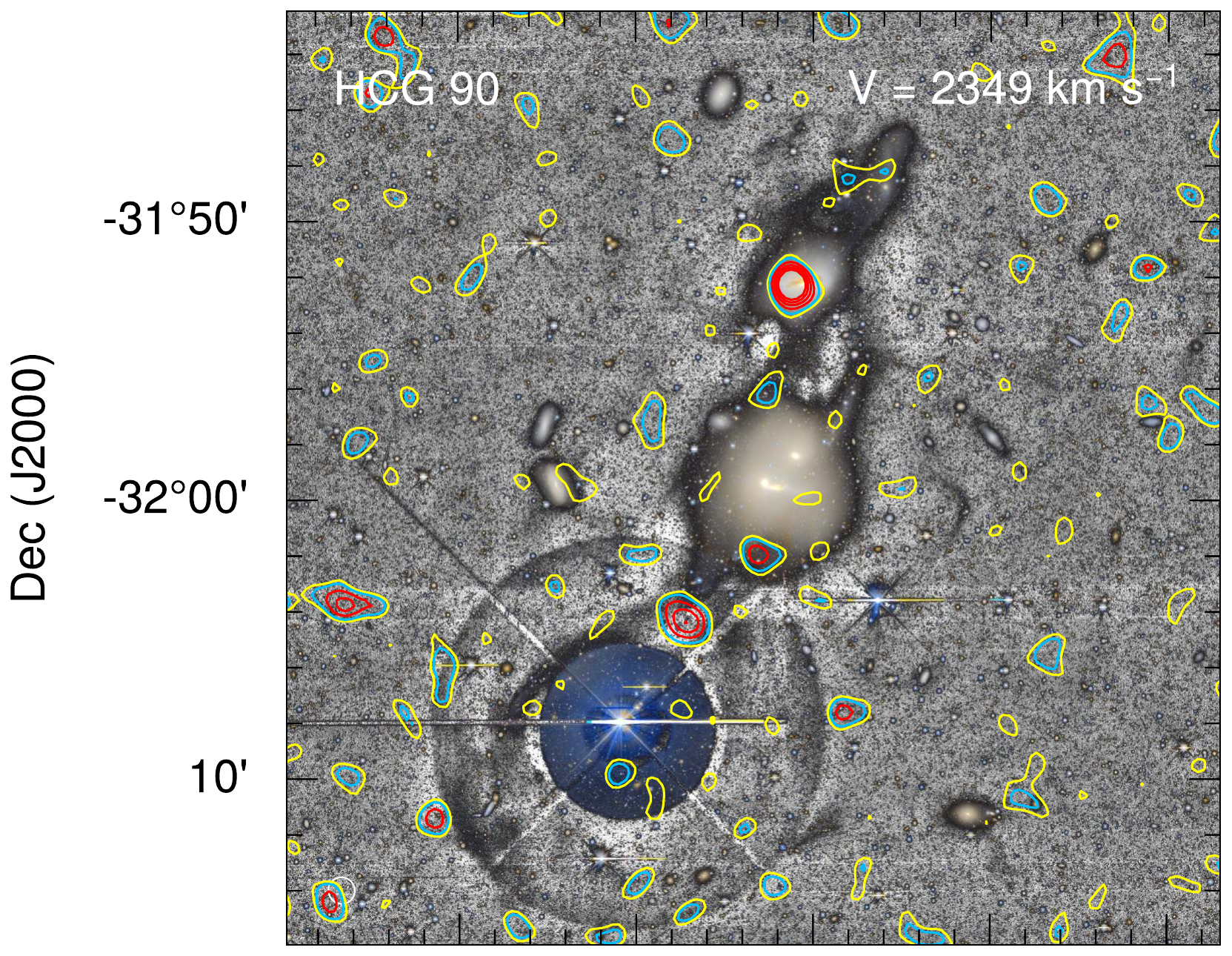} &
          \includegraphics[scale=0.25]{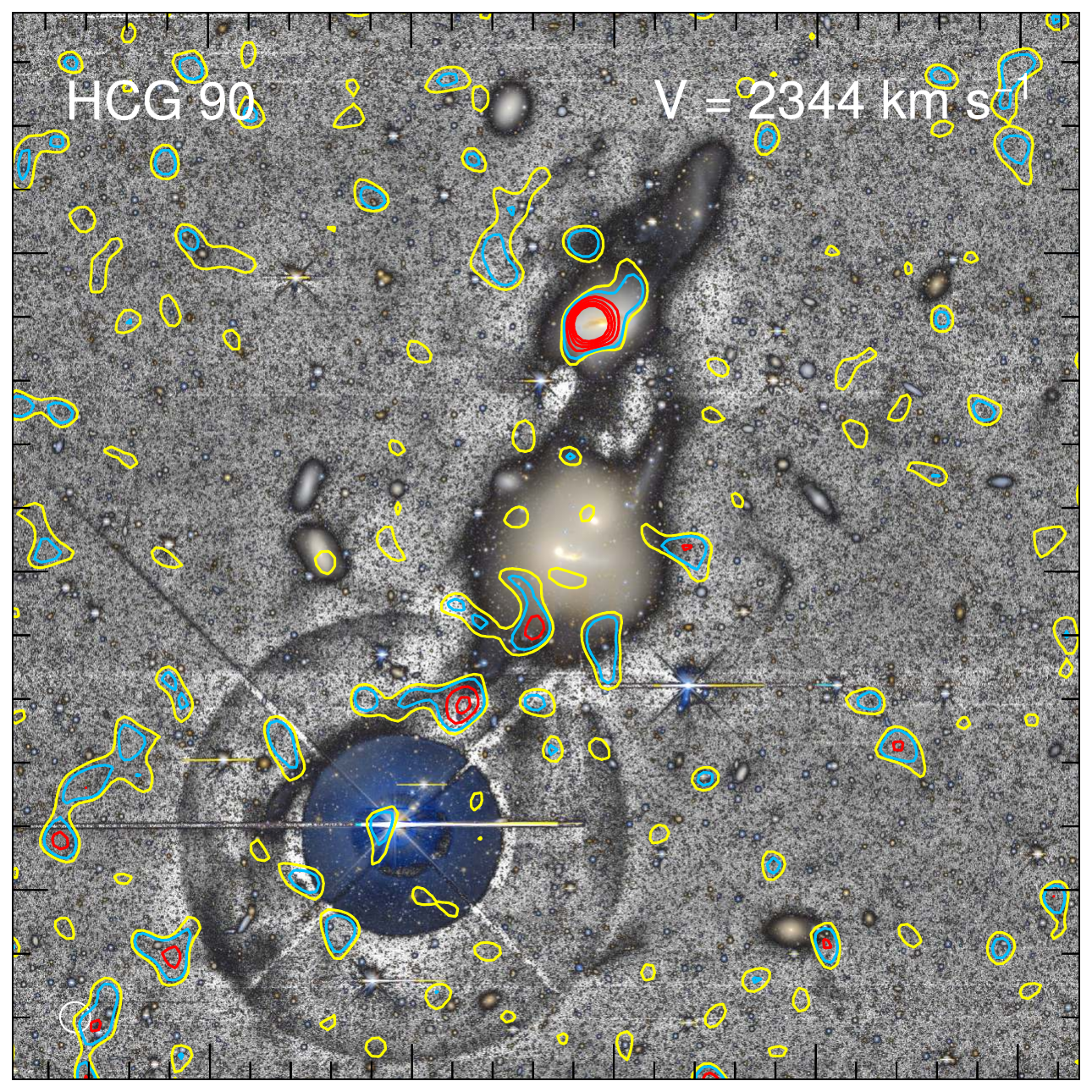} &
          \includegraphics[scale=0.25]{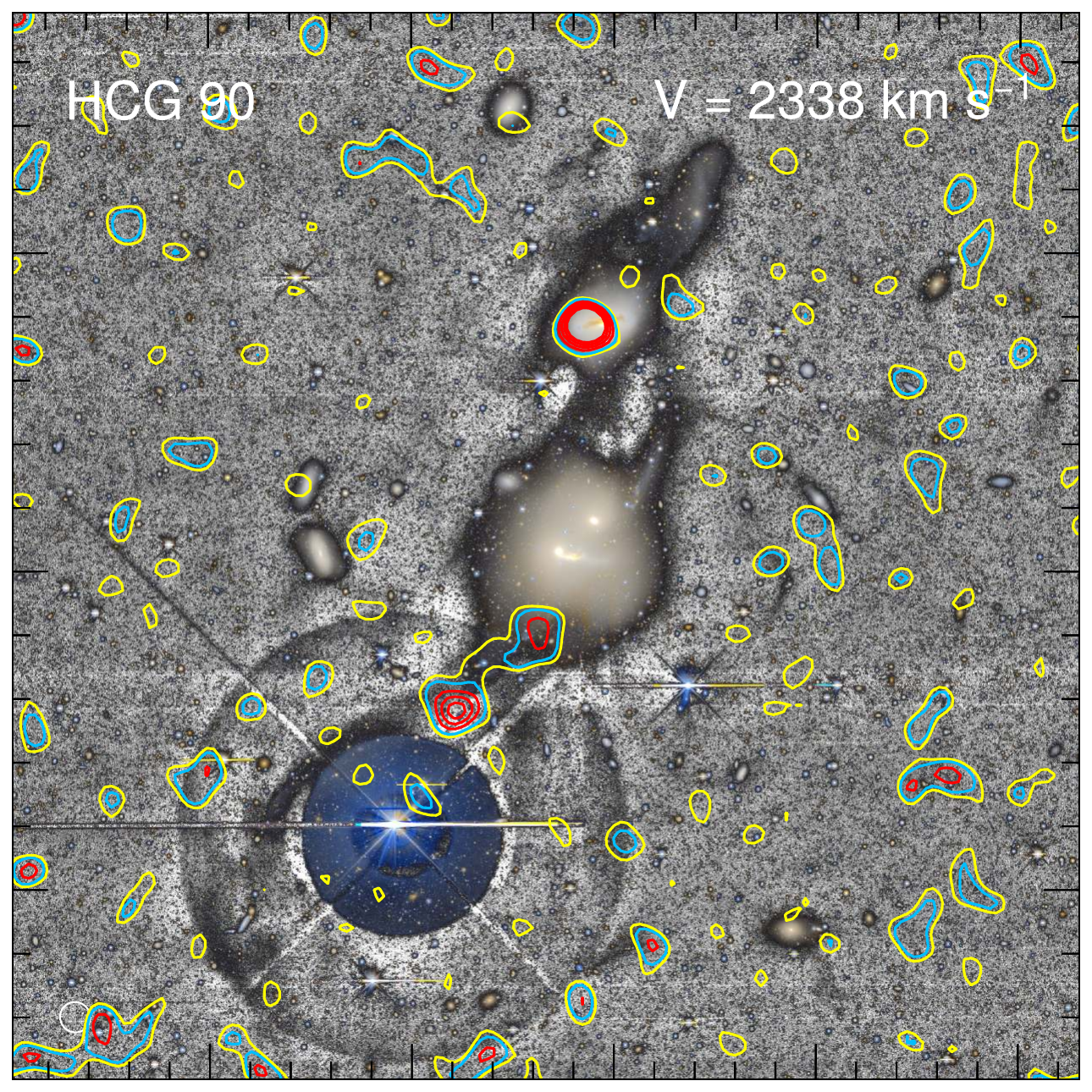} \\[-0.2cm]
          \includegraphics[scale=0.25]{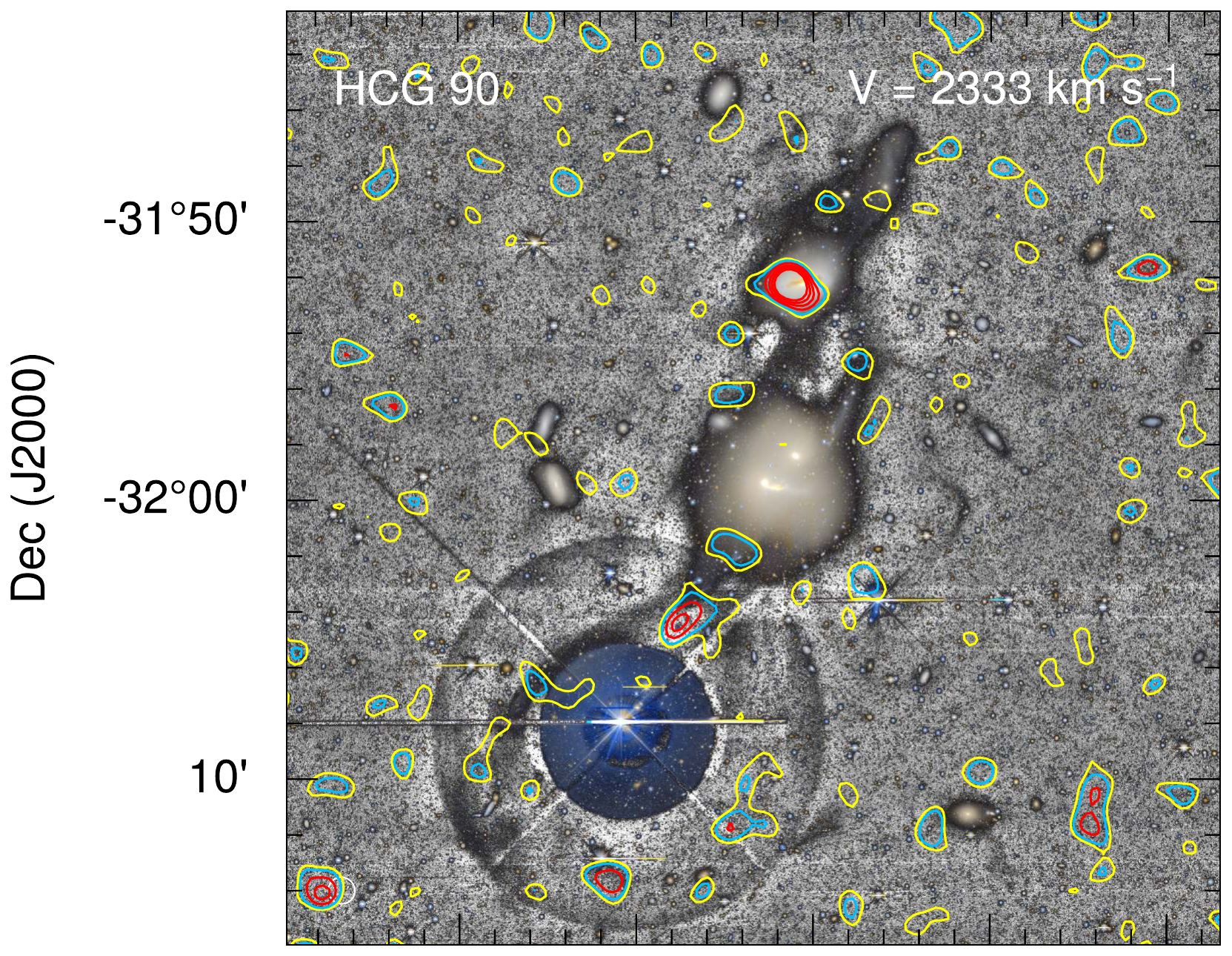} &
          \includegraphics[scale=0.25]{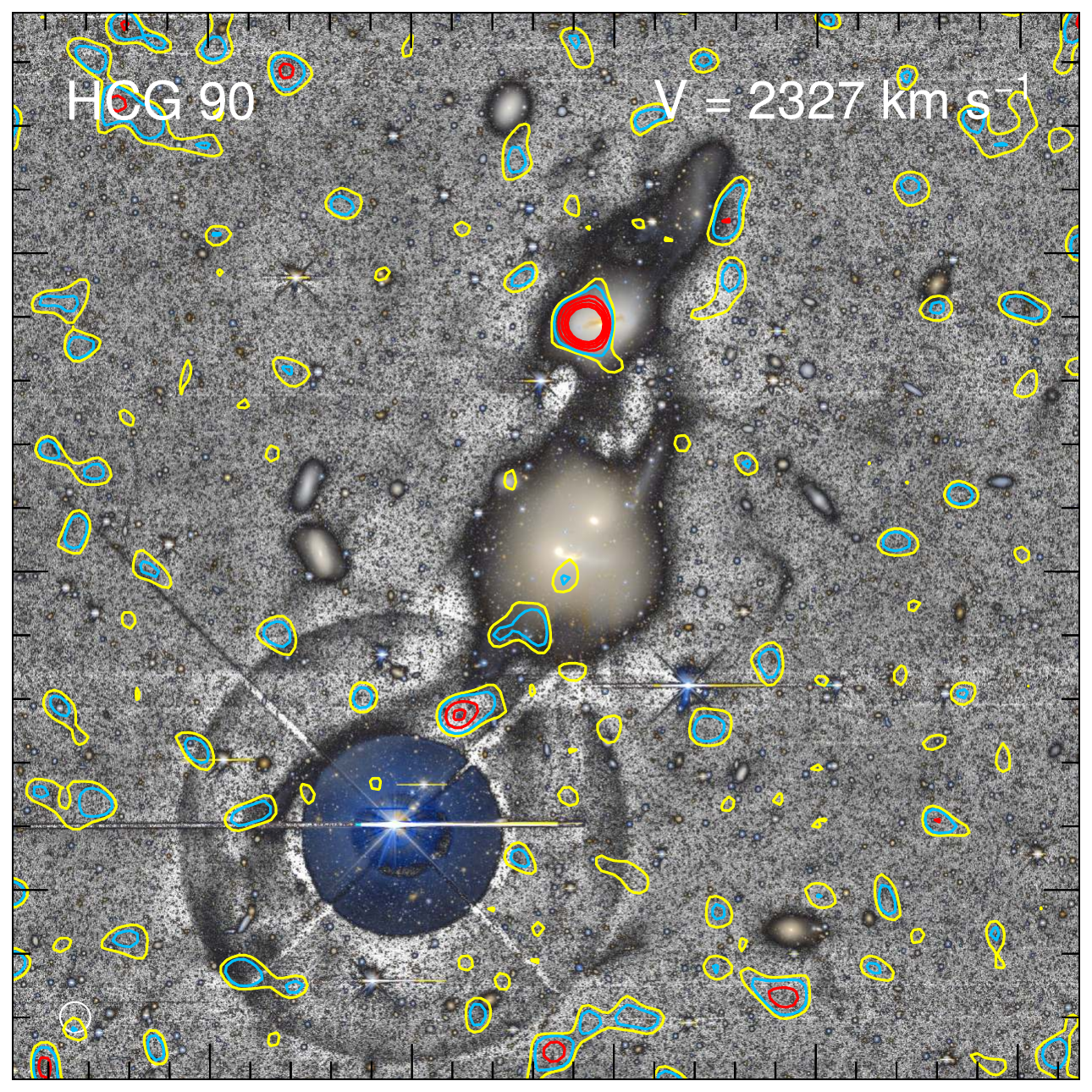} &
          \includegraphics[scale=0.25]{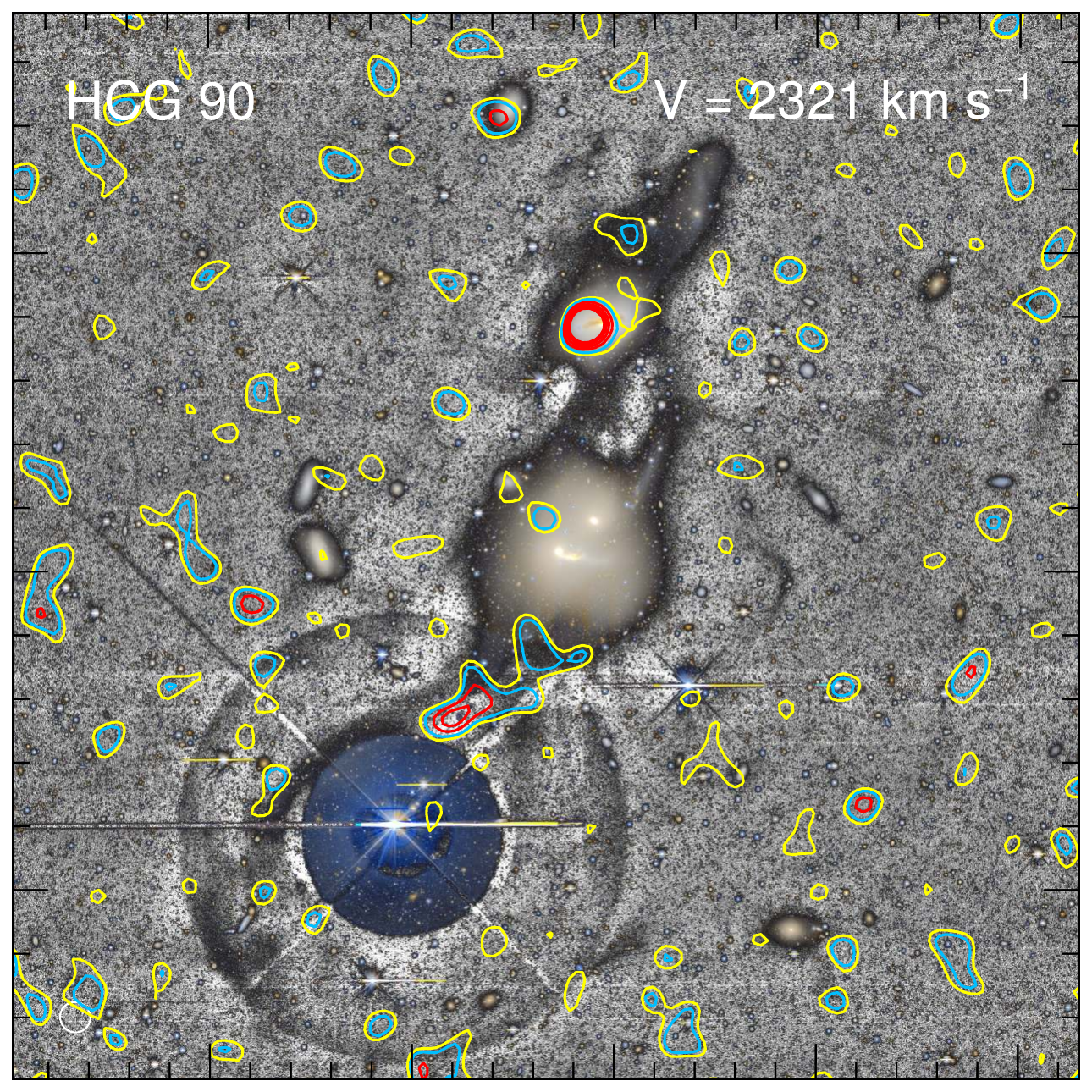} \\[-0.2cm]
          \includegraphics[scale=0.25]{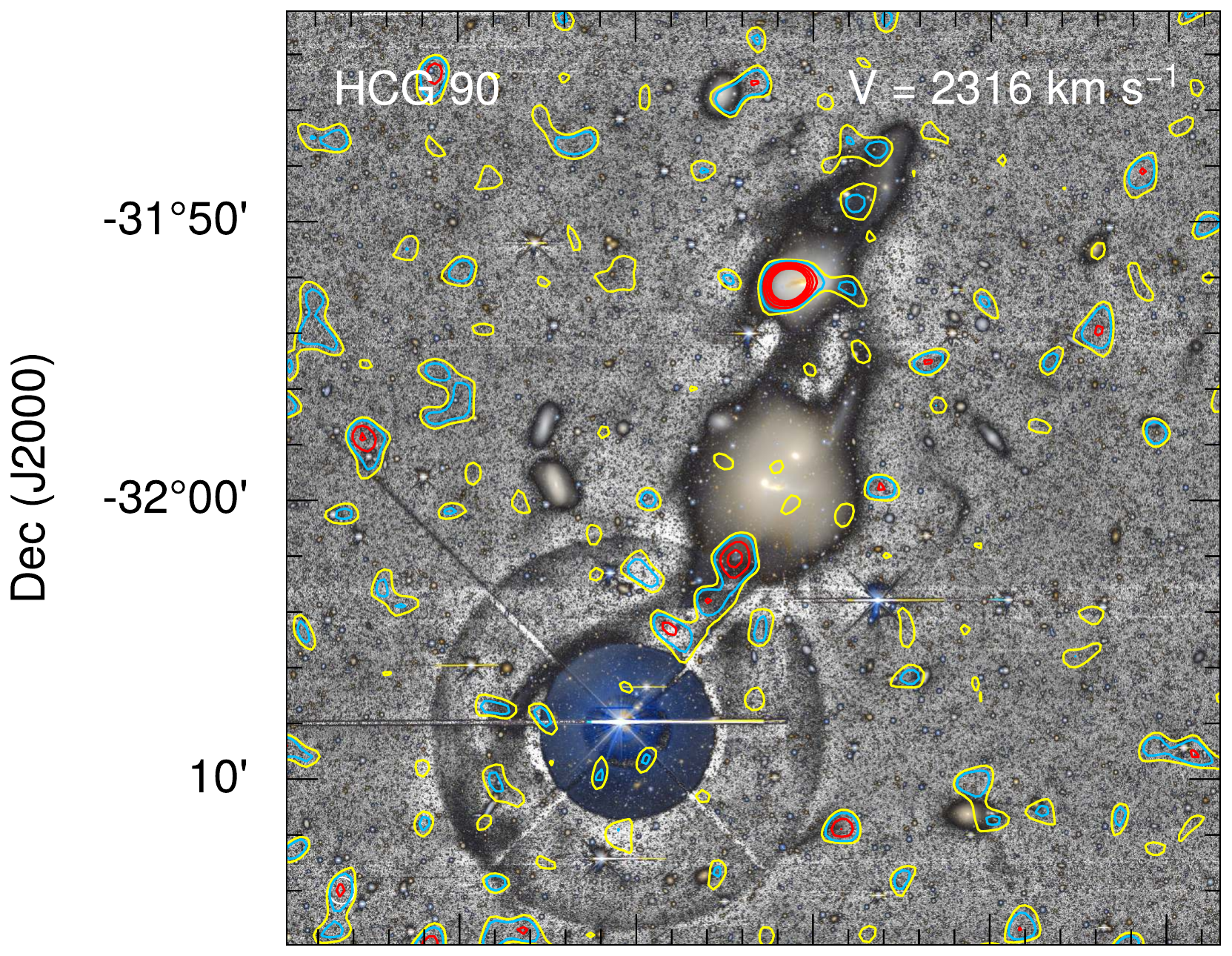} &
          \includegraphics[scale=0.25]{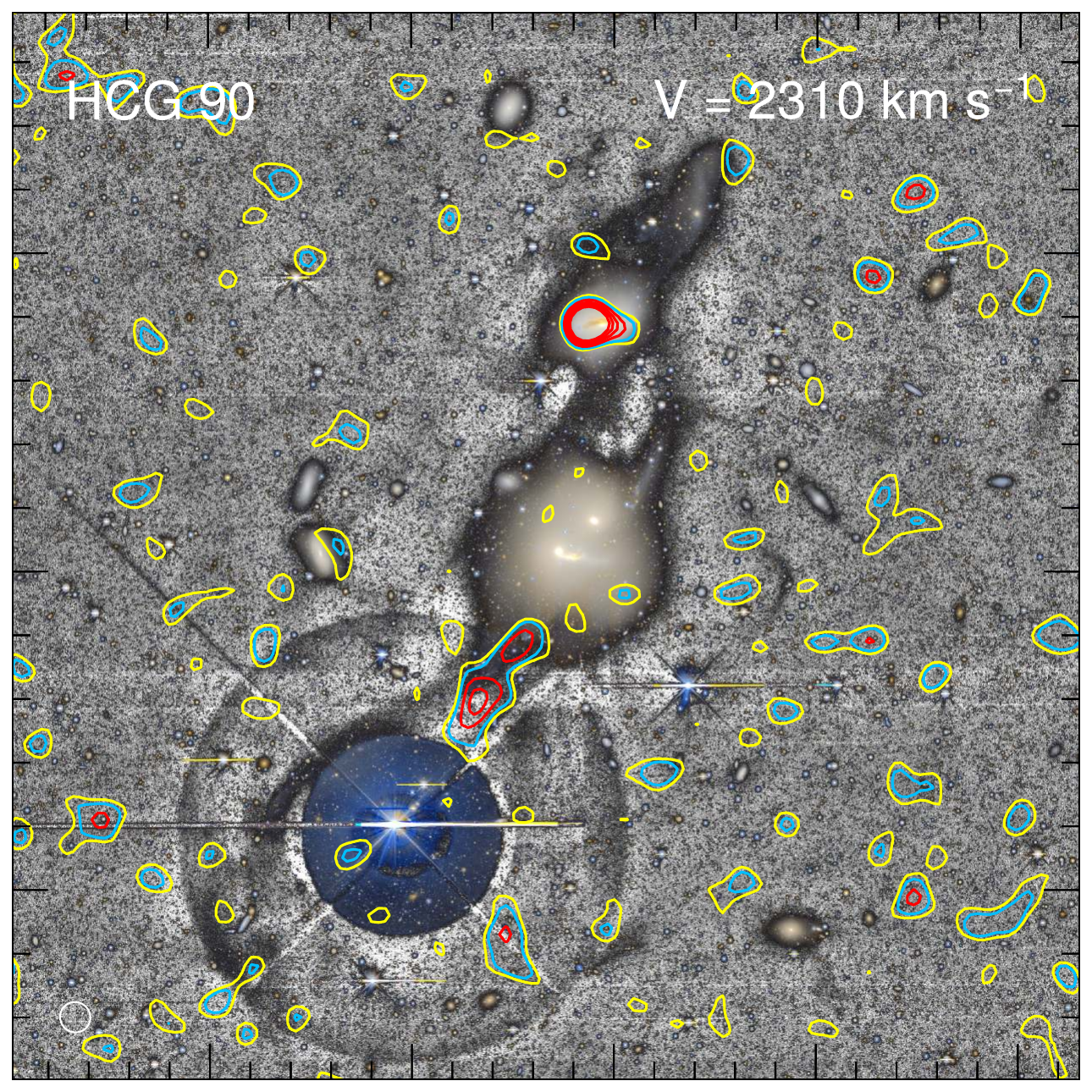} &
          \includegraphics[scale=0.25]{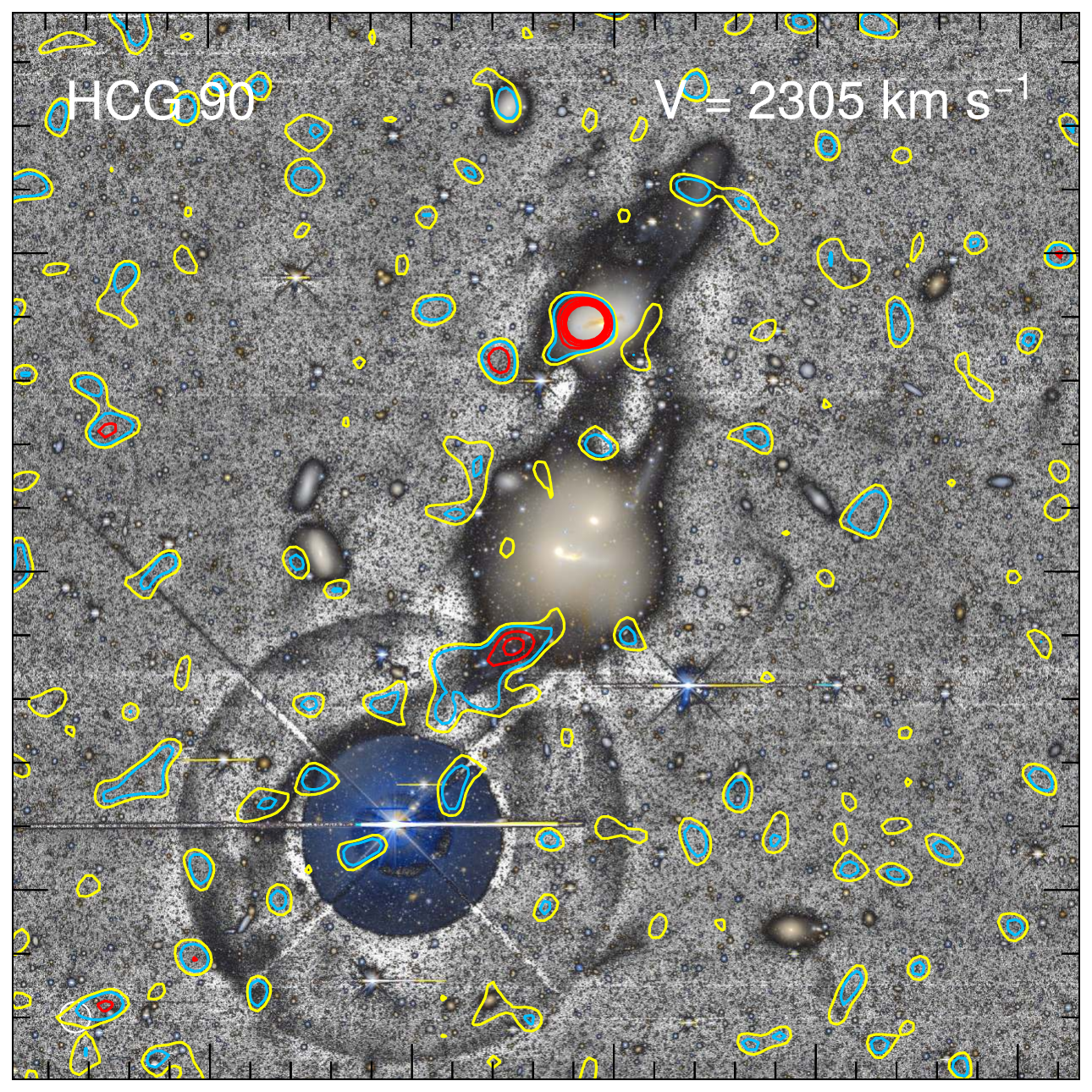} \\[-0.2cm]
          \includegraphics[scale=0.25]{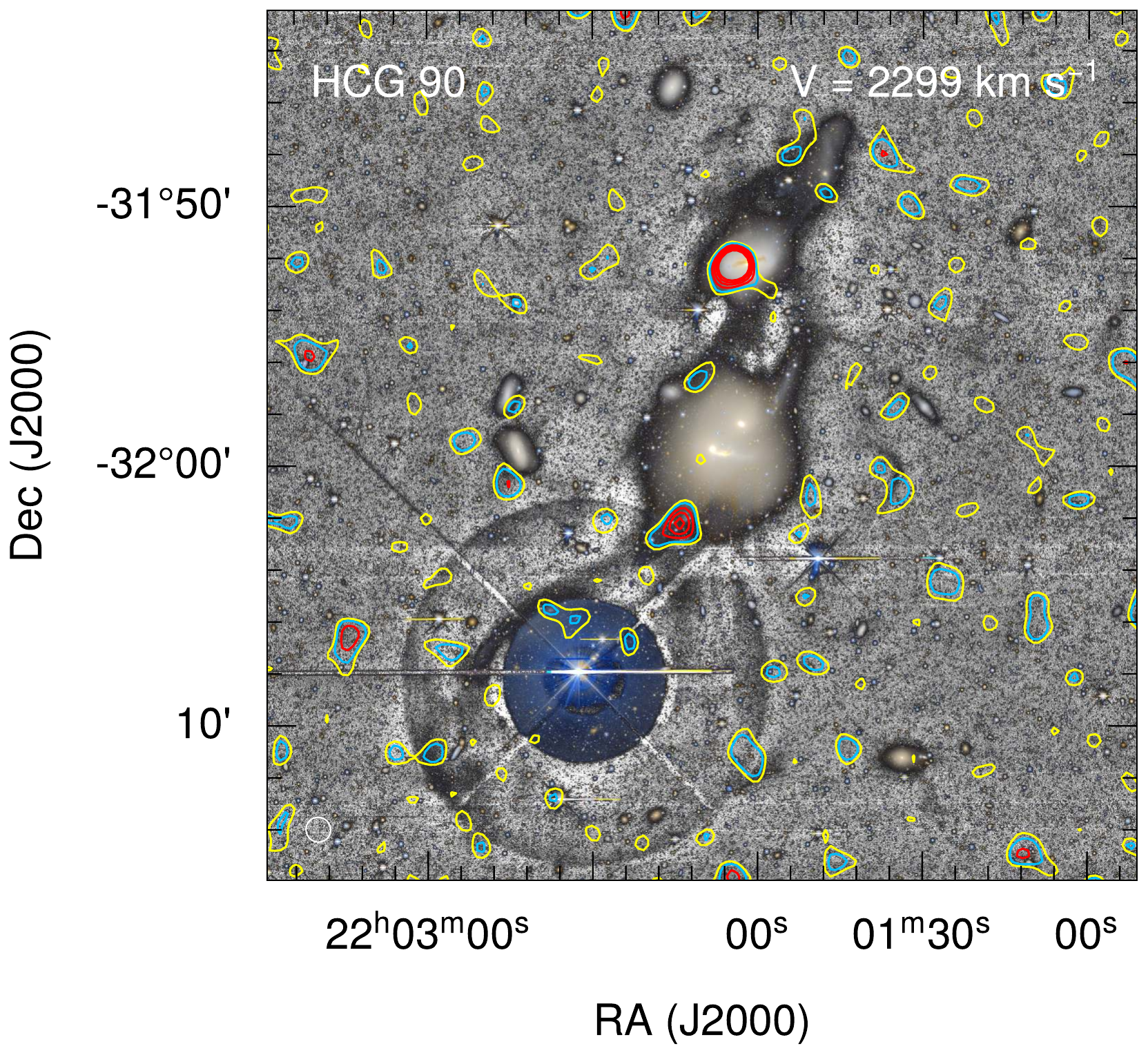} &
          \includegraphics[scale=0.25]{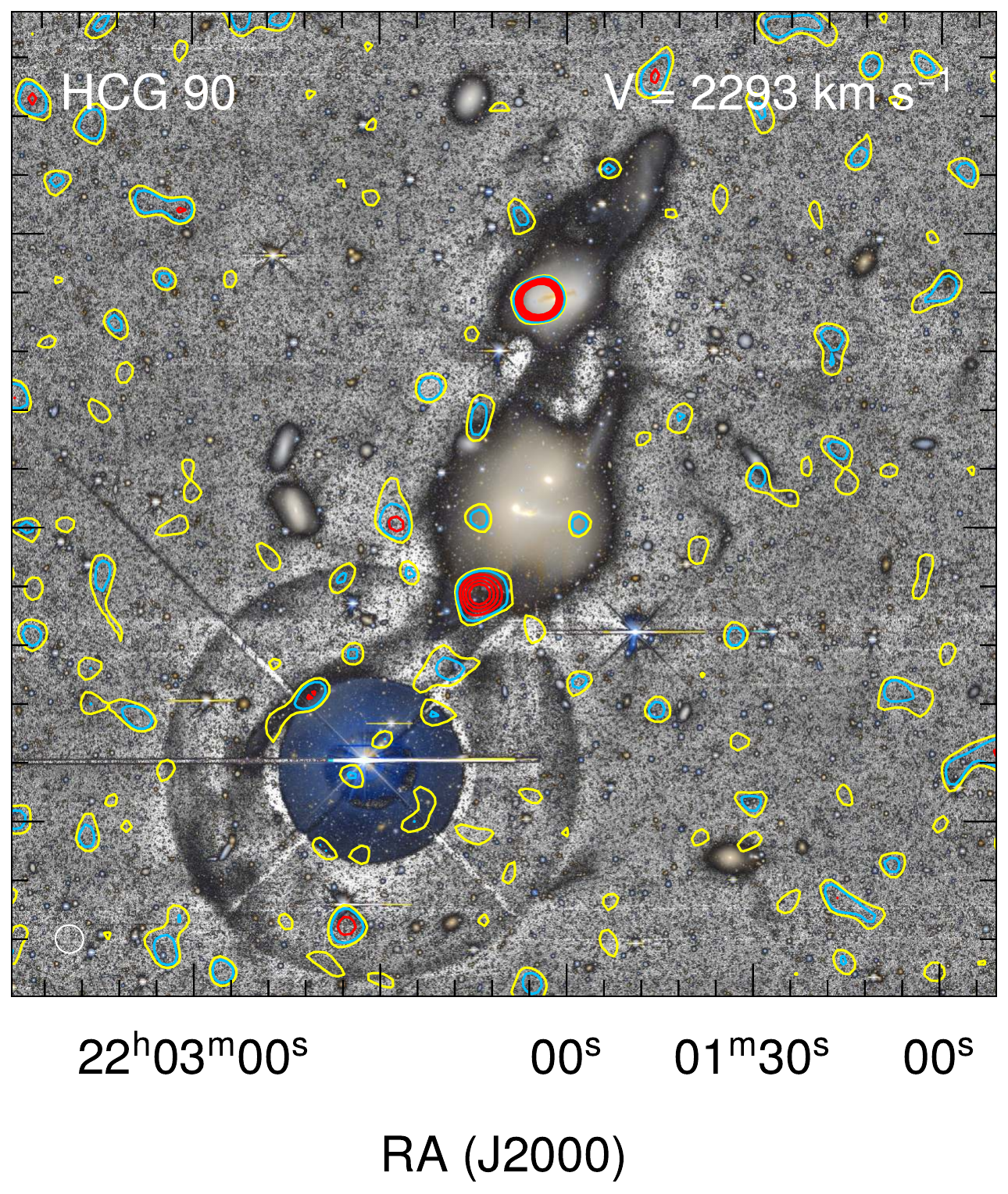} &
          \includegraphics[scale=0.25]{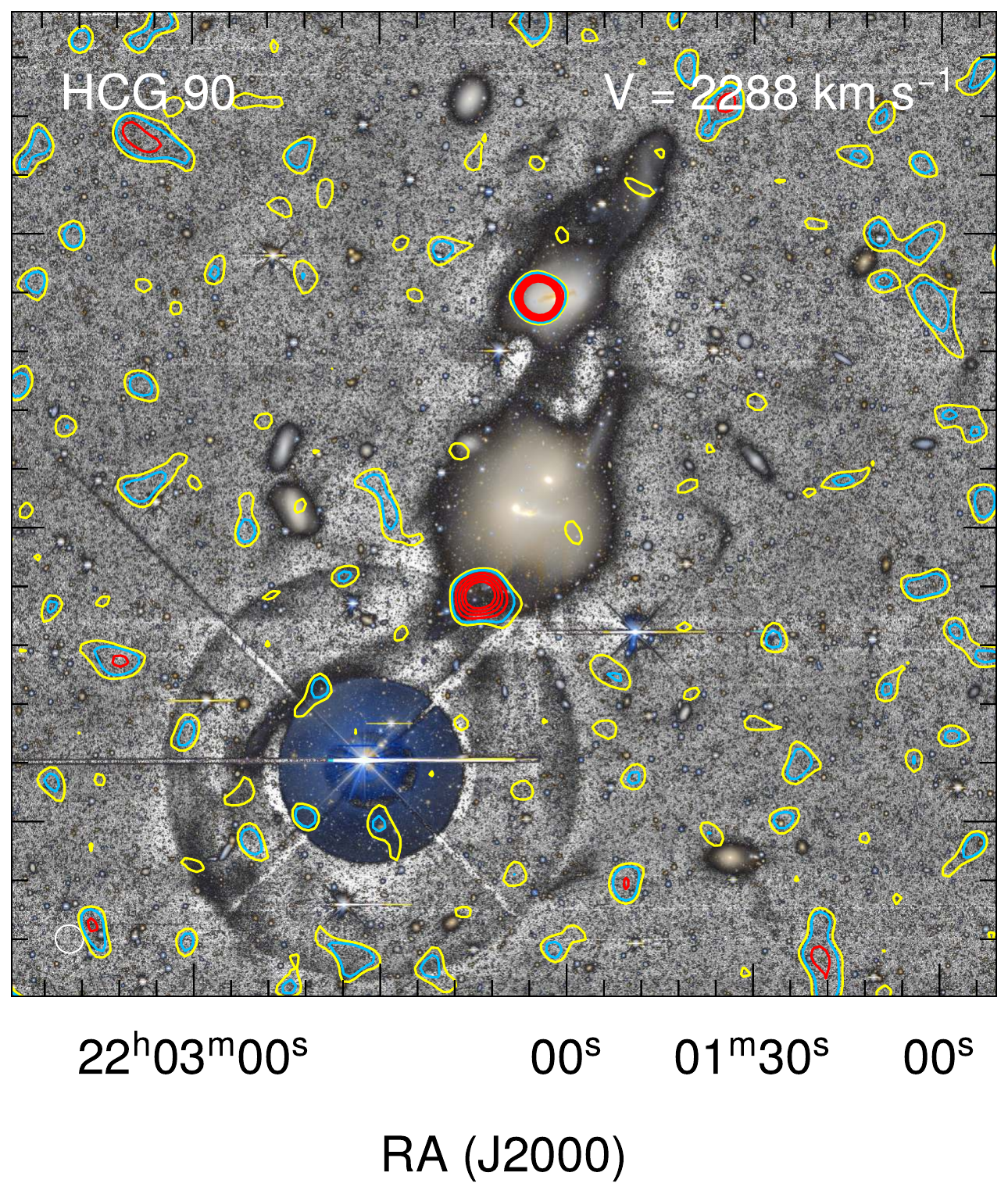} 
        \end{tabular}
    \caption{Example channel maps of the primary beam corrected cube of HCG 90 overlaid on enhanced DECaLS optical images. 
    To improve the signal-to-noise ratio of the optical data and highlight the optical tail, we have added G-band and R-band images and aligned the pixel scale to 8$\times$8 
    (0.27$\times$8 arcsec). Contour levels are (1.5, 2, 3, 4, 5, 6) times 
        the median noise level in the cube (0.66 $\mathrm{mJy~beam{-1}}$). The yellow and green colours show contour levels below 3$\sigma$. The red colours 
        represent contour levels at 3$\sigma$ or higher. Additional figures can be downloaded \href{https://zenodo.org/records/14856489}{here}.}
        \label{fig:hcg90_chanmap}
       \end{figure*}

Figure~\ref{fig:hcg90_chanmap} displays representative channel maps from the primary beam corrected data cube of HCG~90, 
superimposed on enhanced DECaLS optical images. To enhance faint optical features, G- and R-band images were combined, and pixel scaling was adjusted. 
The rest of the channel maps can be found \href{https://zenodo.org/records/14856489}{here}.
\subsection{Moment maps}
\begin{figure*}
  \begin{tabular}{l l}
      \includegraphics[scale=0.27]{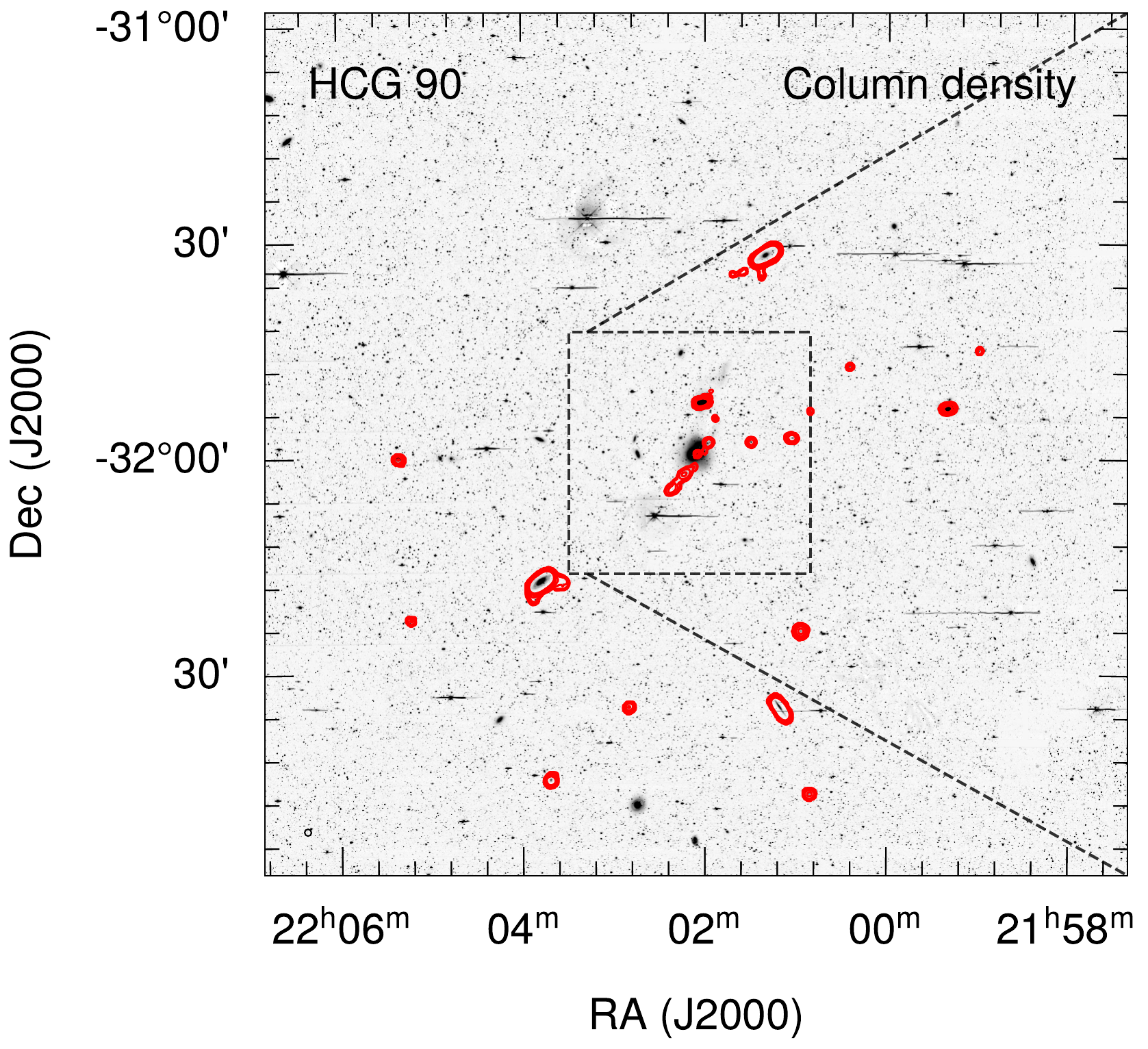}
      & \includegraphics[scale=0.27]{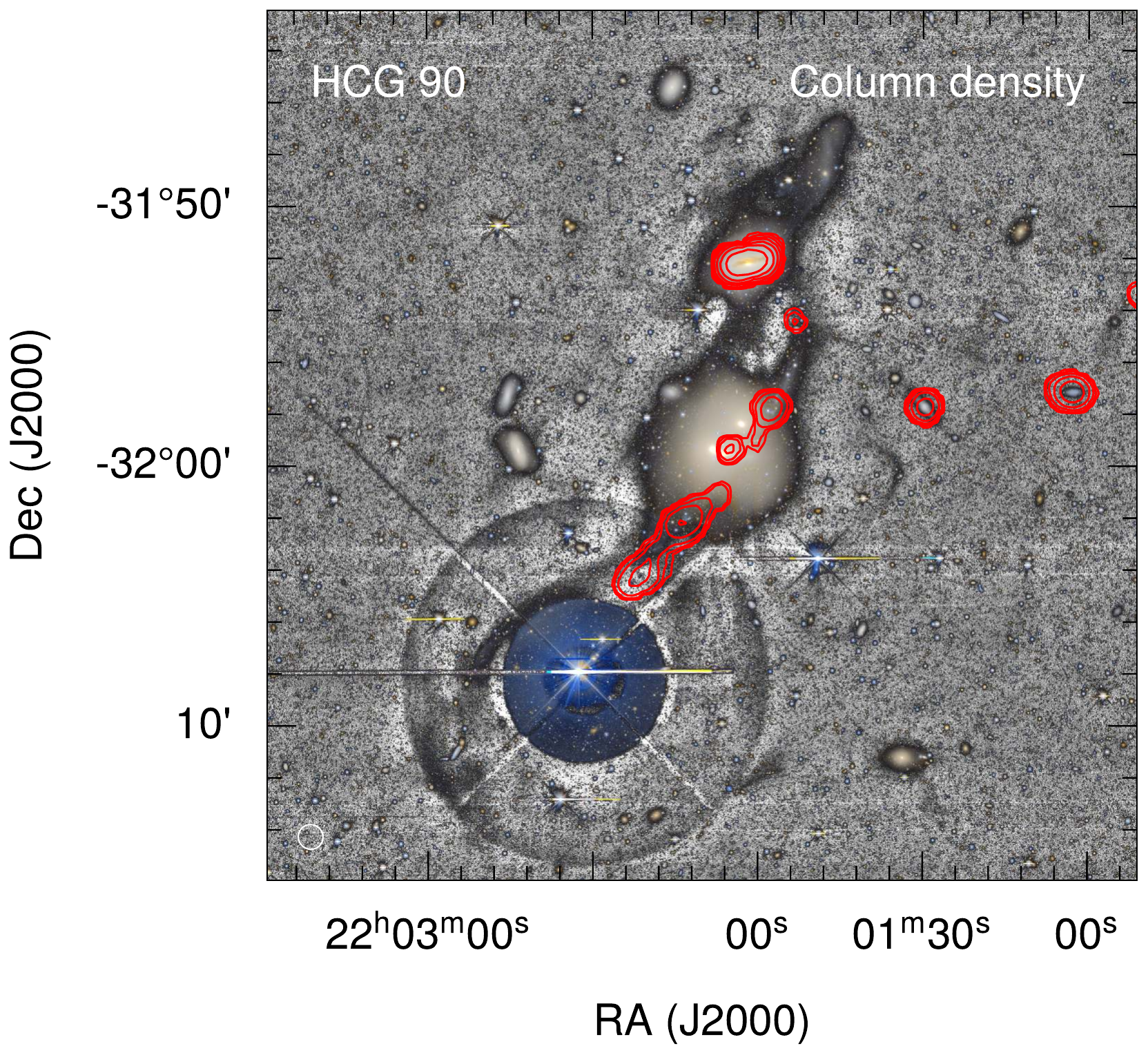}\\
      \includegraphics[scale=0.27]{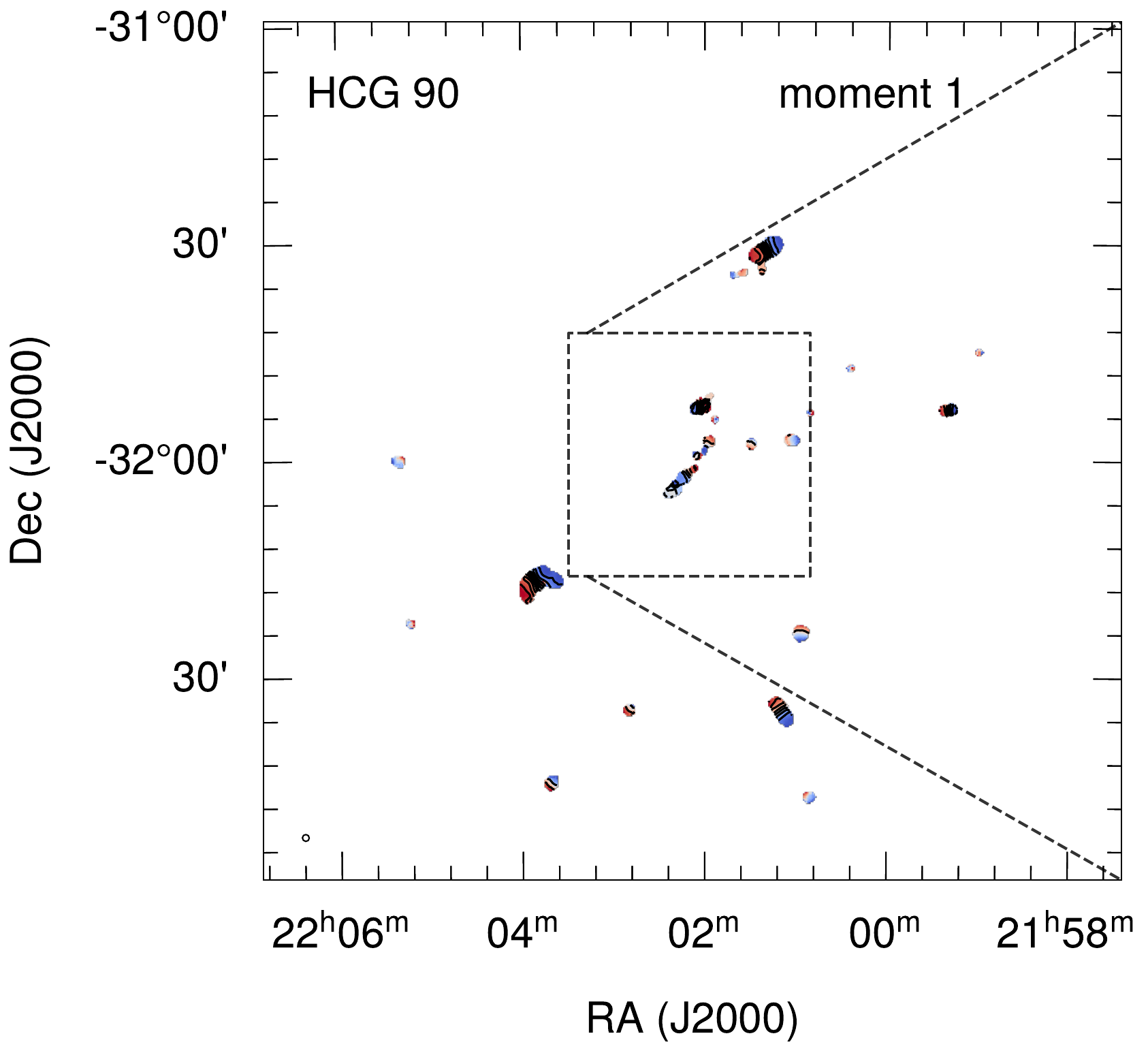} &
      \includegraphics[scale=0.27]{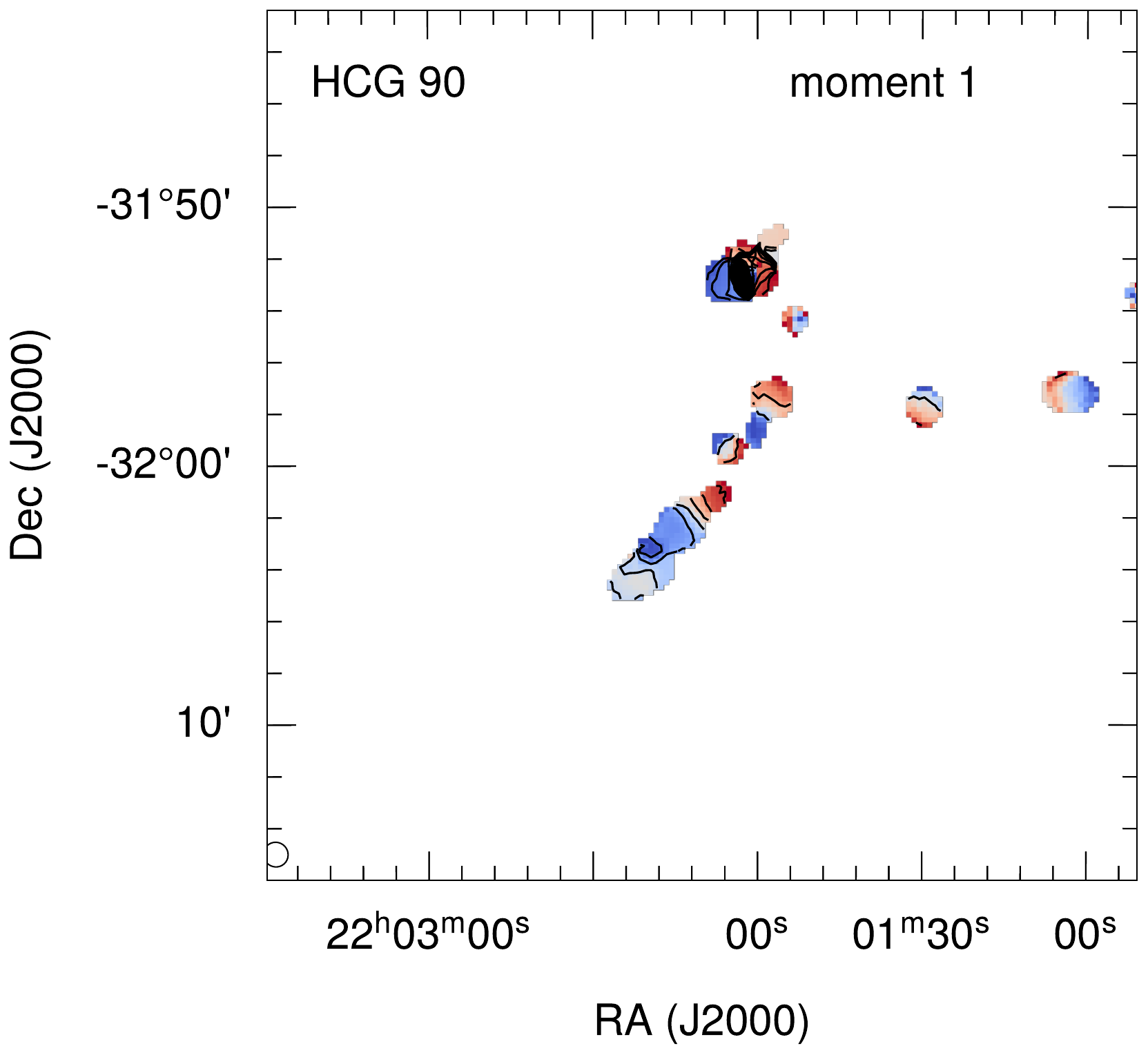}
  \end{tabular}
  \caption{\HI\ Moment maps of HCG 90. Left panels show all sources detected by SoFiA. The right panels show sources within the rectangular box shown on the left to better show the central 
  part of the group. The top panels show the column density maps with contour levels of
          ($\mathrm{3.9~\times~10^{18}}$, $\mathrm{7.7~\times~10^{18}}$, $\mathrm{1.5~\times~10^{19}}$, $\mathrm{3.1~\times~10^{19}}$, 
          $\mathrm{6.2~\times~10^{19}}$, $\mathrm{1.2~\times~10^{20}}$, $\mathrm{2.5~\times~10^{20}}$) $\mathrm{cm^{-2}}$. The contours are overlaid on DECaLS optical images. 
          The bottom panels show the moment one map. Each individual source has its own colour scaling and contour levels to highlight any rotational component.}
  \label{fig:hcg90_mom}
  \end{figure*}

Figure~\ref{fig:hcg90_mom} shows the column density and moment one (velocity field) maps of HCG~90. The left panels present all sources detected by SoFiA, 
whereas the right panels focus on the central region. The top panels overlay \HI\ column density contours on DECaLS optical images. 
The bottom panels display the velocity field with individual colour scalings for each source to highlight the rotational patterns of the surrounding galaxies.  
\subsection{\HI\ tails in HCG~90}
\begin{figure*}
      \setlength{\tabcolsep}{0pt}
      \begin{tabular}{ccc} 
              \includegraphics[scale=0.24]{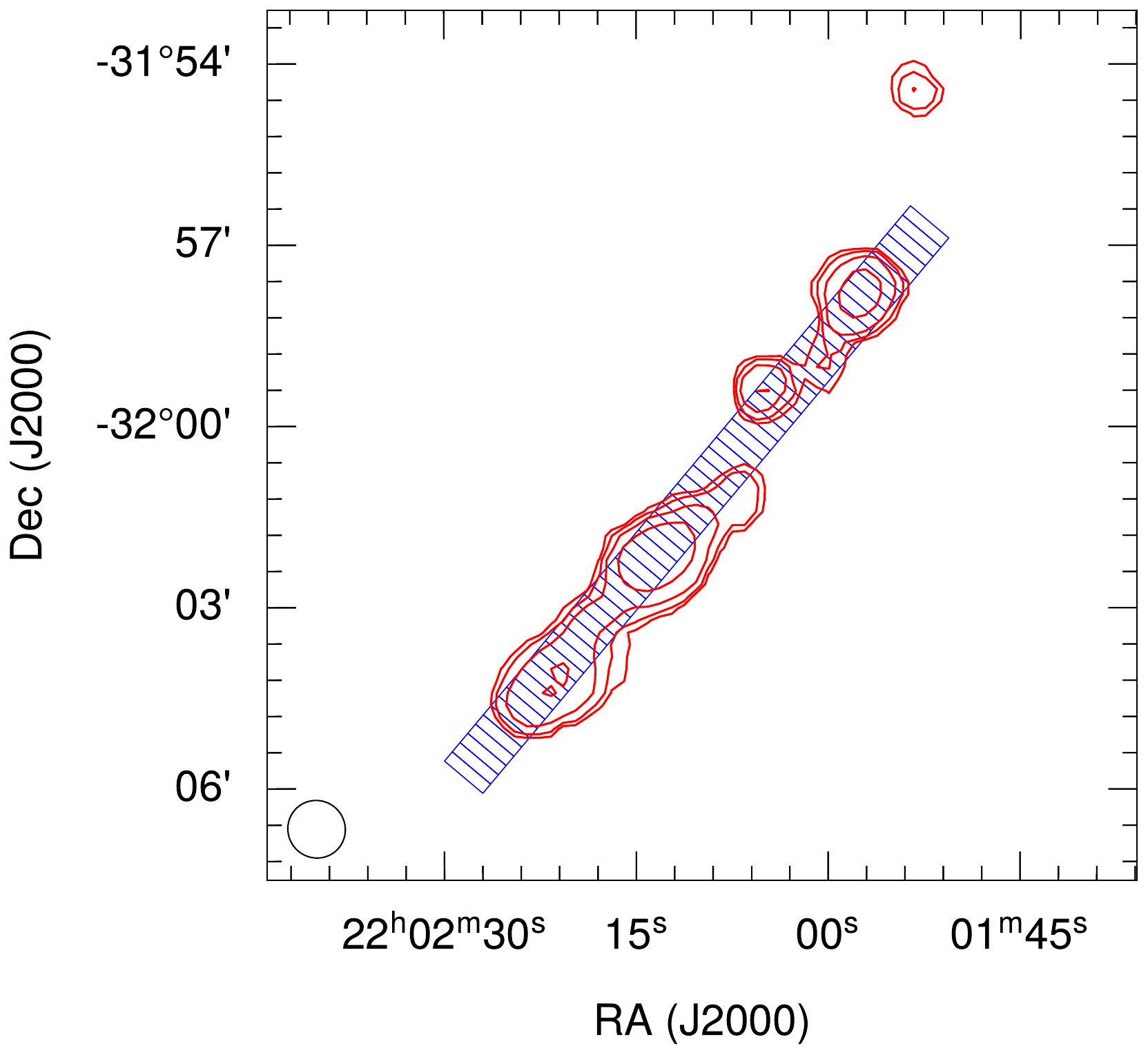} &
              \includegraphics[scale=0.24]{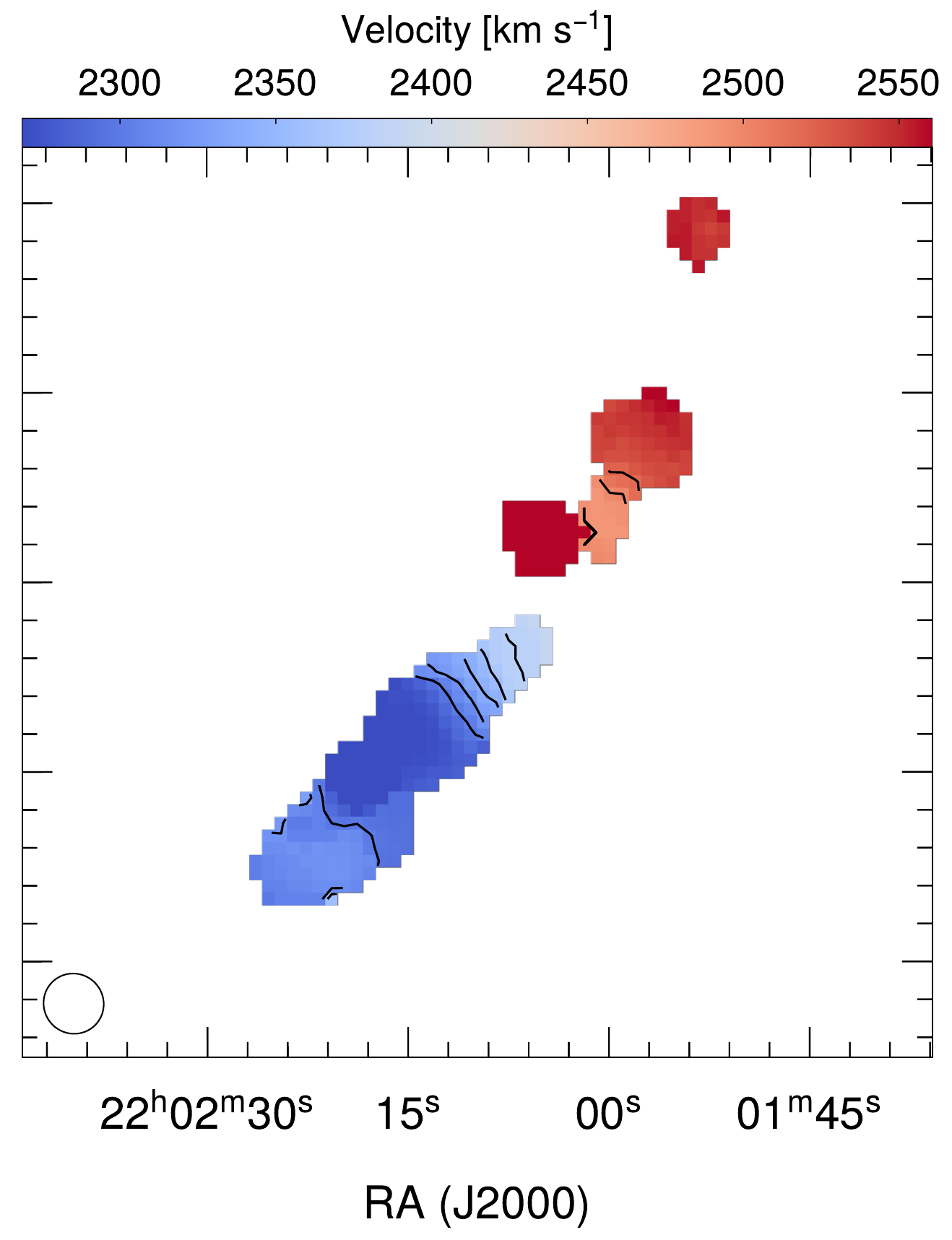} &
              \includegraphics[scale=0.24]{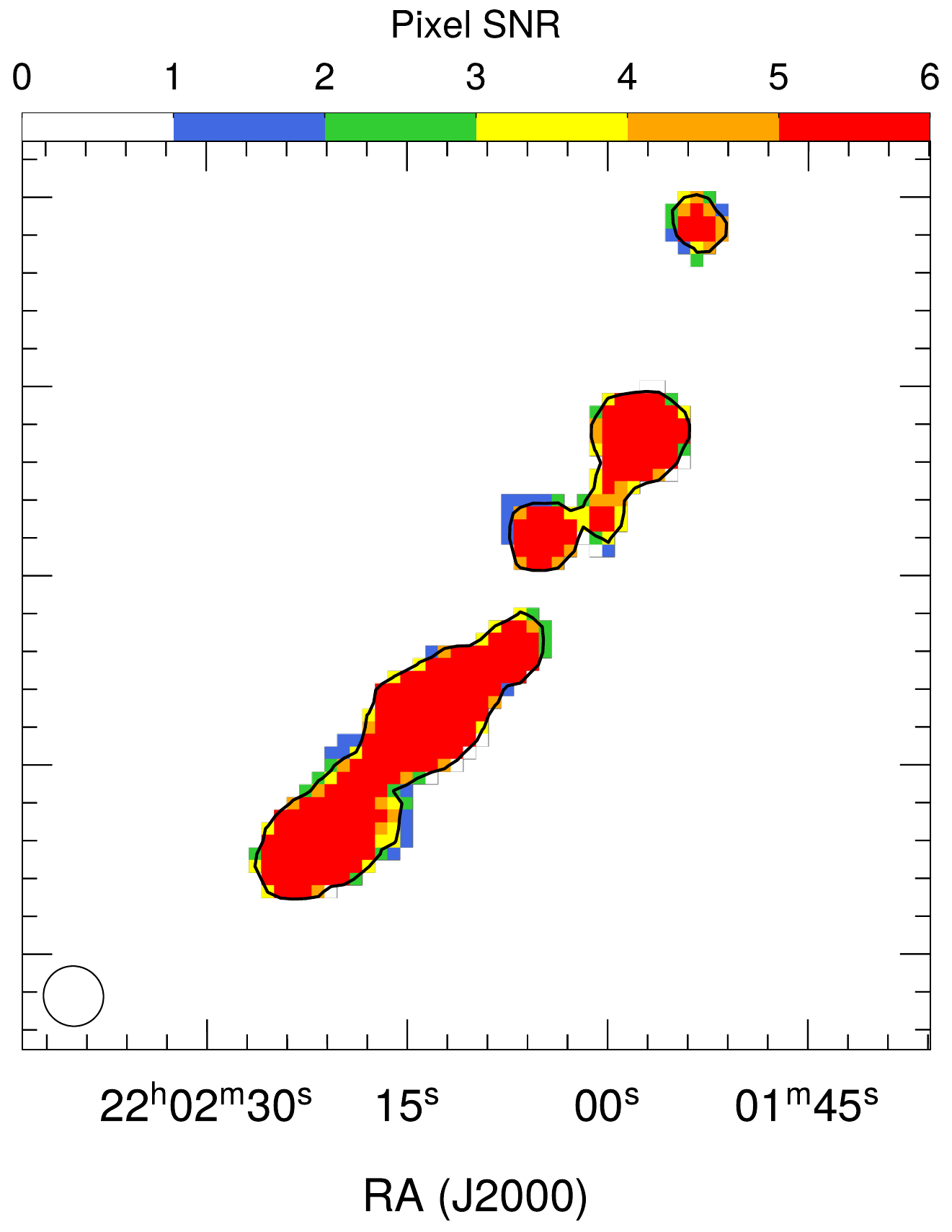} \\
          \begin{minipage}{0.33\textwidth}
              \includegraphics[scale=0.24]{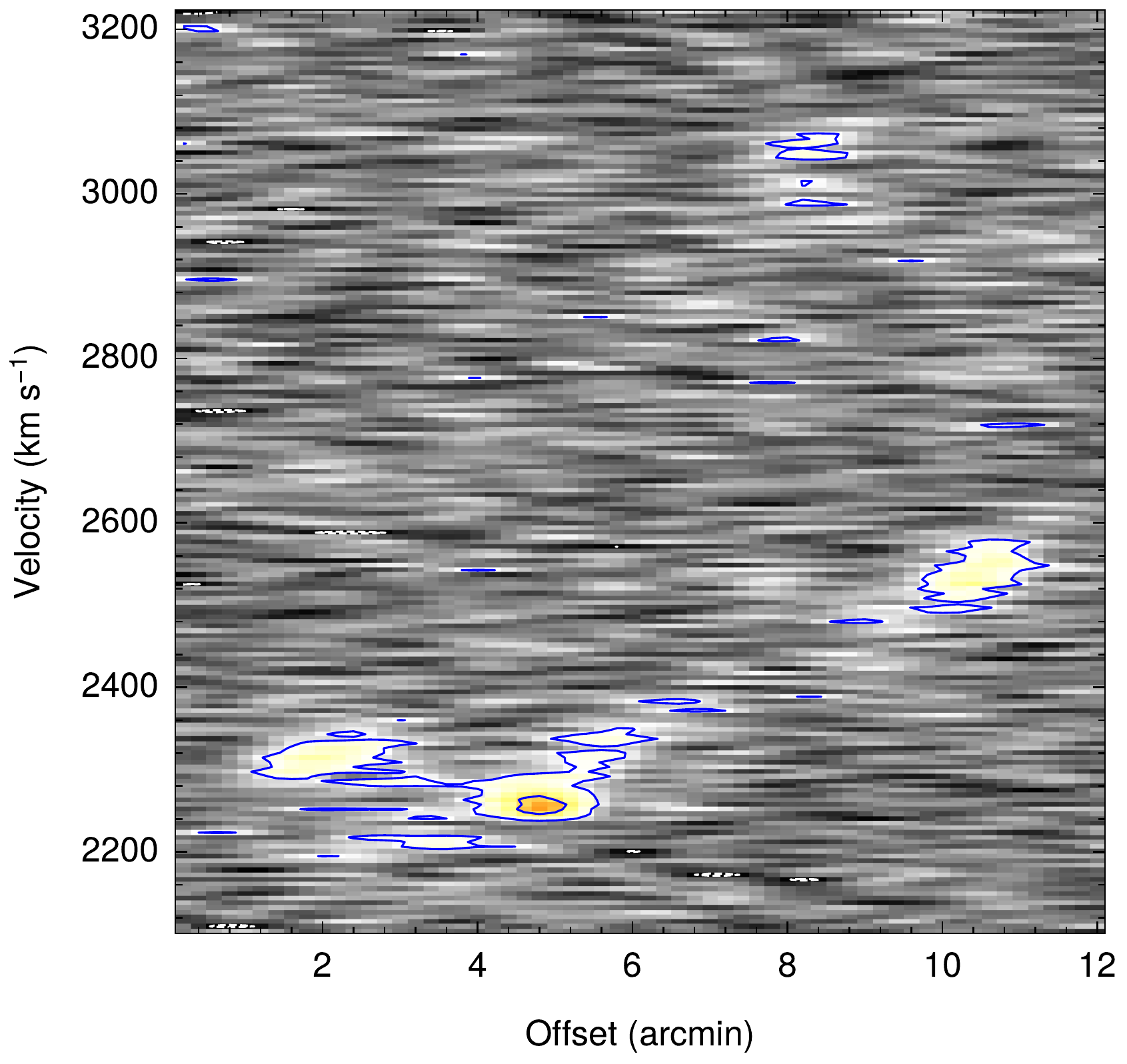}
          \end{minipage} &
          \begin{minipage}{0.33\textwidth}
              \hspace*{0.4cm}
              \includegraphics[scale=0.24]{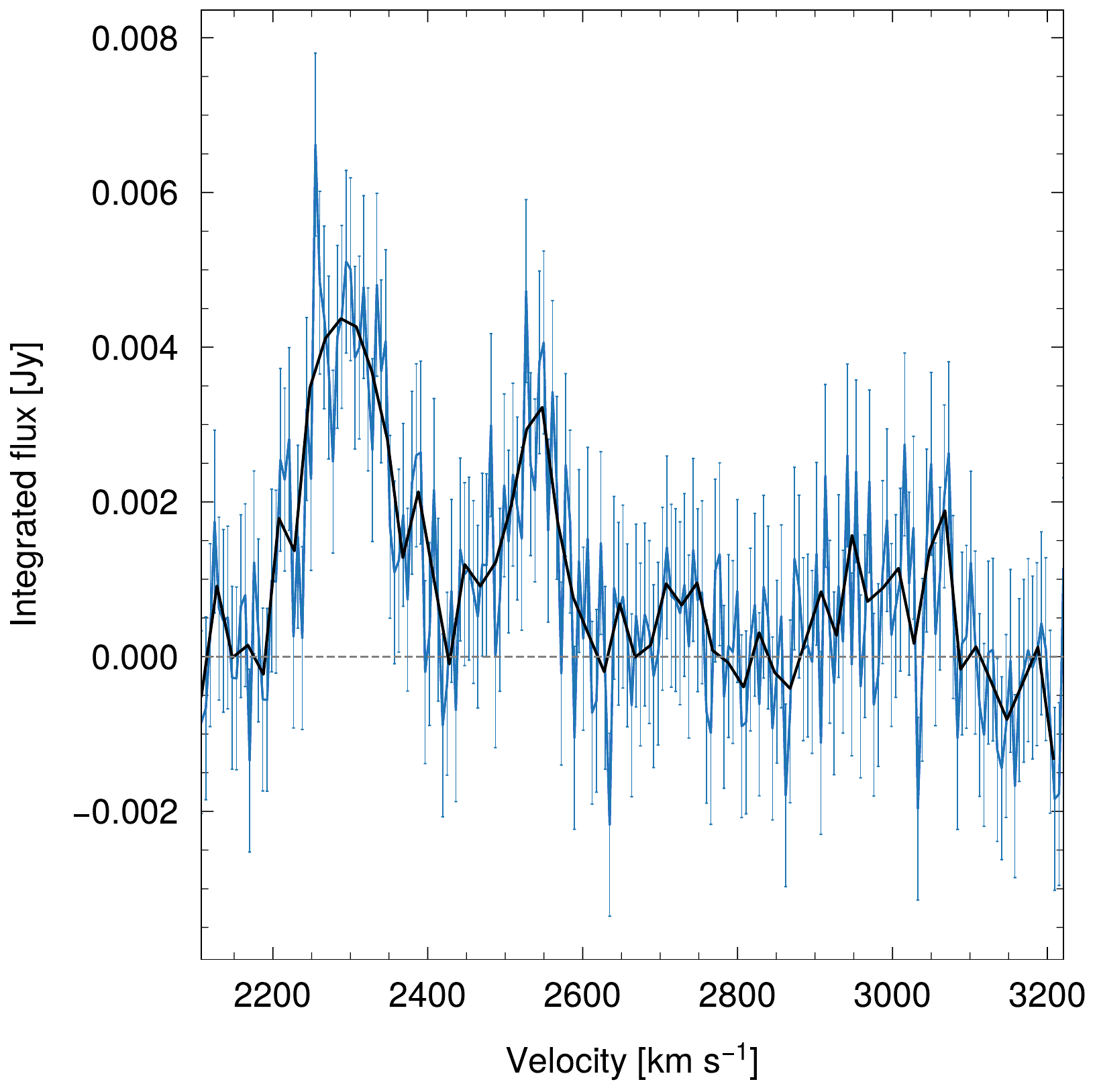}
          \end{minipage} &
      \end{tabular}
        \caption{Top left: \HI\ column density map of the tails of HCG~90. The contour levels are 
        (0.35, 0.69, 1.38, 2.77)~$\times~10^{19}~\mathrm{cm^{-2}}$. The blue area indicates the slice from which 
        the position-velocity diagram shown at the bottom panel was derived. Top centre: moment-1 map, 
        the contour levels are (2300, 2320, 2340, 2360, 2380, 2400, 2420, 2440, 2460, 2480, 2500, 2520)~$\mathrm{km~s^{-1}}$.  
        Top right: signal-to-noise ratio map. Pixels above 6-$\sigma$ are all shown in red. The black contour represent the lowest column density 
        contour plotted at the top left panel of the figure. 
        Bottom left: position-velocity diagram taken from the slice shown at the top right panel. Bottom right: global \HI\ profile. 
        The black line is a smoothed version of the profile using at 20 $\mathrm{km~s^{-1}}$ velocity resolution.}
        \label{fig:hcg90_tail}
      \end{figure*}

Figure~\ref{fig:hcg90_tail} presents an overview plots of the \HI\ tails in HCG~90. The top-left panel displays the HI column density map, 
with the blue region marking the slice used to derive the position-velocity 
diagram in the bottom-left panel of the figure. The top-centre panel shows the velocity field (moment-1 map), 
and the top-right panel presents the signal-to-noise ratio map. The bottom panels show a position-velocity diagram and a (smoothed) global 
profile of the tails.
\subsection{3D visualisation}
\begin{figure*}
        \setlength{\tabcolsep}{0pt}
        \begin{tabular}{c}
        \includegraphics[scale=0.345]{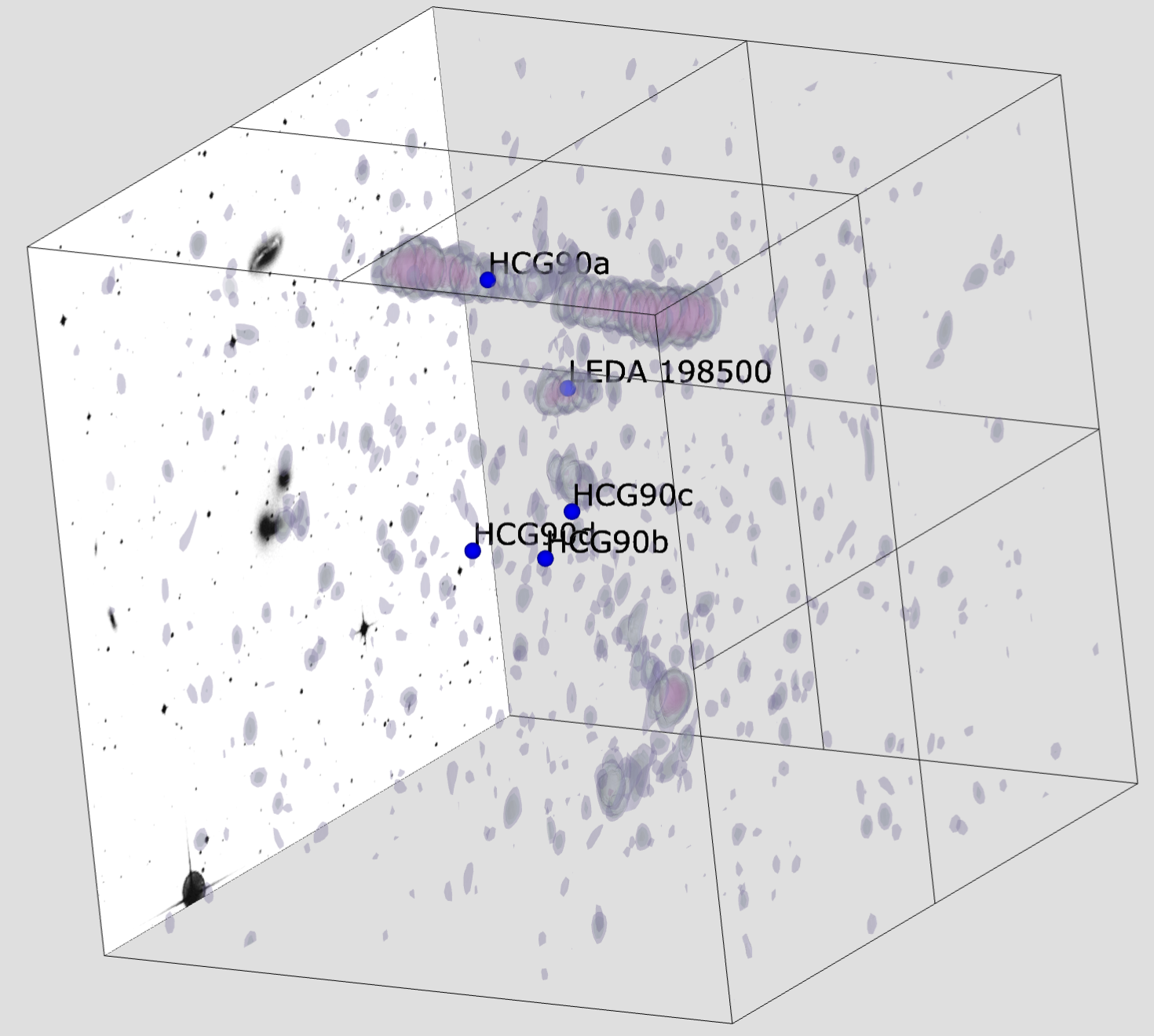}
        \end{tabular}
        \caption{3D visualisation of HCG 90. The blue circles indicate the position of the member galaxies. The 2D grayscale image is a DeCaLS R-band optical image of the group.
        The online version of the cubes are available at \href{https://amiga.iaa.csic.es/x3d-menu/}{https://amiga.iaa.csic.es/x3d-menu/}.}
      \label{fig:hcg90_3dvis}
     \end{figure*}

Figure~\ref{fig:hcg90_3dvis} presents a 3D visualisation of HCG~90, where the blue circles mark the positions of the member galaxies. 
The background image is a DeCaLS R-band optical image of the group. An interactive version of these 3D cubes is available online at \href{https://amiga.iaa.csic.es/x3d-menu/}{https://amiga.iaa.csic.es/x3d-menu/}.    
\section{Additional figures of HCG~97}
\subsection{Noise properties and global profiles}
\begin{figure*}
  \setlength{\tabcolsep}{0pt}
  \begin{tabular}{l l l}
      \includegraphics[scale=0.215]{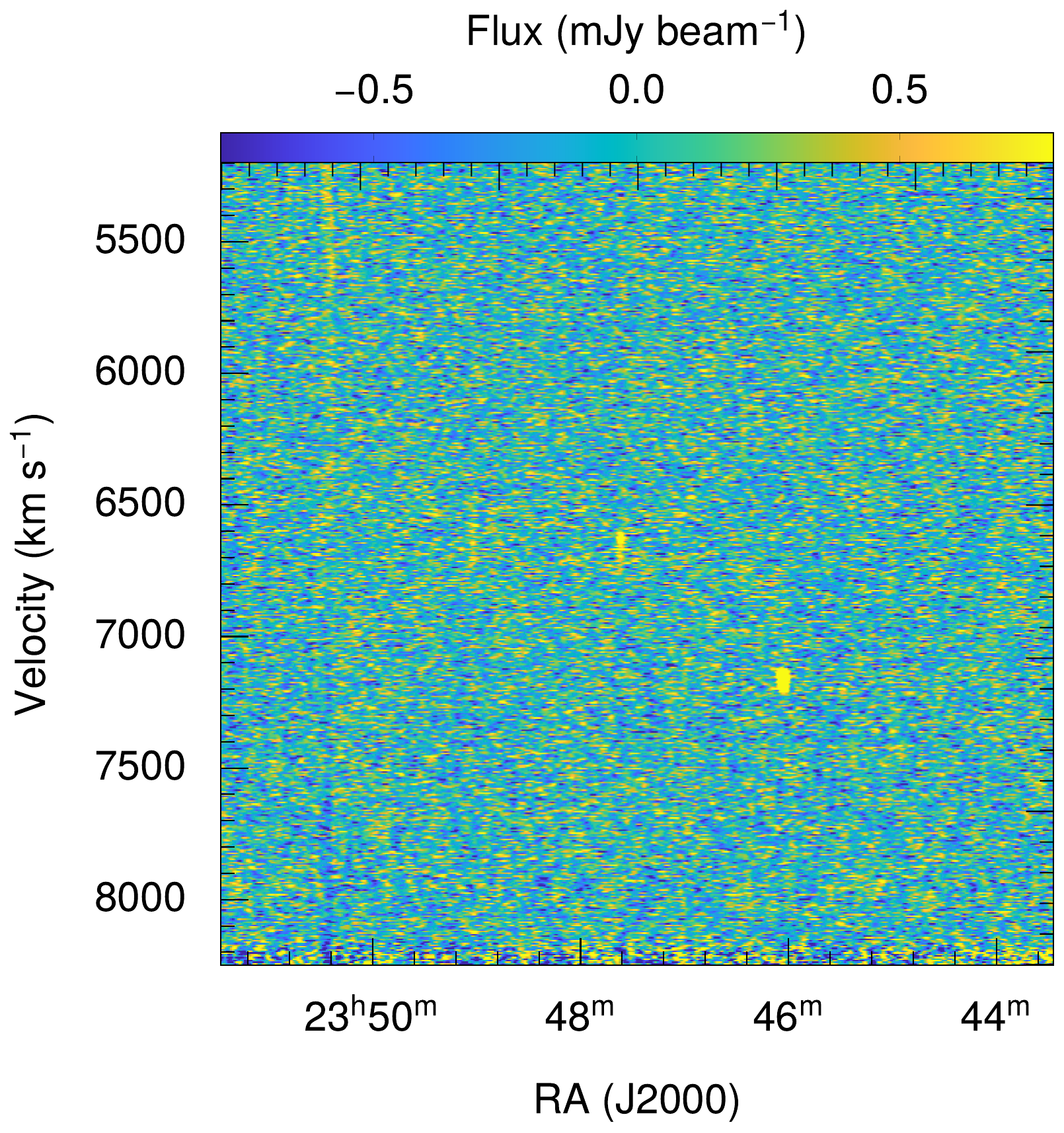} &  
      \includegraphics[scale=0.215]{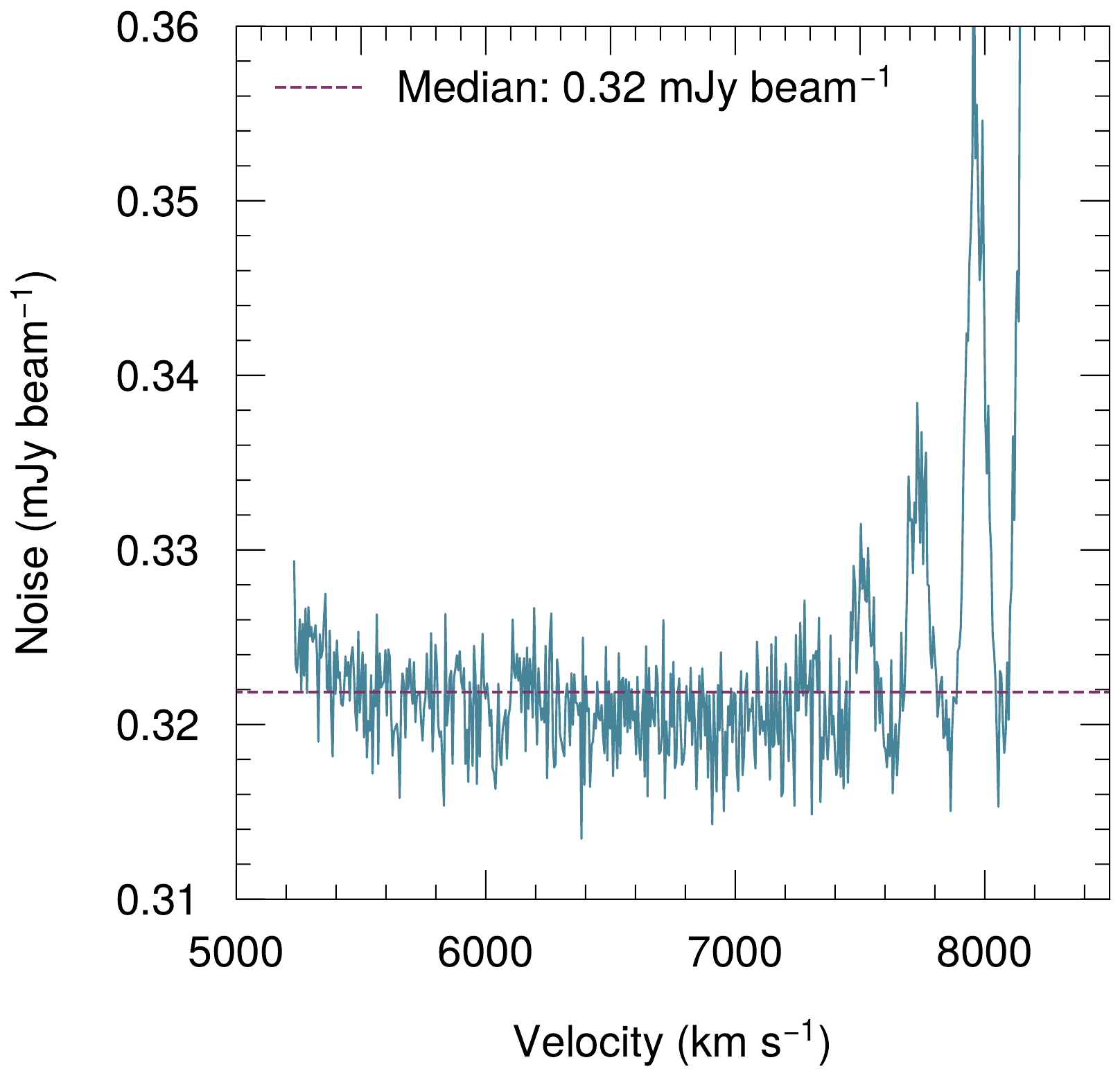} &
      \includegraphics[scale=0.215]{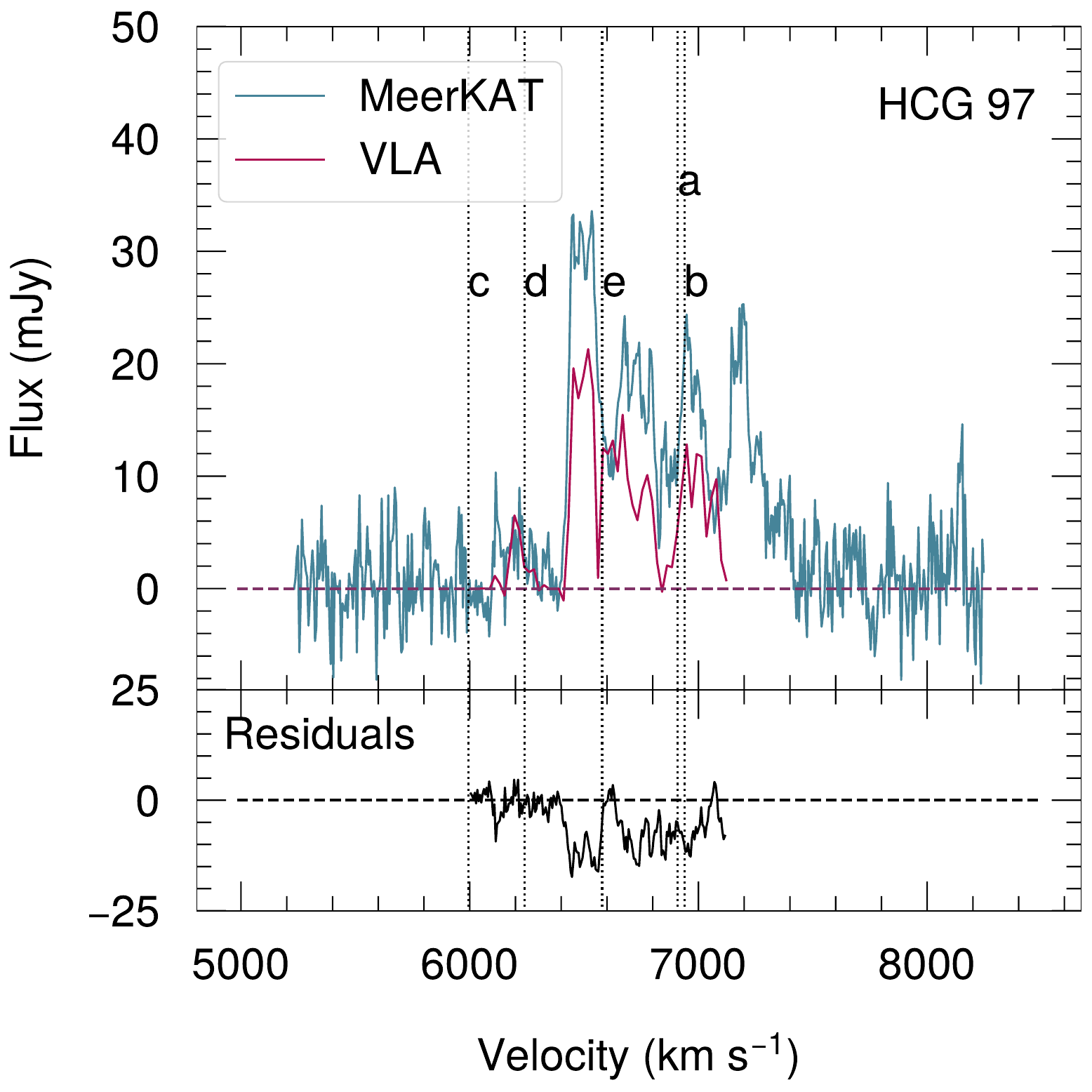}
    \end{tabular}
    \caption{Left panel: velocity vs right ascension of HCG~97. Middle panel: median noise values of each RA-DEC slice of the non-primary beam corrected 60\arcsec\ data cube of 
    HCG 97 as a function of velocity. The horizontal dashed line indicates the median of all the noise values from each slice. Right panel: the blue solid lines indicates the 
    MeerKAT integrated spectrum of HCG~97; the red solid line indicates VLA integrated spectrum of the group derived by \citep{2023A&A...670A..21J}. 
    The vertical dotted lines indicate the velocities of the galaxies in the core of the group. The spectra have been extracted from areas containing only genuine \HI\ emission. }
    \label{fig:hcg97_noise}
   \end{figure*}

Figure~\ref{fig:hcg97_noise} shows the RA-velocity plot of HCG~97 (left), the median noise level across the velocity channels (middle), and the MeerKAT and VLA integrated spectra (right). 

\subsection{Moment maps}
\begin{figure*}
  \begin{tabular}{l l}
      \includegraphics[scale=0.27]{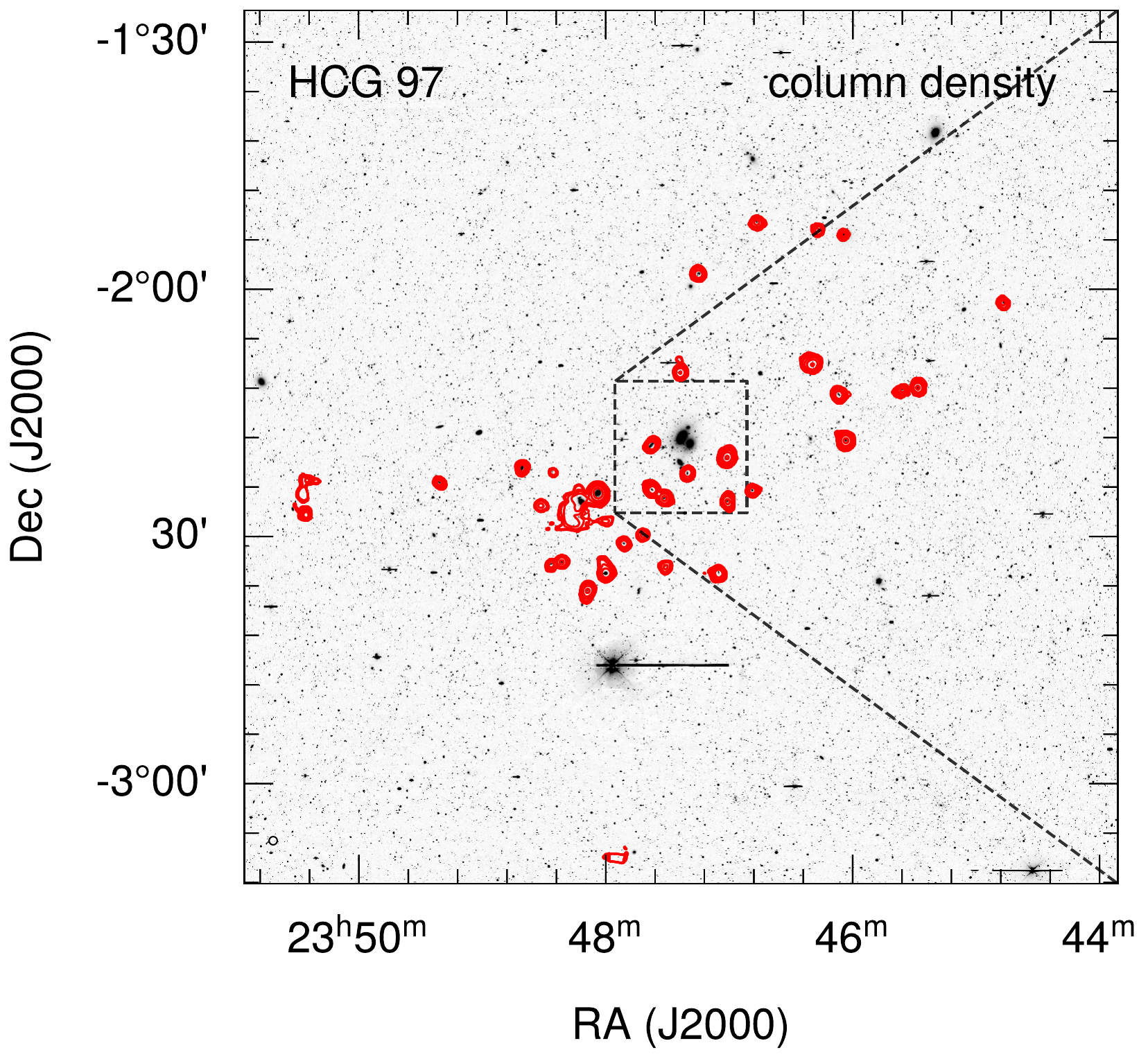}
      & \includegraphics[scale=0.27]{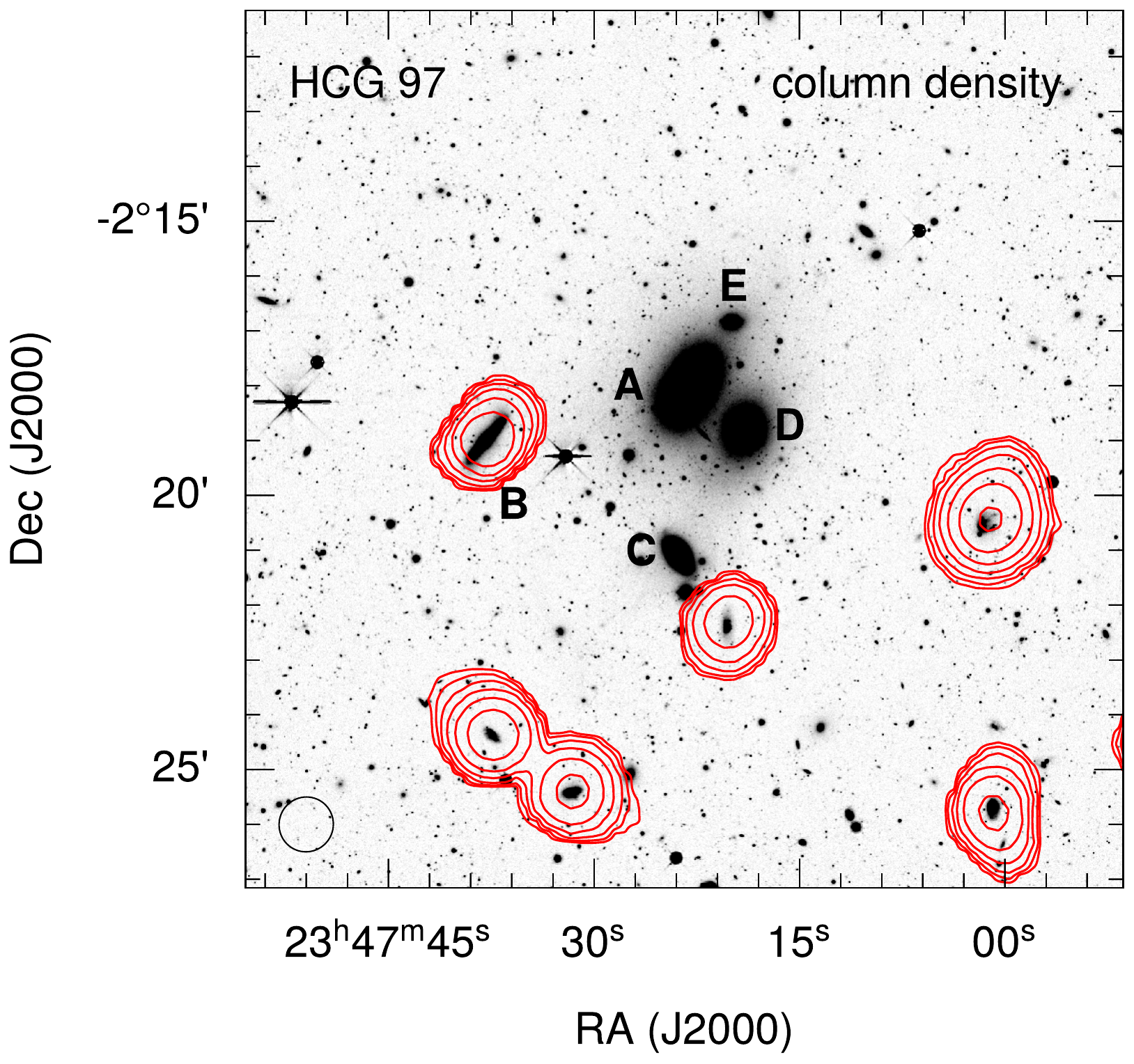}\\
      \includegraphics[scale=0.27]{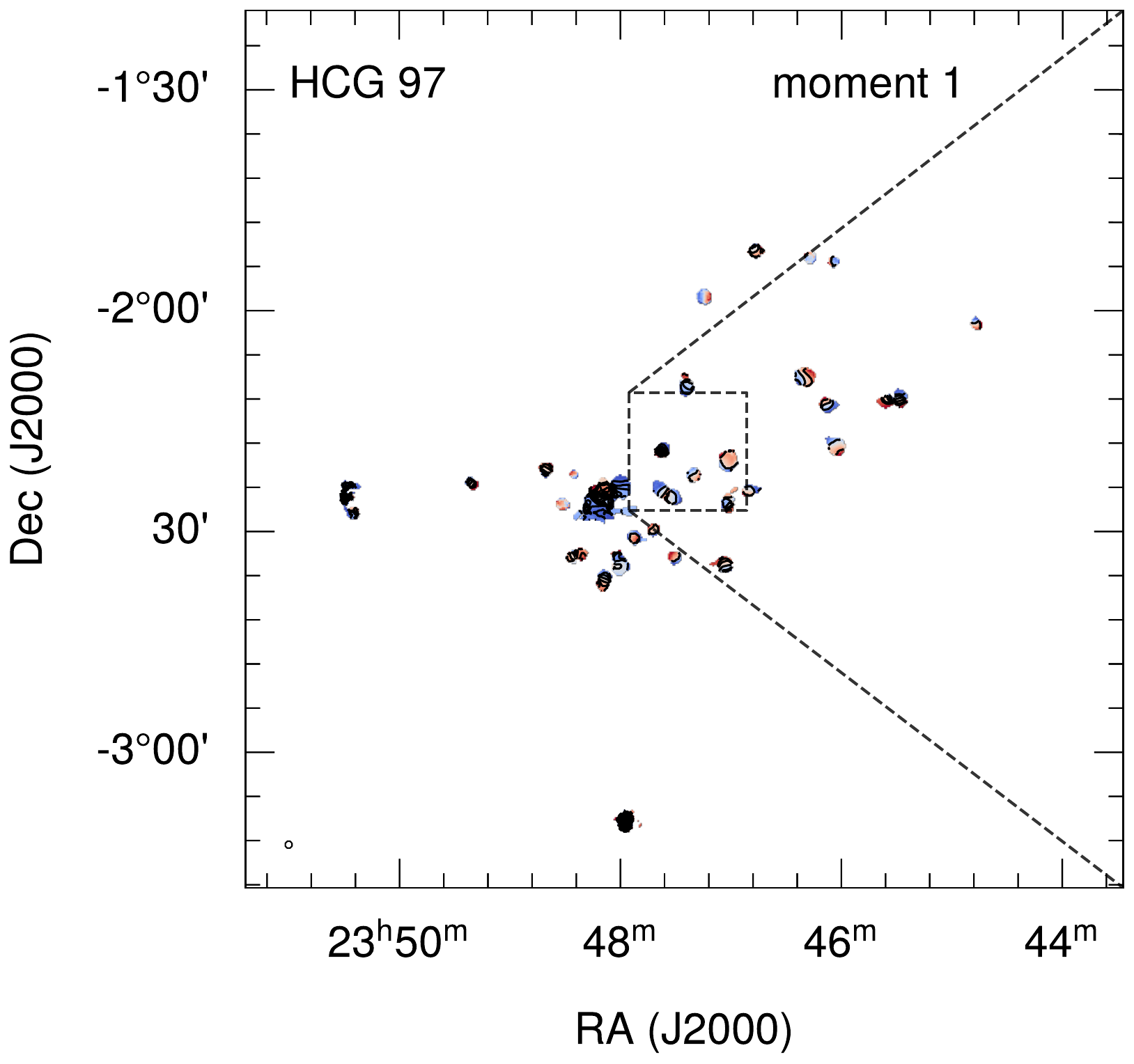} &
      \includegraphics[scale=0.27]{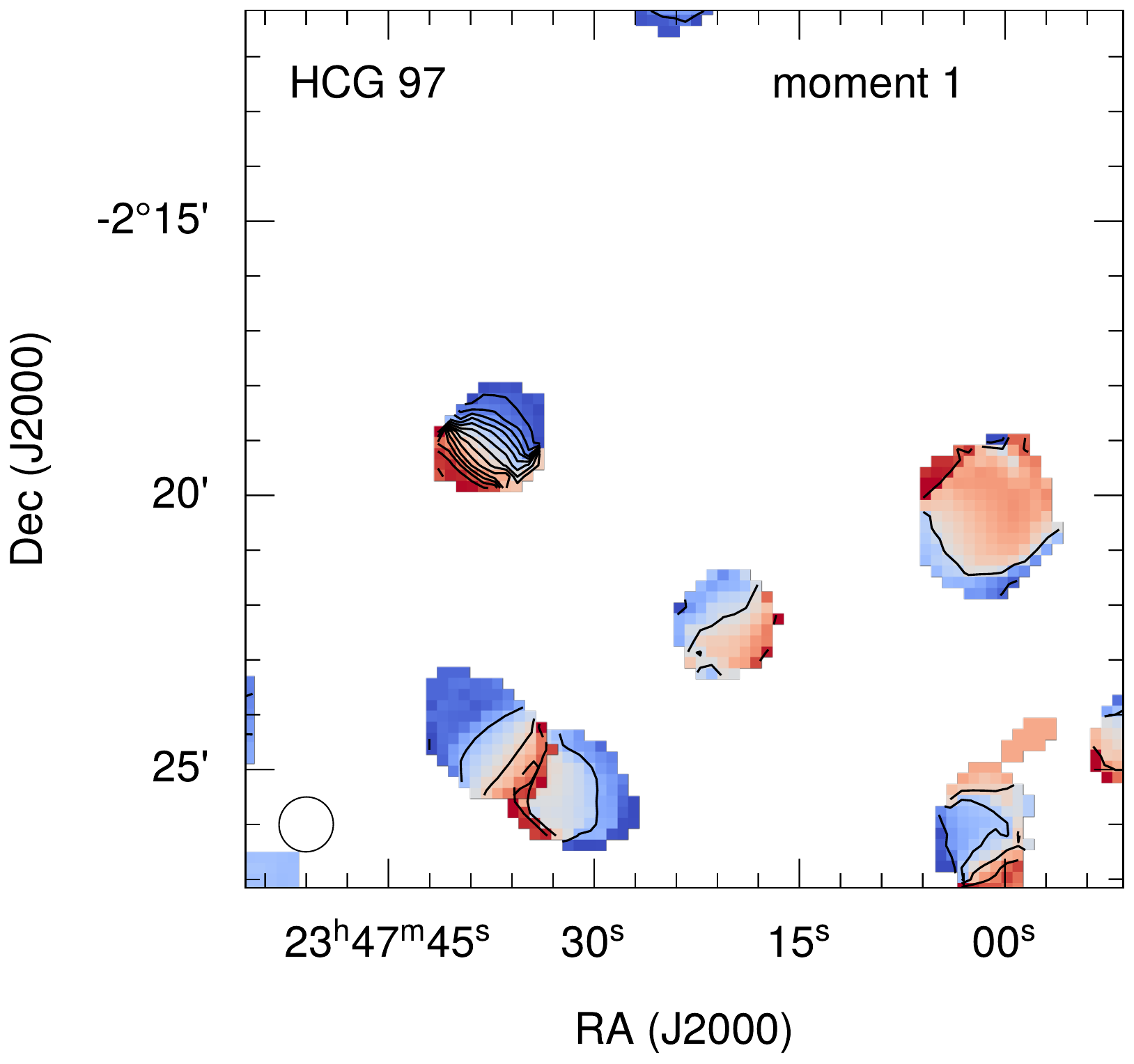}
  \end{tabular}
  \caption{\HI\ Moment maps of HCG~97. Left panels show all sources detected by SoFiA. The right panels show sources within the rectangular 
  box shown on the left to better show the central part of the group. The top panels represent the column density map with contour levels of
          ($\mathrm{3.5~\times~10^{18}}$, $\mathrm{7.0~\times~10^{18}}$, $\mathrm{1.4~\times~10^{19}}$, $\mathrm{2.8~\times~10^{19}}$,
          $\mathrm{5.6~\times~10^{19}}$, $\mathrm{1.1~\times~10^{20}}$, $\mathrm{2.2~\times~10^{20}}$) $\mathrm{cm^{-2}}$. 
          The contours are overlaid on DECaLS R-band optical images. The bottom panels show the moment one map. Each individual 
          source has its own colour scaling and contour levels to highlight any rotational component.}
  \label{fig:hcg97_mom}
  \end{figure*}

Figure~\ref{fig:hcg97_mom} presents the column density maps (top panels) and moment one (bottom panels) maps of HCG~97. The left panels show all sources detected by SoFiA within the 
MeerKAT field of view, whereas the right panels present zoomed-in views of the group’s centre. The grayscale images are DECaLS R-band optical data.
\subsection{\HI\ in HCG~97b}  
\begin{figure*}
      \setlength{\tabcolsep}{1.2pt}
      \begin{tabular}{c c c}
          \includegraphics[scale=0.29]{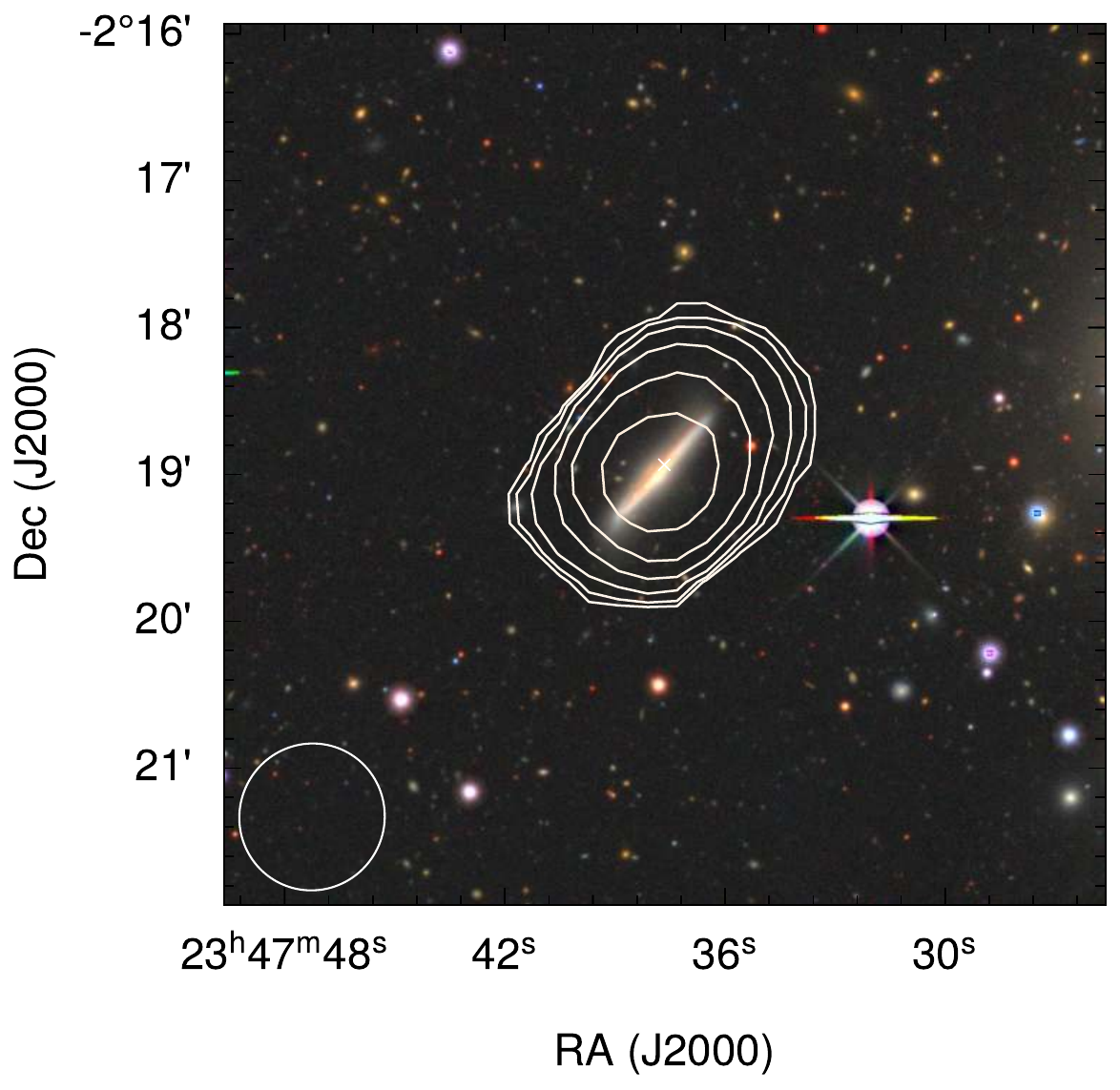} &
          \includegraphics[scale=0.29]{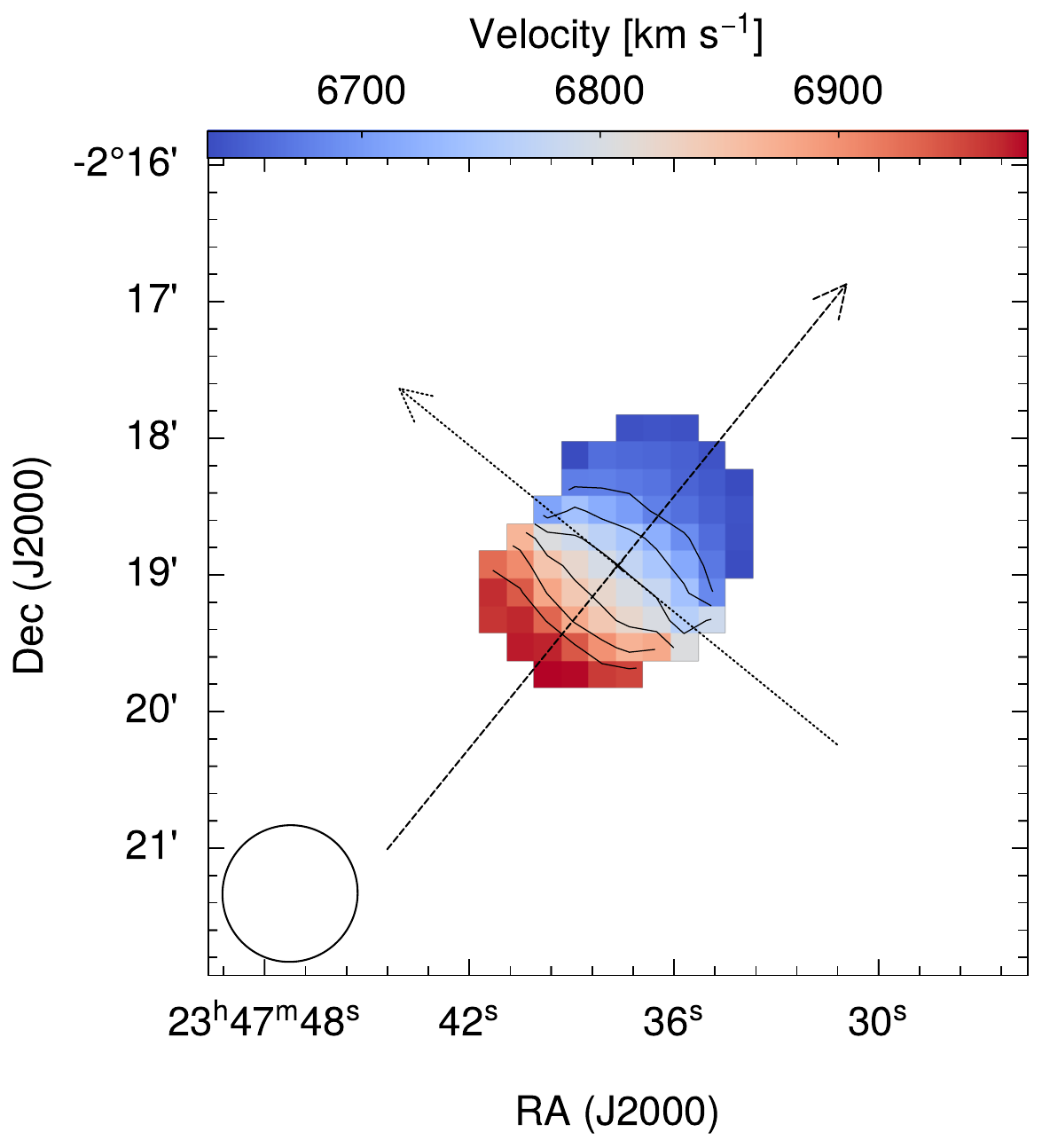} &
          \includegraphics[scale=0.29]{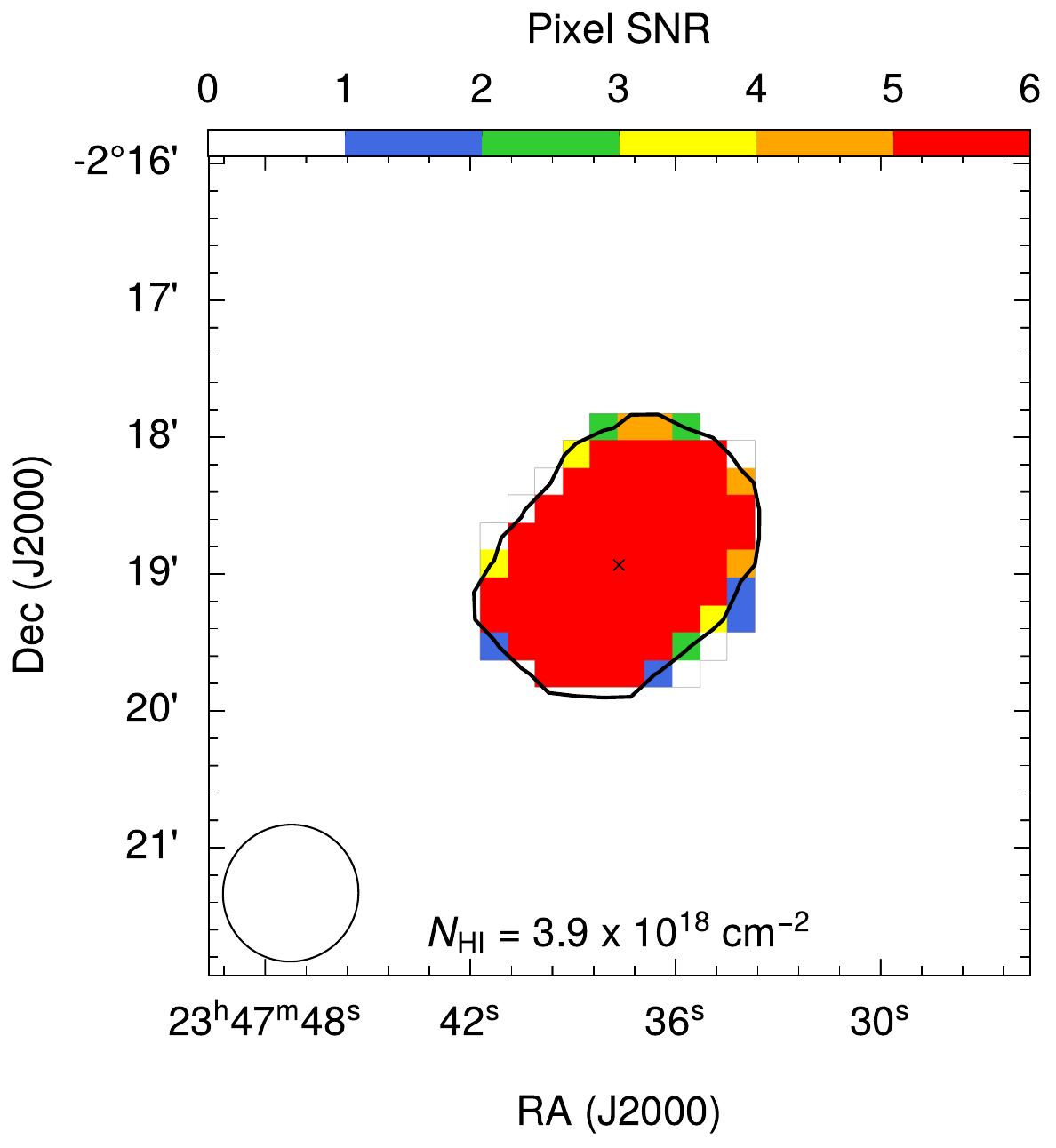}\\ 
          \includegraphics[scale=0.29]{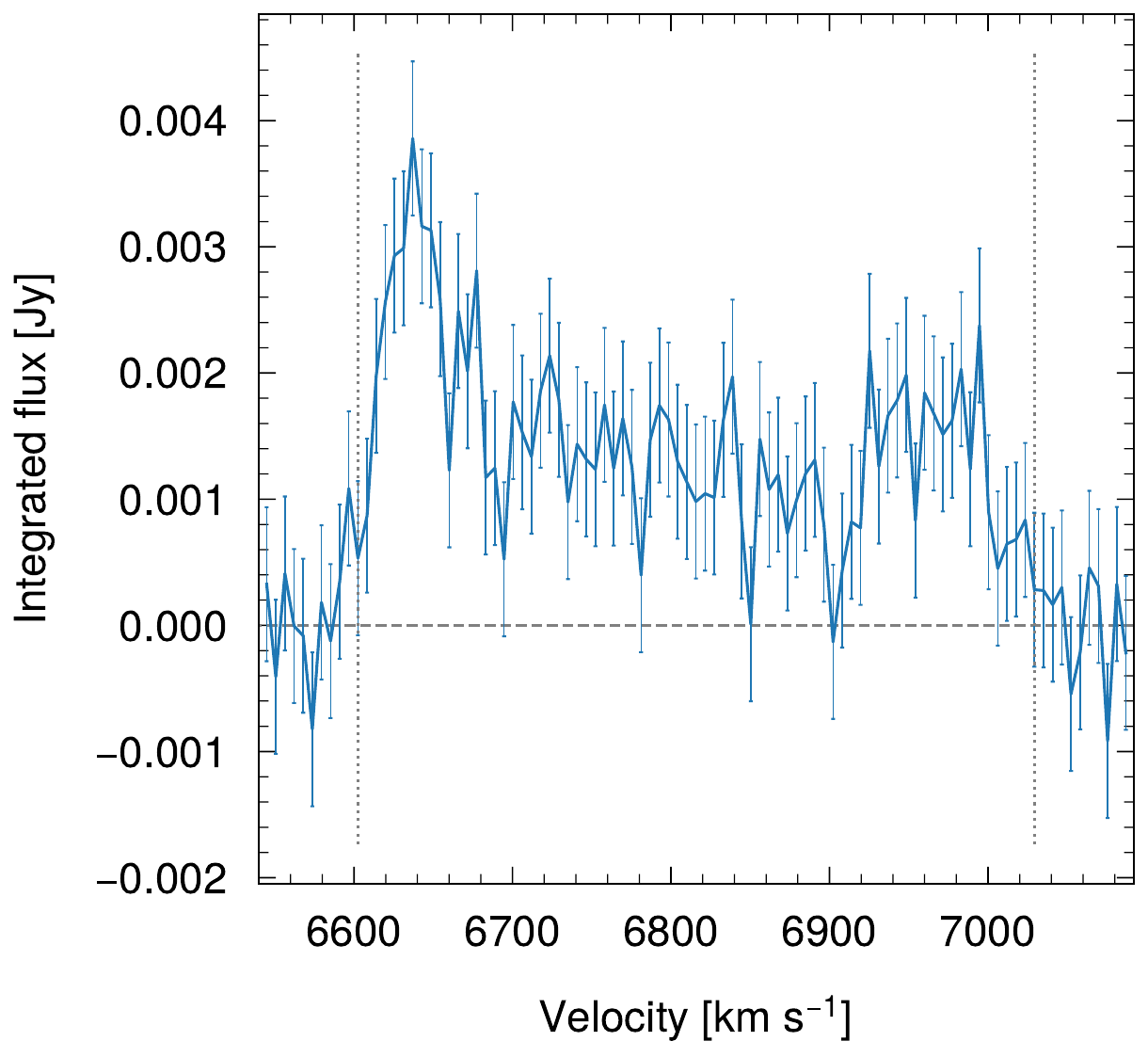} &
          \includegraphics[scale=0.29]{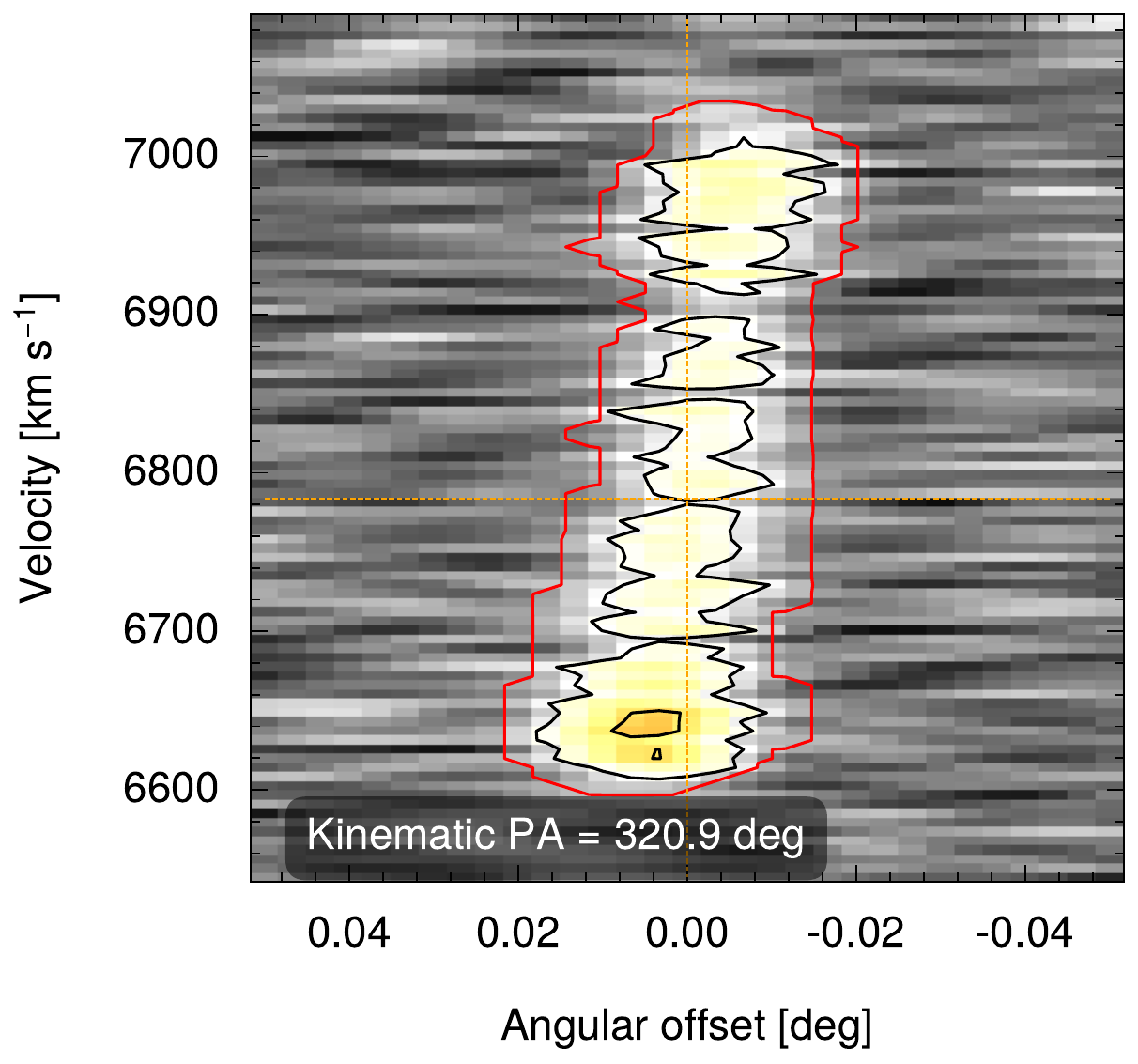}&
          \includegraphics[scale=0.29]{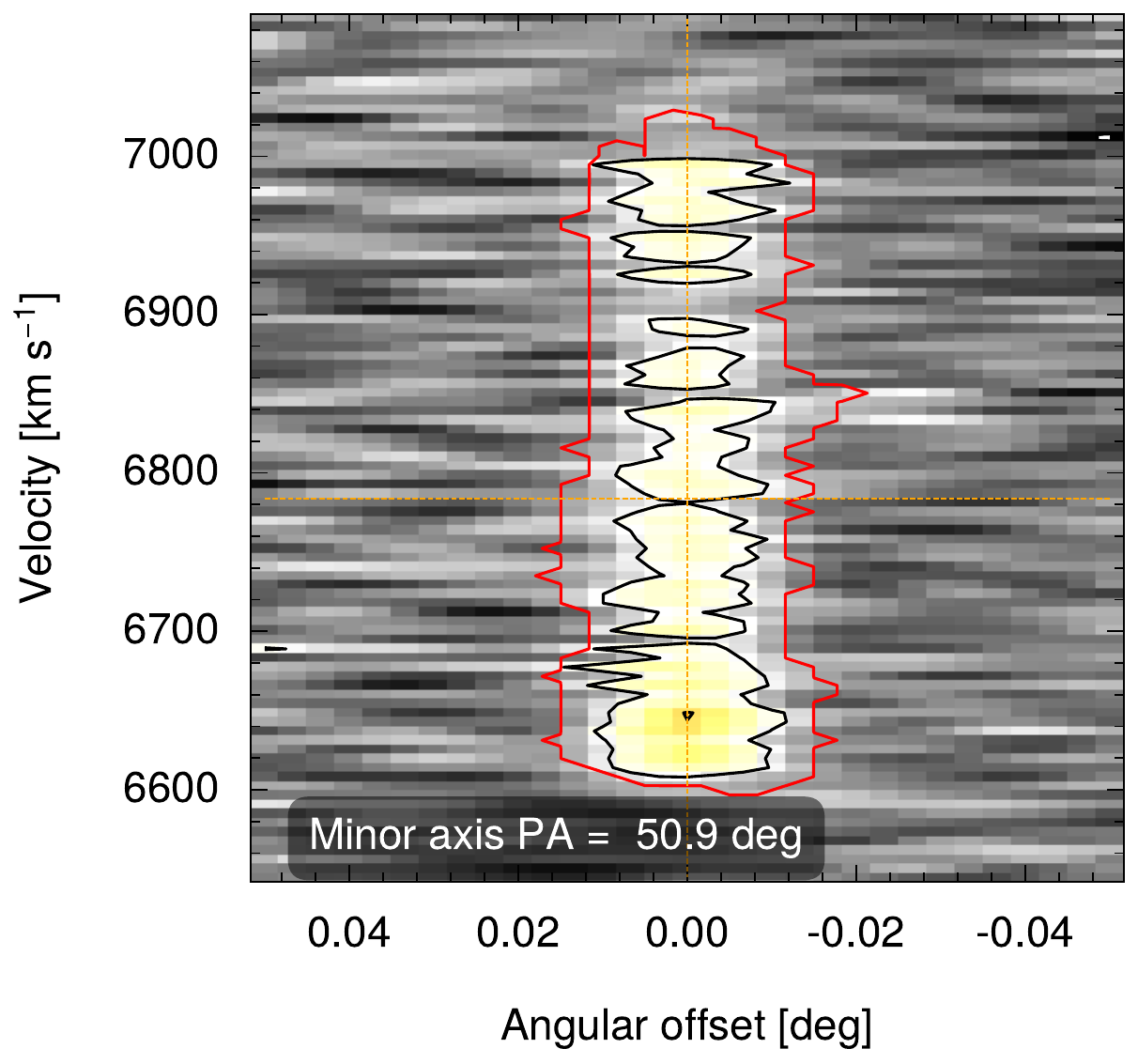}
        \end{tabular}
        \caption{Top left: \HI\ column density map overlaid on DECaLS optical image of HCG~97b. The contour levels are (2.10, 4.20, 8.41)~$\times~\mathrm{10^{18}~cm^{-2}}$. 
        Top centre: moment-1 map. The arrows indicate the slices from which the position-velocity diagrams shown at the bottom panels were derived. Top right: signal-to-noise ratio map. 
        Bottom left: global \HI\ profile. Bottom centre: major axis position-velocity diagram. Bottom right: minor axis position-velocity diagram.}
        \label{fig:hcg97b}
       \end{figure*}

Overview plots of the \HI\ in HCG~97b are shown in Figure~\ref{fig:hcg97b}. These images were generated by SIP, showing moment maps, signal-to-noise ratio, position-velocity diagrams 
and an integrated spectrum.  
\subsection{3D visualisation}
\begin{figure*}
        \setlength{\tabcolsep}{0pt}
        \begin{tabular}{c}
        \includegraphics[scale=0.345]{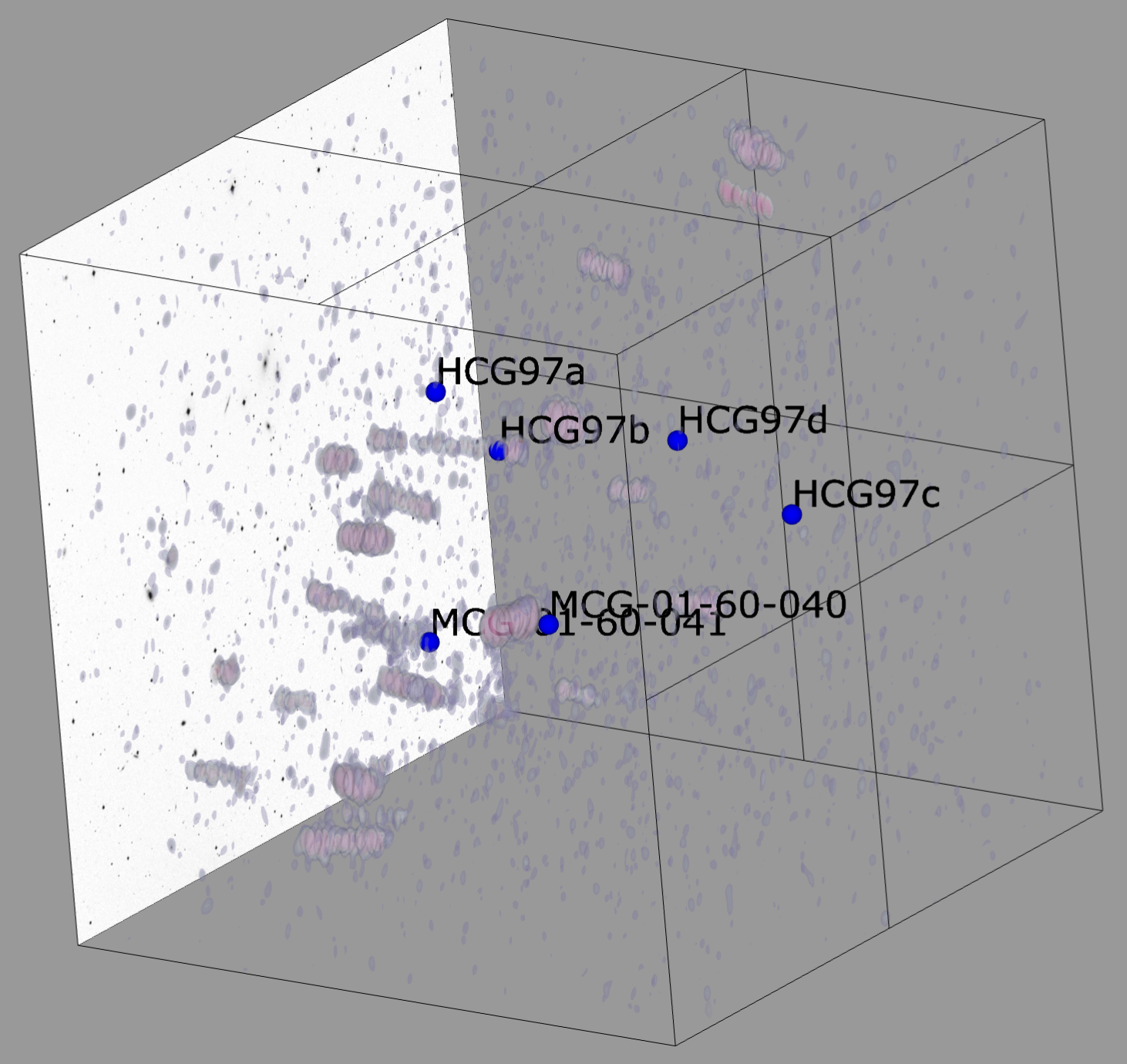}
        \end{tabular}
        \caption{3D visualisation of HCG 97. The blue circles indicate the position of the member galaxies. 
        The 2D grayscale image is a DeCaLS R-band optical image of the group. The online version of the cubes are available at \href{https://amiga.iaa.csic.es/x3d-menu/}{https://amiga.iaa.csic.es/x3d-menu/}.}
      \label{fig:hcg97_3dvis}
     \end{figure*}

Figure~\ref{fig:hcg97_3dvis} shows a 3D visualisation of HCG~97. The blue circles mark the member galaxies and the 2D grayscale background image is a
DECaLS R-band optical data. An interactive version of the data cubes is also accessible at \href{https://amiga.iaa.csic.es/x3d-menu/}{https://amiga.iaa.csic.es/x3d-menu/}.
\end{appendix}
\end{document}